\documentclass{article}
\usepackage{amsmath}
\usepackage{amsfonts}
\usepackage{mathrsfs}
\usepackage {amssymb}
\usepackage {amsmath}
\usepackage {amsthm}
\usepackage{graphicx,color}
\usepackage {amscd}
\usepackage{geometry}
\usepackage{hyperref}
\usepackage{color}
\usepackage{epsfig}

\usepackage{tikz}

\usepackage{bbm}
\usepackage{graphics}

\title{Formation of Trapped Surfaces from Past Null Infinity}
\date{\today}

\author{Xinliang An  \footnote{\href{mailto:An@math.princeton.edu}{\textcolor[rgb]{0.00,0.00,1.00}{Email: xan@math.princeton.edu, xan@math.rutgers.edu}}}}
\theoremstyle{definition}
\newtheorem{lemma}{Lemma}[section]

\newtheorem{proposition}[lemma]{Proposition}
\newtheorem{theorem}[lemma]{Theorem}

\numberwithin{equation}{section}

\begin{document}

\newcommand{\ub}{\underline{u}}
\newcommand{\Cb}{\underline{C}}
\newcommand{\Lb}{\underline{L}}
\newcommand{\Lh}{\hat{L}}
\newcommand{\Lbh}{\hat{\Lb}}
\newcommand{\phib}{\underline{\phi}}
\newcommand{\Phib}{\underline{\Phi}}
\newcommand{\Db}{\underline{D}}
\newcommand{\Dh}{\hat{D}}
\newcommand{\Dbh}{\hat{\Db}}
\newcommand{\omb}{\underline{\omega}}
\newcommand{\omh}{\hat{\omega}}
\newcommand{\ombh}{\hat{\omb}}
\newcommand{\Pb}{\underline{P}}
\newcommand{\chib}{\underline{\chi}}
\newcommand{\chih}{\hat{\chi}}
\newcommand{\chibh}{\hat{\chib}}

\newcommand{\alb}{\underline{\alpha}}
\newcommand{\zeb}{\underline{\zeta}}
\newcommand{\beb}{\underline{\beta}}
\newcommand{\etb}{\underline{\eta}}
\newcommand{\Mb}{\underline{M}}
\newcommand{\oth}{\hat{\otimes}}

\def\a {\alpha}
\def\b {\beta}
\def\ab {\alphab}
\def\bb {\betab}
\def\nab {\nabla}

\def\ub {\underline{u}}
\def\th {\theta}
\def\Lb {\underline{L}}
\def\Hb {\underline{H}}
\def\chib {\underline{\chi}}
\def\chih {\hat{\chi}}
\def\chibh {\hat{\underline{\chi}}}
\def\omegab {\underline{\omega}}
\def\etab {\underline{\eta}}
\def\betab {\underline{\beta}}
\def\alphab {\underline{\alpha}}
\def\Psib {\underline{\Psi}}
\def\hot{\widehat{\otimes}}
\def\Phib {\underline{\Phi}}
\def\thb {\underline{\theta}}
\def\t {\tilde}
\def\st {\tilde{s}}

\def\d {\delta}
\def\f {\frac}
\def\i {\infty}
\def\l {\bigg(}
\def\r {\bigg)}
\def\S {S_{u,\underline{u}}}
\def\o{\omega}
\def\O{\Omega}
\def\be{\begin{equation}\begin{split}}
\def\en{\end{split}\end{equation}}

\def\od{\omega^{\dagger}}
\def\ombd{\underline{\omega}^{\dagger}}
\def\K{K-\frac{1}{|u|^2}}
\def\ut{\frac{1}{|u|^2}}
\def\Kb{K-\frac{1}{(u+\underline{u})^2}}
\def\M{\mathcal}
\def\p{\psi}

\def\D{\Delta}
\def\T{\Theta}
\def\s{S_{u',\underline{u}'}}
\def\Hu{H_u^{(0,\underline{u})}}
\def\Hbu{\underline{H}_{\underline{u}}^{(u_{\infty},u)}}
\def\ee{(\eta,\underline{\eta})}

\def\at{\delta a^{\f12}}
\def\sigmac{\check{\sigma}}
\def\p{\psi}
\def\q{\underline{\psi}}
\def\ls{\leq}
\def\de{\delta}
\def\ls{\lesssim}
\def\oo{\Omega\mbox{tr}\chib-\frac{2}{u}}

\renewcommand{\div}{\mbox{div }}
\newcommand{\curl}{\mbox{curl }}
\newcommand{\trchb}{\mbox{tr} \chib}
\def\trch{\mbox{tr}\chi}

\newcommand{\Ls}{{\mathcal L} \mkern-10mu /\,}
\newcommand{\eps}{{\epsilon} \mkern-8mu /\,}

\newcommand{\tr}{\mbox{tr}}

\newcommand{\xib}{\underline{\xi}}
\newcommand{\psib}{\underline{\psi}}
\newcommand{\rhob}{\underline{\rho}}
\newcommand{\thetab}{\underline{\theta}}
\newcommand{\gammab}{\underline{\gamma}}
\newcommand{\nub}{\underline{\nu}}
\newcommand{\lb}{\underline{l}}
\newcommand{\mub}{\underline{\mu}}
\newcommand{\Xib}{\underline{\Xi}}
\newcommand{\Thetab}{\underline{\Theta}}
\newcommand{\Lambdab}{\underline{\Lambda}}
\newcommand{\vphb}{\underline{\varphi}}

\newcommand{\ih}{\hat{i}}

\newcommand{\tcL}{\widetilde{\mathscr{L}}}

\newcommand{\sRic}{Ric\mkern-19mu /\,\,\,\,}
\newcommand{\sL}{{\cal L}\mkern-10mu /}
\newcommand{\sLh}{\hat{\sL}}
\newcommand{\sg}{g\mkern-9mu /}
\newcommand{\seps}{\epsilon\mkern-8mu /}
\newcommand{\sd}{d\mkern-10mu /}
\newcommand{\sR}{R\mkern-10mu /}
\newcommand{\snab}{\nabla\mkern-13mu /}
\newcommand{\sdiv}{\mbox{div}\mkern-19mu /\,\,\,\,}
\newcommand{\scurl}{\mbox{curl}\mkern-19mu /\,\,\,\,}
\newcommand{\slap}{\mbox{$\triangle  \mkern-13mu / \,$}}
\newcommand{\sGamma}{\Gamma\mkern-10mu /}
\newcommand{\somega}{\omega\mkern-10mu /}
\newcommand{\somb}{\omb\mkern-10mu /}
\newcommand{\spi}{\pi\mkern-10mu /}
\newcommand{\sJ}{J\mkern-10mu /}
\renewcommand{\sp}{p\mkern-9mu /}
\newcommand{\su}{u\mkern-8mu /}

\maketitle

 \large

\begin{abstract}
In this paper, we extend the results of Klainerman and Rodnianski in \cite{KR:Trapped}, which were obtained for a finite region, by showing similar results from past null infinity.
This allows us to recover and extend the results in the work of Christodoulou \cite{Chr:book} from past null infinity. 
\end{abstract}

\section{INTRODUCTION}

\subsection{Background and Goals}

In this paper, we study the formation of trapped surfaces, which is due to focusing of incoming radiation in a spacetime $(M,g)$ satisfying the Einstein vacuum equation
\begin{equation}\label{1.1}
Ric_{\mu\nu}=0.
\end{equation}

A celebrated solution to (\ref{1.1})
is the Schwarzschild metric:
$$g=-\l1-\f{2M}{r}\r dt^2+\l1-\f{2M}{r}\r^{-1} dt^2+r^2 dS,$$
where $dS$ denotes the standard metric on a unit sphere and parameters $M$, $r$ represent mass and radial coordinate, respectively. 

In the Schwarzschild spacetime, a three dimensional hypersurface, where $r=2M$, represents the boundary of an interior region.
Every timelike or null geodesic $\gamma(s)$ starting from the interior region ($r<2M$) is confined in the region $r\leq2M$. 
The interior region of Schwarzschild spacetime is the most famous example of a black hole. Moreover, each causal geodesic $\gamma(s)$
is future incomplete, because $\gamma(s)$ will reach $r=0$, where
$R^{\a\b\gamma\d}R_{\a\b\gamma\d}=\infty$. Therefore, \textit{the Schwarzschild spacetime is future causally geodesically incomplete}. 

Indeed, the Schwarzschild metric is a very special solution to Einstein vacuum equations.
Thus, a fundamental question arises: do generic solutions to Einstein vacuum equations possess any of the singular
features of the Schwarzschild metric?  


To some extend, this question was answered by Penrose in \cite{Penrose} with
 \begin{theorem}(\textbf{Penrose's Incompleteness Theorem})
 Let $(M,g)$ satisfy reasonable topological conditions ($M$ is globally hyperbolic with a noncompact Cauchy hypersurface) and physical conditions ($M$ satisfies $Ric(V,V) \geq 0$ for all null $V$).
 It follows that if $M$ contains a closed trapped surface, then it is future causally geodescially incomplete.
 \end{theorem}

By definition, 2-sphere $S$ is called trapped if both its future expansions are negative.
Let $L,\Lb$ be null vector fields. $L$ is the outgoing vectorfield and $\Lb$ is the incoming vectorfield.
Define $\chi,\chib$ to be the null second fundamental forms of the hypersurfaces generated by $L,\Lb$.
If both $\tr\chi<0$ and $\tr\chib<0$ hold pointwise, then the 2-sphere $S$ is called a \textbf{trapped surface}.
As a comparison, a standard 2-sphere $S$ with radius $r$ in Minkowski space possesses 
$\tr\chi={2}/{r}>0$ and $\tr\chib=-{2}/{r}<0$.

Thus, following Penrose's Incompleteness Theorem, trapped surface formation implies geodesically incomplete.
Hence, one may formally equate the existence of a trapped surface with the existence of a black hole.
This is very useful because trapped surfaces are local and concrete objects.

However, Penrose's Incompleteness Theorem does not answer whether a trapped surface can form dynamically from initial data free of trapped surfaces.
Answering this question requires a good understanding of the solution to the EVE and it requires an understanding of the dynamics of the Einstein vacuum equation (\ref{1.1}) in some large data regime. 


Indeed, this problem was open for a long time until a recent breakthrough by Christodoulou in his  589-page monumental work \cite{Chr:book} \textit{The Formation of Black Holes in General Relativity}. 
In \cite{Chr:book}, Christodoulou studied the characteristic initial value problem with data posed on a truncated incoming cone $\Hb_0$ and a truncated outgoing cone $H_{u_{\infty}}$, which intersect at a 2-sphere $S_{u_{\infty},0}$ (See the figure below). 
The data on $\Hb_0$ are prescribed to coincide with a backward light cone in Minkowski space such that the sphere $S_{u_{\infty},0}$ is the standard 2-sphere with radius $|u_{\infty}|$.
Along $H_{u_{\infty}}$, the data are given in a region with a short characteristic length $\ub\leq \d$ and we further require the traceless part of the null second fundamental form $\chih$ to be large in terms of $\d$. 

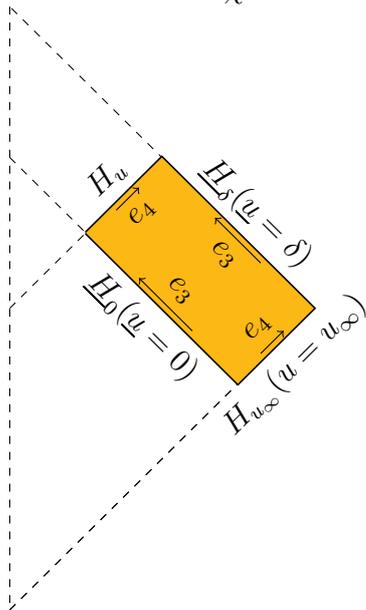
\begin{minipage}[!t]{0.4\textwidth}
\begin{tikzpicture}
\draw [white](3,-1)-- node[midway, sloped, below,black]{$H_{u_{\infty}}(u=u_{\infty})$}(4,0);
\draw [white](1,1)-- node[midway,sloped,above,black]{$H_u$}(2,2);
\draw [white](2,2)--node [midway,sloped,above,black] {$\Hb_{\delta}(\ub=\delta)$}(4,0);
\draw [white](1,1)--node [midway,sloped, below,black] {$\Hb_{0}(\ub=0)$}(3,-1);
\draw [dashed] (0, 4)--(0, -4);
\draw [dashed] (0, -4)--(4,0)--(0,4);
\draw [dashed] (0,0)--(2,2);
\draw [dashed] (0,-4)--(4,0);
\draw [dashed] (0,2)--(3,-1);
\draw [very thick] (1,1)--(3,-1)--(4,0)--(2,2)--(1,1);
\fill[yellow!70!red] (1,1)--(3,-1)--(4,0)--(2,2)--(1,1);
\draw [->] (3.3,-0.6)-- node[midway, sloped, above,black]{$e_4$}(3.6,-0.3);
\draw [->] (1.4,1.3)-- node[midway, sloped, below,black]{$e_4$}(1.7,1.6);
\draw [->] (3.3,0.6)-- node[midway, sloped, below,black]{$e_3$}(2.7,1.2);
\draw [->] (2.4,-0.3)-- node[midway, sloped, above,black]{$e_3$}(1.7,0.4);
\end{tikzpicture}
\end{minipage}
\hspace{0.02\textwidth} 
\begin{minipage}[!t]{0.5\textwidth}
This special form of initial data was termed a ``short pulse" by Christodoulou. In particular, the short pulse allows one to consider a hierarchy of large and small quantities, parametrized by the smallness parameter $\d$, whose sizes are preserved by the nonlinear evolution. Therefore, despite being a problem in a large data regime, a long time existence theorem can be established.
With this \textit{short pulse ansatz}, Christodoulou identified an open set of regular initial conditions. These initial conditions lead to the formation of a trapped surface in the future of the prescribed characteristic initial data along $\Hb_0$ and $H_{u_\infty}$.
\end{minipage}


In order to construct a spacetime such that a trapped surface is formed by the focusing of gravitational waves from past null infinity, the parameter $u_{\infty}$ needs to be passed to infinity. Therefore, in \cite{Chr:book} Christodoulou carefully estimated the decay rates of all curvature components and Ricci coefficients towards past null infinity.

In a subsequent paper \cite{KR:Trapped}, Klainerman and Rodnianski simplified and extended Christodoulou's result in a finite region. 
Based on a novel scaling (parabolic scaling) with respect to the short pulse parameter $\delta$, they assigned numbers (signatures) to various geometric quantities. These signatures come from scaling and associate specific behaviors in power of $\delta$ to these geometric quantities. With these signatures, they further defined scale invariant norms. This procedure allows Klainerman and Rodnianski to show that under the scale invariant norms, except for few exceptions (called anomalies) , all nonlinear terms are small and proportional to  $\delta^{\frac 12}$. Hence, they were left to analyze these anomalies.  This method offers a systematic approach to derive estimates and reduces the workload significantly.

In our paper we extend the results in \cite{KR:Trapped}, which is done for a finite region, by showing a similar result from past null infinity. With the spirit of \cite{KR:Trapped} we complement their scale invariant norms (only involving $\delta$ weights) with a new scaling corresponding to the powers of $u$, which describe the decay 
rates near past null infinity. We will thus see two different hierarchies corresponding to $\delta$ and $u$, respectively. To propagate these two hierarchies, we will encounter some anomalous terms.
In addition to the anomalies in \cite{KR:Trapped}, we will come across additional anomalies corresponding to the decay rate, which lead to  \textit{borderline terms}. In our paper, we will give a systematic way to treat these borderline terms as well. Moreover, based on this systematic way and an observation about structures of Einstein vacuum equation, we give a more direct and intuitive approach to doing energy estimates in an infinite region by integrating the null Bianchi equations. We do not use the Bel-Robinson tensor or Lie derivatives.

Given that the estimates in \cite{KR:Trapped} are more systematic and thus easier to implement, it is natural to ask whether the results of \cite{KR:Trapped} can be used as a stepping stone to derive the results of \cite{Chr:book}. 
In our paper we show that this is indeed the case.
More precisely, the results of \cite{KR:Trapped} hold true for a larger class of data than that of \cite{Chr:book}. The spacetime estimates derived from these data are however weaker.
Nevertheless, we show that starting with Christodoulou's data the estimates derived in \cite{KR:Trapped} can be indeed improved. Those improved results are consistent with Christodoulou's in \cite{Chr:book}.






\subsection{Other Related Works}
For a finite region, regarding other related results, we also point out the treatment of \cite{R-T},  \cite{Yu1}, \cite{Yu2}, \cite{Luk}, \cite{KR:Scarred}. Combining the estimates in \cite{Chr:book} and the Covino-Schoen gluing method, Li and Yu in \cite{L-Y} constructed a class of Cauchy data such that a trapped surface is guaranteed to form in the future. With the existence result in \cite{Chr:book}, Klainerman, Luk, Rodnianski in \cite{K-L-R} offered a fully anisotropic mechanism for formation of trapped surfaces in vacuum. 

While all of the aforementioned works aim to obtain a trapped surface of radius one, in a more result work \cite{A-L} An and Luk studied the ``minimal requirement" on the incoming radiation that guarantees a trapped surface to form in vacuum. They extended the region of existence in \cite{Chr:book}, and proved that a trapped surface of radius $\delta$ (scale invariant) will arise under their assumptions. 

Finally, we refer the interested readers to the beautiful exposition \cite{Dafermos} for more background on the problem and a further discussion on the original work of Christodoulou.

\subsection{Heuristic Argument}

In this subsection, we demonstrate the heuristic argument for trapped surfaces formation. We consider a region $D=D(u,\ub)$ of a vacuum spacetime $(\M M, g)$ generated by optical functions
$(u,\ub)$, which are increasing toward the future, where $u_{\infty}\leq u \leq -c \leq -1$ and $0 \leq \ub \leq \d$. Here $u_{\infty}$ and $c$ are fixed constants and $\d$ is to be determined.
Our results are independent of $u_{\infty}$. Later we will set $u_{\infty}$ to go to $-\infty$ and obtain theorems from
past null infinity. We denote 
by $H_u$ the outgoing null hypersurfaces generated by the level surfaces of $u$, and by $\Hb_{\ub}$ 
the incoming null hypersurfaces generated by the level surfaces of $\ub$. Hence $S_{u,\ub}=H_u \cap \Hb_{\ub}$
is a 2-sphere.\\

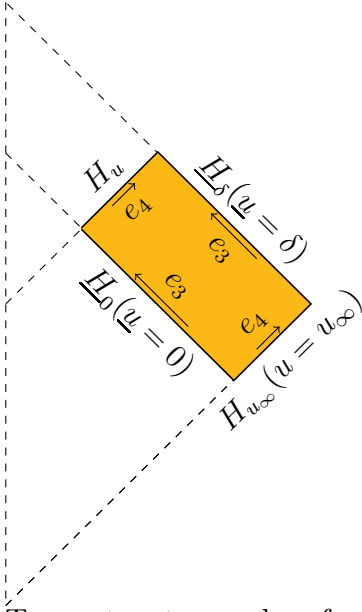
\begin{minipage}[!t]{0.4\textwidth}
\begin{tikzpicture}
\draw [white](3,-1)-- node[midway, sloped, below,black]{$H_{u_{\infty}}(u=u_{\infty})$}(4,0);
\draw [white](1,1)-- node[midway,sloped,above,black]{$H_u$}(2,2);
\draw [white](2,2)--node [midway,sloped,above,black] {$\Hb_{\delta}(\ub=\delta)$}(4,0);
\draw [white](1,1)--node [midway,sloped, below,black] {$\Hb_{0}(\ub=0)$}(3,-1);
\draw [dashed] (0, 4)--(0, -4);
\draw [dashed] (0, -4)--(4,0)--(0,4);
\draw [dashed] (0,0)--(2,2);
\draw [dashed] (0,-4)--(4,0);
\draw [dashed] (0,2)--(3,-1);
\draw [very thick] (1,1)--(3,-1)--(4,0)--(2,2)--(1,1);
\fill[yellow!70!red] (1,1)--(3,-1)--(4,0)--(2,2)--(1,1);
\draw [->] (3.3,-0.6)-- node[midway, sloped, above,black]{$e_4$}(3.6,-0.3);
\draw [->] (1.4,1.3)-- node[midway, sloped, below,black]{$e_4$}(1.7,1.6);
\draw [->] (3.3,0.6)-- node[midway, sloped, below,black]{$e_3$}(2.7,1.2);
\draw [->] (2.4,-0.3)-- node[midway, sloped, above,black]{$e_3$}(1.7,0.4);
\end{tikzpicture}
\end{minipage}
\hspace{0.02\textwidth} 
\begin{minipage}[!t]{0.5\textwidth}
\begin{itemize}
      \item $D(u,\ub)$ is the colored region on the left.
      \item The Optical functions $(u,\ub)$ satisfy

      $g^{\mu\nu}\partial_{\mu}u\partial_{\nu}u=0,$

      $g^{\mu\nu}\partial_{\mu}\ub\partial_{\nu}\ub=0.$
      \item One point stands for a 2-sphere.
      \item $(e_a)_{a=1,2}$ is a frame tangent to the 2-sphere $\S$.
      \item $g(e_a,e_b)=\delta_{ab}$ for $a,b=1,2.$
      \item $e_3,e_4$ are a null pair.
      \item $g(e_3,e_4)=-2.$\\
\end{itemize}
\end{minipage}

To create a trapped surface $\S$, we need that both $\tr\chi<0$ and $\tr\chib<0$ hold pointwise on $\S$. 
For the initial data along $\Hb_0$, on each $S_{u,0}$ we have
$$\tr\chib(u,0)=-\frac{2}{|u|}, \quad \quad \tr\chi(u,0)=\frac{2}{|u|}.$$
For the initial data along $H_{u_{\infty}}$, we have 
$$\tr\chib(u_{\infty},\ub)=-\frac{2}{|u_{\infty}|}+l.o.t.<0, \quad \quad \tr\chi(u_{\infty},\ub)=\frac{2}{|u_{\infty}|}+l.o.t.>0.$$
In the colored region $D(u,\ub)$, $\tr\chib<0$ is always true due to the \textbf{Raychaudhuri Equation}:
$$\nab_3\tr\chib=-\f12(\tr\chib)^2-|\chibh|^2+l.o.t.$$
For $\chi$, we have the following transport equations:
\begin{equation}\label{trapped surface 1}
\nabla_4\tr\chi+\frac{1}{2}(\tr\chi)^2=-|\hat{\chi}|^{2}+l.o.t.,
\end{equation}
and
\begin{equation}\label{trapped surface 2}
\nabla_3\hat{\chi}+\frac{1}{2}\tr{\underline{\chi}}\hat{{\chi}}=l.o.t.
\end{equation}
Here $\chih$ denotes the traceless part of $\chi$.

Employing \eqref{trapped surface 1}, we derive
             $$\nab_3 \tr\chi\leq -|\chih|^2.$$
Hence, it follows that
$$\tr\chi(u,\ub)\leq \tr\chi(u,0)-\int_0^{\ub}|\chih|^2(u,{\ub}')d{\ub}'=\frac{2}{|u|}-\int_0^{\ub}|\chih|^2(u,{\ub}')d{\ub}'.$$

Using the fact that $\tr\chib=-{2}/{|u|}+l.o.t.$ in $D(u,\ub)$ as well as \eqref{trapped surface 2}, we obtain
             
$$|u|^2|\chih|^2(u,\ub)= |u_{\infty}|^2|\chih|^2(u_{\infty},\ub)+l.o.t.$$
Combining these together, along $H_{-c}$ we have
$$\tr\chi(-c,\ub)\leq \tr\chi(-c,0)-\int_0^{\ub}|\chih|^2(-c,{\ub}')d{\ub}'=\frac{2}{|c|}-\frac{|u_{\infty}|^2}{|c|^2}\int_0^{\ub}|\chih|^2(u_{\infty},{\ub}')d{\ub}'+l.o.t.$$
In order to create a trapped surface along the hypersurface $H_c^{(0,\d)}$,  and to avoid trapped surfaces in the initial hypersurface ${H_{u_{\infty}}^{(0,\d)}}$, it is sufficient to require
$$\frac{4c}{|u_{\infty}|^2}<\int_0^{\ub}|\chih|^2(u_{\infty},{\ub}')d{\ub}'<\frac{1}{|u_{\infty}|}.$$
Thus, we expect 
$$|u_{\infty}|\|\chih\|_{L^{\infty}(S_{u_{\infty},\ub})} \approx \delta^{-\frac12},$$ 
which is very large.

To rigorously verify this heuristic argument, we encounter two main difficulties.

\textbf{1.} With arbitrary dispersed initial data at past null infinity, we employ a focusing mechanism to create a
trapped surface of radius $c$. To enable the gravitational radiation to go sufficiently far inside from past null infinity,
we will essentially need a semi-global existence result for the Einstein vacuum equations without symmetry assumption.
And we know that
\begin{itemize}
 \item with no symmetry assumptions, Einstein vacuum equations are energy supercritical.
 \item This is a large data problem, since small data will lead to the stability of Minkowski spacetime (see \cite{Chr-Kl}). 
\end{itemize}

To deal with these difficulties, in \cite{Chr:book} Christodoulou introduced the so called ``short pulse ansatz" and the corresponding hierarchy. 
Based on the smallness assumption on $\delta$, he established the initial data hierarchy for various geometric quantities at $H_{u_{\infty}}$.
Then he proved preservation of this hierarchy in the whole colored region $D(u,\ub)$ starting from $H_{u_{\infty}}$. This enables him to obtain the desired
semi-global existence result. 

In our paper, we also use the short pulse method, but with a different initial data hierarchy, and we will
employ a more direct and intuitive approach to the energy estimates. 

In the proof, we will obtain results independent of $u_{\infty}$. Thus, in the end we will set
$u_{\infty}$ to go to $-\infty$ to obtain \textit{formation of trapped surfaces from past null infinity}.

\textbf{2.} Explicitly referring to  heuristic argument, we need to make sure that all of the lower order terms are truly negligible compared with the main terms.
Since Einstein vacuum equations are a coupled system of many geometric quantities, this requires the understanding of detailed information about all of the geometric quantities and their interactions. 

By following and extending the ideas of Klainerman and Rodnianski in \cite{KR:Trapped}, 
we introduce the notion of \textit{signatures}. We associate a pair of numbers $(s_1, s_2)$ to all the quantities of interest,
where $s_1$ and $s_2$ encode the information about short pulse and decay rates, respectively.
This approach allows for a systematic treatment of many terms and it significantly simplifies the proof. 
There are few terms which do not fall into this framework and we are left to tracked them carefully. 


\subsection{Main Results}

\textbf{Signature and Scale Invariant Norms}

We now turn to the explicit definition of signature and associate norms. 
We assign to each geometric quantity $\phi$ two numbers $s_1(\phi)$ and $s_2(\phi)$. The former is introduced by Klainerman and Rodnianski in \cite{KR:Trapped}. 
The latter is called the \textit{signature for decay rates}. Using them we define the scale invariant norms:

$$\|\phi\|_{L^{\infty}_{sc}(\S)}=\d^{s_1(\phi)-\f12}|u|^{2s_2(\phi)+1}\|\phi\|_{L^{\infty}(\S)},$$

$$\|\phi\|_{L^{2}_{sc}(\S)}=\d^{s_1(\phi)-1}|u|^{2s_2(\phi)}\|\phi\|_{L^{2}(\S)}.$$

\noindent Two identities follow from these definitions:

$$s_1(\phi_1\cdot\phi_2)=s_1(\phi_1)+s_1(\phi_2),$$
$$s_2(\phi_1\cdot\phi_2)=s_2(\phi_1)+s_2(\phi_2).$$

\noindent With these definitions, we therefore obtain H\"older's inequality in scale invariant norms:

$$\|\phi_1\cdot\phi_2\|_{L^2_{sc}(\S)}\leq \f{\d^{\f12}}{|u|}\|\phi_1\|_{L^{\infty}_{sc}(\S)}\|\phi_2\|_{L^{2}_{sc}(\S)}.$$

This inequality tells us that if all the terms are normal, then the nonlinear interactions can be treated as lower order terms. 
Hence, only rare anomalous terms are left for further analysis. 
By adapting the techniques in \cite{Chr-Kl} and \cite{KNI:book},
we deal with all of the borderline terms for decay rate without encountering a logarithmic divergence.

\textbf{Energy Estimates without Bel-Robinson Tensor}

Based on the relation between the new signatures and the coefficients in front of the borderline terms,
with the methods used by Holzegel in \cite{Hol} and Luk-Rodnianski in \cite{L-R:Propagation},  we employ a more direct and intuitive approach to establish energy estimates without using the \textit{Bel-Robinson tensor}.

Denote $\Psi$ to be Curvature component and $\psi$ to be Ricci coefficient. We observe that by separating $\Psi$ 
into proper pairs,  the pair $\Psi^{({s,s'})}$ and $\Psi^{({s-\frac 12,s'+\frac 12})}$ satisfy

$$
\nab_3 \Psi^{({s,s'})}+({\frac 12+s'})\tr\chib\Psi^{{(s,s')}}=\nab\Psi^{({s-\frac12,s'+\frac12})}+\sum_{s_1+s_2=s, \atop s'_1+s'_2=s'+1}\psi^{(s_1,s'_1)}\cdot\Psi^{(s_2,s'_2)},
$$
and
$$
\nab_4 \Psi^{({s-\frac 12,s'+\frac 12})}=\nab\Psi^{({s,s'})}+\sum_{\st_1+\st_2=s+\frac 12, \atop \st'_1+\st'_2=s'+\frac 12}\psi^{(\st_1,\st'_1)}\cdot\Psi^{(\st_2,\st'_2)}.
$$

Here $\Psi^{(s,s')}$ and $\psi^{(s,s')}$ stand for S-tangent tensors $\Psi$ and $\psi$ with signatures $s_1(\Psi)=s$, $s_2(\Psi)=s'$ and $s_1(\psi)=s$, $s_2(\psi)=s'$, respectively.
In above equations, $(1/2+s')\tr\chib\Psi^{(s,s')}$ is the borderline term. Taking the fact $\tr\chib=-{2}/{|u|}+l.o.t$ in $D(u,\ub)$ and using the connection between the coefficient $1/2+s'$ and $\Psi^{(s,s')}$'s signature
for decay rates $s_2(\Psi^{(s,s')})=s'$, we will cancel this borderline 
term in our newly defined scale invariant norms. This observation avoids employing the Bel-Robinson tensor and gives us a more direct and intuitive approach for energy estimates.

\textbf{Retrieving Christodoulou's Estimates}

The norms we exploit in this paper are consistent with the norms in \cite{KR:Trapped}, which are weaker than the norms in \cite{Chr:book}.
With these weaker norms, we encounter fewer borderline terms and it is less difficult to establish the semi-global existence result.
Furthermore, based on the existence results obtained in weak norms together with initial data in strong norms, we can improve the estimates derived in weak norms to strong norms and thus retrieve Christodoulou's estimates in \cite{Chr:book}.

Let $\chi$ be the second fundamental form for $S_{u,\ub}$ with respect to $H_u$  and let $\chih$ be the traceless part
of $\chi$. In this paper, we re-prove the main theorem in Christodoulou's monograph: 




\begin{theorem}\label{TSEVE}(Christodoulou \cite{Chr:book}, 2008; A.)

Given $c$ and $B$, there exists $\delta_0=\delta_0(c, B)$ sufficiently small, such that for $0<\delta<\delta_0$, with initial data:

    \begin{itemize}
            \item $\sum_{i\leq 5, k\leq 3}\delta^{\frac12+k}|u_{\infty}|^{1+i}\|\nab^{k}_4\nab^{i}\chih_{\infty}\|_{L^{\infty}(S_{u_{\infty},\ub})}\leq B$ along $u=u_{\infty}$
            \item Minkowskian initial data \quad \quad \quad \quad \quad \quad \quad \quad \quad \quad along $\ub=0$
            \item $\int_0^{\delta}u^2_{\infty}|\chih_{\infty}|^2\geq 4c$ for every direction \quad \quad \quad \quad \quad along $u=u_{\infty}$
    \end{itemize}

\noindent we have that $S_{c,\delta}$ is a trapped surface.

\end{theorem}

Moreover, all of our proofs are independent of $u_{\infty}$. Letting $u_{\infty}$ go to $-\infty$, we obtain 

\begin{theorem}(Christodoulou \cite{Chr:book}, 2008; A.)

Trapped surfaces can form dynamically for Einstein vacuum equations with initial data which are dispersed at past null infinity.
\end{theorem}

\subsection{Structure of the Paper}

The structure of this paper is as following:

\begin{itemize}
\item In Section 2, we demonstrate setting, equations and notations.  
\item In Section 3, we state the main theorem. 
\item In Section 4, we offer preliminary estimates. 
\item In Section 5- Section 6, we derive estimates for the zeroth and first derivatives of Ricci coefficients. 
\item In Section 7, we derive elliptic estimates for the third derivatives of Ricci coefficients.  
\item In Section 8, we derive energy estimates. 
\item In Section 9, we prove formation of trapped surfaces. 
\item In Section 10, we outline how to pursuit Christodoulou's results.  
\end{itemize}

\subsection{Notation}

We collect much of the notation that is introduced throughout the article. 
\begin{itemize}
\item We denote $\sup_{u,\ub}$ to be the supremum over all values of $u,\ub$, where $u_{\infty}\leq u\leq -c$ and $0\leq \ub \leq \d$. 
\item If $A$ and $B$ are two quantities, we often write $A\lesssim B$ to mean that there exists a constant $C>0$ such that $A\leq CB$. Whenever there is no danger of confusion, we substitute $\leq$ for $\lesssim$ for convenience. 
\item When deriving equations for higher order derivatives or using these equations, the coefficients on the left hand side are precise. And when it will not cause confusion, the coefficients of nonborderline terms on the right hand side are allowed to vary up to a nonzero constant.
\item We will employ brackets to denote sum of all terms, which have one of the components in the brackets. For instance, the notation $\phi_1(\phi_2, \phi_3)$ denotes 
the sum of all terms of the form $\phi_1\phi_2$ or $\phi_1\phi_3$.

\end{itemize}

{\bf Acknowledgments.} The author is grateful to
 his adviser Sergiu Klainerman for encouragement and constant support. The author thanks Jonathan Luk for many enlightening discussions. The author thanks Mihalis Dafermos and Jeremie Szeftel for valuable suggestions.


\section{Setting, Equations and Notations}

\subsection{Definitions.} \label{Definitions}

We will work on a characteristic initial value problem. And the initial data are given on the two characteristic hypersurfaces $H_{u_{\infty}}$ and $\Hb_0$. These two hypersurfaces intersect at the sphere $S_{u_{\infty},0}$. The spacetime under consideration will be a solution to the Einstein equations constructed in a neighborhood of $H_{u_{\infty}}$ and $\Hb_0$ containing $S_{u_{\infty},0}$.

\subsection{Double Null Foliation}\label{secdnf}
We define a double null foliation in a neighborhood of $S_{u_{\infty},0}$ in the following: 

\begin{minipage}[!t]{0.4\textwidth}
\begin{tikzpicture}
\draw [white](3,-1)-- node[midway, sloped, below,black]{$H_{u_{\infty}}(u=u_{\infty})$}(4,0);

\draw [white](2,2)--node [midway,sloped,above,black] {$\Hb_{\delta}(\ub=\delta)$}(4,0);
\draw [white](1,1)--node [midway,sloped, below,black] {$\Hb_{0}(\ub=0)$}(3,-1);
\draw [dashed] (0, 4)--(0, -4);
\draw [dashed] (0, -4)--(4,0)--(0,4);
\draw [dashed] (0,0)--(2,2);
\draw [dashed] (0,-4)--(2,-2);
\draw [dashed] (0,2)--(3,-1);
\draw [very thick] (1,1)--(3,-1)--(4,0)--(2,2)--(1,1);
\fill[yellow!70!red] (1,1)--(3,-1)--(4,0)--(2,2)--(1,1);
\draw [white](1,1)-- node[midway,sloped,above,black]{$H_{u}$}(2,2);
\draw [->] (3.3,-0.6)-- node[midway, sloped, above,black]{$\Lb$}(3.6,-0.3);
\draw [->] (1.4,1.3)-- node[midway, sloped, below,black]{$\Lb$}(1.7,1.6);
\draw [->] (3.3,0.6)-- node[midway, sloped, below,black]{$L$}(2.7,1.2);
\draw [->] (2.4,-0.3)-- node[midway, sloped, above,black]{$L$}(1.7,0.4);
\end{tikzpicture}
\end{minipage}
\hspace{0.02\textwidth} 
\begin{minipage}[!t]{0.5\textwidth}

Denote $u$ and $\ub$ be solutions to the eikonal equations
$$g^{\mu\nu}\partial_\mu u\partial_\nu u=0,\quad g^{\mu\nu}\partial_\mu\ub\partial_\nu \ub=0.$$
They are increasing towards the future and satisfy the initial conditions $u=u_{\infty}$ $(u_{\infty}<<-1)$ on $H_{u_{\infty}}$ and $\ub=0$ on $\Hb_0$. \\

Let
$L'^\mu=-2g^{\mu\nu}\partial_\nu u,\quad \Lb'^\mu=-2g^{\mu\nu}\partial_\nu \ub$ 
be null and geodesic vector fields. \\

Define $\f12\Omega^2=-g(L',\Lb')^{-1}$. Throughout this paper we work with the normalized null pair $(e_3, e_4)$:
$$e_3=\Omega\Lb', \quad e_4=\Omega L', \quad g(e_3, e_4)=-2.$$

And we denote $\Lb=-\Omega^2\Lb', \quad L=\Omega^2 L'$ as equivariant vector filed. 

\end{minipage}



Moreover, for the characteristic initial data, we choose the following gauge:
$$\Omega=1 \quad\mbox{on $H_{u_{\infty}}$ and $\Hb_0$}.$$


We denote by $H_u$ the outgoing null hyper surfaces generated by the level surfaces of $u$ and by $\Hb_{\ub}$ the incoming null hypersurfaces generated by the level hypersurface  of $\ub$. We write $S_{u,\ub}=H_u \bigcap \Hb_{\ub}$, which is a topologically 2-sphere.

\subsection{The Coordinate System}\label{coordinates}

In this paper, we will use a coordinate system $(u,\ub, \theta^1, \theta^2)$. Here $u$ and $\ub$ are solutions to the eikonal equations. To get $(\theta^1, \theta^2)$ on $\S$, we follow the approach in Chapter 1 of \cite{Chr:book}: we first define a coordinate system $(\theta^1, \theta^2)$ on $S_{u_{\infty},0}$. Then we extend this coordinate system to $\Hb_0$ and $H_{u_{\infty}}$ by solving
$$\frac{\partial}{\partial u}\th^A=0\mbox{ on $\Hb_0$, and }\frac{\partial}{\partial \ub}\th^A=0\mbox{ on $H_{u_{\infty}}$}.$$
We then further extend this coordinate system to the whole spacetime under consideration by requiring
$$\Ls_L \th^A=0.$$ 
Here $\Ls_L$ is the restriction of the Lie derivative to $TS_{u,\ub}$ 

Thus we have established a coordinate system in a neighborhood of $S_{u_{\infty}, 0}$. With this coordinate system, we can rewrite $e_3$ and $e_4$ as 
$$e_3=\Omega^{-1}\left(\frac{\partial}{\partial u}+b^A\frac{\partial}{\partial \th^A}\right), e_4=\Omega^{-1}\frac{\partial}{\partial \ub}.$$
We require $b^A$ to satisfy $b^A=0$ on $\Hb_0$.



\subsection{Equations}\label{seceqn}

We decompose the curvature components and Ricci coefficients with respect to a null frame $e_3, e_4$ and a farme $e_1, e_2$ tangent to the 2-sphere $S_{u,\ub}$.

Denote the indices $A,B$ to be $1,2$. With frame $e_3, e_4, e_A,  e_B$, we define the null curvature components:
 \begin{equation}
\begin{split}
\a_{AB}&=R(e_A, e_4, e_B, e_4),\quad \, \,\,   \ab_{AB}=R(e_A, e_3, e_B, e_3),\\
\b_A&= \frac 1 2 R(e_A,  e_4, e_3, e_4) ,\quad \bb_A =\frac 1 2 R(e_A,  e_3,  e_3, e_4),\\
\rho&=\frac 1 4 R(e_4,e_3, e_4,  e_3),\quad \sigma=\frac 1 4  \,^*R(e_4,e_3, e_4,  e_3).
\end{split}
\end{equation}
Here $\, ^*R$ stands for the Hodge dual of $R$.

Denote $D_A:=D_{e_{A}}$. We introduce the following Ricci coefficients. 

 \begin{equation}
\begin{split}
&\chi_{AB}=g(D_A e_4,e_B),\, \,\, \quad \chib_{AB}=g(D_A e_3,e_B),\\
&\eta_A=-\frac 12 g(D_3 e_A,e_4),\quad \etab_A=-\frac 12 g(D_4 e_A,e_3),\\
&\omega=-\frac 14 g(D_4 e_3,e_4),\quad\,\,\, \omegab=-\frac 14 g(D_3 e_4,e_3),\\
&\zeta_A=\frac 1 2 g(D_A e_4,e_3),
\end{split}
\end{equation}

{\bf Remark:} We further decompose $\chi$ and $\chib$ into trace and traceless part. Denote $\chih$ and $\chibh$ be the traceless part of $\chi$ and $\chib$ respectively.

We also name the induced covariant derivative operator on $S_{u,\ub}$ as $\nab$ and
the projections of covariant derivatives $D_3$ and $D_4$ to $S_{u,\ub}$ as $\nab_3$ and $\nab_4$  respectively.  
(A detailed definition could be found in \cite{KNI:book}.)

{\bf Remark:} For Ricci coefficients, the following equalities also hold from definitions:

\begin{equation}
\begin{split}
&\omega=-\frac 12 \nab_4 (\log\Omega),\qquad \omegab=-\frac 12 \nab_3 (\log\Omega),\\
&\eta_A=\zeta_A +\nab_A (\log\Omega),\quad \etab_A=-\zeta_A+\nab_A (\log\Omega).
\end{split}
\end{equation}

We further define different contractions between tensors.  Let
$$(\phi^{(1)}\hot\phi^{(2)})_{AB}:=\phi^{(1)}_A\phi^{(2)}_B+\phi^{(1)}_B\phi^{(2)}_A-\delta_{AB}(\phi^{(1)}\cdot\phi^{(2)}) \quad\mbox{for one forms $\phi^{(1)}_A$, $\phi^{(2)}_A$,}$$
$$(\phi^{(1)}\wedge\phi^{(2)})_{AB}:=\eps^{AB}(\gamma^{-1})^{CD}\phi^{(1)}_{AC}\phi^{(2)}_{BD}\quad\mbox{for symmetric two tensors $\phi^{(1)}_{AB}$, $\phi^{(2)}_{AB}$},$$
where $\eps$ is the volume form associated to the metric $\gamma$. For simplicity, we also use  $\phi^{(1)}\cdot\phi^{(2)}$ as an arbitrary contraction of the tensor product of $\phi^{(1)}$ and $\phi^{(2)}$ with respect to the metric $\gamma$.

We will also employ $\div$, $\curl$ and $\tr$ operators. For totally symmetric tensors, we define these operators through
$$(\div\phi)_{A_1...A_r}:=\nabla^B\phi_{BA_1...A_r},$$
$$(\curl\phi)_{A_1...A_r}:=\eps^{BC}\nabla_B\phi_{CA_1...A_r},$$
$$(\mbox{tr}\phi)_{A_1...A_{r-1}}:=(\gamma^{-1})^{BC}\phi_{BCA_1...A_{r-1}}.$$


We are ready to state the transport equations for curvature components and Ricci coefficients. Rewrite the second Bianchi equations with null frame, we arrive at 

\begin{equation}
\label{eq:null.Bianchi}
\begin{split}
&\nab_3\alpha+\frac 12 \trchb \alpha=\nabla\hot \beta+ 4\omegab\alpha-3(\chih\rho+^*\chih\sigma)+
(\zeta+4\eta)\hot\beta,\\
&\nab_4\beta+2\trch\beta = \div\alpha - 2\omega\beta +  \eta \alpha,\\
&\nab_3\beta+\trchb\beta=\nabla\rho + 2\omegab \beta +^*\nabla\sigma +2\chih\cdot\betab+3(\eta\rho+^*\eta\sigma),\\
&\nab_4\sigma+\frac 32\trch\sigma=-\div^*\beta+\frac 12\chibh\cdot ^*\alpha-\zeta\cdot^*\beta-2\etab\cdot
^*\beta,\\
&\nab_3\sigma+\frac 32\trchb\sigma=-\div ^*\betab+\frac 12\chih\cdot ^*\alphab-\zeta\cdot ^*\betab-2\eta\cdot 
^*\betab,\\
&\nab_4\rho+\frac 32\trch\rho=\div\beta-\frac 12\chibh\cdot\alpha+\zeta\cdot\beta+2\etab\cdot\beta,\\
&\nab_3\rho+\frac 32\trchb\rho=-\div\betab- \frac 12\chih\cdot\alphab+\zeta\cdot\betab-2\eta\cdot\betab,\\
&\nab_4\betab+\trch\betab=-\nabla\rho +^*\nabla\sigma+ 2\omega\betab +2\chibh\cdot\beta-3(\etab\rho-^*\etab\sigma),\\
&\nab_3\betab+2\trchb\, \betab=-\div\alphab-2\omegab\betab+\etab \cdot\alphab,\\
&\nab_4\alphab+\frac 12 \trch\alphab=-\nabla\hot \betab+ 4\omega\alphab-3(\chibh\rho-^*\chibh\sigma)+
(\zeta-4\etab)\hot \betab.
\end{split}
\end{equation}
Here $^*$ denotes the Hodge dual on $S_{u,\ub}$. These transport equations for curvature components are called null Bianchi equations. 

We then rewrite $Ric_{\mu\nu}=0$ in null frames.  For $\chi$ and $\chib$ we have 
\begin{equation}
\label{null.str1}
\begin{split}
\nab_4 \trch+\frac 12 (\trch)^2&=-|\chih|^2-2\omega \trch,\\
\nab_4\chih+\trch \chih&=-2 \omega \chih-\alpha,\\
\nab_3 \trchb+\frac 12 (\trchb)^2&=-2\omegab \trchb-|\chibh|^2,\\
\nab_3\chibh + \trchb\,  \chibh&= -2\omegab \chibh -\alphab,\\
\nab_4 \trchb+\frac1 2 \trch \trchb &=2\omega \trchb +2\rho- \chih\cdot\chibh +2\div \etab +2|\etab|^2,\\
\nab_4\chibh +\frac 1 2 \trch \chibh&=\nab\widehat{\otimes} \etab+2\omega \chibh-\frac 12 \trchb \chih +\etab\widehat{\otimes} \etab,\\
\nab_3 \trch+\frac1 2 \trchb \trch &=2\omegab \trch+2\rho- \chih\cdot\chibh+2\div \eta+2|\eta|^2,\\
\nab_3\chih+\frac 1 2 \trchb \chih&=\nab\widehat{\otimes} \eta+2\omegab \chih-\frac 12 \trch \chibh +\eta\widehat{\otimes} \eta.
\end{split}
\end{equation}
For the remaining Ricci coefficients, we arrive at
\begin{equation}
\label{null.str2}
\begin{split}
\nabla_4\eta&=-\chi\cdot(\eta-\etab)-\b,\\
\nabla_3\etab &=-\chib\cdot (\etab-\eta)+\bb,\\
\nabla_4\omegab&=2\omega\omegab+\frac 34 |\eta-\etab|^2-\frac 14 (\eta-\etab)\cdot (\eta+\etab)-
\frac 18 |\eta+\etab|^2+\frac 12 \rho,\\
\nabla_3\omega&=2\omega\omegab+\frac 34 |\eta-\etab|^2+\frac 14 (\eta-\etab)\cdot (\eta+\etab)- \frac 18 |\eta+\etab|^2+\frac 12 \rho.\\
\end{split}
\end{equation}
These transport equations for Ricci coefficients are call null structure equations.

We also rewrite Gauss-Codazzi equations in null frames. 
Denote $K$ to be the Gauss curvature of the spheres $S_{u,\ub}$. 
And we come to the following constraint equations:
\begin{equation}
\label{null.str3}
\begin{split}
\div\chih&=\frac 12 \nabla \trch - \frac 12 (\eta-\etab)\cdot (\chih -\frac 1 2 \trch) -\beta,\\
\div\chibh&=\frac 12 \nabla \trchb + \frac 12 (\eta-\etab)\cdot (\chibh-\frac 1 2   \trchb) +\betab,\\
\curl\eta &=-\curl\etab=\sigma +\frac 1 2\chibh \wedge\chih,\\
K&=-\rho+\frac 1 2 \chih\cdot\chibh-\frac 1 4 \trch \trchb.
\end{split}
\end{equation}

\subsection{Integration}

Denote $U$ to be a coordinate patch on $\S$. Let $p_U$ be the corresponding partition of unity.
For a function $\phi$,  we define its integration on $S_{u,\ub}$, $H_u$ and $\Hb_{\ub}$ through 

$$\int_{S_{u,\ub}} \phi :=\sum_U \int_{-\infty}^{\infty}\int_{-\infty}^{\infty}\phi p_U\sqrt{\det\gamma}d\th^1 d\th^2,$$
$$\int_{H_{u}} \phi :=\sum_U \int_0^{\delta}\int_{-\infty}^{\infty}\int_{-\infty}^{\infty}\phi2 p_U\Omega\sqrt{\det\gamma}d\th^1 d\th^2d\ub',$$
$$\int_{H_{\ub}} \phi :=\sum_U \int_{u_{\infty}}^{-c}\int_{-\infty}^{\infty}\int_{-\infty}^{\infty}\phi2p_U\Omega\sqrt{\det\gamma}d\th^1 d\th^2du'.$$

Let $D_{u ,\ub}$ be the region $u_{\infty}\leq u'\leq u$, $0\leq \ub'\leq \ub$. We define the integration of $\phi$ in $D_{u,\ub}$ as
\begin{equation*}
\begin{split}
\int_{D_{u,\ub}} \phi :=&\sum_U \int_{u_{\infty}}^u\int_0^{\ub}\int_{-\infty}^{\infty}\int_{-\infty}^{\infty}\phi p_U\sqrt{-\det g}d\th^1 d\th^2 du' d\ub'.\\
\end{split}
\end{equation*}

For $1\leq p < \infty$, we further define the $L^p$ norms ($1\leq p<\infty$) for an arbitrary tensorfield $\phi$:
$$||\phi||_{L^p(S_{u,\ub})}^p:=\int_{S_{u,\ub}} <\phi,\phi>_\gamma^{p/2},$$
$$||\phi||_{L^p(H_u)}^p:=\int_{H_{u}} <\phi,\phi>_\gamma^{p/2},$$
$$||\phi||_{L^p(\Hb_{\ub})}^p:=\int_{\Hb_{\ub}} <\phi,\phi>_\gamma^{p/2}.$$
When $p=\infty$, we define $L^\infty$ norm by
$$||\phi||_{L^\infty(S_{u,\ub})}:=\sup_{\th\in S_{u,\ub}} <\phi,\phi>_\gamma^{1/2}(\th).$$

The following mixed type of norms are also extensively employed in this paper: 
$$||\phi||_{L^2_{\ub}L^\infty_u L^p(\S)}=\left(\int_0^{\d}(\sup_{u_{\infty}\leq u \leq -c}||\phi||_{L^p(S_{u,\ub'})})^2d\ub'\right)^{\frac{1}{2}},$$
$$||\phi||_{L^2_{u}L^\infty_{\ub} L^p(\S)}=\left(\int_{u_{\infty}}^{-c}(\sup_{0\leq \ub \leq \d}||\phi||_{L^p(S_{u',\ub})})^2du'\right)^{\frac{1}{2}}.$$

{\bf Remark:} In this paper we will frequently use the following Minkowski's inequality
$$||\cdot||_{L^\infty_{\ub}L^2_{u} L^p(\S)}\leq||\cdot||_{L^2_{u}L^\infty_{\ub} L^p(\S)}.$$  


\subsection{Signatures} \label{Signatures}

To capture the structure of Einstein's equation, we introduce the following definitions below.

\textbf{Definition of signatures}

To $\phi\in \{\a,\beta,\rho,\sigma, K, \beb,\ab, \chi,\chib,\zeta,\eta,\etb,\omega,\omb, \gamma\}$, 
we assign signatures $s(\phi)$ according to the following rules:
\begin{equation*}\label{signature}
s(\phi):=(s_1(\phi),s_2(\phi)),
\end{equation*}
where
\begin{equation*}\label{signature 1}
s_1(\phi):=1\cdot{N_4}(\phi)+\frac{1}{2}\cdot{N_a}(\phi)+0\cdot{N_3}(\phi)-1,
\end{equation*}
and
\begin{equation*}\label{signature 2}
s_2(\phi):=0\cdot{N_4}(\phi)+\frac{1}{2}\cdot{N_a}(\phi)+1\cdot{N_3}(\phi)-1.
\end{equation*}

$N_4(\phi)$ is the number of times $e_4$ appears in the definition of $\phi$. 
Similarly we define $N_3(\phi)$ and $N_a(\phi)$ where $a=1,2$.

By the definition above, we have

\textbf{Signature table}

\begin{tabular}{|r|r|r|r|r|r|r|r|r|r|r|r|r|r|r|r|r|}
  \hline
     & $\alpha$ & $\beta$ & $\rho$ & $\sigma$ & $K$ & $\underline{\beta}$ & $\underline{\alpha}$ & $\chi$ & $\omega$ & $\zeta$ & $\eta$ & $\underline{\eta}$ & $\mbox{tr}\underline{\chi}$ & $\hat{\underline{\chi}}$ & $\underline{\omega}$ & $\gamma$ \\
  $s_1$ & 2 & 1.5 & 1 & 1 &1 & 0.5 & 0 & 1 & 1 & 0.5 & 0.5 & 0.5 & 0 & 0 & 0 & 0 \\
  $s_2$ & 0 & 0.5 & 1 & 1 & 1 & 1.5 & 2 & 0 & 0 & 0.5 & 0.5 & 0.5 & 1 & 1 & 1 & 0 \\
  \hline
\end{tabular} \\

\textbf{Properties of signatures}
\begin{equation*}\label{signature prop 1}
s_1(\nabla_4{\phi})=s_1(\phi)+1, \quad s_2(\nabla_4{\phi})=s_2(\phi),
\end{equation*}

\begin{equation*}\label{signature prop 2}
s_1(\nabla{\phi})=s_1(\phi)+\frac{1}{2}, \quad s_2(\nabla{\phi})=s_2(\phi)+\frac{1}{2},
\end{equation*}

\begin{equation*}\label{signature prop 3}
s_1(\nabla_3{\phi})=s_1(\phi),\quad s_2(\nabla_3{\phi})=s_2(\phi)+1.
\end{equation*}

\textbf{Conservation of signatures}
\begin{equation*}\label{signature conservation 1}
s_1(\phi_1\cdot{\phi_2})=s_1(\phi_1)+s_1(\phi_2),
\end{equation*}

\begin{equation*}\label{signature conservation 2}
s_2(\phi_1\cdot{\phi_2})=s_2(\phi_1)+s_1(\phi_2),
\end{equation*}

\begin{equation*}\label{signature conservation 3}
s(\phi_1\cdot{\phi_2})=(s_1(\phi_1\cdot{\phi_2}),s_2(\phi_1\cdot{\phi_2}))=(s_1(\phi_1)+s_1(\phi_2),s_2(\phi_1)+s_1(\phi_2))=s(\phi_1)+s(\phi_2). 
\end{equation*}

{\bf Remark}: $s_1$ is the same signature used in \cite{KR:Trapped} and $s_2$ is introduced to study the decay rate near past null infinity. \\

\subsection{Scale Invariant Norms} \label{Scale invariant norms}

For any horizontal tensor-field $\phi$ with signature $s(\phi)=(s_1(\phi),s_2(\phi))$, we give\\
\textbf{Definition of scale invariant norms on $S_{u,\ub}$}
\begin{equation*}\label{An1}
\|\phi\|_{L_{sc}^{\infty}(\S)}:=\delta^{s_1(\phi)-\frac{1}{2}}|u|^{2s_2(\phi)+1}\|\phi\|_{L^{\infty}(\S)},
\end{equation*}

\begin{equation*}\label{An1}
\|\phi\|_{L_{sc}^{4}(\S)}:=\delta^{s_1(\phi)-\frac{3}{4}}|u|^{2s_2(\phi)+\frac{1}{2}}\|\phi\|_{L^{4}(\S)},
\end{equation*}

\begin{equation*}\label{An1}
\|\phi\|_{L_{sc}^{2}(\S)}:=\delta^{s_1(\phi)-1}|u|^{2s_2(\phi)}\|\phi\|_{L^{2}(\S)},
\end{equation*}

\begin{equation*}\label{An1}
\|\phi\|_{L_{sc}^{1}(\S)}:=\delta^{s_1(\phi)-\f32}|u|^{2s_2(\phi)-1}\|\phi\|_{L^{1}(\S)}.
\end{equation*}

More generally, for $1\leq p \leq \infty$, we define,

\begin{equation*}\label{An1}
\|\phi\|_{L_{sc}^{p}(\S)}:=\delta^{s_1(\phi)-\frac12-\frac{1}{p}}|u|^{2s_2(\phi)+1-\frac{2}{p}}\|\phi\|_{L^{p}(\S)}.
\end{equation*}

In scale invariant norms, we have\\
\textbf{H\"{o}lder's inequalities}
\begin{equation*}\label{An1}
\|\phi_1\cdot\phi_2\|_{L_{sc}^{1}(\S)}\leq\frac{\delta^{\frac{1}{2}}}{|u|}\|\phi_1\|_{L_{sc}^{\infty}(S)}\|\phi_2\|_{L_{sc}^{1}(\S)},
\end{equation*}

\begin{equation*}\label{An1}
\|\phi_1\cdot\phi_2\|_{L_{sc}^{1}(\S)}\leq\frac{\delta^{\frac{1}{2}}}{|u|}\|\phi_1\|_{L_{sc}^{2}(S)}\|\phi_2\|_{L_{sc}^{2}(\S)},
\end{equation*}

\begin{equation*}\label{An1}
\|\phi_1\cdot\phi_2\|_{L_{sc}^{2}(\S)}\leq\frac{\delta^{\frac{1}{2}}}{|u|}\|\phi_1\|_{L_{sc}^{\infty}(S)}\|\phi_2\|_{L_{sc}^{2}(\S)},
\end{equation*}

\begin{equation*}\label{An1}
\|\phi_1\cdot\phi_2\|_{L_{sc}^{2}(\S)}\leq\frac{\delta^{\frac{1}{2}}}{|u|}\|\phi_1\|_{L_{sc}^{4}(S)}\|\phi_2\|_{L_{sc}^{4}(\S)},
\end{equation*}

\begin{equation*}\label{An1}
\|\phi_1\cdot\phi_2\|_{L_{sc}^{4}(\S)}\leq\frac{\delta^{\frac{1}{2}}}{|u|}\|\phi_1\|_{L_{sc}^{\infty}(S)}\|\phi_2\|_{L_{sc}^{4}(\S)}.
\end{equation*}

For convenience, along the null hypersurfaces $H_u^{(0,\ub)}$ and $\Hb_{\ub}^{(u_{\infty},u)}$ we also define\\ 
\textbf{Scale invariant norms along a hypersurface}
\begin{equation*}\label{An1}
\|\phi\|^2_{L^2_{sc}(H_u^{(0,\underline{u})})}:=
\delta^{-1}\int_0^{\underline{u}}\|\phi\|^2_{L^2_{sc}(S_{u,\underline{u}'})}d\underline{u}',
\end{equation*}

\begin{equation*}\label{An1}
\|\phi\|^2_{L^2_{sc}(\underline{H}_{\underline{u}}^{(u_{\infty},u)})}:=
\int_{u_{\infty}}^{u}{\frac{1}{|u'|^2}}\|\phi\|^2_{L^2_{sc}(S_{u',\underline{u}})}du'.
\end{equation*}


\subsection{Norms}
In the rest of this paper, we use the following norms:

\textbf{Ricci coefficient norms:}

For any $\S$, we introduce $\mathcal{O}_{s,p}(u,\ub)$:

\begin{equation*}\label{O_{0,infty} sc}
\begin{split}
\mathcal{O}_{0,\infty}(u,\underline{u}):=&\|\omega\|_{L^{\infty}_{sc}(\S)}+\|\chih\|_{L^{\infty}_{sc}(\S)}+\|\tr\chi\|_{L^{\infty}_{sc}(\S)}+\|\eta\|_{L^{\infty}_{sc}(\S)}\\
&+\|\underline \eta\|_{L^{\infty}_{sc}(\S)}+\frac{1}{|u|}\|\hat{\underline{\chi}}\|_{L^{\infty}_{sc}(\S)}+\frac{\delta^{\frac 12}}{|u|^2}\|\tr{\underline{\chi}}\|_{L^{\infty}_{sc}(\S)}+\|\underline{\omega}\|_{L^{\infty}_{sc}(\S)},
\end{split}
\end{equation*}

\begin{equation*}\label{O_(0,4) sc}
\begin{split}
\mathcal{O}_{0,4}(u,\underline{u}):=&\|\omega\|_{L^{4}_{sc}(\S)}+\delta^{\frac{1}{4}}\|\chih\|_{L^{4}_{sc}(\S)}+\|\tr\chi\|_{L^{4}_{sc}(\S)}+\|\eta\|_{L^{4}_{sc}(\S)}\\
&+\|\underline \eta\|_{L^{4}_{sc}(\S)}+\frac{{\delta}^{\frac 14}}{|u|}\|\hat{\underline{\chi}}\|_{L^{4}_{sc}(\S)}+\frac{\delta^{\frac 34}}{|u|^2}\|\tr{\underline{\chi}}\|_{L^{4}_{sc}(\S)}+\|\underline{\omega}\|_{L^{4}_{sc}(\S)},
\end{split}
\end{equation*}



\begin{equation*}\label{O_{1,4} sc}
\begin{split}
\mathcal{O}_{1,4}(u,\underline{u}):=&\|\nabla\omega\|_{L^{4}_{sc}(\S)}+\|\nabla\chih\|_{L^{4}_{sc}(\S)}+\|\nabla \tr\chi\|_{L^{4}_{sc}(\S)}\\
&+\|\nabla\eta\|_{L^{4}_{sc}(\S)}+\|\nabla{\underline{\eta}}\|_{L^{4}_{sc}(\S)}+\frac{1}{|u|}\|\nabla\hat{\underline{\chi}}\|_{L^{4}_{sc}(\S)}\\
&+\|\nabla \tr{\underline{\chi}}\|_{L^{4}_{sc}(\S)}+\|\nabla \underline{\omega}\|_{L^{4}_{sc}(\S)},
\end{split}
\end{equation*}

\begin{equation*}\label{O_{2,4} sc}
\begin{split}
\mathcal{O}_{2,4}(u,\underline{u}):=&\|\nabla^2\omega\|_{L^{4}_{sc}(\S)}+\|\nabla^2\chih\|_{L^{4}_{sc}(\S)}+\|\nabla^2 \tr\chi\|_{L^{4}_{sc}(\S)}\\
&+\|\nabla^2\eta\|_{L^{4}_{sc}(\S)}+\|\nabla^2{\underline{\eta}}\|_{L^{4}_{sc}(\S)}+\frac{1}{|u|}\|\nabla^2\hat{\underline{\chi}}\|_{L^{4}_{sc}(\S)}\\
&+\|\nabla^2 \tr{\underline{\chi}}\|_{L^{4}_{sc}(\S)}+\|\nabla^2 \underline{\omega}\|_{L^{4}_{sc}(\S)},
\end{split}
\end{equation*}

\begin{equation*}\label{O_{3,2} sc}
\begin{split}
\mathcal{O}_{3,2}(u,\underline{u}):=&\|\nabla^3\omega\|_{L^{2}_{sc}(\S)}+\|\nabla^3\chih\|_{L^{2}_{sc}(\S)}+\|\nabla^3 \tr\chi\|_{L^{2}_{sc}(\S)}\\
&+\|\nabla^3\eta\|_{L^{2}_{sc}(\S)}+\|\nabla^3{\underline{\eta}}\|_{L^{2}_{sc}(\S)}+\frac{1}{|u|}\|\nabla^3\hat{\underline{\chi}}\|_{L^{2}_{sc}(\S)}\\
&+\|\nabla^3 \tr{\underline{\chi}}\|_{L^{2}_{sc}(\S)}+\|\nabla^3 \underline{\omega}\|_{L^{2}_{sc}(\S)}.
\end{split}
\end{equation*}

We denote $\mathcal{O}_{0,4}$, $\mathcal{O}_{0,\infty}$,  $\mathcal{O}_{1,4}$, $\mathcal{O}_{2,4}$ and $\mathcal{O}_{3,2}$ to be the supremum of the corresponding norms over all values of $u,\ub$, where $u_{\infty}\leq u \leq -c$ and $0\leq\ub\leq \d$. Finally, we define the total Ricci norm $\mathcal{O}$
$$\mathcal{O}:=\mathcal{O}_{0,4}+\mathcal{O}_{0,\infty}+\mathcal{O}_{1,4}+\mathcal{O}_{2,4}+\mathcal{O}_{3,2}$$
and let $\mathcal{O}^{(0)}$ be the corresponding norm of the initial hypersurface $H_{u_{\infty}}$.\\

\textbf{Curvature norms:}

Along the null hypersurfaces $H=H_u^{(0,\ub)}$ and $\Hb=\Hb_{\ub}^{(u_{\infty},u)}$, we introduce

\begin{equation*}\label{R_0(H) sc}
\mathcal{R}_0(u,\underline{u}):=\delta^{\frac{1}{2}}\|\alpha\|_{L^{2}_{sc}(H)}+\|\beta\|_{L^{2}_{sc}(H)}+\|\rho\|_{L^{2}_{sc}(H)}+\|\sigma\|_{L^{2}_{sc}(H)}+\|\underline{\beta}\|_{L^{2}_{sc}(H)},
\end{equation*}

\begin{equation*}\label{R_0(Hb) sc}
\mathcal{\underline{R}}_0(u,\underline{u}):=\delta^{\frac{1}{2}}\|\beta\|_{L^{2}_{sc}(\underline{H})}+\|\rho\|_{L^{2}_{sc}(\underline{H})}+\|\sigma\|_{L^{2}_{sc}(\underline{H})}+\|\underline{\beta}\|_{L^{2}_{sc}(\underline{H})}+\|\underline{\alpha}\|_{L^{2}_{sc}(\underline{H})},
\end{equation*}

\begin{equation*}\label{R_1(H) sc}
\mathcal{R}_1(u,\underline{u}):=\|\nabla{\alpha}\|_{L^{2}_{sc}(H)}+\|\nabla{\beta}\|_{L^{2}_{sc}(H)}+\|\nabla{\rho}\|_{L^{2}_{sc}(H)}+\|\nabla{\sigma}\|_{L^{2}_{sc}(H)}+\|\nabla{\underline{\beta}}\|_{L^{2}_{sc}(H)},
\end{equation*}

\begin{equation*}\label{R_1(Hb) sc}
\mathcal{\underline{R}}_1(u,\underline{u}):=\|\nabla{\beta}\|_{L^{2}_{sc}(\underline{H})}+\|\nabla{\rho}\|_{L^{2}_{sc}(\underline{H})}+\|\nabla{\sigma}\|_{L^{2}_{sc}(\underline{H})}+\|\nabla{\underline{\beta}}\|_{L^{2}_{sc}(\underline{H})}+\|\nabla{\underline{\alpha}}\|_{L^{2}_{sc}(\underline{H})},
\end{equation*}

\begin{equation*}\label{R_1(H) sc}
\mathcal{R}_2(u,\underline{u}):=\|\nabla^2{\alpha}\|_{L^{2}_{sc}(H)}+\|\nabla^2{\beta}\|_{L^{2}_{sc}(H)}+\|\nabla^2{\rho}\|_{L^{2}_{sc}(H)}+\|\nabla^2{\sigma}\|_{L^{2}_{sc}(H)}+\|\nabla^2{\underline{\beta}}\|_{L^{2}_{sc}(H)},
\end{equation*}

\begin{equation*}\label{R_1(Hb) sc}
\mathcal{\underline{R}}_2(u,\underline{u}):=\|\nabla^2{\beta}\|_{L^{2}_{sc}(\underline{H})}+\|\nabla^2{\rho}\|_{L^{2}_{sc}(\underline{H})}+\|\nabla^2{\sigma}\|_{L^{2}_{sc}(\underline{H})}+\|\nabla^2{\underline{\beta}}\|_{L^{2}_{sc}(\underline{H})}+\|\nabla^2{\underline{\alpha}}\|_{L^{2}_{sc}(\underline{H})}.
\end{equation*}

We set $\mathcal{R}_0$, $\mathcal{R}_1$, $\mathcal{R}_2$ to be the supremum over $u,\ub$ in our spacetime slab of $\mathcal{R}_0(u,\ub)$, $\mathcal{R}_1(u,\ub)$ and $\mathcal{R}_2(u,\ub)$, respectively. Similarly, we define $\underline{\mathcal{R}}_0$, $\underline{\mathcal{R}}_1$ and $\underline{\mathcal{R}}_2$. 
We write $\mathcal{R}:=\mathcal{R}_0+\mathcal{R}_1+\mathcal{R}_2$ and $\underline{\mathcal{R}}:=\underline{\mathcal{R}}_0+\underline{\mathcal{R}}_1+\underline{\mathcal{R}}_2$. Finally, we denote $\mathcal{R}^{(0)}$ as the initial value for the norm $\mathcal{R}$, i.e.,
$$\mathcal{R}^{(0)}:=\sup_{0\leq \ub \leq \delta} \l\mathcal{R}_0(u_{\infty},\ub)+\mathcal{R}_1(u_{\infty},\ub)+\mathcal{R}_2(u_{\infty},\ub)\r.$$

%
%
%
%
%

\textbf{Initial data assumptions:}\\

We define the initial data quantity
$$\mathcal{I}^{(0)}:=\sup_{0\leq \ub \leq \delta}\mathcal{I}^{(0)}(\ub),$$
where
\begin{equation*}
\begin{split}
\mathcal{I}^{(0)}(\ub):=& \delta^{\frac 12}|u_{\infty}|\|\chih_0\|_{L^{\infty}(S_{u_{\infty},\ub})}+\sum_{0\leq k \leq 2}\delta^{\frac 12}\|(\delta\nab_4)^k\chih_0\|_{L^2(S_{u_{\infty},\ub})}\\
&+\sum_{0\leq k \leq 1,}\sum_{1\leq m \leq 4}\delta^{\frac 12}\|(\delta^{\frac 12}|u_{\infty}|\nab)^{m-1}(\delta\nab_4)^k\nab\chih_0\|_{L^2(S_{u_{\infty},\ub})}.
\end{split}
\end{equation*}

Here $\chih_0$ denotes $\chih$ along $H_{u_{\infty}}^{(0,\ub)}$.\\





\section{Statement of main theorem}

We are now ready to state our main theorem
\begin{theorem} (Main Theorem)\label{main.thm}

Consider the following characteristic initial value problem for the Einstein vacuum equations. The initial incoming hypersurface $\Hb_0$ is required to coincide with a backwards light cone in Minkowski space with $u_{\infty}\leq u \leq 0$. On the initial outgoing hypersurface $H_{u_{\infty}}$, 
the data are smooth and $\M I^{(0)}(\ub)$ is bounded by an arbitrary constant $\M I^{(0)}$ uniformly. Given $\M I^{(0)}$ and another arbitrary positive constant $c$, there exists a sufficiently small $\d=\d(\M I^{(0)}, c)>0$ such that in the region $u_{\infty}\leq u \leq -c$, $ 0 \leq \ub \leq \d$, we have

$$\M R+\underline{\M R}+\M O \ls \M I^{(0)}.$$

%
\end{theorem}

{\bf Remark:} In the following, we will only prove the a priori estimates for $\mathcal O$, $\tilde{\M O}_{5,2}$ and $\M R$. The existence and uniqueness of solution and the propagation of regularity follow from standard arguments (see for example \cite{Chr:book}).

{\bf Remark:} Following \cite{Chr:book}, one can solve the constraint ODEs and obtain bounds for the initial data on $H_{u_{\infty}}$ from that of the initial shear. In particular, under the assumption of Theorem \ref{main.thm}, we have the following initial bounds for the Ricci coefficients and curvature components
%
$$\M R^{(0)}+\M O^{(0)}\ls \M I^{(0)}.$$

Once the existence theorem is established, the actual formation of trapped surfaces follows from a simple ODE argument as in \cite{Chr:book}:

\begin{theorem}(Formation of Trapped Surfaces from Past Null Infinity)\label{main.thm2}

Given $\M I^{(0)}$ and $c$, there exist $\d_0=\d_0(\M I^{(0)}, c)$ sufficiently large, such that for $0<\d<\d_0$, with initial data:
   \begin{itemize}
            \item $\sum_{i\leq 5, k\leq 3}\delta^{k} \d^{\frac12}\|\nab^{k}_4(|u_{\infty}|\nab)^{i}\chih_0\|_{L^{\infty}(S_{u_{\infty},\ub})}\leq \M I^{(0)}$ \quad \quad along $u=u_{\infty}$
            \item $\sum_{2\leq j\leq 7}\delta^{\frac12}\|(\delta^{\frac12}|u_{\infty}|\nab)^j\chih_0\|_{L^2(S_{u_{\infty},\ub})}\leq \epsilon$ \quad \quad \quad \quad \quad along $u=u_{\infty}$
            \item Minkowskian initial data \quad \quad \quad \quad \quad \quad \quad \quad \quad \quad \quad \quad along $\ub=0$
            \item $\int_0^{\delta}|u_{\infty}|^2|\chih_0|^2(u_{\infty}, \ub')d\ub'\geq 4c$ for every direction  \quad \quad along $u=u_{\infty}$
    \end{itemize}
Then $S_{c,\delta}$ is a trapped surface.
\end{theorem}

\section{The Preliminary Estimates}\label{secbasic}

We will make the following bootstrap assumptions on the Ricci coefficients:
\begin{equation}\label{BA.0}
 \M O_{0,\infty}\leq \D_0,
\end{equation}

\begin{equation}\label{BA.1}
 \M O_{0,4}\leq \D_1,
\end{equation}

\begin{equation}\label{BA.2}
\M O_{1,4}\leq \D_2.
\end{equation}


Moreover, we set a bootstrap assumption on curvature components and their derivatives:

\begin{equation}\label{BA.4}
 \M R +\underline{\M R} \leq R.
\end{equation}

\subsection{Estimates for Metric Components}\label{metric}
We derive bound for $\Omega$ first:
\begin{proposition}\label{Omega}
Under the assumptions of Theorem \ref{main.thm} and bootstrap assumption \eqref{BA.0}, we have
$$\|\Omega-1\|_{L^\i(\S)}\ls \f{\d^{\f12}}{|u|}\D_0.$$
\end{proposition}
\begin{proof}
For $\Omega$ we have equation
\begin{equation}\label{Omegatransport}
 \omega=-\frac{1}{2}\nabla_4\log\Omega=\frac{1}{2}\Omega\nabla_4\Omega^{-1}=\frac{1}{2}\frac{\partial}{\partial \ub}\Omega^{-1}.
\end{equation}
We integrate this equation.  On $\Hb_0$ we have $\Omega^{-1}=1$ and this leads to
$$||\Omega^{-1}-1||_{L^\infty(S_{u,\ub})}\ls \int_0^{\ub}||\omega||_{L^\infty(S_{u,\ub'})}d\ub'\ls \f{\d^{\f12}}{|u|}\D_0,$$
where we have used the bootstrap assumption \eqref{BA.0}. Finally, notice that 
$$\|\Omega-1\|_{L^\i(\S)}=\|\Omega\|_{L^\i(\S)}\|\Omega^{-1}-1\|_{L^\i(\S)}\ls \f{\d^{\f12}}{|u|}\D_0.$$
\end{proof}

We then control $\gamma$ under the bootstrap assumption (\ref{BA.0}):
\begin{proposition}\label{gamma}
Under the assumptions of Theorem \ref{main.thm} and the bootstrap assumptions \eqref{BA.0}, \eqref{BA.1} and \eqref{BA.2}, for metric $\gamma$ we have
$$c'\leq \det\gamma\leq C'. $$
Moreover, in $D$
$$|\gamma_{AB}|,|(\gamma^{-1})^{AB}|\leq C'.$$
Here $C'$ and $c'$ are constants depending only on initial data.
\end{proposition}
\begin{proof}
We employ the first variation formula
$$\Ls_L\gamma=2\Omega\chi.$$
With coordinates, this gives
$$\frac{\partial}{\partial \ub}\gamma_{AB}=2\Omega\chi_{AB}.$$
Hence we derive that 
$$\frac{\partial}{\partial \ub}\log(\det\gamma)=\Omega\trch.$$
Let $\gamma_0(u,\ub,\th^1,\th^2)=\gamma(u,0,\th^1,\th^2)$. Then it follows
\begin{equation}\label{detgaper}
|\det\gamma-\det(\gamma_0)|\leq \int_0^{\ub}|\trch|d\ub'\leq \f{\d^{\f12}}{|u|}\Delta_0 .
\end{equation}
Thus, we derive the lower and upper bound for $\det \gamma$.
For $\gamma$, denote $\Lambda$ to be the greater eigenvalue. We have
\begin{equation*}\label{La}
\Lambda\leq\sup_{A,B=1,2}\gamma_{AB},
\end{equation*}

$$\sum_{A,B=1,2}|\chi_{AB}|^2\leq\Lambda ||\chi||_{L^\infty(S_{u,\ub})},$$

$$|\gamma_{AB}-(\gamma_0)_{AB}|\leq \int_0^{\ub}|\chi_{AB}|d\ub'\leq\Lambda\f{\d^{\f12}}{|u|}\D_0.$$
Using bootstrap assumption (\ref{La}), we thus bound $|\gamma_{AB}|$ from above. We further bound $|(\gamma^{-1})^{AB}|$ from above by using the upper bound for $|\gamma_{AB}|$ and the lower bound for $\det\gamma$.
\end{proof}

This proposition implies an estimate on the surface area of the two sphere $S_{u,\ub}$.
\begin{proposition}\label{area}
In $D$ we have
$$\sup_{\ub}|\mbox{Area}(S_{u,\ub})-\mbox{Area}(S_{u,0})|\leq \Delta^{\f12}_0\f{\d^{\f14}}{|u|^{\f12}}|u|^2.$$
\end{proposition}
\begin{proof}
This follows from \eqref{detgaper}.
\end{proof}


\subsection{Estimates for Transport Equations}\label{transportsec}
In latter sections of the paper, we will employ the following propositions for transport equations:


\begin{proposition}
Under the assumptions of Theorem \ref{main.thm} and the bootstrap assumptions \eqref{BA.0}, \eqref{BA.1} and \eqref{BA.2}, it follows that
\begin{equation}\label{transport1}
 ||\phi||_{L^2(S_{u,\ub})}\ls ||\phi||_{L^2(S_{u,\ub'})}+\int_{\ub'}^{\ub} ||\nabla_4\phi||_{L^2(S_{u,\ub''})}d{\ub''},
\end{equation}

\begin{equation}\label{transport2}
 ||\phi||_{L^4(S_{u,\ub})}\ls ||\phi||_{L^4(S_{u,\ub'})}+\int_{\ub'}^{\ub} ||\nabla_4\phi||_{L^4(S_{u,\ub''})}d{\ub''}
\end{equation}
for an $S_{u,\ub}$ tangent tensor $\phi$ of arbitrary rank.
\end{proposition}

\begin{proof}
For any scalar $f$, we have the first variation of area formula:
\[
 \frac{d}{d\ub}\int_{\S} f=\int_{\S} \left(\frac{df}{d\ub}+\Omega \trch f\right)=\int_{\S} \Omega\left(e_4(f)+ \trch f\right).
\]
Hence taking $f=|\phi|_{\gamma}^2$, we have
\begin{equation*}\label{Lptransport}
\begin{split}
 ||\phi||^2_{L^2(S_{u,\ub})}=&||\phi||^2_{L^2(S_{u,\ub'})}+\int_{\ub'}^{\ub}\int_{S_{u,\ub''}} 2\Omega\left(<\phi,\nabla_4\phi>_\gamma+ \frac{1}{2}\trch |\phi|^2_{\gamma}\right)d{\ub''}.
\end{split}
\end{equation*}
The inequality (\ref{transport1})  follows by employing Cauchy-Schwarz inequality on the sphere and the $L^\infty$ bounds for $\Omega$ and $\trch$  provided by Proposition \ref{Omega} and the bootstrap assumption \eqref{BA.0}.
With the same method taking $f=|\phi|_{\gamma}^4$, we obtain inequality (\ref{transport2}).
\end{proof}

\begin{proposition}
Under the assumptions of Theorem \ref{main.thm} and the bootstrap assumptions \eqref{BA.0}, \eqref{BA.1} and \eqref{BA.2}, we have
\begin{equation}\label{transport3}
 ||\phi||_{L^2(S_{u,\ub})}\ls ||\phi||_{L^2(S_{u',\ub})}+\int_{u'}^{u} ||\nabla_3\phi||_{L^2(S_{u'',\ub})}d{u''},
\end{equation}

\begin{equation}\label{transport4}
 ||\phi||_{L^4(S_{u,\ub})}\ls ||\phi||_{L^4(S_{u',\ub})}+\int_{u'}^{u} ||\nabla_3\phi||_{L^4(S_{u'',\ub})}d{u''}
\end{equation}
for an $S_{u,\ub}$ tangent tensor $\phi$ of arbitrary rank.
\end{proposition}

\begin{proof}
For scalar $f$, we use the first variation of area formula:
\[
 \frac{d}{du}\int_{\S} f=\int_{\S} \left(\frac{df}{du}+\Omega \tr\chib f\right)=\int_{\S} \Omega\left(e_3(f)+ \tr\chib f\right).
\]
Hence, taking $f=|\phi|_{\gamma}^2$, we have
\begin{equation*}\label{Lptransport}
\begin{split}
 ||\phi||^2_{L^2(S_{u,\ub})}=&||\phi||^2_{L^2(S_{u_{\infty},\ub})}+\int_{u'}^{u}\int_{S_{u'',\ub}} 2\Omega\left(<\phi,\nabla_3\phi>_\gamma+ \frac{1}{2}\tr\chib |\phi|^2_{\gamma}\right)d u''.
\end{split}
\end{equation*}
The inequality (\ref{transport3}) can be concluded using Cauchy-Schwarz on the sphere together with the help of $L^\infty$ control for $\tr\chib$  and $\Omega$, which are provided by Proposition \ref{Omega} and the bootstrap assumption \eqref{BA.0} respectively.
With the same method taking $f=|\phi|_{\gamma}^4$, we obtain inequality (\ref{transport4}).
\end{proof}

We rewrite the above inequalities in scale invariant norms as follows:

\begin{proposition}\label{transport}
For an $\S$ tangent tensor $\phi$ of arbitrary rank, we have

\begin{equation*}
\|\phi\|_{L^2_{sc}(S_{u,\underline{u}})}\leq
\|\phi\|_{L^2_{sc}(S_{u,0})}+\int_0^{\underline{u}}\delta^{-1}\|\nabla_4\phi\|_{L^2_{sc}(S_{u,\underline{u}'})}d\underline{u}',
\end{equation*}

\begin{equation*}
\|\phi\|_{L^2_{sc}(S_{u,\underline{u}})}\leq
\|\phi\|_{L^2_{sc}(S_{u_{\infty},\underline{u}})}+\int_{u_{\infty}}^{u}\frac{1}{|u'|^2}\|\nabla_3\phi\|_{L^2_{sc}(S_{u',\underline{u}})}du',
\end{equation*}

\begin{equation*}
\|\phi\|_{L^4_{sc}(S_{u,\underline{u}})}\leq
\|\phi\|_{L^4_{sc}(S_{u,0})}+\int_0^{\underline{u}}\delta^{-1}\|\nabla_4\phi\|_{L^4_{sc}(S_{u,\underline{u}'})}d\underline{u}',
\end{equation*}

\begin{equation*}
\|\phi\|_{L^4_{sc}(S_{u,\underline{u}})}\leq
\|\phi\|_{L^4_{sc}(S_{u_{\infty},\underline{u}})}+\int_{u_{\infty}}^{u}\frac{1}{|u'|^2}\|\nabla_3\phi\|_{L^4_{sc}(S_{u',\underline{u}})}du'.
\end{equation*}
\end{proposition}

For the $\nab_3$ equation, we can get more precise estimates by incorporating the weights in the norms. These weights depend on the coefficients in front of the linear term with a $\trchb$ factor. The main observation is that under the bootstrap assumption \eqref{BA.0}, $\trchb$ can be viewed essentially as $-\f2{|u|}$. More clearly, we have
\begin{proposition}[Evolution Lemma] \label{evolution lemma}
We continue to work under the assumptions of Theorem \ref{main.thm} and the bootstrap assumptions \eqref{BA.0}, \eqref{BA.1} and \eqref{BA.2}. Moreover, assume that $\|\tr\chib+\f{2}{|u|}\|_{L^{\infty}(\S)}\leq \f{\d^{\f14}}{|u|^2}$ holds in $D_{u,\ub}$.  Let $\phi$ and $F$ be $\S$-tangent tensor fields of rank $k$ satisfying the following transport equation:
\begin{equation*}
\nab_3 \phi_{A_1...A_k}+\lambda_0{\tr\underline{\chi}}\phi_{A_1...A_k}=F_{A_1...A_k}.
\end{equation*}
Set $p\in\{2,4\}$. Denoting $\lambda_1=2(\lambda_0-\frac{1}{p})$, for $\phi$ we have
\begin{equation*}
|u|^{\lambda_1}\|\phi\|_{L^{p}(\S)}\lesssim
|u_{\infty}|^{\lambda_1}\|\phi\|_{L^{p}(S_{u_{\infty},\underline{u}})}+\int_{u_{\infty}}^u|u'|^{\lambda_1}\|F\|_{L^{p}(S_{u',\underline{u}})}du',
\end{equation*}

where the implicit constant is allowed to depend on $\lambda_0$.
\end{proposition}
\begin{proof}
To begin, we have the following identity for any scalar function $f$:
\[
 \frac{d}{du}\int_{\S} f=\int_{\S} \left(\frac{df}{du}+\Omega \trchb f\right)=\int_{\S} \Omega\left(e_3(f)+ \trchb f\right).
\]
Using this identity, we obtain
\begin{equation*}
\begin{split}
&\frac{d}{du}(\int_{\S}|u|^{\lambda_1 p}|\phi|^{p})\\
=&\int_{\S}\Omega\l -\lambda_1 p|u|^{\lambda_1 p-1}(e_3 u)|\phi|^{p}+p|u|^{\lambda_1 p}<\phi^{p-1},\nab_3\phi>+ \tr\underline{\chi}|u|^{\lambda_1 p}|\phi|^{p}\r\\
=&\int_{\S}\Omega\l p|u|^{\lambda_1 p}<\phi^{p-1}, \nab_3\phi+{\lambda_0}\trchb\phi>\r\\
&+\int_{\S}\Omega |u|^{\lambda_1 p}\l -\f{\lambda_1 p(e_3u)}{|u|}+(1-\lambda_0 p)\trchb\r|\phi|^p.
\end{split}
\end{equation*}
Observe that we have
\begin{equation*}
\begin{split}
&-\f{\lambda_1 p(e_3u)}{|u|}+(1-\lambda_0 p)\trchb\\
= &-\f{\lambda_1 p\Omega^{-1}}{|u|}+(1-\lambda_0 p)\trchb\\
= &-\f{\lambda_1 p(\Omega^{-1}-1)}{|u|}+(1-\lambda_0 p)(\trchb+\f{2}{|u|})-\f{\lambda_1 p+2-2\lambda_0 p}{|u|}\\
\ls &\f{\d^{\f14}}{|u|^2}\D_0.
\end{split}
\end{equation*}
For the last inequality, we employ Proposition \ref{Omega} and assumption $\|\tr\chib+\f{2}{|u|}\|_{L^{\infty}(\S)}\leq \f{\d^{\f14}}{|u|^2}$ together with the chosen parameters to satisfy $\lambda_1 p+2-2\lambda_0 p=0$.

Therefore, 
\begin{equation*}
\begin{split}
|\frac{d}{du}(\int_{\S}|u|^{\lambda_1 p}|\phi|^p)|
\ls &\int_{\S}\l p|u|^{\lambda_1 p}|\phi|^{p-1} |F| +|u|^{2\lambda_1-2} \d^{\f14}\D_0|\phi|^p\r.
\end{split}
\end{equation*}
Using Cauchy-Schwarz for the first term and applying Gronwall's inequality for the second term, we obtain
\begin{equation*}
\begin{split}
&|u|^{\lambda_1}\|\phi\|_{L^p(\S)}\\
\ls &e^{\d^{\f14}\D_0\|u^{-2}\|_{L^1_u}}\l|u_{\infty}|^{\lambda_1}\|\phi\|_{L^p(S_{u_{\infty},\underline{u}})}+\int_{u_{\infty}}^u |u'|^{\lambda_1}\|F\|_{L^p(S_{u',\underline{u}})}du'\r\\
\ls &|u_{\infty}|^{\lambda_1}\|\phi\|_{L^p(S_{u_{\infty},\underline{u}})}+\int_{u_{\infty}}^u |u'|^{\lambda_1}\|F\|_{L^p(S_{u',\underline{u}})}du'.
\end{split}
\end{equation*}
since $\de^{\f14}\D_0\|u^{-2}\|_{L^1_u}\ls 1$, when $\d^{\f14}\D_0$ is small.

\end{proof}

{\bf Remark:} Later in the remark of Proposition \ref{O0infty trchib} we will show that under the bootstrap assumptions \eqref{BA.0}, \eqref{BA.1} and \eqref{BA.2}, we have 
$$\|\tr\chib+\f{2}{|u|}\|_{L^{\infty}(\S)}\leq \f{\d^{\f12}}{|u|^2}(\M I^{(0)}+\M R+ \underline{\M R})^5+\f{\d^{\f12}}{|u|^2}\D_0.$$
By choosing $\d$ small enough, $\|\tr\chib+\f{2}{|u|}\|_{L^{\infty}(\S)}\leq \f{\d^{\f14}}{|u|^2}$ is verified.

\subsection{Sobolev Embedding}\label{Embedding}

Since we have the control of volume form on both sides, we have the following Sobolev embedding theorems (see also \cite{L-R:Propagation}).

\begin{proposition}\label{L4}
There exist  $\d_0=\d_0(\Delta_0)$, such that whenever $\d \leq \d_0$, 
in $D$ for any horizontal tensor $\phi$ we have
$$||\phi||_{L^4(S_{u,\ub})}\leq   \|\phi\|^{\frac{1}{2}}_{L^2(\S)}\|\nabla{\phi}\|^{\frac{1}{2}}_{L^2(\S)}+\frac{1}{|u|^{\frac{1}{2}}}\|\phi\|_{L^2(\S)}, $$
\end{proposition}

and in scale invariant norms

\begin{equation*}\label{An1}
\|\phi\|_{L^4_{sc}(S_{u,\underline{u}})}\leq
\|\phi\|^{\frac{1}{2}}_{L^2_{sc}(S_{u,\underline{u}})}\|\nabla{\phi}\|^{\frac{1}{2}}_{L^2_{sc}(S_{u,\underline{u}})}+\delta^{\frac{1}{4}}\|\phi\|_{L^2_{sc}(S_{u,\underline{u}})}.
\end{equation*}

Similarly, for $L^\infty$ norm we obtain
\begin{proposition}\label{Linfty}
There exist  $\d_0=\d_0(\Delta_0)$, such that whenever $\d \leq \d_0$, 
in $D$ for any horizontal tensor $\phi$ we have
$$\|\phi\|_{L^{\infty}(\S)}\leq 
\|\phi\|^{\frac{1}{2}}_{L^4(\S)}\|\nabla{\phi}\|^{\frac{1}{2}}_{L^4(\S)}+\frac{1}{|u|^{\frac{1}{2}}}\|\phi\|_{L^4(\S)}, $$
\end{proposition}

and in scale invariant norms

\begin{equation*}\label{An1}
\|\phi\|_{L^{\infty}_{sc}(S_{u,\underline{u}})}\leq
\|\phi\|^{\frac{1}{2}}_{L^4_{sc}(S_{u,\underline{u}})}\|\nabla{\phi}\|^{\frac{1}{2}}_{L^4_{sc}(S_{u,\underline{u}})}+\delta^{\frac{1}{4}}\|\phi\|_{L^4_{sc}(S_{u,\underline{u}})}.
\end{equation*}



In the same manner, we derive

\begin{proposition}
Let ${^{\d}S}_{u,\ub}\subset S_{u,\ub}$ denote a disk of radius $\d^{\f12}|u|$ relative to either $\theta$ or $\underline{\theta}$ coordinate 
system. There exist  $\d_0=\d_0(\Delta_0)$, such that whenever $\d \leq \d_0$, 
in $D$ for any horizontal tensor $\phi$ we have


\begin{equation*}
\|\phi\|_{L^{\infty}}(S_{u,\ub})\leq \sup_{{^{\d}S}_{u,\ub}\subset S_{u,\ub}}\l\d^{\f14}|u|^{\f12}\|\nab \phi\|_{L^4({^{\d} S}_{u,\ub})}+\frac{\d^{-\f14}}{|u|^{\f12}}\|\phi\|_{L^4({^{\d}S}_{u,\ub})}\r, 
\end{equation*}
and in scale invariant norm
\begin{equation*}
\|\phi\|_{L_{sc}^{\infty}}(S_{u,\ub})\leq \sup_{{^{\d}S}_{u,\ub}\subset S_{u,\ub}}\l\|\nab \phi\|_{L_{sc}^4({^{\d}S}_{u,\ub})}+\|\phi\|_{L_{sc}^4({^{\d}S}_{u,\ub})}\r.
\end{equation*}
\end{proposition}

\subsection{Commutation Formula}\label{commutation}
We list the following formula from \cite{KNI:book}:


\begin{proposition}\label{commute0}
For a scalar function $f$, we have

$$[\nab_4,\nab]f=\frac 12 (\eta+\etb)D_4f-\chi\cdot\nab f,$$
$$[\nab_3,\nab]f=\frac 12 (\eta+\etb)D_3f-\chib\cdot\nab f.$$
\end{proposition}

\begin{proposition}\label{commute1}
For a 1-form $U_b$ tangent to $S_{u,\ub}$, we have

$$[D_4,\nab_a]U_b=-\chi_{ac}\nab_cU_b+\epsilon_{ac}{^*\b_b} U_c+\frac 12(\eta_a+\etb_a)D_4U_b-\chi_{ac}\etb_bU_c+\chi_{ab}\etb\cdot U,$$ 
$$[D_3,\nab_a]U_b=-\chib_{ac}\nab_cU_b+\epsilon_{ac}{^*\beb_b} U_c+\frac 12(\eta_a+\etb_a)D_3U_b-\chib_{ac}\eta_bU_c+\chib_{ab}\eta\cdot U.$$ 
\end{proposition}

\begin{proposition}\label{commute2}
For a 2-form $V_{bc}$ tangent to $\S$, we have
\begin{equation*}
\begin{split}
[D_4,\nab_a]V_{bc}=&\frac 12(\eta_a+\etb_a)D_4V_{bc}-\etb_bV_{dc}\chi_{ad}-\etb_cV_{bd}\chi_{ad}-\epsilon_{bd}{^*\b_a}V_{dc}-\epsilon_{cd}{^*\b_c}V_{bd}\\
&+\chi_{ac}V_{bd}\etb_d+\chi_{ab}V_{dc}\etb_d-\chi_{ad}\nab_dV_{bc},
\end{split}
\end{equation*}

\begin{equation*}
\begin{split}
[D_3,\nab_a]V_{bc}=&\frac 12(\eta_a+\etb_a)D_3V_{bc}-\eta_bV_{dc}\chib_{ad}-\eta_cV_{bd}\chib_{ad}+\epsilon_{bd}{^*\beb_a}V_{dc}+\epsilon_{cd}{^*\beb_c}V_{bd}\\
&+\chib_{ac}V_{bd}\eta_d+\chib_{ab}V_{dc}\eta_d-\chib_{ad}\nab_dV_{bc}.
\end{split}
\end{equation*}
\end{proposition}



We further have a more general formula through mathematical induction (see also \cite{{L-R:Propagation}}):
\begin{proposition}\label{commute3}
Assume $\nabla_4\phi=F_0$. Let $\nabla_4\nabla^i\phi=F_i$.
Then we have
\begin{equation*}
\begin{split}
F_i= &\sum_{i_1+i_2+i_3=i}\nabla^{i_1}(\eta+\underline{\eta})^{i_2}\nabla^{i_3} F_0+\sum_{i_1+i_2+i_3+i_4=i-1} \nabla^{i_1}(\eta+\underline{\eta})^{i_2}\nabla^{i_3}\beta\nabla^{i_4} \phi\\
&+\sum_{i_1+i_2+i_3+i_4=i}\nabla^{i_1}(\eta+\underline{\eta})^{i_2}\nabla^{i_3}\chi\nabla^{i_4} \phi.
\end{split}
\end{equation*}

Similarly, assume $\nabla_3\phi=G_{0}$. Let $\nabla_3\nabla^i\phi=G_{i}$.
We get
\begin{equation*}
\begin{split}
G_{i}-\frac{i}{2}\tr\chib \nab^i \phi= &\sum_{i_1+i_2+i_3=i}\nabla^{i_1}(\eta+\underline{\eta})^{i_2}\nabla^{i_3} G_{0}\\
&+\sum_{i_1+i_2+i_3+i_4=i-1} \nabla^{i_1}(\eta+\underline{\eta})^{i_2}\nabla^{i_3}\underline{\beta}\nabla^{i_4} \phi\\
&+\sum_{i_1+i_2+i_3+i_4=i}\nabla^{i_1}(\eta+\underline{\eta})^{i_2}\nabla^{i_3}(\chibh,\tr\chib+\f{2}{u})\nabla^{i_4} \phi.
\end{split}
\end{equation*}

\end{proposition}


\subsection{General elliptic estimates for Hodge systems}\label{elliptic}

We now prove elliptic estimates for general Hodge systems. 

\begin{proposition}\label{ellipticthm}
We continue to work under the assumptions of Theorem \ref{main.thm} and the bootstrap assumptions \eqref{BA.0}, \eqref{BA.1} and \eqref{BA.2}. Let $\phi$ be a $r+1$ covariant tensorfield on $\S$ and be totally symmetric satisfying
$$\div\phi=f,\quad \curl\phi=g,\quad \mbox{tr}\phi=h.$$
Then we have
\begin{equation*}
\begin{split}
&\|\nab^3\phi\|_{L_{sc}^2(\S)}\\
\ls& \|\nab^2 f\|_{L^2_{sc}(\S)}+\|\nab^2 g\|_{L^2_{sc}(\S)}+\d\|\nab\phi,\nab h, f\|_{L^2_{sc}(\S)}\\
&+\d^{\f12}\|\nab f, \nab g, \nab^2 h\|_{L^2_{sc}(\S)}+\d^{\f32}\|\phi, h\|_{L^2_{sc}(\S)}\\
&+\f{\d^{\f12}}{|u|}\|\nab (\K)\|_{L^2_{sc}(\S)}\|\phi,h\|_{L^{\infty}_{sc}(\S)}\\
&+\f{\d^{\f12}}{|u|}\|\K\|_{L^4_{sc}(\S)}\|\nab\phi,\nab h, f\|_{L^4_{sc}(\S)}\\
&+\f{\d^{\f18}}{|u|^{\f12}}\|\K\|^{\f12}_{L^4_{sc}(\S)}\|\nab f, \nab g, \nab^2 h\|_{L^4_{sc}(\S)}\\
&+\f{\d^{\f58}}{|u|^{\f32}}\|\K\|^{\f32}_{L^4_{sc}(\S)}\|\phi, h\|_{L^{\infty}_{sc}(\S)}.\\
\end{split}
\end{equation*}


\end{proposition}
\begin{proof}
Recall the following identity from Chapter 7 in \cite{Chr:book} that for $\phi$, $f$, $g$ and $h$ as above, we have
\begin{equation*}\label{basic.L2}
\int_{\S} \l |\nab\phi|^2+(r+1) K |\phi|^2\r = \int_{\S} \l|f|^2+|g|^2+K |h|^2\r.
\end{equation*}
To get higher order elliptic estimates, we recall again from \cite{Chr:book} that the symmetrized angular derivative of $\phi$ defined by 
$$(\nab\phi)^s_{BA_1...A_{r+1}}:= \f{1}{r+2}(\nab_B\phi_{A_1...A_r}+\sum_{i=1}^{r+1}\nab_{A_i}\phi_{A_1...<A_i>B...A_{r+1}})$$
satisfies the div-curl system
\begin{equation*}
\begin{split}
\div(\nab\phi)^s=&(\nab f)^s-\f{1}{r+2}(^*\nab g)^s+(r+1)K\phi-\f{2K}{r+1}(\gamma\otimes^s h),\\
\curl(\nab\phi)^s=&\f{r+1}{r+2}(\nab g)^s+(r+1)K(^*\phi)^s,\\
\mbox{tr }(\nab\phi)^s=&\f{2}{r+2}f+\f{r}{r+2}(\nab h)^s,
\end{split}
\end{equation*}
where
$$(\gamma\otimes^s h)_{A_1...A_{r+1}}:=\gamma_{A_iA_j}\sum_{i<j=1,...,r+1}h_{A_1...<A_i>...<A_j>...A_{r+1}}$$
and
$$(^*\phi)^s_{A_1...A_{r+1}}:=\f{1}{r+1}\sum_{i=1}^{r+1}\eps_{A_i}{ }^B\phi_{A_1...<A_i>B...A_r}.$$
Using \eqref{basic.L2}, we obtain 
\begin{equation*}
\begin{split}
&\|\nab^2 \phi\|_{L^2(\S)}^2\\
\ls &\|\nab f\|_{L^2(\S)}^2+\|\nab g\|_{L^2(\S)}^2+\|K(|\nab\phi|^2+|f|^2+|\nab h|^2)\|_{L^1(\S)}\\
&+\|K\phi\|_{L^2(\S)}^2+\|K h\|_{L^2(\S)}^2.
\end{split}
\end{equation*}
Iterating the procedure and applying \eqref{basic.L2}, we thus derive

\begin{equation*}
\begin{split}
&\|\nab^3\phi\|^2_{L^2(\S)}\\
\ls& \|\nab^2 f\|^2_{L^2(\S)}+\|\nab^2 g\|^2_{L^2(\S)}+\|\nab K (\phi,h)\|^2_{L^2(\S)}+\|K (\nab\phi,\nab h)\|^2_{L^2(\S)}\\
&+\|K f\|^2_{L^2(\S)}+\|K(\nab f)^2\|_{L^1(\S)}+\|K(\nab g)^2\|_{L^1(\S)}+\|K(\nab^2\phi)^2\|_{L^1(\S)}\\
&+\|K(\nab^2 h)^2\|_{L^1(\S)}+\|K^3(\phi^2, h^2)\|_{L^1(\S)}.
\end{split}
\end{equation*}

By decomposing $K=\K+\f{1}{|u|^2}$, we deduce

\begin{equation*}
\begin{split}
&\|\nab^3\phi\|^2_{L^2(\S)}\\
\ls& \|\nab^2 f\|^2_{L^2(\S)}+\|\nab^2 g\|^2_{L^2(\S)}+\|\nab (\K) (\phi,h)\|^2_{L^2(\S)}\\
&+\|(\K) (\nab\phi,\nab h)\|^2_{L^2(\S)}+\|\f{1}{|u|^2} (\nab\phi,\nab h)\|^2_{L^2(\S)}\\
&+\|(\K) f\|^2_{L^2(\S)}+\|(\K)(\nab f)^2\|_{L^1(\S)}+\|(\K)(\nab^2\phi, \nab g)^2\|_{L^1(\S)}\\
&+\|\f{1}{|u|^2} f\|^2_{L^2(\S)}+\|\f{1}{|u|^2}(\nab f)^2\|_{L^1(\S)}+\|\f{1}{|u|^2}(\nab^2\phi, \nab g)^2\|_{L^1(\S)}\\
&+\|(\K)(\nab^2 h)^2\|_{L^1(\S)}+\|(\K)^3(\phi^2, h^2)\|_{L^1(\S)}\\
&+\|\f{1}{|u|^2}(\nab^2 h)^2\|_{L^1(\S)}+\|\f{1}{|u|^6}(\phi^2, h^2)\|_{L^1(\S)}.
\end{split}
\end{equation*}

With H\"older's inequality, we obtain 

\begin{equation*}
\begin{split}
&\|\nab^3\phi\|^2_{L^2(\S)}\\
\ls& \|\nab^2 f\|^2_{L^2(\S)}+\|\nab^2 g\|^2_{L^2(\S)}+\f{1}{|u|^4}\|\nab\phi,\nab h\|^2_{L^2(\S)}\\
&+\f{1}{|u|^2}\|\nab^2\phi, \nab f, \nab g, \nab^2 h\|^2_{L^2(\S)}+\f{1}{|u|^6}\|\phi, h\|^2_{L^2(\S)}+\f{1}{|u|^4}\|f\|^2_{L^2(\S)}\\
&+\|\nab (\K)\|^2_{L^2(\S)}\|(\phi,h)\|^2_{L^{\infty}(\S)}+\|\K\|^2_{L^4(\S)}\|\nab\phi,\nab h, f\|^2_{L^4(\S)}\\
&+|u|^{\f12}\|\K\|_{L^4(\S)}\|\nab^2\phi, \nab f, \nab g, \nab^2 h\|^2_{L^4(\S)}\\
&+|u|^{\f12}\|\K\|^3_{L^4(\S)}\|\phi, h\|^2_{L^{\infty}(\S)}.\\
\end{split}
\end{equation*}

In scale invariant norms, we deduce

\begin{equation*}
\begin{split}
&\|\nab^3\phi\|^2_{L_{sc}^2(\S)}\\
\ls& \|\nab^2 f\|^2_{L^2_{sc}(\S)}+\|\nab^2 g\|^2_{L^2_{sc}(\S)}+\d^2\|\nab\phi,\nab h, f\|^2_{L^2_{sc}(\S)}\\
&+\d\|\nab^2\phi, \nab f, \nab g, \nab^2 h\|^2_{L^2_{sc}(\S)}+\d^3\|\phi, h\|^2_{L^2_{sc}(\S)}\\
&+\f{\d}{|u|^2}\|\nab (\K)\|^2_{L^2_{sc}(\S)}\|\phi,h\|^2_{L^{\infty}_{sc}(\S)}\\
&+\f{\d}{|u|^2}\|\K\|^2_{L^4_{sc}(\S)}\|\nab\phi,\nab h, f\|^2_{L^4_{sc}(\S)}\\
&+\f{\d^{\f14}}{|u|}\|\K\|_{L^4_{sc}(\S)}\|\nab^2\phi, \nab f, \nab g, \nab^2 h\|^2_{L^4_{sc}(\S)}\\
&+\f{\d^{\f54}}{|u|^3}\|\K\|^3_{L^4_{sc}(\S)}\|\phi, h\|^2_{L^{\infty}_{sc}(\S)},\\
\end{split}
\end{equation*}

which is equivalent to

\begin{equation*}
\begin{split}
&\|\nab^3\phi\|_{L_{sc}^2(\S)}\\
\ls& \|\nab^2 f\|_{L^2_{sc}(\S)}+\|\nab^2 g\|_{L^2_{sc}(\S)}+\d\|\nab\phi,\nab h, f\|_{L^2_{sc}(\S)}\\
&+\d^{\f12}\|\nab^2\phi, \nab f, \nab g, \nab^2 h\|_{L^2_{sc}(\S)}+\d^{\f32}\|\phi, h\|_{L^2_{sc}(\S)}\\
&+\f{\d^{\f12}}{|u|}\|\nab (\K)\|_{L^2_{sc}(\S)}\|\phi,h\|_{L^{\infty}_{sc}(\S)}\\
&+\f{\d^{\f12}}{|u|}\|\K\|_{L^4_{sc}(\S)}\|\nab\phi,\nab h, f\|_{L^4_{sc}(\S)}\\
&+\f{\d^{\f18}}{|u|^{\f12}}\|\K\|^{\f12}_{L^4_{sc}(\S)}\|\nab^2\phi, \nab f, \nab g, \nab^2 h\|_{L^4_{sc}(\S)}\\
&+\f{\d^{\f58}}{|u|^{\f32}}\|\K\|^{\f32}_{L^4_{sc}(\S)}\|\phi, h\|_{L^{\infty}_{sc}(\S)}.\\
\end{split}
\end{equation*}

\end{proof}

If furthermore a symmetric 2-tensor $\phi$ is traceless, to derive elliptic estimates we only need information from $\div\phi$.

\begin{proposition}\label{ellipticthm2}
Let $\phi$ be a symmetric traceless 2-tensor satisfying
$$\div\phi=f.$$
Then under the assumptions of Theorem \ref{main.thm} and the bootstrap assumptions \eqref{BA.0}, \eqref{BA.1} and \eqref{BA.2}, we have

\begin{equation*}
\begin{split}
&\|\nab^3\phi\|_{L_{sc}^2(\S)}\\
\ls& \|\nab^2 f\|_{L^2_{sc}(\S)}+\d\|\nab\phi,f\|_{L^2_{sc}(\S)}+\d^{\f12}\|\nab^2\phi, \nab f\|_{L^2_{sc}(\S)}\\
&+\d^{\f32}\|\phi\|_{L^2_{sc}(\S)}+\f{\d^{\f12}}{|u|}\|\nab (\K)\|_{L^2_{sc}(\S)}\|\phi\|_{L^{\infty}_{sc}(\S)}\\
&+\f{\d^{\f12}}{|u|}\|\K\|_{L^4_{sc}(\S)}\|\nab\phi, f\|_{L^4_{sc}(\S)}+\f{\d^{\f18}}{|u|^{\f12}}\|\K\|^{\f12}_{L^4_{sc}(\S)}\|\nab^2\phi, \nab f\|_{L^4_{sc}(\S)}\\
&+\f{\d^{\f58}}{|u|^{\f32}}\|\K\|^{\f32}_{L^4_{sc}(\S)}\|\phi\|_{L^{\infty}_{sc}(\S)}.\\
\end{split}
\end{equation*}

\end{proposition}
\begin{proof}
When $\phi$ is both symmetric and traceless, 
$\div\phi=f$ implies $\curl\phi=^*f$. 
Using Proposition \ref{ellipticthm}, the desired estimate follows.
\end{proof}

\section {$\mathcal
{O}_{0,4}(u,\underline{u})$ AND $\mathcal
{O}_{1,4}(u,\underline{u})$ ESTIMATES}

In this section, we prove estimates for the Ricci coefficients and their first angular derivatives in $L^4(\S).$

\subsection{$\M O_{0,4}$ Estimates.}
We start with the $L^4(\S)$ estimate for the Ricci coefficients.
We first notice a fact that 
for all null structure equations the Ricci coefficients $\psi$ with signature $(s,s')$ satisfies either transport equation
\begin{equation}\label{tran4}
\nab_4\psi^{(s,s')}=\sum_{s_1+s_2=s+1, \atop s'_1+s'_2=s'}\psi^{(s_1,s'_1)}\cdot\psi^{(s_2,s'_2)}+\Psi^{(s+1,s')},
\end{equation}
or
\begin{equation}\label{tran3}
\nab_3\psi^{(s,s')}+\lambda[\psi^{(s,s')}]\tr\chib\psi^{(s,s')}=\sum_{s_1+s_2=s, \atop s'_1+s'_2=s'+1}\psi^{(s_1,s'_1)}\cdot\psi^{(s_2,s'_2)}+\Psi^{(s,s'+1)}.
\end{equation}

In this section, $\psi^{(s,s')}$ and $\Psi^{(s,s')}$ stand for arbitrary Ricci coefficient with signature $(s,s')$ and null curvature component with signature $(s,s')$, respectively.
And $\lambda[\psi^{(s,s')}]$ is a constant depending on $\psi^{(s,s')}$. 

For Ricci coefficients $\psi$ satisfying (\ref{tran4}), we have 

\begin{proposition}\label{O04 4sc}
 \begin{equation*}
\begin{split}
&\|\psi^{(s,s')}\|_{L^4_{sc}(S_{u,\ub})}\\
\leq& \|\psi^{(s,s')}\|_{L^4_{sc}(S_{u,0})}+\sup_{0\leq\ub'\leq\delta}\sum_{s_1+s_2=s+1, \atop s'_1+s'_2=s'}\frac{\delta^{\frac 12}}{|u|}\|\psi^{(s_1,s'_1)}\|_{L^{\infty}_{sc}(S_{u,\ub'})}\|\psi^{(s_2,s'_2)}\|_{L^4_{sc}(S_{u,\ub'})}\\
&+\|\Psi^{(s+1,s')}\|^{\frac 12}_{L^2_{sc}(H_u^{(0,\ub)})}\|\nab\Psi^{(s+1,s')}\|^{\frac 12}_{L^2_{sc}(H_u^{(0,\ub)})}+\delta^{\frac 14}\|\Psi^{(s+1,s')}\|_{L^2_{sc}(H_u^{(0,\ub)})}.
\end{split}
\end{equation*}
\end{proposition}

\begin{proof}
Using Proposition \ref{transport}, we have  
\begin{equation*}
\begin{split}
&\|\psi^{(s,s')}\|_{L^4_{sc}(S_{u,\ub})}\\
\leq& \|\psi^{(s,s')}\|_{L^4_{sc}(S_{u,0})}+\int_0^{\ub}\delta^{-1}\|\nab_4\psi^{(s,s')}\|_{L^4_{sc}(S_{u,\ub'})}d\ub'\\
\leq& \|\psi^{(s,s')}\|_{L^4_{sc}(S_{u,0})}+\int_0^{\ub}\delta^{-1}\sum_{s_1+s_2=s+1, \atop s'_1+s'_2=s'}\|\psi^{(s_1,s'_1)}\cdot\psi^{(s_2,s'_2)}\|_{L^4_{sc}(S_{u,\ub'})}d\ub'\\
&+\int_0^{\ub}\delta^{-1}\|\Psi^{(s+1,s')}\|_{L^4_{sc}(S_{u,\ub'})}d\ub' \\
\leq& \|\psi^{(s,s')}\|_{L^4_{sc}(S_{u,0})}+\int_0^{\ub}\delta^{-1}\sum_{s_1+s_2=s+1, \atop s'_1+s'_2=s'}\frac{\delta^{\frac 12}}{|u|}\|\psi^{(s_1,s'_1)}\|_{L^{\infty}_{sc}(S_{u,\ub'})}\|\psi^{(s_2,s'_2)}\|_{L^4_{sc}(S_{u,\ub'})}d\ub'\\
&+\|\Psi^{(s+1,s')}\|^{\frac 12}_{L^2_{sc}(H_u^{(0,\ub)})}\|\nab\Psi^{(s+1,s')}\|^{\frac 12}_{L^2_{sc}(H_u^{(0,\ub)})}+\delta^{\frac 14}\|\Psi^{(s+1,s')}\|_{L^2_{sc}(H_u^{(0,\ub)})}\\
\leq& \|\psi^{(s,s')}\|_{L^4_{sc}(S_{u,0})}+\sup_{0\leq \ub' \leq \delta}\sum_{s_1+s_2=s+1, \atop s'_1+s'_2=s'}\frac{\delta^{\frac 12}}{|u|}\|\psi^{(s_1,s'_1)}\|_{L^{\infty}_{sc}(S_{u,\ub'})}\|\psi^{(s_2,s'_2)}\|_{L^4_{sc}(S_{u,\ub'})}\\
&+\|\Psi^{(s+1,s')}\|^{\frac 12}_{L^2_{sc}(H_u^{(0,\ub)})}\|\nab\Psi^{(s+1,s')}\|^{\frac 12}_{L^2_{sc}(H_u^{(0,\ub)})}+\delta^{\frac 14}\|\Psi^{(s+1,s')}\|_{L^2_{sc}(H_u^{(0,\ub)})},
\end{split}
\end{equation*}
where we employ Proposition \ref{L4} and H\"older's inequality.

\end{proof}

As a consequence, we can prove
\begin{proposition}\label{O04 4norm}
 Under the assumptions of Theorem \ref{main.thm} and the bootstrap assumptions \eqref{BA.0}, we have
\begin{equation*}
 \|\eta\|_{L^4_{sc}(\S)}\leq \M R^{\f12}_0 \M R^{\f12}_1+\d^{\f14} \M R_0+\f{\d^{\f12}}{|u|}\D_0 \M O_{0,4},
\end{equation*}

\begin{equation*}
 \|\omb\|_{L^4_{sc}(\S)}\leq \M R^{\f12}_0 \M R^{\f12}_1+\d^{\f14} \M R_0+\f{\d^{\f12}}{|u|}\D_0 \M O_{0,4},
\end{equation*}

\begin{equation*}
  \|\tr\chi\|_{L^4_{sc}(\S)}\leq  \|\tr\chi\|_{L^4_{sc}(S_{u,0})}+\f{\d^{\f14}}{|u|}\D_0[\chih]\M O_{0,4}[\chih]+\f{\d^{\f12}}{|u|}\D_0 \M O_{0,4},
\end{equation*}

\begin{equation*}
 \d^{\f14}\|\chih\|_{L^4_{sc}(\S)}\leq \M R^{\f12}_0[\a] \M R^{\f12}_1[\a]+\M R_0[\a]+\f{\d^{\f34}}{|u|}\D_0 \M O_{0,4}.
\end{equation*}
\end{proposition}

\begin{proof}
The null Ricci coefficients $\eta$ and $\omb$ satisfy
\begin{equation*}
 \nab_4\eta=-\chi\cdot(\eta-\etb)-\b,
\end{equation*}
and
\begin{equation*}
 \nab_4\omb=2\o\omb+\f12\rho+(\eta,\etb)(\eta,\etb).
\end{equation*}
Since there are no anomalous terms, the conclusion is guaranteed by Proposition \ref{O04 4sc}.

The null Ricci coefficient $\tr\chi$ obeys 
\begin{equation*}
\nab_4 \tr\chi=-\f12(\tr\chi)^2-|\chih|^2-2\o \tr\chi.
\end{equation*}
With Proposition \ref{O04 4sc}, we deduce
 \begin{equation*}
\begin{split}
&\|\tr\chi\|_{L^4_{sc}(\S)}\\
\leq& \|\tr\chi\|_{L^4_{sc}(S_{u,0})}+\sup_{0\leq \ub' \leq \delta}\frac{\delta^{\frac 14}}{|u|}\|\chih\|_{L^{\infty}_{sc}(S_{u,\ub'})}\d^{\f14}\|\chih\|_{L^4_{sc}(S_{u,\ub'})}\\
&+\sup_{0\leq \ub' \leq \delta}\frac{\delta^{\frac 12}}{|u|}\|\tr\chi,\o\|_{L^{\infty}_{sc}(S_{u,\ub'})}\|\tr\chi\|_{L^4_{sc}(S_{u,\ub'})}\\
\leq& \|\tr\chi\|_{L^4_{sc}(S_{u,0})}+\f{\d^{\f14}}{|u|}\D_0[\chih]\M O_{0,4}[\chih]+\f{\d^{\f12}}{|u|}\D_0 \M O_{0,4}.
\end{split}
\end{equation*}

For $\chih$, we have 
\begin{equation*}
 \nab_4\chih+\tr\chi\chih=-2\o\chih-\a.
\end{equation*}
Proposition \ref{O04 4sc} implies
 \begin{equation*}
\begin{split}
&\d^{\f14}\|\chih\|_{L^4_{sc}(\S)}\\
\leq& \sup_{\ub'\in [0,\d]}\frac{\delta^{\frac 34}}{|u|}\|\chih\|_{L^{\infty}_{sc}(S_{u,\ub'})}\|\tr\chi,\o\|_{L^4_{sc}(S_{u,\ub'})}\\
&+\d^{\f14}\|\a\|^{\frac 12}_{L^2_{sc}(H_u^{(0,\ub)})}\|\nab\a\|^{\frac 12}_{L^2_{sc}(H_u^{(0,\ub)})}+\delta^{\frac 12}\|\a\|_{L^2_{sc}(H_u^{(0,\ub)})}\\
\leq&\M R^{\f12}_0[\a] \M R^{\f12}_1[\a]+\M R_0[\a]+\f{\d^{\f34}}{|u|}\D_0 \M O_{0,4}.
\end{split}
\end{equation*}
\end{proof}

All the other Ricci coefficients $\psi^{(s,s')}$ satisfy
\begin{equation*}
\nab_3\psi^{(s,s')}+\lambda[\psi^{(s,s')}]\tr\chib\psi^{(s,s')}=\sum_{s_1+s_2=s, \atop s'_1+s'_2=s'+1}\psi^{(s_1,s'_1)}\cdot\psi^{(s_2,s'_2)}+\Psi^{(s,s'+1)},
\end{equation*}
and for these $\psi^{(s,s')}$ we have
\begin{proposition}
 \begin{equation*}\label{O04 3}
\begin{split}
&|u|^{2\lambda-2s'-1}\|\psi^{(s,s')}\|_{L^4_{sc}(S_{u,\ub})}\\
\leq& |u_{\infty}|^{2\lambda-2s'-1}\|\psi^{(s,s')}\|_{L^4_{sc}(S_{u_{\infty},\ub})}+\int_{u_{\infty}}^{u}|u'|^{2\lambda-2s'-3}\|\Psi^{(s,s'+1)}\|_{L^4_{sc}(S_{u',\ub})}du'\\
&+\int_{u_{\infty}}^{u}|u'|^{2\lambda-2s'-3}\sum_{s_1+s_2=s, \atop s'_1+s'_2=s'+1}\|\psi^{(s_1,s'_1)}\cdot\psi^{(s_2,s'_2)}\|_{L^4_{sc}(S_{u',\ub})}du'.
\end{split}
\end{equation*}
\end{proposition}

\begin{proof}
With the help of Proposition \ref{evolution lemma}, for
$$p=4,\lambda_1=2\lambda-\frac 12,$$
we derive
\begin{equation*}
\begin{split}
&|u|^{2\lambda-\frac 12}\|\psi^{(s,s')}\|_{L^4(S_{u,\ub})}\\
\leq& |u_{\infty}|^{2\lambda-\frac 12}\|\psi^{(s,s')}\|_{L^4(S{u_{\infty},\ub})}+\int_{u_{\infty}}^{u}|u'|^{2\lambda-\frac 12}\|\nab_3\psi^{(s,s')}\|_{L^4(S_{u',\ub})}du'\\
\leq& |u_{\infty}|^{2\lambda-\frac 12}\|\psi^{(s,s')}\|_{L^4(S{u_{\infty},\ub})}+\int_{u_{\infty}}^{u}|u'|^{2\lambda-\frac 12}\|\Psi^{(s,s'+1)}\|_{L^4(S_{u',\ub})}du'\\
&+\int_{u_{\infty}}^{u}|u'|^{2\lambda-\frac12}\sum_{s_1+s_2=s, \atop s'_1+s'_2=s'+1}\|\psi^{(s_1,s'_1)}\cdot\psi^{(s_2,s'_2)}\|_{L^4(S_{u',\ub})}du'.
\end{split}
\end{equation*}
In views of the definition
\begin{equation*}
\|\psi^{(s,s')}\|_{L^4_{sc}(S_{u,\ub})}=\d^{s-\f34}|u|^{2s'+\f12}\|\psi^{(s,s')}\|_{L^4(\S)}, 
\end{equation*}
we rewrite the estimate above in scale invariant norms:
\begin{equation*}\label{4.11}
\begin{split}
&|u|^{2\lambda-2s'-1}\|\psi^{(s,s')}\|_{L^4_{sc}(S_{u,\ub})}\\
\leq& |u_{\infty}|^{2\lambda-2s'-1}\|\psi^{(s,s')}\|_{L^4_{sc}(S_{u_{\infty},\ub})}+\int_{u_{\infty}}^{u}|u'|^{2\lambda-2s'-3}\|\Psi^{(s,s'+1)}\|_{L^4_{sc}(S_{u',\ub})}du'\\
&+\int_{u_{\infty}}^{u}|u'|^{2\lambda-2s'-3}\sum_{s_1+s_2=s, \atop s'_1+s'_2=s'+1}\|\psi^{(s_1,s'_1)}\cdot\psi^{(s_2,s'_2)}\|_{L^4_{sc}(S_{u',\ub})}du'.
\end{split}
\end{equation*}
\end{proof}

Furthermore, it is observed that for $\psi^{(s,s')}\in \{\o,\chibh,\etb, \tr\chib\}$ in the null structure equations $\nab_3\psi^{(s,s')}$, we have $\lambda[\psi^{(s,s')}]\leq s'$. With this observation, we derive
\begin{proposition}
Under the assumption $\lambda \leq s'$, for $\psi^{(s,s')}$ satisfying equation (\ref{tran3}), we have
\begin{equation*}\label{O04 3sc1}
\begin{split}
&\|\psi^{(s,s')}\|_{L^4_{sc}(S_{u,\ub})}\\
\leq&\|\psi^{(s,s')}\|_{L^4_{sc}(S_{u_{\infty},\ub})}+\f{1}{|u|^{\f12}}\|\Psi^{(s,s'+1)}\|^{\f12}_{L^2_{sc}(\Hb_{\ub}^{(u_{\infty,u})})}\|\nab\Psi^{(s,s'+1)}\|^{\f12}_{L^2_{sc}(\Hb_{\ub}^{(u_{\infty,u})})}\\
&+\f{\d^{\f14}}{|u|^{\f12}}\|\Psi^{(s,s'+1)}\|_{L^2_{sc}(\Hb_{\ub}^{(u_{\infty,u})})}+\int_{u_{\infty}}^{u}|u'|^{-2}\sum_{s_1+s_2=s, \atop s'_1+s'_2=s'+1}\|\psi^{(s_1,s'_1)}\cdot\psi^{(s_2,s'_2)}\|_{L^4_{sc}(S_{u',\ub})}du',
\end{split}
\end{equation*}
or
\begin{equation*}\label{O04 3sc2}
\begin{split}
&\f{1}{|u|}\|\psi^{(s,s')}\|_{L^4_{sc}(S_{u,\ub})}\\
\leq&\f{1}{|u_{\infty}|}\|\psi^{(s,s')}\|_{L^4_{sc}(S_{u_{\infty},\ub})}+\f{1}{|u|^{\f32}}\|\Psi^{(s,s'+1)}\|^{\f12}_{L^2_{sc}(\Hb_{\ub}^{(u_{\infty,u})})}\|\nab\Psi^{(s,s'+1)}\|^{\f12}_{L^2_{sc}(\Hb_{\ub}^{(u_{\infty,u})})}\\
&+\f{\d^{\f14}}{|u|^{\f32}}\|\Psi^{(s,s'+1)}\|_{L^2_{sc}(\Hb_{\ub}^{(u_{\infty,u})})}+\int_{u_{\infty}}^{u}|u'|^{-3}\sum_{s_1+s_2=s, \atop s'_1+s'_2=s'+1}\|\psi^{(s_1,s'_1)}\cdot\psi^{(s_2,s'_2)}\|_{L^4_{sc}(S_{u',\ub})}du'.
\end{split}
\end{equation*}
\end{proposition}

\begin{proof}
With the aid of assumption $\lambda[\psi^{(s,s')}]\leq s'$,
we multiply $|u|^{-2\lambda+2s'+1}$ on both sides of Proposition \ref{O04 3}. Employing the fact $1\leq |u| \leq |u'| \leq |u_{\infty}|$, we conclude
\begin{equation*}
\begin{split}
&\|\psi^{(s,s')}\|_{L^4_{sc}(S_{u,\ub})}\\
\leq&\|\psi^{(s,s')}\|_{L^4_{sc}(S_{u_{\infty},\ub})}+\int_{u_{\infty}}^{u}|u'|^{-2}\|\Psi^{(s,s'+1)}\|_{L^4_{sc}(S_{u',\ub})}du'\\
&+\int_{u_{\infty}}^{u}|u'|^{-2}\sum_{s_1+s_2=s, \atop s'_1+s'_2=s'+1}\|\psi^{(s_1,s'_1)}\cdot\psi^{(s_2,s'_2)}\|_{L^4_{sc}(S_{u',\ub})}du'\\
\leq&\|\psi^{(s,s')}\|_{L^4_{sc}(S_{u_{\infty},\ub})}+\int_{u_{\infty}}^{u}|u'|^{-2}\|\Psi^{(s,s'+1)}\|^{\f12}_{L^2_{sc}(S_{u',\ub})}\|\nab\Psi^{(s,s'+1)}\|^{\f12}_{L^2_{sc}(S_{u',\ub})}du'\\
&+\int_{u_{\infty}}^{u}|u'|^{-2}\d^{\f14}\|\Psi^{(s,s'+1)}\|_{L^2_{sc}(S_{u',\ub})}du'\\
&+\int_{u_{\infty}}^{u}|u'|^{-2}\sum_{s_1+s_2=s, \atop s'_1+s'_2=s'+1}\|\psi^{(s_1,s'_1)}\cdot\psi^{(s_2,s'_2)}\|_{L^4_{sc}(S_{u',\ub})}du'\\
\leq&\|\psi^{(s,s')}\|_{L^4_{sc}(S_{u_{\infty},\ub})}+\f{1}{|u|^{\f12}}\|\Psi^{(s,s'+1)}\|^{\f12}_{L^2_{sc}(\Hb_{\ub}^{(u_{\infty,u})})}\|\nab\Psi^{(s,s'+1)}\|^{\f12}_{L^2_{sc}(\Hb_{\ub}^{(u_{\infty,u})})}\\
&+\f{\d^{\f14}}{|u|^{\f12}}\|\Psi^{(s,s'+1)}\|_{L^2_{sc}(\Hb_{\ub}^{(u_{\infty,u})})}+\int_{u_{\infty}}^{u}|u'|^{-2}\sum_{s_1+s_2=s, \atop s'_1+s'_2=s'+1}\|\psi^{(s_1,s'_1)}\cdot\psi^{(s_2,s'_2)}\|_{L^4_{sc}(S_{u',\ub})}du',
\end{split}
\end{equation*}
where we employ Proposition \ref{L4} and H\"older's inequality.

Similarly, if we multiply $|u|^{-2\lambda+2s'}$ on both sides of Proposition \ref{O04 3}, we derive
\begin{equation*}
\begin{split}
&\f{1}{|u|}\|\psi^{(s,s')}\|_{L^4_{sc}(S_{u,\ub})}\\
\leq&\f{1}{|u_{\infty}|}\|\psi^{(s,s')}\|_{L^4_{sc}(S_{u_{\infty},\ub})}+\f{1}{|u|^{\f32}}\|\Psi^{(s,s'+1)}\|^{\f12}_{L^2_{sc}(\Hb_{\ub}^{(u_{\infty,u})})}\|\nab\Psi^{(s,s'+1)}\|^{\f12}_{L^2_{sc}(\Hb_{\ub}^{(u_{\infty,u})})}\\
&+\f{\d^{\f14}}{|u|^{\f32}}\|\Psi^{(s,s'+1)}\|_{L^2_{sc}(\Hb_{\ub}^{(u_{\infty,u})})}+\int_{u_{\infty}}^{u}|u'|^{-3}\sum_{s_1+s_2=s, \atop s'_1+s'_2=s'+1}\|\psi^{(s_1,s'_1)}\cdot\psi^{(s_2,s'_2)}\|_{L^4_{sc}(S_{u',\ub})}du'.
\end{split}
\end{equation*}
\end{proof}

Consequently, we obtain
\begin{proposition}\label{O04 3norm}
 Under the assumptions of Theorem \ref{main.thm} and bootstrap assumption \eqref{BA.0}, the following inequalities hold
\begin{equation*}
 \|\o\|_{L^4_{sc}(\S)}\leq \|\o\|_{L^4_{sc}(S_{u_{\infty},\ub})}+\f{1}{|u|^{\f12}}\underline{\M R}^{\f12}_0 \underline{\M R}^{\f12}_1+\f{\d^{\f14}}{|u|^{\f12}} \underline{\M R}_0+\f{\d^{\f12}}{|u|^2}\D_0 \M O_{0,4},
\end{equation*}

\begin{equation*}
 \f{\d^{\f14}}{|u|}\|\chibh\|_{L^4_{sc}(\S)}\leq \f{\d^{\f14}}{|u_{\infty}|}\|\chibh\|_{L^4_{sc}(S_{u_{\infty},\ub})}+\f{\d^{\f14}}{|u|^{\f32}}\underline{\M R}^{\f12}_0 \underline{\M R}^{\f12}_1+\f{\d^{\f12}}{|u|^{\f32}} \underline{\M R}_0+\f{\d^{\f34}}{|u|^2}\D_0 \M O_{0,4},
\end{equation*}

\begin{equation*}
 \|\etb\|_{L^4_{sc}(\S)}\leq \|\etb\|_{L^4_{sc}(S_{u_{\infty},\ub})}+\M R^{\f12}_0 \M R^{\f12}_1+\d^{\f14} \M R_0+\f{1}{|u|^{\f12}}\underline{\M R}^{\f12}_0 \underline{\M R}^{\f12}_1+\f{\d^{\f14}}{|u|^{\f12}} \underline{\M R}_0+\f{\d^{\f12}}{|u|}\D_0 \M O_{0,4},
\end{equation*}

\begin{equation*}
\f{\d^{\f34}}{|u|^2}\|\tr\chib\|_{L^4_{sc}(\S)}\leq \f{\d^{\f34}}{|u_{\infty}|^2}\|\tr\chib\|_{L^4_{sc}(\S)}+\f{\d^{\f34}}{|u|}\D_0 \M O_{0,4}.
\end{equation*}
\end{proposition}

\begin{proof}
The null Ricci coefficient $\o$ satisfies
 \begin{equation*}
 \nab_3\o=2\o\omb+\f12\rho+(\eta,\etb)(\eta,\etb).
 \end{equation*}
Since no anomalous term appears in the equation above. The conclusion follows by adopting Proposition \ref{O04 3sc1}.

The null Ricci coefficient $\chibh$ obeys
\begin{equation*}
 \nab_3\chibh+\tr\chib\,\chibh=-2\omb\chibh-\ab.
\end{equation*}
With Proposition \ref{O04 3sc2}, we derive 
\begin{equation*}
\begin{split}
&\f{1}{|u|}\|\chibh\|_{L^4_{sc}(S_{u,\ub})}\\
\leq&\f{1}{|u_{\infty}|}\|\chibh\|_{L^4_{sc}(S_{u_{\infty},\ub})}+\f{1}{|u|^{\f32}}\|\ab\|^{\f12}_{L^2_{sc}(\Hb_{\ub}^{(u_{\infty,u})})}\|\ab\|^{\f12}_{L^2_{sc}(\Hb_{\ub}^{(u_{\infty,u})})}\\
&+\f{\d^{\f14}}{|u|^{\f32}}\|\ab\|_{L^2_{sc}(\Hb_{\ub}^{(u_{\infty,u})})}+\int_{u_{\infty}}^{u}|u'|^{-3}\|\omb\chibh\|_{L^4_{sc}(S_{u',\ub})}du'\\
\leq&\f{1}{|u_{\infty}|}\|\chibh\|_{L^4_{sc}(S_{u_{\infty},\ub})}+\f{1}{|u|^{\f32}}\|\ab\|^{\f12}_{L^2_{sc}(\Hb_{\ub}^{(u_{\infty,u})})}\|\ab\|^{\f12}_{L^2_{sc}(\Hb_{\ub}^{(u_{\infty,u})})}\\
&+\f{\d^{\f14}}{|u|^{\f32}}\|\ab\|_{L^2_{sc}(\Hb_{\ub}^{(u_{\infty,u})})}+\int_{u_{\infty}}^{u}\f{\d^{\f12}}{|u'|^{3}}\|\omb\|_{L^4_{sc}(S_{u',\ub})}\f{1}{|u'|}\|\chibh\|_{L^{\infty}_{sc}(S_{u',\ub})}du'\\
\leq&\f{1}{|u_{\infty}|}\|\chibh\|_{L^4_{sc}(S_{u_{\infty},\ub})}+\f{1}{|u|^{\f32}}\|\ab\|^{\f12}_{L^2_{sc}(\Hb_{\ub}^{(u_{\infty,u})})}\|\nab\ab\|^{\f12}_{L^2_{sc}(\Hb_{\ub}^{(u_{\infty,u})})}\\
&+\f{\d^{\f14}}{|u|^{\f32}}\|\ab\|_{L^2_{sc}(\Hb_{\ub}^{(u_{\infty,u})})}+\f{\d^{\f12}}{|u|^{2}}\D_0[\chibh]\M O_{0,4}[\omb]du'.\\
\end{split}
\end{equation*}
Multiplying $\d^{\f14}$ on both sides, we obtain the desired estimate.

For $\etb$, we have null structure equation
\begin{equation*}
 \nab_3\etb+\f12 \tr\chib \, \etb= \f12 \tr\chib \eta-\chibh\cdot(\eta-\etb)+\beb.
\end{equation*}
There is a borderline term $\tr\chib\etb$. With Proposition \ref{evolution lemma} 
and $$\lambda[\etb]=s_2(\etb)=\f12,$$ we obtain
\begin{equation*}
\begin{split}
&|u|^{-1}\|\etb\|_{L^4_{sc}(S_{u,\ub})}\\
\leq& |u_{\infty}|^{-1}\|\etb\|_{L^4_{sc}(S_{u_{\infty},\ub})}+\int_{u_{\infty}}^{u}|u'|^{-3}\|\beb\|_{L^4_{sc}(S_{u',\ub})}du'\\
&+\int_{u_{\infty}}^{u}|u'|^{-3}\|\chibh\cdot(\eta-\etb)\|_{L^4_{sc}(S_{u',\ub})}du'\\
&+\int_{u_{\infty}}^{u}|u'|^{-3}\|\tr\chib\eta\|_{L^4_{sc}(S_{u',\ub})}du'\\
\leq& |u_{\infty}|^{-1}\|\etb\|_{L^4_{sc}(S_{u_{\infty},\ub})}+\f{1}{|u|^{\f32}}\|\beb\|^{\f12}_{L^2_{sc}(\Hb_{\ub}^{(u_{\infty,u})})}\|\nab\beb\|^{\f12}_{L^2_{sc}(\Hb_{\ub}^{(u_{\infty,u})})}\\
&+\f{\d^{\f14}}{|u|^{\f32}}\|\beb\|_{L^2_{sc}(\Hb_{\ub}^{(u_{\infty,u})})}+\f{\d^{\f12}}{|u|^2}\D_0[\chibh]\M O_{0,4}[\eta,\etb]+\f{1}{|u|}\M O_{0,4}[\eta].
\end{split}
\end{equation*}
Multiply $|u|$ on both sides, we deduce 
\begin{equation*}
 \|\etb\|_{L^4_{sc}(\S)}\leq \|\etb\|_{L^4_{sc}(S_{u_{\infty},\ub})}+\M O_{0,4}(\eta)+\f{1}{|u|^{\f12}}\underline{\M R}^{\f12}_0 \underline{\M R}^{\f12}_1+\f{\d^{\f14}}{|u|^{\f12}} \underline{\M R}_0+\f{\d^{\f12}}{|u|}\D_0 \M O_{0,4}.
\end{equation*}
In Proposition \ref{O04 4norm}, we have proved
\begin{equation*}
 \|\eta\|_{L^4_{sc}(\S)}\leq \M R^{\f12}_0 \M R^{\f12}_1+\d^{\f14} \M R_0+\f{\d^{\f12}}{|u|}\D_0 \M O_{0,4}.
\end{equation*}
Putting these together, we get
\begin{equation*}
 \|\etb\|_{L^4_{sc}(\S)}\leq \|\etb\|_{L^4_{sc}(S_{u_{\infty},\ub})}+\M R^{\f12}_0 \M R^{\f12}_1+\d^{\f14} \M R_0+\f{1}{|u|^{\f12}}\underline{\M R}^{\f12}_0 \underline{\M R}^{\f12}_1+\f{\d^{\f14}}{|u|^{\f12}} \underline{\M R}_0+\f{\d^{\f12}}{|u|}\D_0 \M O_{0,4}.
\end{equation*}

The null Ricci coefficient $\tr\chib$ obeys
\begin{equation*}
 \nab_3 \tr\chib+\f12 \tr\chib \tr\chib=-2\omb \tr\chib-|\chibh|^2.
\end{equation*}
With Proposition \ref{evolution lemma} and $$\lambda[\tr\chib]=\f12, \quad \quad s_2(\tr\chib)=1,$$ we derive
\begin{equation*}
\begin{split}
&|u|^{-2}\|\tr\chib\|_{L^4_{sc}(S_{u,\ub})}\\
\leq& |u_{\infty}|^{-2}\|\tr\chib\|_{L^4_{sc}(S_{u_{\infty},\ub})}+\int_{u_{\infty}}^{u}|u'|^{-4}\|\chibh\,\chibh\|_{L^4_{sc}(S_{u',\ub})}du'\\
&+\int_{u_{\infty}}^{u}|u'|^{-4}\|\tr\chib\omb\|_{L^4_{sc}(S_{u',\ub})}du'\\
\leq& |u_{\infty}|^{-2}\|\tr\chib\|_{L^4_{sc}(S_{u_{\infty},\ub})}+\int_{u_{\infty}}^{u}\f{\d^{\f14}}{|u'|^{2}}\f{1}{|u'|}\|\chibh\|_{L^{\infty}_{sc}(S_{u',\ub})}\f{\d^{\f14}}{|u'|}\|\chibh\|_{L^{4}_{sc}(S_{u',\ub})}du'\\
&+\int_{u_{\infty}}^{u}|u'|^{-3}\f{\d^{\f12}}{|u'|^2}\|\tr\chib\|_{L^{\infty}_{sc}(S_{u',\ub})}\|\omb\|_{L^4_{sc}(S_{u',\ub})}du'\\
\leq& |u_{\infty}|^{-2}\|\tr\chib\|_{L^4_{sc}(S_{u_{\infty},\ub})}+\f{\d^{\f14}}{|u|}\D_0[\chibh]\M O_{0,4}[\chibh]+\f{1}{|u|^2}\D_0[\tr\chib]\M O_{0,4}[\omb].
\end{split}
\end{equation*}
Multiplying $\d^{\f34}$ on both sides, we finish the proof. 
\end{proof}

Collecting all the estimates in this section, we conclude
\begin{proposition}\label{O04}
Assume
$$\M R<\infty, \quad \quad \underline{\M R}<\infty, \quad \quad \M O_{0,\infty}<\infty,$$
then there exist $\d_0=\d_0(\M I^{(0)}, \M R^{(0)}, \M R, \underline{\M R}, \M O_{0,\infty})$, such that whenever $\d\leq\d_0$, in $D$ we have
\begin{equation*}
\M O_{0,4}\leq C\l \M I^{(0)}+\M R+\underline{\M R}\r,
\end{equation*}
where $C$ is a large universal constant.
\end{proposition}
\begin{proof}
As a consequence of the estimates above, by taking $\d_0$ sufficient small we derive
 \begin{equation*}
\M O_{0,4}\leq \M I^{(0)}+\M R_0^{\f12}\M R_1^{\f12}+\M R_0+\f{1}{|u|^{\f12}}\underline{\M R}_0^{\f12}\underline{\M R}_1^{\f12}+\f{\d^{\f14}}{|u|}\underline{\M R}_0.
\end{equation*}
By setting 
$$\D_1\gg \M I^{(0)}+\M R_0^{\f12}\M R_1^{\f12}+\M R_0+\f{1}{|u|^{\f12}}\underline{\M R}_0^{\f12}\underline{\M R}_1^{\f12}+\f{\d^{\f14}}{|u|}\underline{\M R}_0,$$
we have improved bootstrap assumption \eqref{BA.1}. 
\end{proof}

\subsection{$\M O_{1,4}$ Estimates.}
We now move to the $L^4(\S)$ estimate for first angular derivatives of the Ricci coefficients.
With the help of Propositions \ref{commute0}, \ref{commute1} and \ref{commute2}, it is easy to verify that
the null Ricci coefficient $\psi$ with signature $(s,s')$ satisfies either transport equation
\begin{equation}\label{tran44}
\begin{split}
&\nab_4\nab\psi^{(s,s')}\\
=&\sum_{s_1+s_2=s+1, \atop s'_1+s'_2=s'}\nab\psi^{(s_1,s'_1)}\cdot\psi^{(s_2,s'_2)}+\nab\Psi^{(s+1,s')}\\
&+\sum_{s_3+s_4=s+\f32, \atop s'_3+s'_4=s'+\f12}\psi^{(s_3,s'_3)}\cdot\Psi^{(s_4,s'_4)}+\sum_{s_5+s_6+s_7=s+\f32, \atop s'_5+s'_6+s'_7=s'+\f12}\psi^{(s_5, s'_5)}\cdot\psi^{(s_6, s'_6)}\cdot\psi^{(s_7, s'_7)},
\end{split}
\end{equation}
or
\begin{equation}\label{tran33}
\begin{split}
&\nab_3\nab\psi^{(s,s')}+\l\lambda[\psi^{(s,s')}]+\f12\r \tr\chib\nab\psi^{(s,s')}\\
=&\sum_{s_1+s_2=s, \atop s'_1+s'_2=s'+1}\nab\psi^{(s_1,s'_1)}\cdot\psi^{(s_2,s'_2)}+\nab\Psi^{(s,s'+1)}\\
&+\sum_{s_3+s_4=s+\f12, \atop s'_3+s'_4=s'+\f32}\psi^{(s_3,s'_3)}\cdot\Psi^{(s_4,s'_4)}+\sum_{s_5+s_6+s_7=s+\f12, \atop s'_5+s'_6+s'_7=s'+\f32}\psi^{(s_5, s'_5)}\cdot\psi^{(s_6, s'_6)}\cdot\psi^{(s_7, s'_7)}.
\end{split}
\end{equation}

As a consequence, we have
\begin{proposition}\label{O14 4sc}
For $\psi^{(s,s')}$ satisfying (\ref{tran44}), the following inequality holds
\end{proposition}
\begin{equation*}
\begin{split}
&\|\nab\psi^{(s,s')}\|_{L^4_{sc}(S_{u,\ub})}\\
\leq& \|\nab\psi^{(s,s')}\|_{L^4_{sc}(S_{u,0})}+\sup_{0\leq u'\leq\delta}\sum_{s_1+s_2=s+1, \atop s'_1+s'_2=s'}\frac{\delta^{\frac 12}}{|u|}\|\nab\psi^{(s_1,s'_1)}\|_{L^{4}_{sc}(S_{u,\ub'})}\|\psi^{(s_2,s'_2)}\|_{L^{\infty}_{sc}(S_{u,\ub'})}\\
&+\|\nab\Psi^{(s+1,s')}\|^{\frac 12}_{L^2_{sc}(H_u^{(0,\ub)})}\|\nab^2\Psi^{(s+1,s')}\|^{\frac 12}_{L^2_{sc}(H_u^{(0,\ub)})}+\delta^{\frac 14}\|\nab\Psi^{(s+1,s')}\|_{L^2_{sc}(H_u^{(0,\ub)})}\\
&+\f{\d^{\f14}}{|u|}\sup_{0\leq u'\leq\delta}\sum_{s_3+s_4=s+\f32, \atop s'_3+s'_4=s'+\f12}\|\psi^{(s_3,s'_3)}\|_{L_{sc}^{\infty}(S_{u,\ub'})}\d^{\f14}\|\Psi^{(s_4,s'_4)}\|^{\f12}_{L^2_{sc}(H_{u}^{(0,\ub)})}\|\nab\Psi^{(s_4,s'_4)}\|^{\f12}_{L^2_{sc}(H_{u}^{(0,\ub)})}\\
&+\f{\d^{\f14}}{|u|}\sup_{0\leq u'\leq\d}\sum_{s_3+s_4=s+\f32, \atop s'_3+s'_4=s'+\f12}\|\psi^{(s_3,s'_3)}\|_{L_{sc}^{\infty}(S_{u,\ub'})}\d^{\f12}\|\Psi^{(s_4,s'_4)}\|_{L^2_{sc}(H_{u}^{(0,\ub)})}\\
&+\f{\d}{|u|^2}\sup_{0\leq u'\leq \d}\sum_{s_5+s_6+s_7=s+\f32, \atop s'_5+s'_6+s'_7=s'+\f12}\|\psi^{(s_5,s'_5)}\|_{L^{\infty}_{sc}(S_{u,\ub'})}\|\psi^{(s_6,s'_6)}\|_{L^{\infty}_{sc}(S_{u,\ub'})}\|\psi^{(s_7, s'_{7})}\|_{L^4_{sc}(S_{u,\ub'})}.\\
\end{split}
\end{equation*}

\begin{proof}
Adopting Proposition \ref{transport}, we obtain  
\begin{equation}
\begin{split}
&\|\nab\psi^{(s,s')}\|_{L^4_{sc}(S_{u,\ub})}\\
\leq& \|\nab\psi^{(s,s')}\|_{L^4_{sc}(S_{u,0})}+\int_0^{\ub}\delta^{-1}\|\nab_4\nab\psi^{(s,s')}\|_{L^4_{sc}(S_{u,\ub'})}d\ub'\\
\leq& \|\nab\psi^{(s,s')}\|_{L^4_{sc}(S_{u,0})}+\sup_{0\leq u'\leq\delta}\sum_{s_1+s_2=s+1, \atop s'_1+s'_2=s'}\|\nab\psi^{(s_1,s'_1)}\cdot\psi^{(s_2,s'_2)}\|_{L^4_{sc}(S_{u,\ub'})}\\
&+\int_0^{\ub}\delta^{-1}\|\nab\Psi^{(s+1,s')}\|_{L^4_{sc}(S_{u,\ub'})}d\ub'\\
&+\int_0^{\ub}\delta^{-1}\sum_{s_3+s_4=s+\f32, \atop s'_3+s'_4=s'+\f12}\|\psi^{(s_3,s'_3)}\cdot\Psi^{(s_4,s'_4)}\|_{L^4_{sc}(S_{u,\ub'})}d\ub'\\
&+\int_0^{\ub}\delta^{-1}\sum_{s_5+s_6+s_7=s+\f32, \atop s'_5+s'_6+s'_7=s'+\f12}\|\psi^{(s_5,s'_5)}\cdot\psi^{(s_6,s'_6)}\cdot\psi^{(s_7, s'_7)}\|_{L^4_{sc}(S_{u,\ub'})}d\ub'\\
\leq& \|\nab\psi^{(s,s')}\|_{L^4_{sc}(S_{u,0})}+\sup_{0\leq u'\leq\delta}\sum_{s_1+s_2=s+1, \atop s'_1+s'_2=s'}\frac{\delta^{\frac 12}}{|u|}\|\nab\psi^{(s_1,s'_1)}\|_{L^{4}_{sc}(S_{u,\ub'})}\|\psi^{(s_2,s'_2)}\|_{L^{\infty}_{sc}(S_{u,\ub'})}\\
&+\|\nab\Psi^{(s+1,s')}\|^{\frac 12}_{L^2_{sc}(H_u^{(0,\ub)})}\|\nab^2\Psi^{(s+1,s')}\|^{\frac 12}_{L^2_{sc}(H_u^{(0,\ub)})}+\delta^{\frac 14}\|\nab\Psi^{(s+1,s')}\|_{L^2_{sc}(H_u^{(0,\ub)})}\\
&+\f{\d^{\f14}}{|u|}\sup_{0\leq u'\leq\delta}\sum_{s_3+s_4=s+\f32, \atop s'_3+s'_4=s'+\f12}\|\psi^{(s_3,s'_3)}\|_{L_{sc}^{\infty}(S_{u,\ub'})}\d^{\f14}\|\Psi^{(s_4,s'_4)}\|^{\f12}_{L^2_{sc}(H_{u}^{(0,\ub)})}\|\nab\Psi^{(s_4,s'_4)}\|^{\f12}_{L^2_{sc}(H_{u}^{(0,\ub)})}\\
&+\f{\d^{\f14}}{|u|}\sup_{0\leq u'\leq\d}\sum_{s_3+s_4=s+\f32, \atop s'_3+s'_4=s'+\f12}\|\psi^{(s_3,s'_3)}\|_{L_{sc}^{\infty}(S_{u,\ub'})}\d^{\f12}\|\Psi^{(s_4,s'_4)}\|_{L^2_{sc}(H_{u}^{(0,\ub)})}\\
&+\f{\d}{|u|^2}\sup_{0\leq u'\leq \d}\sum_{s_5+s_6+s_7=s+\f32, \atop s'_5+s'_6+s'_7=s'+\f12}\|\psi^{(s_5,s'_5)}\|_{L^{\infty}_{sc}(S_{u,\ub'})}\|\psi^{(s_6,s'_6)}\|_{L^{\infty}_{sc}(S_{u,\ub'})}\|\psi^{(s_7, s'_7)}\|_{L^4_{sc}(S_{u,\ub'})},\\
\end{split}
\end{equation}
where we employ Proposition \ref{L4} and H\"older's inequality. 
\end{proof}

With this preparation, it follows that

\begin{proposition}\label{O14 4norm}
 Under the assumptions of Theorem \ref{main.thm} and the bootstrap assumptions \eqref{BA.0} and \eqref{BA.1}, we have
\begin{equation*}
\begin{split}
 \|\nab\eta, \nab\omb, \nab \tr \chi\|_{L^4_{sc}(\S)}\leq &\M R_1^{\f12} \M R_2^{\f12}+\d^{\f14}\M R_1+\f{\d^{\f12}}{|u|}\D_0 \M R_0^{\f12} \M R_1^{\f12}+\f{\d^{\f34}}{|u|}\D_0 \M R_0\\
&+\f{\d^{\f12}}{|u|}\D_0 \M O_{1,4}+\f{\d}{|u|}\D_0^2 \M O_{0,4},
\end{split}
\end{equation*}



and

\begin{equation*}
\begin{split}
\|\nab\chih\|_{L^4_{sc}(\S)}\leq &\M R_1^{\f12} \M R_2^{\f12}+\d^{\f14}\M R_1+\f{\d^{\f14}}{|u|}\D_0 \M R_0^{\f12} \M R_1^{\f12}+\f{\d^{\f14}}{|u|}\D_0 \M R_0\\
&+\f{\d^{\f12}}{|u|}\D_0 \M O_{1,4}+\f{\d}{|u|}\D_0^2 \M O_{0,4}.
\end{split}
\end{equation*}
\end{proposition}

\begin{proof}
The null Ricci coefficients $\nab\eta, \nab\omb, \nab \tr\chi$ satisfy 

\begin{equation*}
\begin{split}
\nab_4\nab\eta=\nab\b+(\eta,\etb)\b+(\eta,\etb)\nab\chi+\chi(\nab\eta,\nab\etb)+\chi(\eta,\etb)(\eta,\etb),
\end{split}
\end{equation*}

\begin{equation*}
\begin{split}
\nab_4\nab\omb=&\nab\rho+(\eta,\etb)\rho+\omb\nab\o+(\chi,\o)\nab\omb+(\eta,\etb)(\nab\eta,\nab\etb)\\
&+(\eta,\etb)\o\omb+(\eta,\etb)(\eta,\etb)(\eta,\etb),
\end{split}
\end{equation*}

and
\begin{equation*}
\begin{split}
\nab_4\nab \tr\chi=(\o,\chi)\nab \tr\chi+\chih\nab\chih+\tr\chi\nab\o+(\eta,\etb)(\o,\chi)(\o,\chi).
\end{split}
\end{equation*}
There is no anomalous term in these equations. The conclusion follows from Proposition \ref{O14 4sc}.

For $\chih$, we have
\begin{equation*}
\begin{split}
\nab_4\nab \chih=&\nab\a+(\eta,\etb)\a+\chih\b+\chih\nab\tr\chi+(\o,\chi)\nab\chih+\chih\nab\o\\
&+(\eta,\etb)(\chi,\o)\chih.
\end{split}
\end{equation*}
The anomalous term is due to $\a$. And the terms including $\nab\a$ are normal. Recalling that
$$\d^{\f12}\|\a\|_{L^2_{sc}(H)}\leq \M R_0,$$
we obtain

\begin{equation*}
\begin{split}
\|\nab\chih\|_{L^4_{sc}(\S)}\leq &\M R_1^{\f12} \M R_2^{\f12}+\d^{\f14}\M R_1+\f{\d^{\f14}}{|u|}\D_0 \M R_0^{\f12} \M R_1^{\f12}+\f{\d^{\f14}}{|u|}\D_0 \M R_0\\
&+\f{\d^{\f12}}{|u|}\D_0 \M O_{1,4}+\f{\d}{|u|}\D_0^2 \M O_{0,4}.
\end{split}
\end{equation*}
\end{proof}

We still need to derive estimates for the other Ricci coefficients $\psi^{(s,s')}$, and it is 
easy to verify that these $\psi^{(s,s')}$ satisfy
\begin{equation}\label{tran13}
\begin{split}
&\nab_3\nab\psi^{(s,s')}+\l\lambda[\psi^{(s,s')}]+\f12\r \tr\chib\nab\psi^{(s,s')}\\
=&\sum_{s_1+s_2=s, \atop s'_1+s'_2=s'+1}\nab\psi^{(s_1,s'_1)}\cdot\psi^{(s_2,s'_2)}+\nab\Psi^{(s,s'+1)}\\
&+\sum_{s_3+s_4=s+\f12, \atop s'_3+s'_4=s'+\f32}\psi^{(s_3,s'_3)}\cdot\Psi^{(s_4,s'_4)}+\sum_{s_5+s_6+s_7=s+\f12, \atop s'_5+s'_6+s'_7=s'+\f32}\psi^{(s_5, s'_5)}\cdot\psi^{(s_6, s'_6)}\cdot\psi^{(s_7, s'_7)}.
\end{split}
\end{equation}

\begin{proposition}\label{O14 3sc}
For $\psi^{s,s'}$ obeys \eqref{tran13}, we have
\begin{equation*}
\begin{split}
&|u|^{2\lambda-2s'-1}\|\nab\psi^{(s,s')}\|_{L^4_{sc}(S_{u,\ub})}\\
\leq& |u_{\infty}|^{2\lambda-2s'-1}\|\nab\psi^{(s,s')}\|_{L^4_{sc}(S_{u_{\infty},\ub})}+\int_{u_{\infty}}^{u}|u'|^{2\lambda-2s'-3}\|\nab\Psi^{(s,s'+1)}\|_{L^4_{sc}(S_{u',\ub})}du'\\
&+\int_{u_{\infty}}^{u}|u'|^{2\lambda-2s'-3}\sum_{s_1+s_2=s, \atop s'_1+s'_2=s'+1}\|\nab\psi^{(s_1,s'_1)}\cdot\psi^{(s_2,s'_2)}\|_{L^4_{sc}(S_{u',\ub})}du'\\
&+\int_{u_{\infty}}^{u}|u'|^{2\lambda-2s'-3}\sum_{s_3+s_4=s+\f12, \atop s'_3+s'_4=s'+\f32}\|\psi^{(s_3,s'_3)}\cdot\Psi^{(s_4,s'_4)}\|_{L^4_{sc}(S_{u',\ub})}du'\\
&+\int_{u_{\infty}}^{u}|u'|^{2\lambda-2s'-3}\sum_{s_5+s_6+s_7=s+\f12, \atop s'_5+s'_6+s'_7=s'+\f32}\|\psi^{(s_5, s'_5)}\cdot\psi^{(s_6, s'_6)}\cdot\psi^{(s_7, s'_7)}\|_{L^4_{sc}(S_{u',\ub})}du'.
\end{split}
\end{equation*}
\end{proposition}

\begin{proof}
Adopting Proposition \ref{evolution lemma} for
$$p=4,\lambda_1=2\lambda+1-\frac 12,$$
we derive
\begin{equation}
\begin{split}
&|u|^{2\lambda+\frac 12}\|\nab\psi^{(s,s')}\|_{L^4(S_{u,\ub})}\\
\leq& |u_{\infty}|^{2\lambda+\frac 12}\|\nab\psi^{(s,s')}\|_{L^4(S{u_{\infty},\ub}))}+\int_{u_{\infty}}^{u}|u'|^{2\lambda+\frac 12}\|\nab_3\nab\psi^{(s,s')}\|_{L^4(S_{u',\ub})}du'\\
\leq& |u_{\infty}|^{2\lambda+\frac 12}\|\nab\psi^{(s,s')}\|_{L^4(S{u_{\infty},\ub})}+\int_{u_{\infty}}^{u}|u'|^{2\lambda+\frac 12}\|\nab\Psi^{(s,s'+1)}\|_{L^4(S_{u',\ub})}du'\\
&+\int_{u_{\infty}}^{u}|u'|^{2\lambda+\frac12}\sum_{s_1+s_2=s, \atop s'_1+s'_2=s'+1}\|\nab\psi^{(s_1,s'_1)}\cdot\psi^{(s_2,s'_2)}\|_{L^4(S_{u',\ub})}du'\\
&+\int_{u_{\infty}}^{u}|u'|^{2\lambda+\frac12}\sum_{s_3+s_4=s+\f12, \atop s'_3+s'_4=s'+\f32}\|\psi^{(s_3,s'_3)}\cdot\Psi^{(s_4,s'_4)}\|_{L^4(S_{u',\ub})}du'\\
&+\int_{u_{\infty}}^{u}|u'|^{2\lambda+\frac12}\sum_{s_5+s_6+s_7=s+\f12, \atop s'_5+s'_6+s'_7=s'+\f32}\|\psi^{(s_5, s'_5)}\cdot\psi^{(s_6, s'_6)}\cdot\psi^{(s_7, s'_7)}\|_{L^4(S_{u',\ub})}du'.
\end{split}
\end{equation}
With the definition
\begin{equation*}
\|\psi^{(s,s')}\|_{L^4_{sc}(S_{u,\ub})}=\d^{s-\f34}|u|^{2s'+\f12}\|\psi^{(s,s')}\|_{L^4(\S)},
\end{equation*}
we deduce the following estimate in scale invariant norms:
\begin{equation*}
\begin{split}
&|u|^{2\lambda-2s'-1}\|\nab\psi^{(s,s')}\|_{L^4_{sc}(S_{u,\ub})}\\
\leq& |u_{\infty}|^{2\lambda-2s'-1}\|\nab\psi^{(s,s')}\|_{L^4_{sc}(S_{u_{\infty},\ub})}+\int_{u_{\infty}}^{u}|u'|^{2\lambda-2s'-3}\|\nab\Psi^{(s,s'+1)}\|_{L^4_{sc}(S_{u',\ub})}du'\\
&+\int_{u_{\infty}}^{u}|u'|^{2\lambda-2s'-3}\sum_{s_1+s_2=s, \atop s'_1+s'_2=s'+1}\|\nab\psi^{(s_1,s'_1)}\cdot\psi^{(s_2,s'_2)}\|_{L^4_{sc}(S_{u',\ub})}du'\\
&+\int_{u_{\infty}}^{u}|u'|^{2\lambda-2s'-3}\sum_{s_3+s_4=s+\f12, \atop s'_3+s'_4=s'+\f32}\|\psi^{(s_3,s'_3)}\cdot\Psi^{(s_4,s'_4)}\|_{L^4_{sc}(S_{u',\ub})}du'\\
&+\int_{u_{\infty}}^{u}|u'|^{2\lambda-2s'-3}\sum_{s_5+s_6+s_7=s+\f12, \atop s'_5+s'_6+s'_7=s'+\f32}\|\psi^{(s_5, s'_5)}\cdot\psi^{(s_6, s'_6)}\cdot\psi^{(s_7, s'_7)}\|_{L^4_{sc}(S_{u',\ub})}du'.
\end{split}
\end{equation*}
\end{proof}

As a consequence, we conclude
\begin{proposition}\label{O14 3norm}
 Under the assumptions of Theorem \ref{main.thm} and the bootstrap assumptions \eqref{BA.0} and \eqref{BA.1}, the following inequalities hold
\begin{equation*}
\begin{split}
 \|\nab\o\|_{L^4_{sc}(\S)}\leq &\|\nab\o\|_{L^4_{sc}(S_{u_{\infty},\ub})}+\f{1}{|u|^{\f12}}\underline{\M R}_1^{\f12}\underline{\M R}_2^{\f12}+\f{\d^{\f14}}{|u|^{\f12}}\underline{\M R}_1+\f{\d^{\f12}}{|u|^{\f32}}\D_0 \underline{\M R}_0^{\f12}\underline{\M R}_1^{\f12}\\
&+\f{\d^{\f34}}{|u|^{\f32}}\D_0 \underline{\M R}_0+\f{\d^{\f12}}{|u|} \D_0 \M O_{1,4}+\f{\d}{|u|^3}\M O_{0,4}\D_0^2,
\end{split}
\end{equation*}

\begin{equation*}
\begin{split}
 \|\nab\etb\|_{L^4_{sc}(\S)}\leq &\|\nab\etb\|_{L^4_{sc}(S_{u_{\infty},\ub})}+\M R_1^{\f12} \M R_2^{\f12}+\d^{\f14}\M R_1+\f{\d^{\f12}}{|u|}\D_0 \M R_0^{\f12} \M R_1^{\f12}+\f{\d^{\f34}}{|u|}\D_0 \M R_0\\
&+\f{1}{|u|^{\f12}}\underline{\M R}_1^{\f12}\underline{\M R}_2^{\f12}+\f{\d^{\f14}}{|u|^{\f12}}\underline{\M R}_1+\f{\d^{\f12}}{|u|^{\f32}}\D_0 \underline{\M R}_0^{\f12}\underline{\M R}_1^{\f12}\\
&+\f{\d^{\f34}}{|u|^{\f32}}\D_0 \underline{\M R}_0+\f{\d^{\f12}}{|u|} \D_0 \M O_{1,4}+\f{\d^{\f12}}{|u|}\M O_{0,4}\D_0^2,
\end{split}
\end{equation*}


\begin{equation*}
\begin{split}
 \f{1}{|u|}\|\nab\chibh\|_{L^4_{sc}(\S)}\leq& |u_{\infty}|^{-1}\|\nab\chibh\|_{L^4_{sc}(S_{u_{\infty},\ub})}+\f{1}{|u|^{\f32}}\underline{\M R}_1^{\f12}\underline{\M R}_2^{\f12}+\f{\d^{\f14}}{|u|^{\f32}}\underline{\M R}_1+\f{\d^{\f12}}{|u|^{\f32}}\D_0 \underline{\M R}_0^{\f12}\underline{\M R}_1^{\f12}\\
&+\f{\d^{\f34}}{|u|^{\f32}}\D_0 \underline{\M R}_0+\f{\d^{\f12}}{|u|} \D_0 \M O_{1,4}+\f{\d^{\f12}}{|u|}\M O_{0,4}\D_0^2.
\end{split}
\end{equation*}


\end{proposition}

\begin{proof}
The null Ricci coefficient $\o$ satisfies
\begin{equation*}
\begin{split}
\nab_3\nab\o+\f12 \tr\chib\nab\o=&\nab\rho+(\eta,\etb)\rho+\omb\nab\o+\chibh\nab\o+\o\nab\omb+(\eta,\etb)(\nab\eta,\nab\etb)\\
&+(\eta,\etb)\o\omb+(\eta,\etb)(\eta,\etb)(\eta,\etb).
\end{split}
\end{equation*}
The anomalous term on the right hand side is due to $\chibh$. With the fact that
$$\lambda(\o)=s_2(\o)=0,$$
and employing Proposition \ref{O14 3sc}, we deduce
\begin{equation*}
\begin{split}
&|u|^{-1}\|\nab\o\|_{L^4_{sc}(S_{u,\ub})}\\
\leq& |u_{\infty}|^{-1}\|\nab\o\|_{L^4_{sc}(S_{u_{\infty},\ub})}+\int_{u_{\infty}}^{u}|u'|^{-3}\|\nab\Psi^{(1,1)}\|_{L^4_{sc}(S_{u',\ub})}du'\\
&+\int_{u_{\infty}}^{u}|u'|^{-3}\sum_{s_1+s_2=1, \atop s'_1+s'_2=1}\|\nab\psi^{(s_1,s'_1)}\cdot\psi^{(s_2,s'_2)}\|_{L^4_{sc}(S_{u',\ub})}du'\\
&+\int_{u_{\infty}}^{u}|u'|^{-3}\d^{\f12}\|\nab\o\|_{L^4_{sc}(S_{u',\ub})}\f{1}{|u'|}\|\chibh\|_{L^{\infty}_{sc}(S_{u',\ub})}du'\\
&+\int_{u_{\infty}}^{u}|u'|^{-3}\sum_{s_3+s_4=\f32, \atop s'_3+s'_4=\f32}\|\psi^{(s_3,s'_3)}\cdot\Psi^{(s_4,s'_4)}\|_{L^4_{sc}(S_{u',\ub})}du'\\
&+\int_{u_{\infty}}^{u}|u'|^{-3}\sum_{s_5+s_6+s_7=\f32, \atop s'_5+s'_6+s'_7=\f32}\|\psi^{(s_5, s'_5)}\cdot\psi^{(s_6, s'_6)}\cdot\psi^{(s_7, s'_7)}|_{L^4_{sc}(S_{u',\ub})}du'\\
\leq& |u_{\infty}|^{-1}\|\nab\o\|_{L^4_{sc}(S_{u_{\infty},\ub})}+\f{1}{|u|^{\f32}}\underline{\M R}_1^{\f12}\underline{\M R}_2^{\f12}+\f{\d^{\f14}}{|u|^{\f32}}\underline{\M R}_1+\f{\d^{\f12}}{|u|^{\f52}}\D_0 \underline{\M R}_0^{\f12}\underline{\M R}_1^{\f12}\\
&+\f{\d^{\f34}}{|u|^{\f52}}\D_0 \underline{\M R}_0+\f{\d^{\f12}}{|u|^2} \D_0 \M O_{1,4}+\f{\d}{|u|^4}\M O_{0,4}\D_0^2,
\end{split}
\end{equation*}
where we use Proposition \ref{L4} and H\"older's inequality.  
Multiplying $|u|$ on both sides, the conclusion follows.

The null Ricci coefficient $\etb$ obeys
\begin{equation*}
\begin{split}
\nab_3\nab\etb+\tr\chib\nab\etb=&\tr\chib\nab\eta+\nab\beb+(\eta,\etb)\beb+(\eta,\etb)\nab\tr\chib+(\eta,\etb)\nab\chibh\\
&+\chibh\nab(\eta,\etb)+\tr\chib\eta\etb+\chibh\eta\etb.
\end{split}
\end{equation*}
Recalling that 
$$\lambda(\etb)=s_2(\etb)=\f12,$$
adopting Proposition \ref{O14 3sc}, we obtain 
\begin{equation*}
\begin{split}
&|u|^{-1}\|\nab\etb\|_{L^4_{sc}(S_{u,\ub})}\\
\leq& |u_{\infty}|^{-1}\|\nab\etb\|_{L^4_{sc}(S_{u_{\infty},\ub})}+\int_{u_{\infty}}^{u}|u'|^{-3}\|\tr\chib\nab\eta\|_{L^4_{sc}(S_{u',\ub})}du'\\
&+\int_{u_{\infty}}^{u}|u'|^{-3}\|\nab\beb\|_{L^4_{sc}(S_{u',\ub})}du'+\int_{u_{\infty}}^{u}|u'|^{-3}\|(\eta,\etb)\beb\|_{L^4_{sc}(S_{u',\ub})}du'\\
&+\int_{u_{\infty}}^{u}|u'|^{-3}\|(\eta,\etb)\nab\tr\chib+(\eta,\etb)\nab\chibh+\chibh\nab(\eta,\etb)\|_{L^4_{sc}(S_{u',\ub})}du'\\
&+\int_{u_{\infty}}^{u}|u'|^{-3}\|\tr\chib\eta\etb+\chibh\eta\etb\|_{L^4_{sc}(S_{u',\ub})}du'\\
\leq& |u_{\infty}|^{-1}\|\nab\etb\|_{L^4_{sc}(S_{u_{\infty},\ub})}+\int_{u_{\infty}}^{u}\f{1}{|u'|^2}\f{\d^{\f12}}{|u'|^2}\|\tr\chib\|_{L^{\infty}_{sc}(S_{u',\ub})}\|\nab\eta\|_{L^4_{sc}(S_{u',\ub})}du'\\
&+\int_{u_{\infty}}^{u}|u'|^{-3}\|\nab\beb\|_{L^4_{sc}(S_{u',\ub})}du'+\int_{u_{\infty}}^{u}|u'|^{-3}\|(\eta,\etb)\beb\|_{L^4_{sc}(S_{u',\ub})}du'\\
&+\int_{u_{\infty}}^{u}\f{\d^{\f12}}{|u'|^{3}}\|\eta,\etb\|_{L^{\infty}_{sc}(S_{u',\ub})}\f{1}{|u'|}\|\nab\tr\chib, \nab\chibh\|_{L^4_{sc}(S_{u',\ub})}du'\\
&+\int_{u_{\infty}}^{u}\f{\d^{\f12}}{|u'|^{3}}\f{1}{|u'|}\|\chibh\|_{L^{\infty}_{sc}(S_{u',\ub})}\|\nab(\eta,\etb)\|_{L^4_{sc}(S_{u',\ub})}du'\\
&+\int_{u_{\infty}}^{u}\f{\d^{\f12}}{|u'|^{3}}\f{\d^{\f12}}{|u'|^2}\|\tr\chib\|_{L^{\infty}_{sc}(S_{u',\ub})}\|\eta\|_{L^{\infty}_{sc}(S_{u',\ub})}\|\etb\|_{L^4_{sc}(S_{u',\ub})}du'\\
&+\int_{u_{\infty}}^{u}\f{\d}{|u|^4}\f{1}{|u'|}\|\chibh\|_{L^{\infty}_{sc}(S_{u',\ub})}\|\eta\|_{L^{\infty}_{sc}(S_{u',\ub})}\|\etb\|_{L^4_{sc}(S_{u',\ub})}du'\\
\leq& |u_{\infty}|^{-1}\|\nab\etb\|_{L^4_{sc}(S_{u_{\infty},\ub})}+\f{1}{|u|}\M O_{1,4}(\eta)+\f{1}{|u|^{\f32}}\underline{\M R}_1^{\f12}\underline{\M R}_2^{\f12}+\f{\d^{\f14}}{|u|^{\f32}}\underline{\M R}_1+\f{\d^{\f12}}{|u|^{\f52}}\D_0 \underline{\M R}_0^{\f12}\underline{\M R}_1^{\f12}\\
&+\f{\d^{\f34}}{|u|^{\f52}}\D_0 \underline{\M R}_0+\f{\d^{\f12}}{|u|^2} \D_0 \M O_{1,4}+\f{\d^{\f12}}{|u|^2}\M O_{0,4}\D_0^2.
\end{split}
\end{equation*}
Multiplying $|u|$ on both sides, we obtain
\begin{equation*}
\begin{split}
 \|\nab\etb\|_{L^4_{sc}(\S)}\leq &\|\nab\etb\|_{L^4_{sc}(S_{u_{\infty},\ub})}+\M O_{1,4}(\eta)+\f{1}{|u|^{\f12}}\underline{\M R}_1^{\f12}\underline{\M R}_2^{\f12}+\f{\d^{\f14}}{|u|^{\f12}}\underline{\M R}_1+\f{\d^{\f12}}{|u|^{\f32}}\D_0 \underline{\M R}_0^{\f12}\underline{\M R}_1^{\f12}\\
&+\f{\d^{\f34}}{|u|^{\f32}}\D_0 \underline{\M R}_0+\f{\d^{\f12}}{|u|} \D_0 \M O_{1,4}+\f{\d^{\f12}}{|u|}\D_0^2\M O_{0,4}.
\end{split}
\end{equation*}

In Proposition \ref{O14 4norm}, we have proved
\begin{equation*}
\begin{split}
 \|\nab\eta\|_{L^4_{sc}(\S)}\leq &\M R_1^{\f12} \M R_2^{\f12}+\d^{\f14}\M R_1+\f{\d^{\f12}}{|u|}\D_0 \M R_0^{\f12} \M R_1^{\f12}+\f{\d^{\f34}}{|u|}\D_0 \M R_0\\
&+\f{\d^{\f12}}{|u|}\D_0 \M O_{1,4}+\f{\d}{|u|}\D_0^2 \M O_{0,4}.
\end{split}
\end{equation*}
Collecting these together, we derive
\begin{equation*}
\begin{split}
 \|\nab\etb\|_{L^4_{sc}(\S)}\leq &\|\nab\etb\|_{L^4_{sc}(S_{u_{\infty},\ub})}+\M R_1^{\f12} \M R_2^{\f12}+\d^{\f14}\M R_1+\f{\d^{\f12}}{|u|}\D_0 \M R_0^{\f12} \M R_1^{\f12}+\f{\d^{\f34}}{|u|}\D_0 \M R_0\\
&+\f{1}{|u|^{\f12}}\underline{\M R}_1^{\f12}\underline{\M R}_2^{\f12}+\f{\d^{\f14}}{|u|^{\f12}}\underline{\M R}_1+\f{\d^{\f12}}{|u|^{\f32}}\D_0 \underline{\M R}_0^{\f12}\underline{\M R}_1^{\f12}\\
&+\f{\d^{\f34}}{|u|^{\f32}}\D_0 \underline{\M R}_0+\f{\d^{\f12}}{|u|} \D_0 \M O_{1,4}+\f{\d^{\f12}}{|u|}\M O_{0,4}\D_0^2.
\end{split}
\end{equation*}

For $\chibh$, by a straightforward computation we derive

\begin{equation*}
\begin{split}
\nab_3\nab\chibh+\f32\tr\chib\nab\chibh=&\nab\ab+(\eta,\etb)\ab+\chibh\,\beb+\chibh\nab\tr\chib+\chibh\nab\omb+(\omb,\chibh)\nab\chibh\\
&+(\eta,\etb)\tr\chib\,\chibh+(\eta,\etb)\omb\chibh+(\eta,\etb)\chibh\,\chibh.
\end{split}
\end{equation*}

Since
$$\lambda(\chibh)=s_2(\chibh)=1,$$
by employing Proposition \ref{O14 3sc} we deduce
\begin{equation*}
\begin{split}
&|u|^{-1}\|\nab\chibh\|_{L^4_{sc}(S_{u,\ub})}\\
\leq& |u_{\infty}|^{-1}\|\nab\chibh\|_{L^4_{sc}(S_{u_{\infty},\ub})}+\int_{u_{\infty}}^{u}|u'|^{-3}\|\nab\ab\|_{L^4_{sc}(S_{u',\ub})}du'\\
&+\int_{u_{\infty}}^{u}|u'|^{-3}\|(\eta,\etb)\ab\|_{L^4_{sc}(S_{u',\ub})}du'+\int_{u_{\infty}}^{u}|u'|^{-3}\|\chibh\,\beb\|_{L^4_{sc}(S_{u',\ub})}du'\\
&+\int_{u_{\infty}}^{u}|u'|^{-3}\|\chibh\nab\tr\chib+\chibh\nab\omb+(\omb,\chibh)\nab\chibh\|_{L^4_{sc}(S_{u',\ub})}du'\\
&+\int_{u_{\infty}}^{u}|u'|^{-3}\|(\eta,\etb)\tr\chib\,\chibh+(\eta,\etb)\chibh(\chibh,\omb)\|_{L^4_{sc}(S_{u',\ub})}du'\\
\leq& |u_{\infty}|^{-1}\|\nab\chibh\|_{L^4_{sc}(S_{u_{\infty},\ub})}+\int_{u_{\infty}}^{u}\f{1}{|u'|^3}\|\nab\ab\|_{L^4_{sc}(S_{u',\ub})}du'\\
&+\int_{u_{\infty}}^{u}|u'|^{-3}\|(\eta,\etb)\ab\|_{L^4_{sc}(S_{u',\ub})}du'+\int_{u_{\infty}}^{u}|u'|^{-3}\d^{\f12}\|\beb\|_{L^4_{sc}(S_{u',\ub})}\f{1}{|u'|}\|\chibh\|_{L^{\infty}_{sc}(S_{u',\ub})}du'\\
&+\int_{u_{\infty}}^{u}\f{\d^{\f12}}{|u'|^{3}}\f{1}{|u'|}\|\chibh\|_{L^{\infty}_{sc}(S_{u',\ub})}\|\nab\tr\chib, \nab\omb\|_{L^4_{sc}(S_{u',\ub})}du'\\
&+\int_{u_{\infty}}^{u}\f{\d^{\f12}}{|u'|^{2}}\f{1}{|u'|}\|\chibh,\omb\|_{L^{\infty}_{sc}(S_{u',\ub})}\f{1}{|u'|}\|\nab\chibh\|_{L^4_{sc}(S_{u',\ub})}du'\\
&+\int_{u_{\infty}}^{u}\f{\d^{\f12}}{|u'|^{2}}\f{\d^{\f12}}{|u'|^2}\|\tr\chib\|_{L^{\infty}_{sc}(S_{u',\ub})}\f{1}{|u'|}\|\chibh\|_{L^{\infty}_{sc}(S_{u',\ub})}\|\eta,\etb\|_{L^4_{sc}(S_{u',\ub})}du'\\
&+\int_{u_{\infty}}^{u}\f{\d}{|u'|^3}\f{1}{|u'|}\|\chibh\|_{L^{\infty}_{sc}(S_{u',\ub})}\f{1}{|u'|}\|\chibh,\omb\|_{L^{\infty}_{sc}(S_{u',\ub})}\|\eta,\etb\|_{L^4_{sc}(S_{u',\ub})}du'\\
\leq& |u_{\infty}|^{-1}\|\nab\chibh\|_{L^4_{sc}(S_{u_{\infty},\ub})}+\f{1}{|u|^{\f32}}\underline{\M R}_1^{\f12}\underline{\M R}_2^{\f12}+\f{\d^{\f14}}{|u|^{\f32}}\underline{\M R}_1+\f{\d^{\f12}}{|u|^{\f32}}\D_0 \underline{\M R}_0^{\f12}\underline{\M R}_1^{\f12}\\
&+\f{\d^{\f34}}{|u|^{\f32}}\D_0 \underline{\M R}_0+\f{\d^{\f12}}{|u|} \D_0 \M O_{1,4}+\f{\d^{\f12}}{|u|}\M O_{0,4}\D_0^2,
\end{split}
\end{equation*}
which is the desired estimate. 

\end{proof}

We now focus on $\tr\chib$. 
By introducing a new bootstrap assumption
\begin{equation}\label{BA.11}
\f{1}{|u|}\|\Omega\tr\chib-\f{2}{u}\|_{L^{\infty}_{sc}(\S)}\leq \D_4, 
\end{equation}
we can prove
\begin{proposition}\label{O14 trchib}
 \begin{equation*}
\begin{split}
\|\nab\tr\chib\|_{L^4_{sc}(\S)}\leq &\|\nab\tr\chib\|_{L^4_{sc}(S_{u_{\infty},\ub})}+\M R_1^{\f12} \M R_2^{\f12}+\d^{\f14}\M R_1+\f{\d^{\f12}}{|u|}\D_0 \M R_0^{\f12} \M R_1^{\f12}+\f{\d^{\f34}}{|u|}\D_0 \M R_0\\
&+\f{\d^{\f12}}{|u|} \D_0 \M O_{1,4}+\f{\d^{\f12}}{|u|}\M O_{0,4}\D_0^2+{\d^{\f12}} (\D_0+\D_4)(\M O_{0,4}+\M O_{1,4})\\
&+{\d^{\f12}}\M O_{0,4}(\D_0+\D_4)^2.
\end{split}
\end{equation*}
\end{proposition}

\begin{proof}
Thanks to the fact
$$\omb=-\f{1}{2\Omega}\nab_3\Omega,$$
we calculate
\begin{equation*}
\begin{split}
&\nab_3(\Omega \tr\chib-\f{2}{u})\\
=&\nab_3\Omega\tr\chib+\Omega\nab_3\tr\chib-\nab_3(\f{2}{u}) \\
=&-2\Omega\omb\tr\chib+\Omega\l-\f12(\tr\chib)^2-2\omb\tr\chib-|\chibh|^2\r+\f{2\Omega^{-1}}{u^2}\\
=&-4\Omega\omb\tr\chib-\Omega|\chibh|^2-\f{\Omega^{-1}}{2}\l(\Omega\tr\chib)^2-\f{4}{u^2}\r\\
=&-4\Omega\omb\tr\chib-\Omega|\chibh|^2-\f{\Omega^{-1}}{2}(\Omega\tr\chib-\f{2}{u})(\Omega\tr\chib+\f{2}{u}).\\
\end{split}
\end{equation*}
Therefore, we obtain
\begin{equation}\label{Omega trchi 3}
\begin{split}
\nab_3(\Omega\tr\chib-\f{2}{u})+\tr\chib(\Omega\tr\chib-\f{2}{u})=-4\Omega\omb\tr\chib-\Omega|\chibh|^2+\f{\Omega^{-1}}{2}(\Omega\tr\chib-\f{2}{u})^2.\\
\end{split}
\end{equation}


Applying \eqref{Omega trchi 3} and Proposition \ref{commute0}, we compute
\begin{equation*}
\begin{split}
&\nab_3\nab(\Omega \tr\chib-\f{2}{u})+\f32\tr\chib\nab(\Omega \tr\chib-\f{2}{u})\\
=&\chibh\nab(\Omega \tr\chib-\f{2}{u})+\nab\tr\chib(\Omega \tr\chib-\f{2}{u})+\nab\Omega\omb\tr\chib+\Omega\nab\omb\tr\chib+\Omega\omb\nab\tr\chib\\
&+\nab\Omega\chibh\,\chibh+\Omega\chibh\nab\chibh+\nab\Omega\Omega^{-2}(\Omega \tr\chib-\f{2}{u})^2+\Omega^{-1}\nab(\Omega \tr\chib-\f{2}{u})(\Omega \tr\chib-\f{2}{u})\\
&+\ee\tr\chib(\Omega \tr\chib-\f{2}{u})+\ee\Omega\omb\tr\chib+\ee\Omega\chibh\,\chibh+\ee\Omega^{-1}(\Omega \tr\chib-\f{2}{u})^2.
\end{split}
\end{equation*}

Combining the fact
$$\eta_A+\etb_A=2\nab_A (\log \O),$$
Proposition \ref{Omega}, bootstrap assumption \eqref{BA.11}, Proposition \ref{evolution lemma} and Gronwall's inequality, we conclude
\begin{equation*}
\begin{split}
&|u|^{-1}\|\nab(\Omega \tr\chib-\f{2}{u})\|_{L^4_{sc}(S_{u,\ub})}\\
%
\leq& |u_{\infty}|^{-1}\|\nab(\Omega \tr\chib-\f{2}{u_{\infty}})\|_{L^4_{sc}(S_{u_{\infty},\ub})}+\int_{u_{\infty}}^{u}\f{1}{|u'|^{2}}\f{\d^{\f12}}{|u'|^2}\|\tr\chib\|_{L^{\infty}_{sc}(S_{u',\ub})}\|\nab\omb\|_{L^4_{sc}(S_{u',\ub})}du'\\
&+\int_{u_{\infty}}^{u}\f{\d^{\f12}}{|u'|^{3}}\f{1}{|u'|}\|\Omega \tr\chib-\f{2}{u'}, \omb\|_{L^{\infty}_{sc}(S_{u',\ub})}\|\nab\tr\chib\|_{L^4_{sc}(S_{u',\ub})}du'\\
&+\int_{u_{\infty}}^{u}\f{\d^{\f12}}{|u'|^{2}}\f{1}{|u'|}\|\chibh\|_{L^{\infty}_{sc}(S_{u',\ub})}\f{1}{|u'|}\|\nab\chibh\|_{L^4_{sc}(S_{u',\ub})}du'\\
&+\int_{u_{\infty}}^{u}\f{\d^{\f12}}{|u'|^{2}}\f{\d^{\f12}}{|u'|^2}\|\tr\chib\|_{L^{\infty}_{sc}(S_{u',\ub})}\|\eta,\etb\|_{L^{4}_{sc}(S_{u',\ub})}\f{1}{|u'|}\|\Omega \tr\chib-\f{2}{u'}\|_{L^{\infty}_{sc}(S_{u',\ub})}du'\\
&+\int_{u_{\infty}}^{u}\f{\d^{\f12}}{|u'|^{3}}\f{\d^{\f12}}{|u'|^2}\|\tr\chib\|_{L^{\infty}_{sc}(S_{u',\ub})}\|\eta,\etb\|_{L^{\infty}_{sc}(S_{u',\ub})}\|\omb\|_{L^4_{sc}(S_{u',\ub})}du'\\
&+\int_{u_{\infty}}^{u}\f{\d}{|u'|^{2}}\f{1}{|u'|}\|\Omega \tr\chib-\f{2}{u},\chibh\|_{L^{\infty}_{sc}(S_{u',\ub})}\f{1}{|u'|}\|\Omega \tr\chib-\f{2}{u'}, \chibh\|_{L^{\infty}_{sc}(S_{u',\ub})}\|\eta,\etb\|_{L^4_{sc}(S_{u',\ub})}du'\\
\leq& |u_{\infty}|^{-1}\|\nab\tr\chib\|_{L^4_{sc}(S_{u_{\infty},\ub})}+\f{1}{|u|}\M O_{1,4}[\omb]+\f{\d^{\f12}}{|u|} (\D_0+\D_4) \M O_{1,4}+\f{\d^{\f12}}{|u|}\M O_{0,4}(\D_0+\D_4)^2.
\end{split}
\end{equation*}
Multiplying $|u|$ on both sides, we obtain
\begin{equation*}
\|\nab(\Omega\tr\chib)\|_{L^4_{sc}(\S)}\leq \|\nab\tr\chib\|_{L^4_{sc}(S_{u_{\infty},\ub})}+\M O_{1,4}[\omb]+{\d^{\f12}} (\D_0+\D_4) \M O_{1,4}+{\d^{\f12}}\M O_{0,4}(\D_0+\D_4)^2.
\end{equation*}

To get desired bound for $\|\nab\tr\chib\|_{L^4_{sc}(\S)}$, we need derive estimate for $\|\nab\Omega\|_{L^4_{sc}(\S)}$. 
And we have the following lemma:
\begin{proposition}\label{nab Omega}
 $$|u|\|\nab\Omega\|_{L^4_{sc}(\S)} \leq \M O_{1,4}[\omb]+\f{\d^{\f12}}{|u|}\D_0\M O_{0,4}.$$
\end{proposition}
\begin{proof}
For $\Omega$, we have
$$\nab_3\Omega=-2\Omega\omb.$$
Applying Proposition \ref{commute0}, we obtain
$$\nab_3\nab\Omega+\f12\tr\chib\nab\Omega=(\chibh,\omb)\nab\Omega+\Omega\nab\omb+\ee\Omega\omb.$$
With Gronwall's inequality, we get
\begin{equation*}
\begin{split}
|u|^{-1}\|\nab\Omega\|_{L^4_{sc}(\S)}\leq & |u_{\infty}|^{-1}\|\nab\Omega\|_{L^4_{sc}(S_{u_{\infty},\ub})}+\int_{u_{\infty}}^u\f{1}{|u'|^3}\|\nab\omb\|_{L^4_{sc}(S_{u',\ub})}du'\\
&+\int_{u_{\infty}}^u\f{\d^{\f12}}{|u'|^4}\|\eta,\etb\|_{L^4_{sc}(S_{u',\ub})}\|\omb\|_{L^{\infty}_{sc}(S_{u',\ub})}du'\\
\leq&\f{1}{|u|^2}\M O_{1,4}[\omb]+\f{\d^{\f12}}{|u|^3}\D_0\M O_{0,4},
\end{split}
\end{equation*}
where we employ $\nab\Omega=0$ along $H^{(0,\ub)}_{u_{\infty}}$. Multiplying $|u|^2$ on both sides, we derive
$$|u|\|\nab\Omega\|_{L^4_{sc}(\S)} \leq \M O_{1,4}[\omb]+\f{\d^{\f12}}{|u|}\D_0\M O_{0,4}.$$
\end{proof}

Therefore, together with Proposition \ref{nab Omega} we infer
\begin{equation*}
\begin{split}
&\|\nab\tr\chib\|_{L^4_{sc}(\S)}\\
\ls& \|\Omega\nab\tr\chib\|_{L^4_{sc}(\S)}\\
\leq& \|\nab(\Omega\tr\chib)\|_{L^4_{sc}(\S)}+\|(\nab\Omega)\tr\chib\|_{L^4_{sc}(\S)}\\
\leq& \|\nab(\Omega\tr\chib)\|_{L^4_{sc}(\S)}+|u|\|\nab\Omega\|_{L^4_{sc}(\S)}\f{\d^{\f12}}{|u|^2}\|\tr\chib\|_{L^{\infty}_{sc}(\S)}\\
\leq& \|\nab\tr\chib\|_{L^4_{sc}(S_{u_{\infty},\ub})}+\M O_{1,4}[\omb]+{\d^{\f12}} (\D_0+\D_4)(\M O_{0,4}+\M O_{1,4})+{\d^{\f12}}\M O_{0,4}(\D_0+\D_4)^2.\\
\end{split}
\end{equation*}

Recall the estimate obtained in Proposition \ref{O14 4norm}
\begin{equation*}
\begin{split}
 \|\nab\omb\|_{L^4_{sc}(\S)}\leq &\M R_1^{\f12} \M R_2^{\f12}+\d^{\f14}\M R_1+\f{\d^{\f12}}{|u|}\D_0 \M R_0^{\f12} \M R_1^{\f12}+\f{\d^{\f34}}{|u|}\D_0 \M R_0\\
&+\f{\d^{\f12}}{|u|}\D_0 \M O_{1,4}+\f{\d}{|u|}\D_0^2 \M O_{0,4}.
\end{split}
\end{equation*}
Putting these estimates altogether, we derive
\begin{equation*}
\begin{split}
\|\nab\tr\chib\|_{L^4_{sc}(\S)}\leq &\|\nab\tr\chib\|_{L^4_{sc}(S_{u_{\infty},\ub})}+\M R_1^{\f12} \M R_2^{\f12}+\d^{\f14}\M R_1+\f{\d^{\f12}}{|u|}\D_0 \M R_0^{\f12} \M R_1^{\f12}+\f{\d^{\f34}}{|u|}\D_0 \M R_0\\
&+\f{\d^{\f12}}{|u|} \D_0 \M O_{1,4}+\f{\d^{\f12}}{|u|}\M O_{0,4}\D_0^2+{\d^{\f12}} (\D_0+\D_4)(\M O_{0,4}+\M O_{1,4})\\
&+{\d^{\f12}}\M O_{0,4}(\D_0+\D_4)^2.
\end{split}
\end{equation*}
\end{proof}

Thanks to the proposition above, we are ready to derive bound for $\|\Omega\tr\chib-\f{2}{u}\|_{L^{\infty}_{sc}(\S)}$.
\begin{proposition}\label{O0infty trchib}
Under the assumption of Theorem \ref{main.thm} and bootstrap assumptions \eqref{BA.0}-\eqref{BA.2}, we have
$$\f{1}{|u|}\|\Omega\tr\chib-\f{2}{u}\|_{L^{\infty}_{sc}(\S)}\leq C (\M I^{(0)}+\M R+\underline{\M R})^5,$$ 
where $C$ is a universal large constant.
\end{proposition}

\begin{proof}
Recall equation \eqref{Omega trchi 3} in the appendix
\begin{equation*}
\begin{split}
\nab_3(\Omega\tr\chib-\f{2}{u})+\tr\chib(\Omega\tr\chib-\f{2}{u})=-4\Omega\omb\tr\chib-\Omega|\chibh|^2+\f{\Omega^{-1}}{2}(\Omega\tr\chib-\f{2}{u})^2.\\
\end{split}
\end{equation*}
Since
$$\lambda(\Omega\tr\chib-\f{2}{u})=s_2(\Omega\tr\chib-\f{2}{u})=1,$$
with Proposition \ref{Omega}, bootstrap assumptions \eqref{BA.0} and \eqref{BA.11} and Gronwall's inequality, we deduce
\begin{equation*}
\begin{split}
&\f{1}{|u|}\|\Omega\tr\chib-\f{2}{u}\|_{L^4_{sc}(\S)}\\
\leq&\f{1}{|u_{\infty}|}\|\Omega\tr\chib-\f{2}{u_{\infty}}\|_{L^4_{sc}(S_{u_{\infty},\ub})}+\int_{u_{\infty}}^u\f{1}{|u'|^2}\|\omb\|_{L^4_{sc}(S_{u',\ub})}\f{\d^{\f12}}{|u'|^2}\|\tr\chib\|_{L^{\infty}_{sc}(S_{u',\ub})} du'\\
&+\int_{u_{\infty}}^u\f{\d^{\f14}}{|u'|^2}\f{\d^{\f14}}{|u'|}\|\chibh\|_{L^4_{sc}(S_{u',\ub})}\f{1}{|u'|}\|\chibh\|_{L^{\infty}_{sc}(S_{u',\ub})} du'\\
\leq&\f{1}{|u_{\infty}|}\|\Omega\tr\chib-\f{2}{u_{\infty}}\|_{L^4_{sc}(S_{u_{\infty},\ub})}+\f{1}{|u|}\M O_{0,4}[\omb]+\f{\d^{\f14}}{|u|}\D_0\M O_{0,4}.\\
\end{split}
\end{equation*}
Employing conclusions in Proposition \ref{O04} yields
$$\f{1}{|u|}\|\Omega\tr\chib-\f{2}{u}\|_{L^{\infty}_{sc}(\S)}\leq (\M I^{(0)}+\M R+\underline{\M R})^5.$$ 
By choosing $\D_4\gg (\M I^{(0)}+\M R+\underline{\M R})^5$, we have improved bootstrap assumption \eqref{BA.11} and obtain desired estimate.
\end{proof}

{\bf Remark:} Together with Proposition \ref{Omega}, Proposition \ref{O0infty trchib} implies 
$$\|\tr\chib+\f{2}{|u|}\|_{L^{\infty}(\S)}\leq \f{\d^{\f12}}{|u|^2}(\M I^{(0)}+\M R+\underline{\M R})+\f{\d^{\f12}}{|u|^2}\D_0.$$

Gathering all the estimates above, we obtain

\begin{proposition}\label{O14}
Assume
$$\M R<\infty, \quad \quad \underline{\M R}<\infty, \quad \quad \M O_{0,\infty}<\infty,$$
then there exist $\d_0=\d_0(\M I^{(0)}, \M R^{(0)}, \M R, \underline{\M R}, \M O_{0,\infty})$, such that whenever $\d\leq\d_0$, in $D$ we have
\begin{equation*}
\M O_{1,4}\leq C\l \M I^{(0)}+\M R+\underline{\M R}\r,
\end{equation*}
where $C$ is a large universal constant.
\end{proposition}

\begin{proof}
From the estimates above, by taking $\d_0$ sufficient small we have
 \begin{equation*}
\M O_{1,4}\leq \M I^{(0)}+\M R_0^{\f12}\M R_1^{\f12}+\M R_0+\f{1}{|u|^{\f12}}\underline{\M R}_0^{\f12}\underline{\M R}_1^{\f12}+\f{\d^{\f14}}{|u|^{\f12}}\underline{\M R}_0.
\end{equation*}
Let $\D_2\gg \M I^{(0)}+\M R+\underline{\M R}$. We have improved bootstrap assumption \eqref{BA.1}. 
\end{proof}

\section{$\M O_{0,\infty}$ ESTIMATES}
We then show how to bound the $L^{\infty}(\S)$ norm of the Ricci coefficients.
\subsection{$\M O_{0,\infty}$ Estimates for $\o,\tr\chi, \eta,\etb,\omb$}
For $\o,\tr\chi, \eta,\etb,\omb$, we have
\begin{proposition}
\begin{equation*}
\|\o,\tr\chi, \eta,\etb,\omb\|_{L^{\infty}_{sc}({S}_{u,\ub})} \leq C\l \M I^{(0)}+\M R+\underline{\M R}\r,
\end{equation*}
where $C$ is a large universal constant.
\end{proposition}
\begin{proof}
From Proposition \ref{Linfty} for any horizontal tensor $\phi$ in $D$, we have
$$\|\phi\|_{L^{\infty}_{sc}(\S)}\leq \|\phi\|^{\f12}_{L^{4}_{sc}(\S)}\|\nab \phi\|^{\f12}_{L^{4}_{sc}(\S)}+\d^{\f14}\|\phi\|_{L^{4}_{sc}(\S)}.$$
Thanks to this inequality and Propositions \ref{O04} and \ref{O14}, the conclusion follows. 
\end{proof}

\subsection{$\M O_{0,\infty}$ Estimates for $\chih, \chibh$}
We complement the anomalous norms of $\chih,\chibh$ by the local, non-anomalous, scale invariant norms
\begin{equation*}
 \M O_0^{\d}[\chih](u,\ub)=\sup_{{^{\d}S}\subset S}\|\chih\|_{L^4_{sc}({^{\d}S}_{u,\ub})},\quad \quad  \M O_0^{\d}[\chibh](u,\ub)=\sup_{{^{\d}S}\subset S}\|\chibh\|_{L^4_{sc}({^{\d}S}_{u,\ub})}
\end{equation*}
where ${^{\d}S}_{u,\ub}$ is a disk of radius $\d^{\f12}|u|$ obtained by transporting from the initial data embedded in $S_{u,0}$ or $S_{u_{\infty},\ub}$.

With the aid of these norms, we can prove
\begin{proposition}\label{O0infty chih}
\begin{equation*}
\begin{split}
\|\chih\|_{L^{\infty}_{sc}({S}_{u,\ub})}\leq &\|\a\|_{L^{4}_{sc}({^{\d}S}_{u_{\infty},\ub})}+\f{1}{|u|^{\f12}}\underline{\M R}^{\f12}_1\underline{\M R}^{\f12}_2+\f{\d^{\f14}}{|u|^{\f12}}\underline{\M R}_1\\
&+\f{\d^{\f12}}{|u|^{\f32}}\D_0\underline{\M R}^{\f12}_0\underline{\M R}^{\f12}_1+\f{\d^{\f34}}{|u|^{\f32}}\D_0\underline{\M R}_0\\
&+\M R_1^{\f12} \M R_2^{\f12}+\d^{\f14}\M R_1+\f{\d^{\f14}}{|u|}\D_0 \M R_0^{\f12} \M R_1^{\f12}+\f{\d^{\f14}}{|u|}\D_0 \M R_0\\
&+\f{\d^{\f12}}{|u|}\D_0 \M O_{1,4}+\f{\d}{|u|}\D_0^2 \M O_{0,4}.
\end{split}
\end{equation*}

\end{proposition}

\begin{proof}
For $\a$, we have the null Bianchi equation
$$\nab_3\a+\f12\tr\chib\a=\nab\widehat{\otimes}\b+4\omb\a-3(\chih\rho+{^{*}\chih}\sigma)+(\zeta+4\eta)\widehat{\otimes}\b.$$
Applying Gronwall's inequality and Proposition \ref{evolution lemma}, we have
\begin{equation*}
\begin{split}
\|\a\|_{L^{4}_{sc}({^{\d}S}_{u,\ub})}\leq &\|\a\|_{L^{4}_{sc}({^{\d}S}_{u_{\infty},\ub})}+\int_{u_{\infty}}^u\f{1}{|u'|^2}\|\nab\b\|_{L^{4}_{sc}({^{\d}S}_{u',\ub})} du'\\
&+\int_{u_{\infty}}^u\f{1}{|u'|^2}\|\chih\cdot(\rho,\sigma)+(\eta,\etb)\b\|_{L^{4}_{sc}({^{\d}S}_{u',\ub})} du'\\
\leq &\|\a\|_{L^{4}_{sc}({^{\d}S}_{u_{\infty},\ub})}+\int_{u_{\infty}}^u\f{1}{|u'|^2}\|\nab\b\|_{L^{4}_{sc}(S_{u',\ub})} du'\\
&+\int_{u_{\infty}}^u\f{1}{|u'|^2}\|\chih\cdot(\rho,\sigma)+(\eta,\etb)\b\|_{L^{4}_{sc}(S_{u',\ub})} du'\\
\leq &\|\a\|_{L^{4}_{sc}({^{\d}S}_{u_{\infty},\ub})}+\f{1}{|u|^{\f12}}\underline{\M R}^{\f12}_1\underline{\M R}^{\f12}_2+\f{\d^{\f14}}{|u|^{\f12}}\underline{\M R}_1\\
&+\f{\d^{\f12}}{|u|^{\f32}}\D_0\underline{\M R}^{\f12}_0\underline{\M R}^{\f12}_1+\f{\d^{\f34}}{|u|^{\f32}}\D_0\underline{\M R}_0.
\end{split}
\end{equation*}
For $\chih$, we have
$$\nab_4\chih+\tr\chi\chih=-2\o\chih-\a.$$
Using Gronwall's inequality and Proposition \ref{transport}, we derive
\begin{equation*}
\begin{split}
\|\chih\|_{L^{4}_{sc}({^{\d}S}_{u,\ub})}\leq &\int_0^{\ub}\d^{-1}\|\a\|_{L^{4}_{sc}({^{\d}S}_{u,\ub'})} d\ub'.\\
\end{split}
\end{equation*}
Together with the estimate for $\|\a\|_{L^{4}_{sc}({^{\d}S}_{u,\ub})}$ above, we obtain
\begin{equation*}
\begin{split}
\|\chih\|_{L^{4}_{sc}({^{\d}S}_{u,\ub})}\leq &\|\a\|_{L^{4}_{sc}({^{\d}S}_{u_{\infty},\ub})}+\f{1}{|u|^{\f12}}\underline{\M R}^{\f12}_1\underline{\M R}^{\f12}_2+\f{\d^{\f14}}{|u|^{\f12}}\underline{\M R}_1\\
&+\f{\d^{\f12}}{|u|^{\f32}}\D_0\underline{\M R}^{\f12}_0\underline{\M R}^{\f12}_1+\f{\d^{\f34}}{|u|^{\f32}}\D_0\underline{\M R}_0.
\end{split}
\end{equation*}

Employing Proposition \ref{Linfty} leads to
\begin{equation*}
\begin{split}
\|\chih\|_{L^{\infty}_{sc}({S}_{u,\ub})}\leq &\sup_{{^{\d}S}\subset S}\l\|\nab\chih\|_{L^{4}_{sc}({^{2\d}S}_{u,\ub})}+\|\chih\|_{L^{4}_{sc}({^{2\d}S}_{u,\ub})}\r\\
\leq &\M O_{1,4}[\chih]+\|\a\|_{L^{4}_{sc}({^{\d}S}_{u_{\infty},\ub})}+\f{1}{|u|^{\f12}}\underline{\M R}^{\f12}_1\underline{\M R}^{\f12}_2+\f{\d^{\f14}}{|u|^{\f12}}\underline{\M R}_1\\
&+\f{\d^{\f12}}{|u|^{\f32}}\D_0\underline{\M R}^{\f12}_0\underline{\M R}^{\f12}_1+\f{\d^{\f34}}{|u|^{\f32}}\D_0\underline{\M R}_0.
\end{split}
\end{equation*}

The following conclusion in Proposition \ref{O14 4norm}:
\begin{equation*}
\begin{split}
\|\nab\chih\|_{L^4_{sc}(\S)}\leq &\M R_1^{\f12} \M R_2^{\f12}+\d^{\f14}\M R_1+\f{\d^{\f14}}{|u|}\D_0 \M R_0^{\f12} \M R_1^{\f12}+\f{\d^{\f14}}{|u|}\D_0 \M R_0\\
&+\f{\d^{\f12}}{|u|}\D_0 \M O_{1,4}+\f{\d}{|u|}\D_0^2 \M O_{0,4},
\end{split}
\end{equation*}
immediately implies
\begin{equation*}
\begin{split}
\|\chih\|_{L^{\infty}_{sc}({S}_{u,\ub})}\leq &\|\a\|_{L^{4}_{sc}({^{\d}S}_{u_{\infty},\ub})}+\f{1}{|u|^{\f12}}\underline{\M R}^{\f12}_1\underline{\M R}^{\f12}_2+\f{\d^{\f14}}{|u|^{\f12}}\underline{\M R}_1\\
&+\f{\d^{\f12}}{|u|^{\f32}}\D_0\underline{\M R}^{\f12}_0\underline{\M R}^{\f12}_1+\f{\d^{\f34}}{|u|^{\f32}}\D_0\underline{\M R}_0\\
&+\M R_1^{\f12} \M R_2^{\f12}+\d^{\f14}\M R_1+\f{\d^{\f14}}{|u|}\D_0 \M R_0^{\f12} \M R_1^{\f12}+\f{\d^{\f14}}{|u|}\D_0 \M R_0\\
&+\f{\d^{\f12}}{|u|}\D_0 \M O_{1,4}+\f{\d}{|u|}\D_0^2 \M O_{0,4}.
\end{split}
\end{equation*}
\end{proof}

For $\chibh$, we can prove
\begin{proposition}\label{O0infty chibh}
\begin{equation*}
\begin{split}
\f{1}{|u|}\|\chibh\|_{L^{\infty}_{sc}({S}_{u,\ub})}\leq & \f{1}{|u_{\infty}|}\|\nab\chibh\|_{L^4_{sc}(S_{u_{\infty},\ub})}+\f{1}{|u_{\infty}|}\|\chibh\|_{L^{4}_{sc}({^{\d}S}_{u_{\infty},\ub})}+\f{1}{|u|^{\f32}}\underline{\M R}_1^{\f12}\underline{\M R}_2^{\f12}\\
&+\f{\d^{\f14}}{|u|^{\f32}}\underline{\M R}_1+\f{\d^{\f12}}{|u|^{\f32}}\D_0 \underline{\M R}_0^{\f12}\underline{\M R}_1^{\f12}+\f{1}{|u|^{\f32}}\underline{\M R}^{\f12}_0\underline{\M R}^{\f12}_1+\f{\d^{\f14}}{|u|^{\f32}}\underline{\M R}_0\\
&+\f{\d^{\f34}}{|u|^{\f32}}\D_0 \underline{\M R}_0+\f{\d^{\f12}}{|u|} \D_0 \M O_{1,4}+\f{\d^{\f12}}{|u|}\M O_{0,4}\D_0^2.\\
\end{split}
\end{equation*}

\end{proposition}

\begin{proof}
The null coefficient $\chibh$ satisfies
$$\nab_3\chibh+\tr\chib\,\chibh=-2\omb\chibh-\ab.$$
Adopting Gronwall's inequality and Proposition \ref{evolution lemma}, we derive

\begin{equation*}
\begin{split}
\f{1}{|u|}\|\chibh\|_{L^{4}_{sc}({^{\d}S}_{u,\ub})}\leq &\f{1}{|u_{\infty}|}\|\chibh\|_{L^{4}_{sc}({^{\d}S}_{u_{\infty},\ub})}+\int_{u_{\infty}}^u\f{1}{|u'|^3}\|\ab\|_{L^{4}_{sc}({^{\d}S}_{u',\ub})} du'\\
\leq&\f{1}{|u_{\infty}|}\|\chibh\|_{L^{4}_{sc}({^{\d}S}_{u_{\infty},\ub})}+\int_{u_{\infty}}^u\f{1}{|u'|^3}\|\ab\|_{L^{4}_{sc}(S_{u',\ub})} du'\\
\leq &\f{1}{|u_{\infty}|}\|\chibh\|_{L^{4}_{sc}({^{\d}S}_{u_{\infty},\ub})}+\f{1}{|u|^{\f32}}\underline{\M R}^{\f12}_0\underline{\M R}^{\f12}_1+\f{\d^{\f14}}{|u|^{\f32}}\underline{\M R}_0.\\
\end{split}
\end{equation*}

Employing Proposition \ref{Linfty}, we have
\begin{equation*}
\begin{split}
\f{1}{|u|}\|\chibh\|_{L^{\infty}_{sc}({S}_{u,\ub})}\leq &\sup_{{^{\d}S}\subset S}\l\f{1}{|u|}\|\nab\chibh\|_{L^{4}_{sc}({^{2\d}S}_{u,\ub})}+\f{1}{|u|}\|\chibh\|_{L^{4}_{sc}({^{2\d}S}_{u,\ub})}\r\\
\leq &\M O_{1,4}[\chibh]+\f{1}{|u_{\infty}|}\|\chibh\|_{L^{4}_{sc}({^{\d}S}_{u_{\infty},\ub})}+\f{1}{|u|^{\f32}}\underline{\M R}^{\f12}_0\underline{\M R}^{\f12}_1+\f{\d^{\f14}}{|u|^{\f32}}\underline{\M R}_0.
\end{split}
\end{equation*}

Hence the following conclusion in Proposition \ref{O14 3norm}:
\begin{equation*}
\begin{split}
 \f{1}{|u|}\|\nab\chibh\|_{L^4_{sc}(\S)}\leq& |u_{\infty}|^{-1}\|\nab\chibh\|_{L^4_{sc}(S_{u_{\infty},\ub})}+\f{1}{|u|^{\f32}}\underline{\M R}_1^{\f12}\underline{\M R}_2^{\f12}+\f{\d^{\f14}}{|u|^{\f32}}\underline{\M R}_1+\f{\d^{\f12}}{|u|^{\f32}}\D_0 \underline{\M R}_0^{\f12}\underline{\M R}_1^{\f12}\\
&+\f{\d^{\f34}}{|u|^{\f32}}\D_0 \underline{\M R}_0+\f{\d^{\f12}}{|u|} \D_0 \M O_{1,4}+\f{\d^{\f12}}{|u|}\M O_{0,4}\D_0^2,
\end{split}
\end{equation*}
implies
\begin{equation*}
\begin{split}
\f{1}{|u|}\|\chibh\|_{L^{\infty}_{sc}({S}_{u,\ub})}\leq & \f{1}{|u_{\infty}|}\|\nab\chibh\|_{L^4_{sc}(S_{u_{\infty},\ub})}+\f{1}{|u_{\infty}|}\|\chibh\|_{L^{4}_{sc}({^{\d}S}_{u_{\infty},\ub})}+\f{1}{|u|^{\f32}}\underline{\M R}_1^{\f12}\underline{\M R}_2^{\f12}\\
&+\f{\d^{\f14}}{|u|^{\f32}}\underline{\M R}_1+\f{\d^{\f12}}{|u|^{\f32}}\D_0 \underline{\M R}_0^{\f12}\underline{\M R}_1^{\f12}+\f{1}{|u|^{\f32}}\underline{\M R}^{\f12}_0\underline{\M R}^{\f12}_1+\f{\d^{\f14}}{|u|^{\f32}}\underline{\M R}_0\\
&+\f{\d^{\f34}}{|u|^{\f32}}\D_0 \underline{\M R}_0+\f{\d^{\f12}}{|u|} \D_0 \M O_{1,4}+\f{\d^{\f12}}{|u|}\M O_{0,4}\D_0^2.\\
\end{split}
\end{equation*}

\end{proof}

Gathering all the estimates in this section, we obtain

\begin{proposition}\label{O0infty}
Assume
$$\M R<\infty, \quad \quad \underline{\M R}<\infty,$$
then there exist $\d_0=\d_0(\M I^{(0)}, \M R^{(0)}, \M R, \underline{\M R})$, such that whenever $\d\leq\d_0$, in $D$ we have
\begin{equation*}
\M O_{0,\infty}\leq C\l\M I^{(0)}+\M R+\underline{\M R}\r,
\end{equation*}
where $C$ is a universal constant.
\end{proposition}
\begin{proof}
From the estimates above, by taking $\d_0$ sufficient small we have 
\begin{equation*}
\M O_{0,\infty}\leq \M I^{(0)}+\M R+\underline{\M R}.
\end{equation*}
Let $\D_3\gg \M I^{(0)}+\M R+\underline{\M R}$. We have improved bootstrap assumption \eqref{BA.0}. 
\end{proof}


\subsection{Improved $\M O_{0,\infty}$ Estimates for $\tr\chi$}

For $\tr\chi$, we can prove the following improved estimate
\begin{proposition}\label{O0infty trchi}
\begin{equation*}
\begin{split}
\d^{-\f12}\|\tr\chi\|_{L^{\infty}_{sc}({S}_{u,\ub})}\leq &\d^{-\f12}\|\tr\chi\|_{L^{\infty}_{sc}({S}_{u,0})}+\|\a\|_{L^{4}_{sc}({^{\d}S}_{u_{\infty},\ub})}\\
&+\l\f{1}{|u|}\underline{\M R}^{\f12}_1\underline{\M R}^{\f12}_2+\f{\d^{\f14}}{|u|}\underline{\M R}_1
+\f{\d^{\f12}}{|u|^2}\D_0\underline{\M R}^{\f12}_0\underline{\M R}^{\f12}_1+\f{\d^{\f34}}{|u|^2}\D_0\underline{\M R}_0+\M R_1^{\f12} \M R_2^{\f12}\r^2\\
&+\l\d^{\f12}\M R_1+\f{\d^{\f14}}{|u|}\D_0 \M R_0^{\f12} \M R_1^{\f12}+\f{\d^{\f14}}{|u|}\D_0 \M R_0+\f{\d^{\f12}}{|u|}\D_0 \M O_{1,4}+\f{\d}{|u|}\D_0^2 \M O_{0,4}\r^2.
\end{split}
\end{equation*}

\end{proposition}

\begin{proof}
The null coefficient $\tr\chi$ obeys
$$\nab_4\tr\chi+\f12(\tr\chi)^2=-|\chih|^2-2\o\tr\chi.$$
Applying Gronwall's inequality, we obtain

\begin{equation*}
\begin{split}
\d^{-\f12}\|\tr\chi\|_{L^{\infty}_{sc}({S}_{u,\ub})}\leq &\d^{-\f12}\|\tr\chi\|_{L^{\infty}_{sc}({S}_{u,0})}+\d^{-\f12}\int_0^{\ub}\d^{-1}\|\chih\cdot\chih\|_{L^{\infty}_{sc}(S_{u,\ub'})} d\ub'\\
\leq& \d^{-\f12}\|\tr\chi\|_{L^{\infty}_{sc}({S}_{u,0})}+\f{1}{|u|}\M O^2_{0,\infty}[\chih].
\end{split}
\end{equation*}

The following conclusion in Proposition \ref{O0infty chih}:
\begin{equation*}
\begin{split}
\|\chih\|_{L^{\infty}_{sc}({S}_{u,\ub})}\leq &\|\a\|_{L^{4}_{sc}({^{\d}S}_{u_{\infty},\ub})}+\f{1}{|u|^{\f12}}\underline{\M R}^{\f12}_1\underline{\M R}^{\f12}_2+\f{\d^{\f14}}{|u|^{\f12}}\underline{\M R}_1\\
&+\f{\d^{\f12}}{|u|^{\f32}}\D_0\underline{\M R}^{\f12}_0\underline{\M R}^{\f12}_1+\f{\d^{\f34}}{|u|^{\f32}}\D_0\underline{\M R}_0\\
&+\M R_1^{\f12} \M R_2^{\f12}+\d^{\f14}\M R_1+\f{\d^{\f14}}{|u|}\D_0 \M R_0^{\f12} \M R_1^{\f12}+\f{\d^{\f14}}{|u|}\D_0 \M R_0\\
&+\f{\d^{\f12}}{|u|}\D_0 \M O_{1,4}+\f{\d}{|u|}\D_0^2 \M O_{0,4},
\end{split}
\end{equation*}
implies
\begin{equation*}
\begin{split}
\d^{-\f12}\|\tr\chi\|_{L^{\infty}_{sc}({S}_{u,\ub})}\leq &\d^{-\f12}\|\tr\chi\|_{L^{\infty}_{sc}({S}_{u,0})}+\|\a\|^2_{L^{4}_{sc}({^{\d}S}_{u_{\infty},\ub})}\\
&+\l\f{1}{|u|^{\f12}}\underline{\M R}^{\f12}_1\underline{\M R}^{\f12}_2+\f{\d^{\f14}}{|u|^{\f12}}\underline{\M R}_1
+\f{\d^{\f12}}{|u|^{\f32}}\D_0\underline{\M R}^{\f12}_0\underline{\M R}^{\f12}_1+\f{\d^{\f34}}{|u|^{\f32}}\D_0\underline{\M R}_0+\M R_1^{\f12} \M R_2^{\f12}\r^2\\
&+\l\d^{\f14}\M R_1+\f{\d^{\f14}}{|u|}\D_0 \M R_0^{\f12} \M R_1^{\f12}+\f{\d^{\f14}}{|u|}\D_0 \M R_0+\f{\d^{\f12}}{|u|}\D_0 \M O_{1,4}+\f{\d}{|u|}\D_0^2 \M O_{0,4}\r^2.
\end{split}
\end{equation*}

\end{proof}

\section{ELLIPTIC ESTIMATES FOR THE THIRD DERIVATIVES OF RICCI COEFFICIENTS}
We then derive estimates for Ricci coefficients' third angular derivatives. This is achieved by a combination 
of transport estimates and elliptic estimates. To start with, we first prove two propositions for curvature components.

\subsection{$L^4_{sc}(\S)$ Estimates for Curvature Components}
For curvature components $\b,\rho,\sigma, \beb$, we have the following null Bianchi equations
$$\nab_4\b+2\tr\chi\b=\div\a-2\o\b+\eta\cdot\a,$$
$$\nab_4\rho+\f32\tr\chi\rho=\div\b-\f12\chibh\cdot\a+\zeta\cdot\b+2\etb\cdot\b,$$
$$\nab_4\sigma+\f32\tr\chi\sigma=-\div{^{*}\b}+\f12\chibh\cdot{^{*}\a}-\zeta\cdot{^{*}\b}-2\etb\cdot{^{*}\b},$$
$$\nab_4\beb+\tr\chi\beb=-\nab\rho+{^{*}\nab}\sigma+2\o\beb+2\chibh\b-3(\etb\rho-{^{*}\etb}\sigma).$$

With these equations we derive
\begin{proposition}\label{R04}
 \begin{equation*}
 \|\b,\rho,\sigma,\beb\|_{L^4_{sc}(\S)}\leq  \M R^{\f12}_1\M R^{\f12}_2+\d^{\f14}\M R_1+{\d^{\f14}}(\M I^{(0)}+\M R+\underline{\M R})(\M R^{\f12}_1\M R^{\f12}_2+\M R_0).
 \end{equation*}
\end{proposition}

\begin{proof}
For $\Psi^{(s,s')}\in\{\b,\rho,\sigma,\beb\}$, $\Psi^{(s,s')}$ satisfies the following systematical equation
$$\nab_4\Psi^{(s,s')}=\nab\Psi^{(s+\f12,s'-\f12)}+\sum_{s_1+s_2=s+1,\atop s'_1+s'_2=s'}\psi^{(s_1,s'_1)}\cdot\Psi^{(s_2,s'_2)}.$$
Employing Proposition \ref{transport}, we have
\begin{equation*}
\begin{split}
&\|\Psi^{(s,s')}\|_{L^4_{sc}(\S)}\\
\leq& \int_0^{\ub}\d^{-1}\|\nab\Psi^{(s+\f12,s'-\f12)}\|_{L^4_{sc}(S_{u,\ub'})} d\ub'\\
&+\sum_{s_1+s_2=s+1,\atop s'_1+s'_2=s'}\int_0^{\ub}\d^{-1}\f{\d^{\f12}}{|u|}\|\psi^{(s_1,s'_1)}\|_{L^{\infty}_{sc}(S_{u,\ub'})}\|\Psi^{(s_2,s'_2)}\|_{L^{4}_{sc}(S_{u,\ub'})}d \ub'\\
\leq& \int_0^{\ub}\d^{-1}\|\nab\Psi^{(s+\f12,s'-\f12)}\|^{\f12}_{L^2_{sc}(S_{u,\ub'})}\|\nab^2\Psi^{(s+\f12,s'-\f12)}\|^{\f12}_{L^2_{sc}(S_{u,\ub'})}d \ub'\\
&+\int_0^{\ub}\d^{-1}\d^{\f14}\|\nab\Psi^{(s+\f12,s'-\f12)}\|_{L^2_{sc}(S_{u,\ub'})}d\ub'\\
&+\sum_{s_1+s_2=s+1,\atop s'_1+s'_2=s'}\int_0^{\ub}\d^{-1}\f{\d^{\f12}}{|u|}\|\psi^{(s_1,s'_1)}\|_{L^{\infty}_{sc}(S_{u,\ub'})}\|\Psi^{(s_2,s'_2)}\|^{\f12}_{L^{2}_{sc}(S_{u,\ub'})}\|\nab\Psi^{(s_2,s'_2)}\|^{\f12}_{L^{2}_{sc}(S_{u,\ub'})}d \ub'\\
&+\sum_{s_1+s_2=s+1,\atop s'_1+s'_2=s'}\int_0^{\ub}\d^{-1}\f{\d^{\f12}}{|u|}\|\psi^{(s_1,s'_1)}\|_{L^{\infty}_{sc}(S_{u,\ub'})}\d^{\f14}\|\Psi^{(s_2,s'_2)}\|_{L^{2}_{sc}(S_{u,\ub'})}d \ub'\\
\leq&\|\nab\Psi^{(s+\f12,s'-\f12)}\|^{\f12}_{L^2_{sc}(\Hu)}\|\nab^2\Psi^{(s+\f12,s'-\f12)}\|^{\f12}_{L^2_{sc}(\Hu)}+\d^{\f14}\|\nab\Psi^{(s+\f12,s'-\f12)}\|_{L^2_{sc}(\Hu)}\\
&+\sum_{s_1+s_2=s+1,\atop s'_1+s'_2=s'}\d^{\f14}\sup_{u,\ub}\f{1}{|u|}\|\psi^{(s_1,s'_1)}\|_{L^{\infty}_{sc}(S_{u,\ub})}\d^{\f14}\|\Psi^{(s_2,s'_2)}\|^{\f12}_{L^2_{sc}(\Hu)}\|\nab\Psi^{(s_2,s'_2)}\|^{\f12}_{L^2_{sc}(\Hu)}\\
&+\sum_{s_1+s_2=s+1,\atop s'_1+s'_2=s'}\d^{\f14}\sup_{u,\ub}\f{1}{|u|}\|\psi^{(s_1,s'_1)}\|_{L^{\infty}_{sc}(S_{u,\ub})}\d^{\f12}\|\Psi^{(s_2,s'_2)}\|_{L^2_{sc}(\Hu)},
\end{split}
\end{equation*}
where we employ Proposition \ref{L4} and H\"older's inequality.
The anomalous term we encounter are among $\{\eta\cdot\a, \quad \chibh\cdot\a, \quad \chibh\cdot{^{*}\a}, \quad \chibh\cdot\b\}$. We control them
through the estimates above by adding suitable $\delta$ or $|u|$ weights in front of anomalous terms. Hence we obtain
 \begin{equation*}
 \|\b,\rho,\sigma,\beb\|_{L^4_{sc}(\S)}\leq  \M R^{\f12}_1\M R^{\f12}_2+\d^{\f14}\M R_1+{\d^{\f14}}\M O_{0,\infty}\M R^{\f12}_0\M R^{\f12}_1+{\d^{\f14}}\M O_{0,\infty}\M R_0.
 \end{equation*}
Combining the result in Proposition \ref{O0infty}, we derive  
 \begin{equation*}
 \|\b,\rho,\sigma,\beb\|_{L^4_{sc}(\S)}\leq  \M R^{\f12}_1\M R^{\f12}_2+\d^{\f14}\M R_1+{\d^{\f14}}C(\M I^{(0)}+\M R+\underline{\M R})(\M R^{\f12}_0\M R^{\f12}_1+\M R_0).
 \end{equation*}


\end{proof}

For $\K$, we have
\begin{proposition}
 \begin{equation*}
\begin{split}
&\|\K\|_{L^4_{sc}(\S)}\\
\leq&\M R^{\f12}_1\M R^{\f12}_2+\d^{\f14}\M R_1+\f{\d^{\f12}}{|u|}\M O_{0,\infty}\M R^{\f12}_0\M R^{\f12}_1+\f{\d^{\f34}}{|u|}\M O_{0,\infty}\M R_0+\f{\d^{\f12}}{|u|}\M O_{0,\infty}\M O_{1,4}\\
&+\f{\d}{|u|^2}\M O^3_{0,\infty}+\f{\d^{\f14}}{|u|^{\f12}}\M O_{0,\infty}\l1+\M R+\d^{\f14}(\M I^{(0)}+\M R+\underline{\M R})^3\r.
\end{split}
\end{equation*}
\end{proposition}

\begin{proof}
The curvature component $K$ obeys
\begin{equation*}
\begin{split}
\nab_4 K+\tr\chi K=&-\div\b-\zeta\cdot\b-2\etb\cdot\b+\f12\chih\cdot\nab\widehat{\otimes}\etb+\f12\chih\cdot(\etb\widehat{\otimes}\etb)\\
&-\f12\tr\chi\div\etb-\f12\tr\chi|\etb|^2.
\end{split}
\end{equation*}
Employing Proposition \ref{transport} and Gronwall's inequality, we derive
\begin{equation*}
\begin{split}
&\|K\|_{L^4_{sc}(\S)}\\
\leq &\|K\|_{L^4_{sc}(S_{u,0})}+\|\nab\b\|^{\f12}_{L^2_{sc}(\Hu)}\|\nab^2\b\|^{\f12}_{L^2_{sc}(\Hu)}+\d^{\f14}\|\nab\b\|_{L^2_{sc}(\Hu)}\\
&+\f{\d^{\f12}}{|u|}\sup_{u,\ub}\|\eta,\etb\|_{L_{sc}^{\infty}(\S)}\|\b\|^{\f12}_{L^2_{sc}(\Hu)}\|\nab\b\|^{\f12}_{L^2_{sc}(\Hu)}\\
&+\f{\d^{\f34}}{|u|}\sup_{u,\ub}\|\eta,\etb\|_{L_{sc}^{\infty}(\S)}\|\b\|_{L^2_{sc}(\Hu)}+\f{\d^{\f12}}{|u|}\sup_{u,\ub}\|\chi\|_{L_{sc}^{\infty}(\S)}\|\nab\etb\|_{L_{sc}^{4}(\S)}\\
&+\f{\d}{|u|^2}\sup_{u,\ub}\|\chi\|_{L_{sc}^{\infty}(\S)}\|\etb\|^2_{L_{sc}^{\infty}(\S)}.
\end{split}
\end{equation*}
For $\|K\|_{L^4_{sc}(S_{u,0})}$, since $K=\f{1}{|u|^2}$ on $S_{u,0}$, we have
$$\|K\|_{L^4_{sc}(S_{u,0})}=\d^{\f14}|u|.$$
Therefore, combining Propositions \ref{O04}, \ref{O14} and \ref{O0infty}, we get
\begin{equation*}
\begin{split}
\f{1}{|u|}\|K\|_{L^4_{sc}(\S)}\leq & \d^{\f14}+\M R+\d^{\f14}(\M I^{(0)}+\M R+\underline{\M R})^3.
\end{split}
\end{equation*}

This bound implies an improved estimate for $\K$. For $\K$, we have
\begin{equation*}
\begin{split}
\nab_4 (\K)=&-\div\b-\zeta\cdot\b-2\etb\cdot\b+\f12\chih\cdot\nab\widehat{\otimes}\etb+\f12\chih\cdot(\etb\widehat{\otimes}\etb)\\
&-\f12\tr\chi\div\etb-\f12\tr\chi|\etb|^2-\tr\chi K.
\end{split}
\end{equation*}
Applying Proposition \ref{transport} and the estimate above, we derive
\begin{equation*}
\begin{split}
&\|\K\|_{L^4_{sc}(\S)}\\
\leq &\|\nab\b\|^{\f12}_{L^2_{sc}(\Hu)}\|\nab^2\b\|^{\f12}_{L^2_{sc}(\Hu)}+\d^{\f14}\|\nab\b\|_{L^2_{sc}(\Hu)}\\
&+\f{\d^{\f12}}{|u|}\sup_{u,\ub}\|\eta,\etb\|_{L_{sc}^{\infty}(\S)}\|\b\|^{\f12}_{L^2_{sc}(\Hu)}\|\nab\b\|^{\f12}_{L^2_{sc}(\Hu)}\\
&+\f{\d^{\f34}}{|u|}\sup_{u,\ub}\|\eta,\etb\|_{L_{sc}^{\infty}(\S)}\|\b\|_{L^2_{sc}(\Hu)}+\f{\d^{\f12}}{|u|}\sup_{u,\ub}\|\chi\|_{L_{sc}^{\infty}(\S)}\|\nab\etb\|_{L_{sc}^{4}(\S)}\\
&+\f{\d}{|u|^2}\sup_{u,\ub}\|\chi\|_{L_{sc}^{\infty}(\S)}\|\etb\|^2_{L_{sc}^{\infty}(\S)}+\d^{\f12}\sup_{u,\ub}\|\tr\chi\|_{L_{sc}^{\infty}(\S)}\f{1}{|u|}\|K\|_{L_{sc}^{4}(\S)}.
\end{split}
\end{equation*}
Putting this and the estimate for $\|K\|_{L^4_{sc}(\S)}$ together, we deduce
\begin{equation*}
\begin{split}
&\|\K\|_{L^4_{sc}(\S)}\\
\leq&\M R^{\f12}_1\M R^{\f12}_2+\d^{\f14}\M R_1+\f{\d^{\f12}}{|u|}\M O_{0,\infty}\M R^{\f12}_0\M R^{\f12}_1+\f{\d^{\f34}}{|u|}\M O_{0,\infty}\M R_0+\f{\d^{\f12}}{|u|}\M O_{0,\infty}\M O_{1,4}\\
&+\f{\d}{|u|^2}\M O^3_{0,\infty}+\d^{\f12}\M O_{0,\infty}\l \d^{\f14}+\M R+\d^{\f14}(\M I^{(0)}+\M R+\underline{\M R})^3\r.
\end{split}
\end{equation*}
\end{proof}

\subsection{$L^2_{sc}(\S)$ Estimates for First Derivatives of Curvature Components}
For $\nab\b,\nab\rho,\nab\sigma,\nab\beb$, the following estimate holds
\begin{proposition}\label{R12}
 \begin{equation*}
\begin{split}
 \|\nab\b,\nab\rho,\nab\sigma,\nab\beb\|_{L^2_{sc}(\S)}\leq&  \M R_2+\d^{\f14}(\M I^{(0)}+\M R+\underline{\M R})(\M R^{\f12}_0\M R^{\f12}_1+\M R_0)\\
&+{\d^{\f12}}(\M I^{(0)}+\M R+\underline{\M R})\M R_1+\f{\d^{\f12}}{|u|}(\M I^{(0)}+\M R+\underline{\M R})^2\M R_0\\
&+(\M R^{\f12}_0\M R^{\f12}_1+\d^{\f14}\M R_0)^2.
\end{split} 
\end{equation*}
\end{proposition}

\begin{proof}
For $\Psi^{(s,s')}\in\{\b,\rho,\sigma,\beb\}$, $\Psi^{(s,s')}$ satisfies the systematical equation:

\begin{equation*} 
\begin{split}
& \nab_4 \nab\Psi^{(s,s')}\\
=&\nab^2\Psi^{(s+\f12,s'-\f12)}+\sum_{s_1+s_2=s+1, \atop s'_1+s'_2=s'}\nab\psi^{(s_1, s'_1)}\cdot\Psi^{(s_2, s'_2)}+\sum_{s_1+s_2=s+1, \atop s'_1+s'_2=s'}\psi^{(s_1, s'_1)}\cdot\nab\Psi^{(s_2, s'_2)}\\
&+\sum_{s_3+s_4+s_5=s+\f32, \atop s'_3+s'_4+s'_5=s'+\f12}\psi^{(s_3, s'_3)}\cdot\psi^{(s_4, s'_4)}\cdot\Psi^{(s_5,s'_5)}+\sum_{s_6+s_7=s+1, \atop s'_6+s'_7=s'}\Psi^{(s_6,s'_6)}\cdot\nab\Psi^{(s_7,s'_7)}.
\end{split}
\end{equation*}

Employing Propositions \ref{transport} and \ref{L4}, we infer
\begin{equation*}
\begin{split}
&\|\nab\Psi^{(s,s')}\|_{L^2_{sc}(\S)}\\
\leq& \int_0^{\ub}\d^{-1}\|\nab^2\Psi^{(s+\f12,s'-\f12)}\|_{L^2_{sc}(S_{u,\ub'})} d\ub'\\
&+\sum_{s_1+s_2=s+1,\atop s'_1+s'_2=s'}\int_0^{\ub}\d^{-1}\f{\d^{\f12}}{|u|}\|\nab\psi^{(s_1,s'_1)}\|_{L^{4}_{sc}(S_{u,\ub'})}\|\Psi^{(s_2,s'_2)}\|_{L^{4}_{sc}(S_{u,\ub'})}d \ub'\\
&+\sum_{s_1+s_2=s+1,\atop s'_1+s'_2=s'}\int_0^{\ub}\d^{-1}\f{\d^{\f12}}{|u|}\|\psi^{(s_1,s'_1)}\|_{L^{\infty}_{sc}(S_{u,\ub'})}\|\nab\Psi^{(s_2,s'_2)}\|_{L^{2}_{sc}(S_{u,\ub'})}d \ub'\\
&+\sum_{s_3+s_4+s_5=s+\f32, \atop s'_3+s'_4+s'_5=s'+\f12}\int_0^{\ub}\d^{-1}\f{\d}{|u|^2}\|\psi^{(s_3,s'_3)}\|_{L^{\infty}_{sc}(S_{u,\ub'})}\|\psi^{(s_4,s'_4)}\|_{L^{\infty}_{sc}(S_{u,\ub'})}\|\Psi^{(s_5,s'_5)}\|_{L^{2}_{sc}(S_{u,\ub'})}d \ub'\\
&+\sum_{s_6+s_7=s+1, \atop s'_6+s'_7=s'}\int_0^{\ub}\d^{-1}\|\Psi^{(s_6,s'_6)}\|_{L^4_{sc}(S_{u,\ub'})}\|\nab\Psi^{(s_7,s'_7)}\|_{L^4_{sc}(S_{u,\ub'})} d \ub'\\
\end{split}
\end{equation*}

\begin{equation*}
\begin{split}
\leq& \|\nab^2\Psi^{(s+\f12,s'-\f12)}\|_{L^2_{sc}(\Hu)}\\
&+\d^{\f14}\sum_{s_1+s_2=s+1,\atop s'_1+s'_2=s'}\sup_{u,\ub}\f{1}{|u|}\|\nab\psi^{(s_1,s'_1)}\|_{L^{4}_{sc}(S_{u,\ub})}\d^{\f14}\|\Psi^{(s_2,s'_2)}\|^{\f12}_{L^2_{sc}(\Hu)}\|\nab\Psi^{(s_2,s'_2)}\|^{\f12}_{L^2_{sc}(\Hu)}\\
&+\d^{\f14}\sum_{s_1+s_2=s+1,\atop s'_1+s'_2=s'}\sup_{u,\ub}\f{1}{|u|}\|\nab\psi^{(s_1,s'_1)}\|_{L^{4}_{sc}(S_{u,\ub})}\d^{\f12}\|\Psi^{(s_2,s'_2)}\|_{L^2_{sc}(\Hu)}\\
&+\d^{\f12}\sum_{s_1+s_2=s+1,\atop s'_1+s'_2=s'}\sup_{u,\ub}\f{1}{|u|}\|\psi^{(s_1,s'_1)}\|_{L^{\infty}_{sc}(S_{u,\ub})}\|\nab\Psi^{(s_2,s'_2)}\|_{L^2_{sc}(\Hu)}\\
&+\f{\d^{\f12}}{|u|}\sum_{s_3+s_4+s_5=s+\f32, \atop s'_3+s'_4+s'_5=s'+\f12}\sup_{u,\ub}\f{1}{|u|}\|\psi^{(s_3,s'_3)}\|_{L^{\infty}_{sc}(S_{u,\ub})}\|\psi^{(s_4,s'_4)}\|_{L^{\infty}_{sc}(S_{u,\ub})}\d^{\f12}\|\Psi^{(s_5,s'_5)}\|_{L^2_{sc}(\Hu)}\\
&+\sum_{s_6+s_7=s+1, \atop s'_6+s'_7=s'}\l\|\Psi^{(s_6,s'_6)}\|^{\f12}_{L^2_{sc}(\Hu)}\|\nab\Psi^{(s_6,s'_6)}\|^{\f12}_{L^2_{sc}(\Hu)}+\d^{\f14}\|\Psi^{(s_6,s'_6)}\|_{L^2_{sc}(\Hu)}\r\\
&\quad\quad\times\l\|\nab\Psi^{(s_7,s'_7)}\|^{\f12}_{L^2_{sc}(\Hu)}\|\nab^2\Psi^{(s_7,s'_7)}\|^{\f12}_{L^2_{sc}(\Hu)}+\d^{\f14}\|\nab\Psi^{(s_7,s'_7)}\|_{L^2_{sc}(\Hu)}\r.\\
\end{split}
\end{equation*}
The anomalous term we have are among 
$$\{\nab\eta\a,\quad \nab\chibh\a,\quad \nab\chibh\b,\quad \chibh\nab\a,\quad \chibh\nab\b,\quad (\eta,\etb)\eta\a,\quad (\eta,\etb)\chibh\a,\quad (\eta,\etb)\chibh\b\}.$$ 
All of them are controlled through the estimates above. Hence we obtain
 \begin{equation*}
\begin{split}
 \|\nab\b,\nab\rho,\nab\sigma,\nab\beb\|_{L^2_{sc}(\S)}\leq&  \M R_2+\d^{\f14}\M O_{1,4}(\M R^{\f12}_0\M R^{\f12}_1+\M R_0)+{\d^{\f12}}\M O_{0,\infty}\M R_1+\f{\d^{\f12}}{|u|}\M O^2_{0,\infty}\M R_0\\
&+(\M R^{\f12}_0\M R^{\f12}_1+\d^{\f14}\M R_0)^2.
\end{split} 
\end{equation*}
Combining the result in Propositions \ref{O0infty} and \ref{O14}, we demonstrate  
\begin{equation*}
\begin{split}
 \|\nab\b,\nab\rho,\nab\sigma,\nab\beb\|_{L^2_{sc}(\S)}\leq&  \M R_2+\d^{\f14}(\M I^{(0)}+\M R+\underline{\M R})(\M R^{\f12}_0\M R^{\f12}_1+\M R_0)\\
&+{\d^{\f12}}(\M I^{(0)}+\M R+\underline{\M R})\M R_1+\f{\d^{\f12}}{|u|}(\M I^{(0)}+\M R+\underline{\M R})^2\M R_0\\
&+(\M R^{\f12}_0\M R^{\f12}_1+\d^{\f14}\M R_0)^2.
\end{split} 
\end{equation*}

\end{proof}

The curvature component $K$ satisfies 
\begin{proposition}
 \begin{equation*}
\begin{split}
\|\nab K\|_{L^2_{sc}(\S)}\leq& \M R+\delta^{\f14}\l\M I^{(0)}+\M R+\underline{\M R}\r^3.
\end{split} 
\end{equation*}
\end{proposition}

\begin{proof}
For $K$, we have
$$K=-\rho+\f12\chih\cdot\chibh-\f14\tr\chi\tr\chib.$$
This is equivalent to
$$\K=-\rho+\f12\chih\cdot\chibh-\f14\tr\chi(\tr\chib+\f{2}{|u|})+\f{1}{2|u|}(\tr\chi-\f{2}{|u|}).$$
Hence, we get
\begin{equation*}
\begin{split}
&\|\nab K\|_{L^2_{sc}(\S)}\\
=&\|\nab (\K)\|_{L^2_{sc}(\S)}\\
\leq&\|\nab \rho\|_{L^2_{sc}(\S)}+\|\chih\nab\chibh\|_{L^2_{sc}(\S)}+\|\nab\chih\chibh\|_{L^2_{sc}(\S)}\\
&+\|\nab\tr\chi (\tr\chib+\f{2}{|u|})\|_{L^2_{sc}(\S)}+\|\tr\chi\nab\tr\chib\|_{L^2_{sc}(\S)}\\
&+\f{1}{|u|}\|\nab\tr\chi\|_{L^2_{sc}(\S)}\\
\leq&\|\nab \rho\|_{L^2_{sc}(\S)}+\d^{\f14}\d^{\f14}\|\chih\|_{L^{4}_{sc}(\S)}\f{1}{|u|}\|\nab\chibh\|_{L^4_{sc}(\S)}+\d^{\f14}\|\nab\chih\|_{L^4_{sc}(\S)}\f{\d^{\f14}}{|u|}\|\chibh\|_{L^{4}_{sc}(\S)}\\
&+{\d^{\f12}}\|\nab\tr\chi\|_{L^2_{sc}(\S)}\f{1}{|u|}\|\tr\chib+\f{2}{|u|}\|_{L^{\infty}_{sc}(\S)}+\f{\d^{\f12}}{|u|}\|\tr\chi\|_{L^{\infty}_{sc}(\S)}\|\nab\tr\chib\|_{L^2_{sc}(\S)}\\
&+\f{1}{|u|}\|\nab\tr\chi\|_{L^2_{sc}(\S)}.\\
\end{split} 
\end{equation*}

Let $\phi$ be $S_{u,\ub}$-tangent tensor field. With the definition of scale invariant norms and H\"older's inequality, we have
$$\|\phi\|_{L^2_{sc}(\S)}\leq \d^{\f14}\|\phi\|_{L^4_{sc}(\S)}.$$

Therefore, we deduce

\begin{equation*}
\begin{split}
&\|\nab K\|_{L^2_{sc}(\S)}\\
\leq&\|\nab \rho\|_{L^2_{sc}(\S)}+\d^{\f14}\d^{\f14}\|\chih\|_{L^{4}_{sc}(\S)}\f{1}{|u|}\|\nab\chibh\|_{L^4_{sc}(\S)}+\d^{\f14}\|\nab\chih\|_{L^4_{sc}(\S)}\f{\d^{\f14}}{|u|}\|\chibh\|_{L^{4}_{sc}(\S)}\\
&+{\d^{\f34}}\|\nab\tr\chi\|_{L^4_{sc}(\S)}\f{1}{|u|}\|\tr\chib+\f{2}{|u|}\|_{L^{\infty}_{sc}(\S)}+\f{\d^{\f34}}{|u|}\|\tr\chi\|_{L^{\infty}_{sc}(\S)}\|\nab\tr\chib\|_{L^4_{sc}(\S)}\\
&+\f{\d^{\f14}}{|u|}\|\nab\tr\chi\|_{L^4_{sc}(\S)}.\\
\end{split} 
\end{equation*}

From Propositions  \ref{O04}, \ref{O0infty trchib}, \ref{O14}, \ref{O0infty} and \ref{R12}, have
\begin{equation*}
\begin{split}
\|\nab K\|_{L^2_{sc}(\S)}\leq& \M R+\delta^{\f14}\l\M I^{(0)}+\M R+\underline{\M R}\r^3.
\end{split} 
\end{equation*}

\end{proof}

\subsection{$L^2_{sc}(\S)$ Estimates for Third Derivatives of Ricci Coefficients}
To get estimates for the third derivatives of Ricci coefficients,  if we will simply employ transport equations for them. We may encounter some trouble. Because the third derivatives
of curvature components will be involved and it will result in lose of derivatives. To deal with this problem, we define new renormalized quantities $\T^{(s,s')}$ with signature $(s,s')$. And we will derive and use transport equations for $\T^{(s,s')}$ instead. We have
$$\T^{(s,s')}\in \{\nab\tr\chi,\nab\tr\chib, \mu, \underline{\mu}, \kappa, \underline{\kappa}\}.$$
We define $\mu$ and $\underline{\mu}$ through
$$\mu:=-\div\eta-\rho,$$
$$\underline{\mu}:=-\div\etb-\rho.$$
Let $\o^{\dagger}$ solve equation
$$\nab_3\o^{\dagger}=\f12\sigma,$$
with zero initial data on $H_{u_{\infty}}$.
And we define $\kappa$ through $\o^{\dagger}$:
$$\kappa:=\nab\o+{^{*}\nab}\o^{\dagger}-\f12\b.$$
Similarly, we define $\underline{\kappa}$ through:
$$\underline{\kappa}:=-\nab\omb+{^{*}\nab}\omb^{\dagger}-\f12\beb.$$
And $\omb^{\dagger}$ is defined to be the solution to 
$$\nab_4\omb^{\dagger}=\f12\sigma,$$
with zero initial data on $\Hb_0$.

Using null structure equations, null Bianchi equations and Propositions \ref{commute0}-\ref{commute3}, it can be demonstrated that 
$\T^{(s,s')}$ obeys the following systematical transport equations:
\begin{equation*}
\begin{split}
&\nab_4\T^{(s,s')}\\
=&\chi^{(1,0)}\cdot\T^{(s,s')}+\sum_{s_1+s_2=s+\f12, \atop s'_1+s'_2=s'-\f12}\nab\psi^{(s_1,s'_1)}\cdot\psi^{(s_2,s'_2)}\\
&+\sum_{s_3+s_4=s+1, \atop s'_3+s'_4=s'}\psi^{(s_3,s'_3)}\cdot\Psi^{(s_4,s'_4)}+\sum_{s_5+s_6+s_7=s+1, \atop s'_5+s'_6+s'_7=s'}\psi^{(s_5, s'_5)}\cdot\psi^{(s_6, s'_6)}\cdot\psi^{(s_7, s'_7)},
\end{split}
\end{equation*}
or
\begin{equation*}
\begin{split}
&\nab_3\T^{(s,s')}+\lambda[\T^{(s,s')}]\tr\chib\T^{(s,s')}\\
=&\chibh^{(0,1)}\cdot\T^{(s,s')}+\sum_{s_1+s_2=s-\f12, \atop s'_1+s'_2=s'+\f12}\nab\psi^{(s_1,s'_1)}\cdot\psi^{(s_2,s'_2)}\\
&+\sum_{s_3+s_4=s, \atop s'_3+s'_4=s'+1}\psi^{(s_3,s'_3)}\cdot\Psi^{(s_4,s'_4)}+\sum_{s_5+s_6+s_7=s, \atop s'_5+s'_6+s'_7=s'+1}\psi^{(s_5, s'_5)}\cdot\psi^{(s_6, s'_6)}\cdot\psi^{(s_7, s'_7)}.
\end{split}
\end{equation*}
In this section, $\psi^{(s,s')}$ and $\Psi^{(s,s')}$ represent an arbitrary Ricci coefficient with signature $(s,s')$ and a null curvature component with signature $(s,s')$, respectively.
$\lambda[\psi^{(s,s')}]$ is a constant depending on $\psi^{(s,s')}$.

Similarly, we can check $\T^{(s,s')}$ satisfies the following systematical transport equations:
\begin{equation*}
\begin{split}
&\nab_4\nab\T^{(s,s')}\\
=&\chi^{(1,0)}\cdot\nab\T^{(s,s')}+\sum_{s_1+s_2=s+\f12, \atop s'_1+s'_2=s'-\f12}\nab^2\psi^{(s_1,s'_1)}\cdot\psi^{(s_2,s'_2)}\\
&+\sum_{s_1+s_2=s+\f12, \atop s'_1+s'_2=s'-\f12}\nab\psi^{(s_1,s'_1)}\cdot\nab\psi^{(s_2,s'_2)}+\sum_{s_3+s_4=s+1, \atop s'_3+s'_4=s'}\nab\psi^{(s_3,s'_3)}\cdot\Psi^{(s_4,s'_4)}\\
&+\sum_{s_3+s_4=s+1, \atop s'_3+s'_4=s'}\psi^{(s_3,s'_3)}\cdot\nab\Psi^{(s_4,s'_4)}+\sum_{s_5+s_6+s_7=s+1, \atop s'_5+s'_6+s'_7=s'}\nab\psi^{(s_5, s'_5)}\cdot\psi^{(s_6, s'_6)}\cdot\psi^{(s_7, s'_7)}\\
&+\sum_{s_8+s_9+s_{10}=s+\f32, \atop s'_8+s'_9+s'_{10}=s'+\f12}\psi^{(s_8, s'_8)}\cdot\psi^{(s_9, s'_9)}\cdot\Psi^{(s_{10}, s'_{10})}+\sum_{s_{11}+s_{12}=s+\f32, \atop s'_{11}+s'_{12}=s'+\f12}\Psi^{(s_{11},s'_{11})}\cdot\Psi^{(s_{12},s'_{12})}\\
&+\sum_{s_{13}+s_{14}+s_{15}+s_{16}=s+\f32, \atop s'_{13}+s'_{14}+s'_{15}+s'_{16}=s'+\f12}\psi^{(s_{13}, s'_{13})}\cdot\psi^{(s_{14}, s'_{14})}\cdot\psi^{(s_{15}, s'_{15})}\cdot\psi^{(s_{16}, s'_{16})},
\end{split}
\end{equation*}
or

\begin{equation*}
\begin{split}
&\nab_3\nab\T^{(s,s')}+\l \lambda[\T^{(s,s')}]+\f12\r\tr\chib\T^{(s,s')}\\
=&\chibh^{(0,1)}\cdot\nab\T^{(s,s')}+\sum_{s_1+s_2=s-\f12, \atop s'_1+s'_2=s'+\f12}\nab^2\psi^{(s_1,s'_1)}\cdot\psi^{(s_2,s'_2)}\\
&+\sum_{s_1+s_2=s-\f12, \atop s'_1+s'_2=s'+\f12}\nab\psi^{(s_1,s'_1)}\cdot\nab\psi^{(s_2,s'_2)}+\sum_{s_3+s_4=s, \atop s'_3+s'_4=s'+1}\nab\psi^{(s_3,s'_3)}\cdot\Psi^{(s_4,s'_4)}\\
&+\sum_{s_3+s_4=s, \atop s'_3+s'_4=s'+1}\psi^{(s_3,s'_3)}\cdot\nab\Psi^{(s_4,s'_4)}+\sum_{s_5+s_6+s_7=s, \atop s'_5+s'_6+s'_7=s'+1}\nab\psi^{(s_5, s'_5)}\cdot\psi^{(s_6, s'_6)}\cdot\psi^{(s_7, s'_7)}\\
&+\sum_{s_8+s_9+s_{10}=s+\f12, \atop s'_8+s'_9+s'_{10}=s'+\f32}\psi^{(s_8, s'_8)}\cdot\psi^{(s_9, s'_9)}\cdot\Psi^{(s_{10}, s'_{10})}+\sum_{s_{11}+s_{12}=s+\f12, \atop s'_{11}+s'_{12}=s'+\f32}\Psi^{(s_{11},s'_{11})}\cdot\Psi^{(s_{12},s'_{12})}\\
&+\sum_{s_{13}+s_{14}+s_{15}+s_{16}=s+\f12, \atop s'_{13}+s'_{14}+s'_{15}+s'_{16}=s'+\f32}\psi^{(s_{13}, s'_{13})}\cdot\psi^{(s_{14}, s'_{14})}\cdot\psi^{(s_{15}, s'_{15})}\cdot\psi^{(s_{16}, s'_{16})}.
\end{split}
\end{equation*}

And $\T^{(s,s')}$ obeys

\begin{equation}\label{tranLLL}
\begin{split}
&\nab_4\nab^2\T^{(s,s')}\\
=&\chi^{(1,0)}\cdot\nab^2\T^{(s,s')}+\sum_{s_1+s_2=s+\f12, \atop s'_1+s'_2=s'-\f12}\nab^3\psi^{(s_1,s'_1)}\cdot\psi^{(s_2,s'_2)}\\
&+\sum_{s_1+s_2=s+\f12, \atop s'_1+s'_2=s'-\f12}\nab^2\psi^{(s_1,s'_1)}\cdot\nab\psi^{(s_2,s'_2)}+\sum_{s_3+s_4=s+1, \atop s'_3+s'_4=s'}\nab^2\psi^{(s_3,s'_3)}\cdot\Psi^{(s_4,s'_4)}\\
&+\sum_{s_3+s_4=s+1, \atop s'_3+s'_4=s'}\nab\psi^{(s_3,s'_3)}\cdot\nab\Psi^{(s_4,s'_4)}+\sum_{s_5+s_6+s_7=s+1, \atop s'_5+s'_6+s'_7=s'}\nab^2\psi^{(s_5, s'_5)}\cdot\psi^{(s_6, s'_6)}\cdot\psi^{(s_7, s'_7)}\\
&+\sum_{s_3+s_4=s+1, \atop s'_3+s'_4=s'}\psi^{(s_3,s'_3)}\cdot\nab^2\Psi^{(s_4,s'_4)}+\sum_{s_5+s_6+s_7=s+1, \atop s'_5+s'_6+s'_7=s'}\nab\psi^{(s_5, s'_5)}\cdot\nab\psi^{(s_6, s'_6)}\cdot\psi^{(s_7, s'_7)}\\
&+\sum_{s_8+s_9+s_{10}=s+\f32, \atop s'_8+s'_9+s'_{10}=s'+\f12}\nab\psi^{(s_8, s'_8)}\cdot\psi^{(s_9, s'_9)}\cdot\Psi^{(s_{10}, s'_{10})}\\
&+\sum_{s_8+s_9+s_{10}=s+\f32, \atop s'_8+s'_9+s'_{10}=s'+\f12}\psi^{(s_8, s'_8)}\cdot\psi^{(s_9, s'_9)}\cdot\nab\Psi^{(s_{10}, s'_{10})}+\sum_{s_{11}+s_{12}=s+\f32, \atop s'_{11}+s'_{12}=s'+\f12}\nab\Psi^{(s_{11},s'_{11})}\cdot\Psi^{(s_{12},s'_{12})}\\
&+\sum_{s_{13}+s_{14}+s_{15}+s_{16}=s+\f32, \atop s'_{13}+s'_{14}+s'_{15}+s'_{16}=s'+\f12}\nab\psi^{(s_{13}, s'_{13})}\cdot\psi^{(s_{14}, s'_{14})}\cdot\psi^{(s_{15}, s'_{15})}\cdot\psi^{(s_{16}, s'_{16})}\\
&+\sum_{s_{17}+s_{18}+s_{19}+s_{20}=s+2, \atop s'_{17}+s'_{18}+s'_{19}+s'_{20}=s'+1}\psi^{(s_{17},s'_{17})}\cdot\psi^{(s_{18}, s'_{18})}\cdot\psi^{(s_{19}, s'_{19})}\cdot\Psi^{(s_{20}, s'_{20})}\\
&+\sum_{s_{21}+s_{22}+s_{23}=s+2, \atop s'_{21}+s'_{22}+s'_{23}=s'+1}\psi^{(s_{21},s'_{21})}\cdot\Psi^{(s_{22},s'_{22})}\cdot\Psi^{(s_{23},s'_{23})}\\
&+\sum_{s_{24}+s_{25}+s_{26}+s_{27}+s_{28}=s+2, \atop s'_{24}+s'_{25}+s'_{26}+s'_{27}+s'_{28}=s'+1}\psi^{(s_{24},s'_{24})}\cdot\psi^{(s_{25}, s'_{25})}\cdot\psi^{(s_{26}, s'_{26})}\cdot\psi^{(s_{27}, s'_{27})}\cdot\psi^{(s_{28}, s'_{28})},
\end{split}
\end{equation}

or

\begin{equation}\label{tranT3}
\begin{split}
&\nab_3\nab^2\T^{(s,s')}+\l\lambda[\T^{(s,s')}]+1\r\tr\chib\nab^2\T^{(s,s')}\\
=&\chibh^{(0,1)}\cdot\nab^2\T^{(s,s')}+\sum_{s_1+s_2=s-\f12, \atop s'_1+s'_2=s'+\f12}\nab^3\psi^{(s_1,s'_1)}\cdot\psi^{(s_2,s'_2)}\\
&+\sum_{s_1+s_2=s-\f12, \atop s'_1+s'_2=s'+\f12}\nab^2\psi^{(s_1,s'_1)}\cdot\nab\psi^{(s_2,s'_2)}+\sum_{s_3+s_4=s, \atop s'_3+s'_4=s'+1}\nab^2\psi^{(s_3,s'_3)}\cdot\Psi^{(s_4,s'_4)}\\
&+\sum_{s_3+s_4=s, \atop s'_3+s'_4=s'+1}\nab\psi^{(s_3,s'_3)}\cdot\nab\Psi^{(s_4,s'_4)}+\sum_{s_5+s_6+s_7=s, \atop s'_5+s'_6+s'_7=s'+1}\nab^2\psi^{(s_5, s'_5)}\cdot\psi^{(s_6, s'_6)}\cdot\psi^{(s_7, s'_7)}\\
&+\sum_{s_3+s_4=s, \atop s'_3+s'_4=s'+1}\psi^{(s_3,s'_3)}\cdot\nab^2\Psi^{(s_4,s'_4)}+\sum_{s_5+s_6+s_7=s, \atop s'_5+s'_6+s'_7=s'+1}\nab\psi^{(s_5, s'_5)}\cdot\nab\psi^{(s_6, s'_6)}\cdot\psi^{(s_7, s'_7)}\\
&+\sum_{s_8+s_9+s_{10}=s+\f12, \atop s'_8+s'_9+s'_{10}=s'+\f32}\nab\psi^{(s_8, s'_8)}\cdot\psi^{(s_9, s'_9)}\cdot\Psi^{(s_{10}, s'_{10})}\\
&+\sum_{s_8+s_9+s_{10}=s+\f12, \atop s'_8+s'_9+s'_{10}=s'+\f32}\psi^{(s_8, s'_8)}\cdot\psi^{(s_9, s'_9)}\cdot\nab\Psi^{(s_{10}, s'_{10})}+\sum_{s_{11}+s_{12}=s+\f12, \atop s'_{11}+s'_{12}=s'+\f32}\nab\Psi^{(s_{11},s'_{11})}\cdot\Psi^{(s_{12},s'_{12})}\\
&+\sum_{s_{13}+s_{14}+s_{15}+s_{16}=s+\f12, \atop s'_{13}+s'_{14}+s'_{15}+s'_{16}=s'+\f32}\nab\psi^{(s_{13}, s'_{13})}\cdot\psi^{(s_{14}, s'_{14})}\cdot\psi^{(s_{15}, s'_{15})}\cdot\psi^{(s_{16}, s'_{16})}\\
&+\sum_{s_{17}+s_{18}+s_{19}+s_{20}=s+1, \atop s'_{17}+s'_{18}+s'_{19}+s'_{20}=s'+2}\psi^{(s_{17},s'_{17})}\cdot\psi^{(s_{18}, s'_{18})}\cdot\psi^{(s_{19}, s'_{19})}\cdot\Psi^{(s_{20}, s'_{20})}\\
&+\sum_{s_{21}+s_{22}+s_{23}=s+1, \atop s'_{21}+s'_{22}+s'_{23}=s'+2}\psi^{(s_{21},s'_{21})}\cdot\Psi^{(s_{22},s'_{22})}\cdot\Psi^{(s_{23},s'_{23})}\\
&+\sum_{s_{24}+s_{25}+s_{26}+s_{27}+s_{28}=s+1, \atop s'_{24}+s'_{25}+s'_{26}+s'_{27}+s'_{28}=s'+2}\psi^{(s_{24},s'_{24})}\cdot\psi^{(s_{25}, s'_{25})}\cdot\psi^{(s_{26}, s'_{26})}\cdot\psi^{(s_{27}, s'_{27})}\cdot\psi^{(s_{28}, s'_{28})}.
\end{split}
\end{equation}

Now we are ready to estimate the second derivative of $\T^{(s,s')}$.
\begin{proposition}
For $\T^{(s,s')}$ satisfying (\ref{tranLLL}), we have 
\end{proposition}
\begin{equation*}
\begin{split}
\|\nab^2\T^{(s,s')}\|_{L^2_{sc}(\S)}\leq & \int_0^{\ub}\f{\d^{-\f12}}{|u|}\sum_{s_1+s_2=s+\f12, \atop s'_1+s'_2=s'-\f12}\|\nab^3\psi^{(s_1, s'_1)}\|_{L^2_{sc}(S_{u,\ub'})}\|\psi^{(s_2, s'_2)}\|_{L^{\infty}_{sc}(S_{u,\ub'})}d\ub'\\
&+\f{\d^{\f12}}{|u|}\M O_{2,4}\M O_{1,4}+\f{\d^{\f12}}{|u|}\M O_{2,4}\M R_0^{\f12}\M R_1^{\f12}+\f{\d^{\f34}}{|u|}\M O_{2,4}\M R_0\\
&+\f{\d^{\f12}}{|u|}\M O_{1,4}\M R_1^{\f12}\M R_2^{\f12}+\f{\d^{\f34}}{|u|}\M O_{1,4}\M R_1+\f{\d}{|u|}\D_0\M O_{2,4}\M O_{0,4}\\
&+\f{\d}{|u|}\D_0\M O_{1,4}\M O_{1,4}+\f{\d}{|u|}\D_0\M O_{1,4}\M R_1^{\f12}\M R_2^{\f12}+\f{\d^{\f54}}{|u|}\D_0\M O_{1,4}\M R_1\\
&+\f{\d}{|u|}\D_0^2\M R_1+\f{\d^{\f12}}{|u|}\M R_1\M R_2^{\f12}\M R_0^{\f12}+\f{\d^{\f34}}{|u|}\M R_1^{\f12}\M R_2^{\f12}\M R_0+\f{\d^{\f34}}{|u|}\M R_1^{\f32}\M R_0^{\f12}+\f{\d}{|u|}\M R_1\M R_0\\
&+\f{\d^{\f32}}{|u|^3}\D_0^2\M O_{1,4}\M O_{0,4}+\f{\d^{\f32}}{|u|^3}\D_0^3\M R_0+\f{\d}{|u|^2}\D_0\M R_0\M R_1\\
&+\f{\d^{\f54}}{|u|^2}\D_0\M R_0^{\f32}\M R_1^{\f12}+\f{\d^{\f32}}{|u|^2}\D_0\M R_0^2+\f{\d^2}{|u|^4}\D_0^4\M R_0^2.
\end{split}
\end{equation*}

\begin{proof}
Employing Proposition \eqref{transport}, it follows that  
\begin{equation*}
\begin{split}
&\|\nab^2\T^{(s,s')}\|_{L^2_{sc}(\S)}\\
\leq&E_1+E_{2}+E_{3}+E_{4}+E_{5}+E_{6}+E_{7}+E_{8}+E_{9}+E_{10}\\
&++E_{11}+E_{12}+E_{13}+E_{14}
\end{split}
\end{equation*}
where $E_1-E_{14}$ are given as follows:
\begin{equation*}
\begin{split}
E_1=& \int_0^{\ub}\f{\d^{-\f12}}{|u|}\sum_{s_1+s_2=s+\f12, \atop s'_1+s'_2=s'-\f12}\|\nab^3\psi^{(s_1, s'_1)}\|_{L^2_{sc}(S_{u,\ub'})}\|\psi^{(s_2, s'_2)}\|_{L^{\infty}_{sc}(S_{u,\ub'})} d\ub',\\
E_2=&\int_0^{\ub}\f{\d^{-\f12}}{|u|}\sum_{s_1+s_2=s+\f12, \atop s'_1+s'_2=s'-\f12}\|\nab^2\psi^{(s_1,s'_1)}\|_{L^4_{sc}(S_{u,\ub'})}\|\nab\psi^{(s_2,s'_2)}\|_{L^4_{sc}(S_{u,\ub'})} d\ub',\\
E_3=&\int_0^{\ub}\f{\d^{-\f12}}{|u|}\sum_{s_3+s_4=s+1, \atop s'_3+s'_4=s'}\|\nab^2\psi^{(s_3,s'_3)}\|_{L^4_{sc}(S_{u,\ub'})}\|\Psi^{(s_4,s'_4)}\|_{L^4_{sc}(S_{u,\ub'})} d\ub',\\
E_4=&\int_0^{\ub}\f{\d^{-\f12}}{|u|}\sum_{s_3+s_4=s+1, \atop s'_3+s'_4=s'}\|\nab\psi^{(s_3,s'_3)}\|_{L^4_{sc}(S_{u,\ub'})}\|\nab\Psi^{(s_4,s'_4)}\|_{L^4_{sc}(S_{u,\ub'})}d\ub',\\
E_5=&\int_0^{\ub}\f{\d^{-\f12}}{|u|}\sum_{s_5+s_6+s_7=s+1, \atop s'_5+s'_6+s'_7=s'}\|\nab^2\psi^{(s_5, s'_5)}\|_{L^4_{sc}(S_{u,\ub'})}\|\psi^{(s_6, s'_6)}\cdot\psi^{(s_7, s'_7)}\|_{L^4_{sc}(S_{u,\ub'})} d\ub',\\
E_6=&\int_0^{\ub}\f{1}{|u|^2}\sum_{s_5+s_6+s_7=s+1, \atop s'_5+s'_6+s'_7=s'}\|\nab\psi^{(s_5, s'_5)}\|_{L^4_{sc}(S_{u,\ub'})}\|\nab\psi^{(s_6, s'_6)}\|_{L^4_{sc}(S_{u,\ub'})}\|\psi^{(s_7, s'_7)}\|_{L^{\infty}_{sc}(S_{u,\ub'})} d\ub',\\
E_7=&\int_0^{\ub}\f{1}{|u|^2}\sum_{s_8+s_9+s_{10}=s+\f32, \atop s'_8+s'_9+s'_{10}=s'+\f12}\|\nab\psi^{(s_8, s'_8)}\|_{L^4_{sc}(S_{u,\ub'})}\|\psi^{(s_9, s'_9)}\|_{L^{\infty}_{sc}(S_{u,\ub'})}\|\Psi^{(s_{10}, s'_{10})}\|_{L^4_{sc}(S_{u,\ub'})} d\ub',\\
E_8=&\int_0^{\ub}\f{\d^{-\f12}}{|u|}\sum_{s_8+s_9+s_{10}=s+\f32, \atop s'_8+s'_9+s'_{10}=s'+\f12}\|\psi^{(s_8, s'_8)}\cdot\psi^{(s_9, s'_9)}\|_{L^4_{sc}(S_{u,\ub'})}\|\nab\Psi^{(s_{10}, s'_{10})}\|_{L^4_{sc}(S_{u,\ub'})} d\ub',\\
E_9=&\int_0^{\ub}\f{\d^{-\f12}}{|u|}\sum_{s_{11}+s_{12}=s+\f32, \atop s'_{11}+s'_{12}=s'+\f12}\|\nab\Psi^{(s_{11},s'_{11})}\|_{L^4_{sc}(S_{u,\ub'})}\|\Psi^{(s_{12},s'_{12})}\|_{L^4_{sc}(S_{u,\ub'})} d\ub',\\
E_{10}=&\int_0^{\ub}\f{\d^{-\f12}}{|u|}\sum_{s_{13}+s_{14}+s_{15}+s_{16}=s+\f32, \atop s'_{13}+s'_{14}+s'_{15}+s'_{16}=s'+\f12}\|\nab\psi^{(s_{13}, s'_{13})}\|_{L^4_{sc}(S_{u,\ub'})}\\
&\quad\times\|\psi^{(s_{14}, s'_{14})}\cdot\psi^{(s_{15}, s'_{15})}\cdot\psi^{(s_{16}, s'_{16})}\|_{L^4_{sc}(S_{u,\ub'})} d\ub',\\
E_{11}=&\int_0^{\ub}\f{\d^{-\f12}}{|u|}\sum_{s_{17}+s_{18}+s_{19}+s_{20}=s+2, \atop s'_{17}+s'_{18}+s'_{19}+s'_{20}=s'+1}\|\psi^{(s_{17},s'_{17})}\cdot\psi^{(s_{18}, s'_{18})}\cdot\psi^{(s_{19}, s'_{19})}\|_{L^4_{sc}(S_{u,\ub'})}\\
&\quad\times\|\Psi^{(s_{20}, s'_{20})}\|_{L^4_{sc}(S_{u,\ub'})} d\ub', \\
E_{12}=&\int_0^{\ub}\f{1}{|u|^2}\sum_{s_{21}+s_{22}+s_{23}=s+2, \atop s'_{21}+s'_{22}+s'_{23}=s'+1}\|\psi^{(s_{21},s'_{21})}\|_{L^{\infty}_{sc}(S_{u,\ub'})}\|\Psi^{(s_{22},s'_{22})}\|_{L^4_{sc}(S_{u,\ub'})}\\
&\quad\times\|\Psi^{(s_{23},s'_{23})}\|_{L^4_{sc}(S_{u,\ub'})} d\ub',\\
E_{13}=&\int_0^{\ub}\f{\d^{-\f12}}{|u|}\sum_{s_{24}+s_{25}+s_{26}+s_{27}+s_{28}=s+2, \atop s'_{24}+s'_{25}+s'_{26}+s'_{27}+s'_{28}=s'+1}\|\psi^{(s_{24},s'_{24})}\cdot\psi^{(s_{25}, s'_{25})}\cdot\psi^{(s_{26}, s'_{26})}\cdot\psi^{(s_{27}, s'_{27})}\|_{L^{4}_{sc}(S_{u,\ub'})}\\
&\quad\times\|\psi^{(s_{28}, s'_{28})}\|_{L^4_{sc}(S_{u,\ub'})} d\ub',\\
\end{split}
\end{equation*}

\begin{equation*}
\begin{split}
E_{14}=&\int_0^{\ub}\f{\d^{-\f12}}{|u|}\sum_{s_3+s_4=s+1, \atop s'_3+s'_4=s'}\|\psi^{(s_3,s'_3)}\|_{L^{\infty}_{sc}(S_{u,\ub'})}\|\nab^2\Psi^{(s_4,s'_4)}\|_{L^2_{sc}(S_{u,\ub'})}d\ub'. \\
\end{split}
\end{equation*}

Adopting Proposition \ref{L4} and H\"older's inequality, we demonstrate the estimates below:
\begin{equation*}
\begin{split}
E_2\leq{\d^{\f12}}\sum_{s_1+s_2=s+\f12, \atop s'_1+s'_2=s'-\f12}\sup_{u',\ub'}\f{1}{|u'|}\|\nab^2\psi^{(s_1,s'_1)}\|_{L^4_{sc}(S_{u',\ub'})}\|\nab\psi^{(s_2,s'_2)}\|_{L^4_{sc}(S_{u',\ub'})},\\
\end{split}
\end{equation*}

\begin{equation*}
\begin{split}
E_3\leq&{\d^{\f14}}\sum_{s_3+s_4=s+1, \atop s'_3+s'_4=s'}\sup_{u',\ub'}\f{1}{|u'|}\|\nab^2\psi^{(s_3,s'_3)}\|_{L^4_{sc}(S_{u',\ub'})}\d^{\f14}\|\Psi^{(s_4,s'_4)}\|^{\f12}_{L^2_{sc}(\Hu)}\|\nab\Psi^{(s_4,s'_4)}\|^{\f12}_{L^2_{sc}(\Hu)}\\
&+{\d^{\f14}}\sum_{s_3+s_4=s+1, \atop s'_3+s'_4=s'}\sup_{u',\ub'}\f{1}{|u'|}\|\nab^2\psi^{(s_3,s'_3)}\|_{L^4_{sc}(S_{u',\ub'})}\d^{\f12}\|\Psi^{(s_4,s'_4)}\|_{L^2_{sc}(\Hu)},
\end{split}
\end{equation*}

\begin{equation*}
\begin{split}
E_4\leq&{\d^{\f12}}\sum_{s_3+s_4=s+1, \atop s'_3+s'_4=s'}\sup_{u',\ub'}\f{1}{|u'|}\|\nab\psi^{(s_3,s'_3)}\|_{L^4_{sc}(S_{u',\ub'})}\|\nab\Psi^{(s_4,s'_4)}\|^{\f12}_{L^2_{sc}(\Hu)}\|\nab^2\Psi^{(s_4,s'_4)}\|^{\f12}_{L^2_{sc}(\Hu)}\\
&+\f{\d^{\f34}}{|u|}\sum_{s_3+s_4=s+1, \atop s'_3+s'_4=s'}\sup_{u',\ub'}\|\nab\psi^{(s_3,s'_3)}\|_{L^4_{sc}(S_{u',\ub'})}\|\nab\Psi^{(s_4,s'_4)}\|_{L^2_{sc}(\Hu)},
\end{split}
\end{equation*}

\begin{equation*}
\begin{split}
E_5\leq\f{\d}{|u|^2}\sum_{s_5+s_6+s_7=s+1, \atop s'_5+s'_6+s'_7=s'}\sup_{u',\ub'}\|\nab^2\psi^{(s_5, s'_5)}\|_{L^4_{sc}(S_{u',\ub'})}\|\psi^{(s_6, s'_6)}\|_{L^4_{sc}(S_{u',\ub'})}\|\psi^{(s_7, s'_7)}\|_{L^{\infty}_{sc}(S_{u',\ub'})},\\
\end{split}
\end{equation*}

\begin{equation*}
\begin{split}
E_6\leq\f{\d}{|u|^2}\sum_{s_5+s_6+s_7=s+1, \atop s'_5+s'_6+s'_7=s'}\sup_{u',\ub'}\|\nab\psi^{(s_5, s'_5)}\|_{L^4_{sc}(S_{u',\ub'})}\|\nab\psi^{(s_6, s'_6)}\|_{L^4_{sc}(S_{u',\ub'})}\|\psi^{(s_7, s'_7)}\|_{L^{\infty}_{sc}(S_{u',\ub'})},\\
\end{split}
\end{equation*}

\begin{equation*}
\begin{split}
E_7\leq&\f{\d^{\f34}}{|u|}\sum_{s_8+s_9+s_{10}=s+\f32, \atop s'_8+s'_9+s'_{10}=s'+\f12}\sup_{u',\ub'}\f{1}{|u'|}\|\nab\psi^{(s_8, s'_8)}\|_{L^4_{sc}(S_{u',\ub'})}\|\psi^{(s_9, s'_9)}\|_{L^{\infty}_{sc}(S_{u',\ub'})}\\
&\quad\times\l\d^{\f14}\|\Psi^{(s_{10}, s'_{10})}\|^{\f12}_{L^2_{sc}(\Hu)}\|\nab\Psi^{(s_{10}, s'_{10})}\|^{\f12}_{L^2_{sc}(\Hu)}+\d^{\f12}\|\Psi^{(s_{10}, s'_{10})}\|_{L^2_{sc}(\Hu)} \r,\\
\end{split}
\end{equation*}

\begin{equation*}
\begin{split}
E_8\leq&\f{\d}{|u|}\sum_{s_8+s_9+s_{10}=s+\f32, \atop s'_8+s'_9+s'_{10}=s'+\f12}\sup_{u',\ub'}\|\psi^{(s_8, s'_8)}\|_{L^4_{sc}(S_{u',\ub'})}\f{1}{|u'|}\|\psi^{(s_9, s'_9)}\|_{L^{\infty}_{sc}(S_{u',\ub'})}\\
&\quad\times\l\|\nab\Psi^{(s_{10}, s'_{10})}\|^{\f12}_{L^2_{sc}(\Hu)}\|\nab^2\Psi^{(s_{10}, s'_{10})}\|^{\f12}_{L^2_{sc}(\Hu)}+\d^{\f14}\|\nab\Psi^{(s_{10}, s'_{10})}\|_{L^2_{sc}(\Hu)} \r,\\
\end{split}
\end{equation*}

\begin{equation*}
\begin{split}
E_9\leq&\f{\d^{\f12}}{|u|}\sum_{s_{11}+s_{12}=s+\f32, \atop s'_{11}+s'_{12}=s'+\f12}\l\|\nab\Psi^{(s_{11}, s'_{11})}\|^{\f12}_{L^2_{sc}(\Hu)}\|\nab^2\Psi^{(s_{11}, s'_{11})}\|^{\f12}_{L^2_{sc}(\Hu)}\\
&\quad\quad\quad+\d^{\f14}\|\nab\Psi^{(s_{12}, s'_{12})}\|_{L^2_{sc}(\Hu)} \r\\
&\quad\times\l\|\Psi^{(s_{12}, s'_{12})}\|^{\f12}_{L^2_{sc}(\Hu)}\|\nab\Psi^{(s_{12}, s'_{12})}\|^{\f12}_{L^2_{sc}(\Hu)}+\d^{\f14}\|\Psi^{(s_{12}, s'_{12})}\|_{L^2_{sc}(\Hu)} \r,
\end{split}
\end{equation*}

\begin{equation*}
\begin{split}
E_{10}\leq&\f{\d^{\f32}}{|u|^3}\sup_{u',\ub'}\sum_{s_{13}+s_{14}+s_{15}+s_{16}=s+\f32, \atop s'_{13}+s'_{14}+s'_{15}+s'_{16}=s'+\f12}\|\nab\psi^{(s_{13}, s'_{13})}\|_{L^{4}_{sc}(S_{u',\ub'})}\|\psi^{(s_{14}, s'_{14})}\|_{L^4_{sc}(S_{u',\ub'})}\\
&\quad\times\|\psi^{(s_{15}, s'_{15})}\|_{L^{\infty}_{sc}(S_{u',\ub'})}\|\psi^{(s_{16}, s'_{16})}\|_{L^{\infty}_{sc}(S_{u',\ub'})},\\
\end{split}
\end{equation*}

\begin{equation*}
\begin{split}
E_{11}\leq&\f{\d^{\f32}}{|u|^3}\sup_{u', \ub'}\sum_{s_{17}+s_{18}+s_{19}+s_{20}=s+2, \atop s'_{17}+s'_{18}+s'_{19}+s'_{20}=s'+1}\|\psi^{(s_{17},s'_{17})}\|_{L^{\infty}_{sc}(S_{u',\ub'})}\|\psi^{(s_{18}, s'_{18})}\|_{L^{\infty}_{sc}(S_{u',\ub'})}\\
&\quad\times\|\psi^{(s_{19}, s'_{19})}\|_{L^4_{sc}(S_{u',\ub'})}\\
&\quad\times\l\|\Psi^{(s_{20}, s'_{20})}\|^{\f12}_{L^2_{sc}(\Hu)}\|\nab\Psi^{(s_{20}, s'_{20})}\|^{\f12}_{L^2_{sc}(\Hu)}+\d^{\f14}\|\Psi^{(s_{20}, s'_{20})}\|_{L^2_{sc}(\Hu)} \r,\\
\end{split}
\end{equation*}

\begin{equation*}
\begin{split}
E_{12}\leq&\f{\d}{|u|^2}\sup_{u',\ub'}\sum_{s_{21}+s_{22}+s_{23}=s+2, \atop s'_{21}+s'_{22}+s'_{23}=s'+1}\|\psi^{(s_{21},s'_{21})}\|_{L^{\infty}_{sc}(S_{u',\ub'})}\\
&\quad\times\l\|\Psi^{(s_{22}, s'_{22})}\|^{\f12}_{L^2_{sc}(\Hu)}\|\nab\Psi^{(s_{22}, s'_{22})}\|^{\f12}_{L^2_{sc}(\Hu)}+\d^{\f14}\|\Psi^{(s_{22}, s'_{22})}\|_{L^2_{sc}(\Hu)} \r\\
&\quad\times\l\|\Psi^{(s_{23}, s'_{23})}\|^{\f12}_{L^2_{sc}(\Hu)}\|\nab\Psi^{(s_{23}, s'_{23})}\|^{\f12}_{L^2_{sc}(\Hu)}+\d^{\f14}\|\Psi^{(s_{23}, s'_{23})}\|_{L^2_{sc}(\Hu)} \r,\\
\end{split}
\end{equation*}

\begin{equation*}
\begin{split}
E_{13}\leq&\f{\d^{2}}{|u|^4}\sup_{u',\ub'}\sum_{s_{24}+s_{25}+s_{26}+s_{27}+s_{28}=s+2, \atop s'_{24}+s'_{25}+s'_{26}+s'_{27}+s'_{28}=s'+1}\|\psi^{(s_{24},s'_{24})}\|_{L^{\infty}_{sc}(S_{u',\ub'})}\|\psi^{(s_{25}, s'_{25})}\|_{L^{\infty}_{sc}(S_{u',\ub'})}\\
&\quad\times\|\psi^{(s_{26}, s'_{26})}\|_{L^{\infty}_{sc}(S_{u',\ub'})}\|\psi^{(s_{27}, s'_{27})}\|_{L^{4}_{sc}(S_{u',\ub'})}\|\psi^{(s_{28}, s'_{28})}\|_{L^4_{sc}(S_{u',\ub'})},
\end{split}
\end{equation*}

\begin{equation*}
\begin{split}
E_{14}\leq{\d^{\f12}}\sum_{s_3+s_4=s+1, \atop s'_3+s'_4=s'}\sup_{u',\ub'}\f{1}{|u'|}\|\psi^{(s_3,s'_3)}\|_{L^{\infty}_{sc}(S_{u',\ub'})}\|\nab^2\Psi^{(s_4,s'_4)}\|_{L^2_{sc}(\Hu)}.\\
\end{split}
\end{equation*}
\end{proof}

When $\T\in\{\nab\tr\chi, \underline{\kappa}\}$, there is no anomalous terms, which implies

\begin{proposition}
For $\T\in\{\nab\tr\chi, \underline{\kappa}\}$, we have
\begin{equation*}
\begin{split}
\|\nab^2\T^{(s,s')}\|_{L^2_{sc}(\S)}\leq & \int_0^{\ub}\f{\d^{-\f12}}{|u|}\sum_{s_1+s_2=s+\f12, \atop s'_1+s'_2=s'-\f12}\|\nab^3\psi^{(s_1, s'_1)}\|_{L^2_{sc}(S_{u,\ub'})}\|\psi^{(s_2, s'_2)}\|_{L^{\infty}_{sc}(S_{u,\ub'})}d\ub'\\
&+\f{\d^{\f12}}{|u|}\M O_{2,4}\M O_{1,4}+\f{\d^{\f12}}{|u|}\M O_{2,4}\M R_0^{\f12}\M R_1^{\f12}+\f{\d^{\f34}}{|u|}\M O_{2,4}\M R_0\\
&+\f{\d^{\f12}}{|u|}\M O_{1,4}\M R_1^{\f12}\M R_2^{\f12}+\f{\d^{\f34}}{|u|}\M O_{1,4}\M R_1+\f{\d}{|u|}\D_0\M O_{2,4}\M O_{0,4}\\
&+\f{\d}{|u|}\D_0\M O_{1,4}\M O_{1,4}+\f{\d}{|u|}\D_0\M O_{1,4}\M R_1^{\f12}\M R_2^{\f12}+\f{\d^{\f54}}{|u|}\D_0\M O_{1,4}\M R_1\\
&+\f{\d}{|u|^2}\M O_{0,4}\D_0\M R_1^{\f12}\M R_2^{\f12}+\f{\d^{\f54}}{|u|^2}\M O_{0,4}\D_0\M R_1\\
&+\f{\d^{\f34}}{|u|}\M R_1\M R_2^{\f12}\M R_0^{\f12}+\f{\d^{\f34}}{|u|}\M R_1^{\f12}\M R_2^{\f12}\M R_0+\f{\d^{\f12}}{|u|}\M R_1^{\f32}\M R_0^{\f12}+\f{\d}{|u|}\M R_1\M R_0\\
&+\f{\d^{\f32}}{|u|^3}\D_0^2\M O_{1,4}\M O_{0,4}+\f{\d^{\f32}}{|u|^3}\D_0^2\M R_0\M O_{0,4}+\f{\d}{|u|^2}\D_0(\M R_0^{\f12}\M R_1^{\f12}+\d^{\f14}\M R_0)^2\\
&+\f{\d^{\f32}}{|u|^3}\D_0^2\M O_{0,4}(\M R_0^{\f12}\M R_1^{\f12}+\d^{\f14}\M R_0)+\f{\d^2}{|u|^4}\D_0^3\M O^2_{0,4}+\f{\d^{\f12}}{|u|}\D_0\M R_2.
\end{split}
\end{equation*}
\end{proposition}

When $\T^{(s,s')}=\mu$, in the equation of $\nab_4\nab^2\mu$, there are several anomalous terms.
We encounter $\nab\chibh\nab\a$ and it is bounded through estimate for $E_4$. We have $\chibh\nab^2\a$ and it is bounded through estimate for $E_{14}$.
There is $\nab^2\chibh\a$ and we bound it through estimate for $E_3$. We also get $(\nab\eta,\nab\etb)\chibh\a, (\eta,\etb)\nab\chibh\a$ and we treat them through estimate for $E_7$.
The term $(\eta,\etb)\chibh\nab\a$ shows up, and it is bounded through estimate for $E_8$. Therefore, we deduce

\begin{proposition}
For $\T=\mu$, we obtain
\begin{equation*}
\begin{split}
\|\nab^2\T^{(s,s')}\|_{L^2_{sc}(\S)}\leq & \int_0^{\ub}\f{\d^{-\f12}}{|u|}\sum_{s_1+s_2=s+\f12, \atop s'_1+s'_2=s'-\f12}\|\nab^3\psi^{(s_1, s'_1)}\|_{L^2_{sc}(S_{u,\ub'})}\|\psi^{(s_2, s'_2)}\|_{L^{\infty}_{sc}(S_{u,\ub'})}d\ub'\\
&+\f{\d^{\f12}}{|u|}\M O_{2,4}\M O_{1,4}+{\d^{\f14}}\M O_{2,4}\M R_0^{\f12}\M R_1^{\f12}+{\d^{\f14}}\M O_{2,4}\M R_0\\
&+{\d^{\f12}}\M O_{1,4}\M R_1^{\f12}\M R_2^{\f12}+{\d^{\f34}}\M O_{1,4}\M R_1+\f{\d}{|u|}\D_0\M O_{2,4}\M O_{0,4}\\
&+\f{\d}{|u|}\D_0\M O_{1,4}\M O_{1,4}+{\d^{\f34}}\D_0\M O_{1,4}\M R_0^{\f12}\M R_1^{\f12}+{\d^{\f34}}\D_0\M O_{1,4}\M R_0\\
&+\f{\d}{|u|}\D_0^2\M R_1+\f{\d^{\f12}}{|u|}\M R_1\M R_2^{\f12}\M R_0^{\f12}+\f{\d^{\f34}}{|u|}\M R_1^{\f12}\M R_2^{\f12}\M R_0+\f{\d^{\f34}}{|u|}\M R_1^{\f32}\M R_0^{\f12}+\f{\d}{|u|}\M R_1\M R_0\\
&+\f{\d^{\f32}}{|u|^3}\D_0^2\M O_{1,4}\M O_{0,4}+\f{\d^{\f32}}{|u|^3}\D_0^3\M R_0+\f{\d}{|u|^2}\D_0\M R_0\M R_1+\f{\d}{|u|}\D_0\M O_{0,4}\M R^{\f12}_1\M R^{\f12}_2\\
&+\f{\d}{|u|}\D_0\M O_{0,4}\M R_1+\f{\d^{\f54}}{|u|^2}\D_0\M R_0^{\f32}\M R_1^{\f12}+\f{\d^{\f32}}{|u|^2}\D_0\M R_0^2\\
&+\f{\d^2}{|u|^4}\D_0^4\M R_0^2+{\d^{\f12}}\D_0\M R_2.
\end{split}
\end{equation*}
\end{proposition}

In the case $\T^{(s,s')}$ obeys equation \eqref{tranT3},
from Proposition \ref{evolution lemma}, Gronwall's inequality and the definition of scale invariant norms, we obtain the following proposition.
\begin{proposition}
When $\T^{(s,s')}$ satisfies \eqref{tranT3}, we get
\begin{equation*}
\begin{split}
&|u|^{2\lambda-2s'-1}\|\nab^2\T^{(s,s')}\|_{L^2_{sc}(\S)}\\
\leq & \int_{u_{\infty}}^{u}\d^{\f12}|u'|^{2\lambda-2s'-4}\sum_{s_1+s_2=s-\f12, \atop s'_1+s'_2=s'+\f12}\|\nab^3\psi^{(s_1, s'_1)}\|_{L^2_{sc}(S_{u',\ub})}\|\psi^{(s_2, s'_2)}\|_{L^{\infty}_{sc}(S_{u',\ub})} du'\\
&+\int_{u_{\infty}}^{u}\d^{\f12}|u'|^{2\lambda-2s'-4}\sum_{s_1+s_2=s-\f12, \atop s'_1+s'_2=s'+\f12}\|\nab^2\psi^{(s_1,s'_1)}\|_{L^4_{sc}(S_{u',\ub})}\|\nab\psi^{(s_2,s'_2)}\|_{L^4_{sc}(S_{u',\ub})} du'\\
&+\int_{u_{\infty}}^{u}\d^{\f12}|u'|^{2\lambda-2s'-4}\sum_{s_3+s_4=s, \atop s'_3+s'_4=s'+1}\|\nab^2\psi^{(s_3,s'_3)}\|_{L^4_{sc}(S_{u',\ub})}\|\Psi^{(s_4,s'_4)}\|_{L^4_{sc}(S_{u',\ub})} du'\\
&+\int_{u_{\infty}}^{u}\d^{\f12}|u'|^{2\lambda-2s'-4}\sum_{s_3+s_4=s, \atop s'_3+s'_4=s'+1}\|\nab\psi^{(s_3,s'_3)}\|_{L^4_{sc}(S_{u',\ub})}\|\nab\Psi^{(s_4,s'_4)}\|_{L^4_{sc}(S_{u',\ub})}du' \\
&+\int_{u_{\infty}}^{u}\d^{\f12}|u'|^{2\lambda-2s'-4}\sum_{s_3+s_4=s, \atop s'_3+s'_4=s'+1}\|\psi^{(s_3,s'_3)}\|_{L^{\infty}_{sc}(S_{u',\ub})}\|\nab^2\Psi^{(s_4,s'_4)}\|_{L^2_{sc}(S_{u',\ub})}du' \\
&+\int_{u_{\infty}}^{u}\d^{\f12}|u'|^{2\lambda-2s'-4}\sum_{s_5+s_6+s_7=s, \atop s'_5+s'_6+s'_7=s'+1}\|\nab^2\psi^{(s_5, s'_5)}\|_{L^4_{sc}(S_{u',\ub})}\|\psi^{(s_6, s'_6)}\cdot\psi^{(s_7, s'_7)}\|_{L^4_{sc}(S_{u',\ub})} du'\\
&+\int_{u_{\infty}}^{u}\d|u'|^{2\lambda-2s'-5}\sum_{s_5+s_6+s_7=s, \atop s'_5+s'_6+s'_7=s'+1}\|\nab\psi^{(s_5, s'_5)}\|_{L^4_{sc}(S_{u',\ub})}\|\nab\psi^{(s_6, s'_6)}\|_{L^4_{sc}(S_{u',\ub})}\\
&\quad\times\|\psi^{(s_7, s'_7)}\|_{L^{\infty}_{sc}(S_{u',\ub})} du'\\
&+\int_{u_{\infty}}^{u}\d|u'|^{2\lambda-2s'-5}\sum_{s_8+s_9+s_{10}=s+\f12, \atop s'_8+s'_9+s'_{10}=s'+\f32}\|\nab\psi^{(s_8, s'_8)}\|_{L^4_{sc}(S_{u',\ub})}\|\psi^{(s_9, s'_9)}\|_{L^{\infty}_{sc}(S_{u',\ub})}\\
&\quad\times\|\Psi^{(s_{10}, s'_{10})}\|_{L^4_{sc}(S_{u',\ub})} du'\\
&+\int_{u_{\infty}}^{u}\d^{\f12}|u'|^{2\lambda-2s'-4}\sum_{s_8+s_9+s_{10}=s+\f12, \atop s'_8+s'_9+s'_{10}=s'+\f32}\|\psi^{(s_8, s'_8)}\cdot\psi^{(s_9, s'_9)}\|_{L^4_{sc}(S_{u',\ub})}\|\nab\Psi^{(s_{10}, s'_{10})}\|_{L^4_{sc}(S_{u',\ub})} du'\\
&+\int_{u_{\infty}}^{u}\d^{\f12}|u'|^{2\lambda-2s'-4}\sum_{s_{11}+s_{12}=s+\f12, \atop s'_{11}+s'_{12}=s'+\f32}\|\nab\Psi^{(s_{11},s'_{11})}\|_{L^4_{sc}(S_{u',\ub})}\|\Psi^{(s_{12},s'_{12})}\|_{L^4_{sc}(S_{u',\ub})} du'\\
&+\int_{u_{\infty}}^{u}\d^{\f12}|u'|^{2\lambda-2s'-4}\sum_{s_{13}+s_{14}+s_{15}+s_{16}=s+\f12, \atop s'_{13}+s'_{14}+s'_{15}+s'_{16}=s'+\f32}\|\nab\psi^{(s_{13}, s'_{13})}\|_{L^4_{sc}(S_{u',\ub})}\\
&\quad\times\|\psi^{(s_{14}, s'_{14})}\cdot\psi^{(s_{15}, s'_{15})}\cdot\psi^{(s_{16}, s'_{16})}\|_{L^4_{sc}(S_{u',\ub})} du'\\
\end{split}
\end{equation*}

\begin{equation*}
\begin{split}&+\int_{u_{\infty}}^{u}\d^{\f12}|u'|^{2\lambda-2s'-4}\sum_{s_{17}+s_{18}+s_{19}+s_{20}=s+1, \atop s'_{17}+s'_{18}+s'_{19}+s'_{20}=s'+2}\|\psi^{(s_{17},s'_{17})}\cdot\psi^{(s_{18}, s'_{18})}\cdot\psi^{(s_{19}, s'_{19})}\|_{L^4_{sc}(S_{u',\ub})}\\
&\quad\times\|\Psi^{(s_{20}, s'_{20})}\|_{L^4_{sc}(S_{u',\ub})} du' \\
&+\int_{u_{\infty}}^{u}\d|u'|^{2\lambda-2s'-5}\sum_{s_{21}+s_{22}+s_{23}=s+1, \atop s'_{21}+s'_{22}+s'_{23}=s'+2}\|\psi^{(s_{21},s'_{21})}\|_{L^{\infty}_{sc}(S_{u',\ub})}\|\Psi^{(s_{22},s'_{22})}\|_{L^4_{sc}(S_{u',\ub})}\\
&\quad\times\|\Psi^{(s_{23},s'_{23})}\|_{L^4_{sc}(S_{u',\ub})} du'\\
&+\int_{u_{\infty}}^{u}\d^{\f12}|u'|^{2\lambda-2s'-4}\sum_{s_{24}+s_{25}+s_{26}+s_{27}+s_{28}=s+1, \atop s'_{24}+s'_{25}+s'_{26}+s'_{27}+s'_{28}=s'+2}\|\psi^{(s_{28}, s'_{28})}\|_{L^4_{sc}(S_{u',\ub})}\\
&\quad\times\|\psi^{(s_{24},s'_{24})}\cdot\psi^{(s_{25}, s'_{25})}\cdot\psi^{(s_{26}, s'_{26})}\cdot\psi^{(s_{27}, s'_{27})}\|_{L^{\infty}_{sc}(S_{u',\ub})} du'.
\end{split}
\end{equation*}

\end{proposition}

{\bf Remark}: Using the equations in the appendix, it is easy to verify that when $\T\in \{\mub, \kappa\}$, $\nab^2\T$ obeys \eqref{tranT3}.
In the case of $\T=\nab(\oo)$, by employing Proposition \ref{Omega} and the fact that 
$$\eta_A+\etb_A=2\O^{-1}\nab_A\O,$$
it can be verified that $\nab^3(\oo)$ obeys the same bound for $\nab^2\T^{(s,s')}$ above.

For $\T\in\{\nab(\oo), \underline{\mu}, \kappa\}$, we have
$\lambda[\nab(\oo)]=s_2(\nab(\oo))=\f32$, $\lambda[\underline{\mu}]=s_2(\underline{\mu})=1$
and $\lambda[\kappa]=s_2(\kappa)=\f12$.
This implies

\begin{proposition}\label{O32311}
For $\T\in\{\nab(\oo), \underline{\mu}, \kappa\}$, we infer 
\begin{equation*}
\begin{split}
&|u|^{-1}\|\nab^2\T^{(s,s')}\|_{L^2_{sc}(\S)}\\
\leq & \int_{u_{\infty}}^{u}\d^{\f12}|u'|^{-4}\sum_{s_1+s_2=s-\f12, \atop s'_1+s'_2=s'+\f12}\|\nab^3\psi^{(s_1, s'_1)}\|_{L^2_{sc}(S_{u',\ub})}\|\psi^{(s_2, s'_2)}\|_{L^{\infty}_{sc}(S_{u',\ub})} du'\\
&+F_{1}+F_{2}+F_{3}+F_{4}+F_{5}+F_{6}+F_{7}+F_{8}+F_{9}+F_{10}\\
&+F_{11}+F_{12}+F_{13},
\end{split}
\end{equation*}
where $F_1-F_{13}$ are listed below:
\begin{equation*}
\begin{split}
F_1=&\int_{u_{\infty}}^{u}\d^{\f12}|u'|^{-4}\sum_{s_1+s_2=s-\f12, \atop s'_1+s'_2=s'+\f12}\|\nab^2\psi^{(s_1,s'_1)}\|_{L^4_{sc}(S_{u',\ub})}\|\nab\psi^{(s_2,s'_2)}\|_{L^4_{sc}(S_{u',\ub})} du',\\
F_2=&\int_{u_{\infty}}^{u}\d^{\f12}|u'|^{-4}\sum_{s_3+s_4=s, \atop s'_3+s'_4=s'+1}\|\nab^2\psi^{(s_3,s'_3)}\|_{L^4_{sc}(S_{u',\ub})}\|\Psi^{(s_4,s'_4)}\|_{L^4_{sc}(S_{u',\ub})} du',\\
F_3=&\int_{u_{\infty}}^{u}\d^{\f12}|u'|^{-4}\sum_{s_3+s_4=s, \atop s'_3+s'_4=s'+1}\|\nab\psi^{(s_3,s'_3)}\|_{L^4_{sc}(S_{u',\ub})}\|\nab\Psi^{(s_4,s'_4)}\|_{L^4_{sc}(S_{u',\ub})}du', \\
F_4=&\int_{u_{\infty}}^{u}\d^{\f12}|u'|^{-4}\sum_{s_5+s_6+s_7=s, \atop s'_5+s'_6+s'_7=s'+1}\|\nab^2\psi^{(s_5, s'_5)}\|_{L^4_{sc}(S_{u',\ub})}\|\psi^{(s_6, s'_6)}\cdot\psi^{(s_7, s'_7)}\|_{L^4_{sc}(S_{u',\ub})} du',\\
F_5=&\int_{u_{\infty}}^{u}\d|u'|^{-5}\sum_{s_5+s_6+s_7=s, \atop s'_5+s'_6+s'_7=s'+1}\|\nab\psi^{(s_5, s'_5)}\|_{L^4_{sc}(S_{u,\ub'})}\|\nab\psi^{(s_6, s'_6)}\|_{L^4_{sc}(S_{u',\ub})}\\
&\quad\times\|\psi^{(s_7, s'_7)}\|_{L^{\infty}_{sc}(S_{u',\ub})} du',\\
F_6=&\int_{u_{\infty}}^{u}\d|u'|^{-5}\sum_{s_8+s_9+s_{10}=s+\f12, \atop s'_8+s'_9+s'_{10}=s'+\f32}\|\nab\psi^{(s_8, s'_8)}\|_{L^4_{sc}(S_{u',\ub})}\|\psi^{(s_9, s'_9)}\|_{L^{\infty}_{sc}(S_{u',\ub})}\\
&\quad\times\|\Psi^{(s_{10}, s'_{10})}\|_{L^4_{sc}(S_{u',\ub})} du',\\
F_7=&\int_{u_{\infty}}^{u}\d^{\f12}|u'|^{-4}\sum_{s_8+s_9+s_{10}=s+\f12, \atop s'_8+s'_9+s'_{10}=s'+\f32}\|\psi^{(s_8, s'_8)}\cdot\psi^{(s_9, s'_9)}\|_{L^4_{sc}(S_{u',\ub})}\|\nab\Psi^{(s_{10}, s'_{10})}\|_{L^4_{sc}(S_{u',\ub})} du',\\
F_8=&\int_{u_{\infty}}^{u}\d^{\f12}|u'|^{-4}\sum_{s_{11}+s_{12}=s+\f12, \atop s'_{11}+s'_{12}=s'+\f32}\|\nab\Psi^{(s_{11},s'_{11})}\|_{L^4_{sc}(S_{u',\ub})}\|\Psi^{(s_{12},s'_{12})}\|_{L^4_{sc}(S_{u',\ub})} du',\\
F_9=&\int_{u_{\infty}}^{u}\d^{\f12}|u'|^{-4}\sum_{s_{13}+s_{14}+s_{15}+s_{16}=s+\f12, \atop s'_{13}+s'_{14}+s'_{15}+s'_{16}=s'+\f32}\|\nab\psi^{(s_{13}, s'_{13})}\|_{L^4_{sc}(S_{u',\ub})}\\
&\quad\times\|\psi^{(s_{14}, s'_{14})}\cdot\psi^{(s_{15}, s'_{15})}\cdot\psi^{(s_{16}, s'_{16})}\|_{L^4_{sc}(S_{u',\ub})} du',\\
F_{10}=&\int_{u_{\infty}}^{u}\d^{\f12}|u'|^{-4}\sum_{s_{17}+s_{81}+s_{19}+s_{20}=s+1, \atop s'_{17}+s'_{18}+s'_{19}+s'_{20}=s'+2}\|\psi^{s_{17},s'_{17}}\cdot\psi^{(s_{18}, s'_{18})}\cdot\psi^{(s_{19}, s'_{19})}\|_{L^4_{sc}(S_{u',\ub})}\\
&\quad\times\|\Psi^{(s_{20}, s'_{20})}\|_{L^4_{sc}(S_{u',\ub})} du', \\
F_{11}=&\int_{u_{\infty}}^{u}\d|u'|^{-5}\sum_{s_{21}+s_{22}+s_{23}=s+1, \atop s'_{21}+s'_{22}+s'_{23}=s'+2}\|\psi^{(s_{21},s'_{21})}\|_{L^{\infty}_{sc}(S_{u',\ub})}\|\Psi^{(s_{22},s'_{22})}\|_{L^4_{sc}(S_{u',\ub})}\\
&\quad\times\|\Psi^{(s_{23},s'_{23})}\|_{L^4_{sc}(S_{u',\ub})} du',\\
F_{12}=&\int_{u_{\infty}}^{u}\d^{\f12}|u'|^{-4}\sum_{s_{24}+s_{25}+s_{26}+s_{27}+s_{28}=s+1, \atop s'_{24}+s'_{25}+s'_{26}+s'_{27}+s'_{28}=s'+2}\|\psi^{(s_{28}, s'_{28})}\|_{L^4_{sc}(S_{u',\ub})}\\
&\quad\times\|\psi^{(s_{24},s'_{24})}\cdot\psi^{(s_{25}, s'_{25})}\cdot\psi^{(s_{26}, s'_{26})}\cdot\psi^{(s_{27}, s'_{27})}\|_{L^{4}_{sc}(S_{u',\ub})} du'\\
\end{split}
\end{equation*}

\begin{equation*}
\begin{split}
F_{13}=&\int_{u_{\infty}}^{u}\d^{\f12}|u'|^{-4}\sum_{s_3+s_4=s, \atop s'_3+s'_4=s'+1}\|\psi^{(s_3,s'_3)}\|_{L^{\infty}_{sc}(S_{u',\ub})}\|\nab^2\Psi^{(s_4,s'_4)}\|_{L^2_{sc}(S_{u',\ub})}du'. \\
\end{split}
\end{equation*}

\end{proposition}

With the aid of Proposition \ref{L4} and H\"older's inequality, we demonstrate the following estimate.

For $F_1$, we obtain
\begin{equation*}
\begin{split}
F_1\leq\f{\d^{\f12}}{|u|}\sum_{s_1+s_2=s-\f12, \atop s'_1+s'_2=s'+\f12}\sup_{u',\ub'}\f{1}{|u'|}\|\nab^2\psi^{(s_1,s'_1)}\|_{L^4_{sc}(S_{u',\ub'})}\f{1}{|u'|}\|\nab\psi^{(s_2,s'_2)}\|_{L^4_{sc}(S_{u',\ub'})}.\\
\end{split}
\end{equation*}
For $F_2$, we get
\begin{equation*}
\begin{split}
F_2\leq&\f{\d^{\f12}}{|u|^{\f32}}\sum_{s_3+s_4=s, \atop s'_3+s'_4=s'+1}\sup_{u',\ub'}\f{1}{|u'|}\|\nab^2\psi^{(s_3,s'_3)}\|_{L^4_{sc}(S_{u',\ub'})}\\
&\quad\times\l\|\Psi^{(s_4,s'_4)}\|^{\f12}_{L^2_{sc}(\Hbu)}\|\nab\Psi^{(s_4,s'_4)}\|^{\f12}_{L^2_{sc}(\Hbu)}+\d^{\f14}\|\Psi^{(s_4,s'_4)}\|_{L^2_{sc}(\Hbu)}\r.\\
\end{split}
\end{equation*}
The anomalous term of $F_2$ type is $\nab^2\chibh\rho$, which is bounded as above.

We estimate $F_3$ via
\begin{equation*}
\begin{split}
F_3\leq&\f{\d^{\f12}}{|u|^{\f32}}\sum_{s_3+s_4=s, \atop s'_3+s'_4=s'+1}\sup_{u',\ub'}\f{1}{|u'|}\|\nab\psi^{(s_3,s'_3)}\|_{L^4_{sc}(S_{u',\ub'})}\\
&\quad\times\l\|\nab\Psi^{(s_4,s'_4)}\|^{\f12}_{L^2_{sc}(\Hbu)}\|\nab^2\Psi^{(s_4,s'_4)}\|^{\f12}_{L^2_{sc}(\Hbu)}+\d^{\f14}\|\nab\Psi^{(s_4,s'_4)}\|_{L^2_{sc}(\Hbu)}\r.\\
\end{split}
\end{equation*}
The anomalous terms of $F_3$ type are $\nab\chibh\nab\rho$ and $\nab\chibh\nab\b$, which is bounded as above.

For $F_4$, we have
\begin{equation*}
\begin{split}
F_4\leq&\f{\d^{\f14}}{|u|}\sum_{s_5+s_6+s_7=s, \atop s'_5+s'_6+s'_7=s'+1}\sup_{u',\ub'}\|\nab^2\psi^{(s_5, s'_5)}\|_{L^4_{sc}(S_{u',\ub'})}\f{\d^{\f14}}{|u'|}\|\psi^{(s_6, s'_6)}\|_{L^4_{sc}(S_{u',\ub'})}\\
&\quad\times\f{\d^{\f12}}{|u'|^2}\|\psi^{(s_7, s'_7)}\|_{L^{\infty}_{sc}(S_{u',\ub'})}. \\
\end{split}
\end{equation*}
A potential most anomalous term of $F_4$ type is $(\nab^2\eta,\nab^2\etb)\tr\chib\tr\chib$.
This term can be avoided by using equation for $\nab_3(\O\tr\chib-\f{2}{u})$.
The anomalous terms of $F_4$ type left are $(\nab^2\eta,\nab^2\etb)(\O\tr\chib-\f{2}{u})(\O\tr\chib-\f{2}{u}),(\nab^2\eta,\nab^2\etb)\omb\tr\chib, (\nab^2\eta,\nab^2\etb)\chibh\,\chibh$ and $(\nab^2\omb, \nab^2\chib)(\eta,\etb)(\omb,\chib)$.
These terms are bounded through the estimate above.

For $F_5$, we obtain
\begin{equation*}
\begin{split}
F_5\leq&\f{\d^{\f12}}{|u|}\sum_{s_5+s_6+s_7=s, \atop s'_5+s'_6+s'_7=s'+1}\sup_{u',\ub'}\|\nab\psi^{(s_5, s'_5)}\|_{L^4_{sc}(S_{u',\ub'})}\f{1}{|u'|}\|\nab\psi^{(s_6, s'_6)}\|_{L^4_{sc}(S_{u',\ub'})}\\
&\quad\times\f{\d^{\f12}}{|u'|^2}\|\psi^{(s_7, s'_7)}\|_{L^{\infty}_{sc}(S_{u',\ub'})}. \\
\end{split}
\end{equation*}

For $F_6$, we derive
\begin{equation*}
\begin{split}
F_6\leq&\f{\d^{\f12}}{|u|^{\f32}}\sum_{s_8+s_9+s_{10}=s+\f12, \atop s'_8+s'_9+s'_{10}=s'+\f32}\sup_{u',\ub'}\|\nab\psi^{(s_8, s'_8)}\|_{L^4_{sc}(S_{u',\ub'})}\f{\d^{\f12}}{|u'|^2}\|\psi^{(s_9, s'_9)}\|_{L^{\infty}_{sc}(S_{u',\ub'})}\\
&\quad\times\l\|\Psi^{(s_{10},s'_{10})}\|^{\f12}_{L^2_{sc}(\Hbu)}\|\nab\Psi^{(s_{10},s'_{10})}\|^{\f12}_{L^2_{sc}(\Hbu)}+\d^{\f14}\|\Psi^{(s_{10},s'_{10})}\|_{L^2_{sc}(\Hbu)}\r.\\
\end{split}
\end{equation*}
The anomalous term of $F_6$ type are $(\nab\eta,\nab\etb)\chib\rho$ and $(\nab\eta,\nab\etb)\chib\b$, which are bounded through the estimates above. 

For $F_7$, we get
\begin{equation*}
\begin{split}
F_7\leq&\f{\d^{\f12}}{|u|^{\f32}}\sum_{s_8+s_9+s_{10}=s+\f12, \atop s'_8+s'_9+s'_{10}=s'+\f32}\sup_{u',\ub'}\|\psi^{(s_8, s'_8)}\|_{L^4_{sc}(S_{u',\ub'})}\f{\d^{\f12}}{|u'|^2}\|\psi^{(s_9, s'_9)}\|_{L^{\infty}_{sc}(S_{u',\ub'})}\\
&\quad\times\l\|\nab\Psi^{(s_{10},s'_{10})}\|^{\f12}_{L^2_{sc}(\Hbu)}\|\nab^2\Psi^{(s_{10},s'_{10})}\|^{\f12}_{L^2_{sc}(\Hbu)}\\
&\quad\quad\quad\times+\d^{\f14}\|\nab\Psi^{(s_{10},s'_{10})}\|_{L^2_{sc}(\Hbu)}\r.\\
\end{split}
\end{equation*}
The anomalous term of $F_7$ type is $(\eta,\etb)\chib\nab\b$, which is bounded as above. 


For $F_8$, 
when $\T=\nab(\oo)$, there is no $F_8$ term. When $\T=\underline{\mu}$, the $F_8$ terms are $\nab\beb\cdot\rho$ and $\beb\cdot\nab\rho$.
When $\T=\kappa$, the $F_8$ terms are $\beb\nab\b$ and $\b\nab\beb$.
From Proposition \ref{R04}, we have
\begin{equation*}
 \|\rho,\beb\|_{L^4_{sc}(\S)}\leq  \M R^{\f12}_1\M R^{\f12}_2+\d^{\f14}\M R_1+{\d^{\f14}}(\M I^{(0)}+\M R+\underline{\M R})(\M R^{\f12}_1\M R^{\f12}_2+\M R_0).
\end{equation*}
Therefore, we derive
\begin{equation*}
\begin{split}
F_8\leq&\f{\d^{\f12}}{|u|^2}\sum_{s_{11}+s_{12}=s+\f12, \atop s'_{11}+s'_{12}=s'+\f32}
\l\|\nab\Psi^{(s_{11},s'_{11})}\|^{\f12}_{L^2_{sc}(\Hbu)}\|\nab^2\Psi^{(s_{11},s'_{11})}\|^{\f12}_{L^2_{sc}(\Hbu)}\\
&\quad\quad\quad+\d^{\f14}\|\nab\Psi^{(s_{11},s'_{11})}\|_{L^2_{sc}(\Hbu)}\r\\
&\times\l\|\Psi^{(s_{12},s'_{12})}\|^{\f12}_{L^2_{sc}(\Hbu)}\|\nab\Psi^{(s_{12},s'_{12})}\|^{\f12}_{L^2_{sc}(\Hbu)}+\d^{\f14}\|\Psi^{(s_{12},s'_{12})}\|_{L^2_{sc}(\Hbu)}\r.\\
\end{split}
\end{equation*}

For $F_9$ and $F_{10}$, we have

\begin{equation*}
\begin{split}
F_9\leq&\f{\d^{\f12}}{|u|}\sum_{s_{13}+s_{14}+s_{15}+s_{16}=s+\f12, \atop s'_{13}+s'_{14}+s'_{15}+s'_{16}=s'+\f32}\sup_{u',\ub'}\|\nab\psi^{(s_{13}, s'_{13})}\|_{L^4_{sc}(S_{u',\ub'})}\\
&\quad\times\|\psi^{(s_{14}, s'_{14})}\|_{L^4_{sc}(S_{u',\ub'})}\f{\d^{\f12}}{|u'|^2}\|\psi^{(s_{15}, s'_{15})}\|_{L^{\infty}_{sc}(S_{u',\ub'})}\f{\d^{\f12}}{|u'|^2}\|\psi^{(s_{16}, s'_{16})}\|_{L^{\infty}_{sc}(S_{u',\ub'})},\\
\end{split}
\end{equation*}
and
\begin{equation*}
\begin{split}
F_{10}\leq&\f{\d}{|u|^{\f52}}\sum_{s_{17}+s_{18}+s_{19}+s_{20}=s+1, \atop s'_{17}+s'_{18}+s'_{19}+s'_{20}=s'+2}\sup_{u',\ub'}\|\psi^{(s_{17},s'_{17})}\|_{L^{4}_{sc}(S_{u',\ub'})}\|\psi^{(s_{18}, s'_{18})}\|_{L^{\infty}_{sc}(S_{u',\ub'})}\\
&\quad\quad\quad\times\f{\d^{\f12}}{|u'|^2}\|\psi^{(s_{19}, s'_{19})}\|_{L^{\infty}_{sc}(S_{u',\ub'})}\\
&\quad\times\l\|\Psi^{(s_{20},s'_{20})}\|^{\f12}_{L^2_{sc}(\Hbu)}\|\nab\Psi^{(s_{20},s'_{20})}\|^{\f12}_{L^2_{sc}(\Hbu)}+\d^{\f14}\|\Psi^{(s_{20},s'_{20})}\|_{L^2_{sc}(\Hbu)}\r.\\
\end{split}
\end{equation*}

For $F_{11}$, when $\T=\nab(\oo)$, there is no $F_{11}$ term. When $\T=\underline{\mu}$, the $F_{11}$ term is $(\eta,\etb)\beb\rho$.
When $\T=\kappa$, the $F_{11}$ term is $(\eta,\etb)\beb\b$.
Hence, we derive
\begin{equation*}
\begin{split}
F_{11}\leq&\f{\d}{|u|^{3}}\sum_{s_{21}+s_{22}+s_{23}=s+1, \atop s'_{21}+s'_{22}+s'_{23}=s'+2}\sup_{u',\ub'}\|\psi^{(s_{21},s'_{21})}\|_{L^{\infty}_{sc}(S_{u',\ub'})}\\
&\quad\times\l\|\Psi^{(s_{22},s'_{22})}\|^{\f12}_{L^2_{sc}(\Hbu)}\|\nab\Psi^{(s_{22},s'_{22})}\|^{\f12}_{L^2_{sc}(\Hbu)}+\d^{\f14}\|\Psi^{(s_{22},s'_{22})}\|_{L^2_{sc}(\Hbu)}\r\\
&\quad\times\l\|\Psi^{(s_{23},s'_{23})}\|^{\f12}_{L^2_{sc}(\Hbu)}\|\nab\Psi^{(s_{23},s'_{23})}\|^{\f12}_{L^2_{sc}(\Hbu)}+\d^{\f14}\|\Psi^{(s_{23},s'_{23})}\|_{L^2_{sc}(\Hbu)}\r.\\
\end{split}
\end{equation*}

We bound $F_{12}$ through
\begin{equation*}
\begin{split}
F_{12}\leq&\f{\d^{\f32}}{|u|^4}\sum_{s_{24}+s_{25}+s_{26}+s_{27}+s_{28}=s+1, \atop s'_{24}+s'_{25}+s'_{26}+s'_{27}+s'_{28}=s'+2}\sup_{u',\ub'}\|\psi^{(s_{28}, s'_{28})}\|_{L^4_{sc}(S_{u',\ub'})}\|\psi^{(s_{24},s'_{24})}\|_{L^4_{sc}(S_{u',\ub'})}\\
&\quad\times\|\psi^{(s_{25}, s'_{25})}\|_{L^{\infty}_{sc}(S_{u',\ub'})}\|\psi^{(s_{26}, s'_{26})}\|_{L^{\infty}_{sc}(S_{u',\ub'})}\f{\d^{\f12}}{|u'|^2}\|\psi^{(s_{27}, s'_{27})}\|_{L^{\infty}_{sc}(S_{u',\ub'})}.
\end{split}
\end{equation*}

Finally, $F_{13}$ obeys
\begin{equation*}
\begin{split}
F_{13}\leq \f{\d^{\f12}}{|u|^{\f52}}\sup_{s_3+s_4=s, \atop s'_3+s'_4=s'+1}\sup_{u,\ub}\|\psi^{(s_3,s'_3)}\|_{L^{\infty}_{sc}(\S)}\|\nab^2\Psi^{(s_4,s'_4)}\|_{L^2_{sc}(\Hbu)}.
\end{split}
\end{equation*}


Gathering all the estimate, with Propositions \ref{O04}, \ref{O14} and \ref{O0infty} we deduce
\begin{proposition}\label{O3231}
For $\T\in\{\nab(\oo), \underline{\mu}, \kappa\}$, we have
\begin{equation*}
\begin{split}
&|u|^{-1}\|\nab^2\T^{(s,s')}\|_{L^2_{sc}(\S)}\\
\leq & \int_{u_{\infty}}^{u}\d^{\f12}|u'|^{-4}\sum_{s_1+s_2=s-\f12, \atop s'_1+s'_2=s'+\f12}\|\nab^3\psi^{(s_1, s'_1)}\|_{L^2_{sc}(S_{u',\ub})}\|\psi^{(s_2, s'_2)}\|_{L^{\infty}_{sc}(S_{u',\ub})} du'\\
&+\f{\d^{\f12}}{|u|}(\M I^{(0)}+\M R+\underline{\M R})^4.
\end{split}
\end{equation*}
\end{proposition}

In the estimates above we encounter $\M O_{2,4}$, which can be bounded via the inequality
$$\M O_{2,4}\leq \M O_{3,2}+\M O_{1,4}.$$
This is guaranteed by the following Gagliardo-Nirenberg's inequality
$$\|\nab^2 \phi\|_{L^4_{sc}(\S)}\leq \|\nab^3 \phi\|^{\f23}_{L^2_{sc}(\S)}\|\nab \phi\|^{\f13}_{L^4_{sc}(\S)},$$
where $\phi$ is a $S_{u,\ub}$-tangent tensor.

To get the desired $\M O_{3,2}$ estimates, we need to make another bootstrap assumption.

Recall
\begin{equation*}
\begin{split}
&\M O_{3,2}\\
=&\sup_{u,\ub}\l\|\nab^3\chih, \nab^3\tr\chi, \nab^3\o, \nab^3\o^{\dagger}, \nab^3\eta, \nab^3\etb, \nab^3\tr\chib, \nab^3\omb, \nab^3\omb^{\dagger}\|_{L^2_{sc}(\S)}\\
&\quad\quad\quad+\f{1}{|u|}\|\nab^3\chibh\|_{L^2_{sc}(\S)}\r,
\end{split}
\end{equation*}
and we set up a new bootstrap assumption
\begin{equation}\label{BA.5}
\M O_{3,2}\leq \D_5.
\end{equation}

\begin{proposition}
 Under the assumption of Theorem \ref{main.thm} and the bootstrap assumption \eqref{BA.5}, we demonstrate
\begin{equation*}
\begin{split}
\|\nab^3\tr\chi\|_{L^2_{sc}(\S)}\leq\f{\d^{\f12}}{|u|}\D_0\D_5+\d^{\f14}((\M I^{(0)})^3+\M R^3+\underline{\M R}^3),
\end{split}
\end{equation*}
and
\begin{equation*}
\begin{split}
\|\nab^3\chih\|_{L^2_{sc}(\S)}\leq& \|\nab^2\b\|_{L^2_{sc}(\S)}+\f{\d^{\f12}}{|u|}\D_0\D_5+\d^{\f14}(\M I^{(0)}+\M R+\underline{\M R})^5\\
&+\f{\d^{\f12}}{|u|}(\M O_{0,\infty}\M O_{2,2}+\M O^2_{1,4}).
\end{split}
\end{equation*}
\end{proposition}

\begin{proof}
From the derived estimates for $\nab^2\T^{(s,s')}$ , we have
\begin{equation*}
\begin{split}
\|\nab^3\tr\chi\|_{L^2_{sc}(\S)}\leq&\int_0^{\ub}\f{\d^{-\f12}}{|u|}\|\chi,\o\|_{L^{\infty}_{sc}(S_{u,\ub'})}\|\nab^3\chi,\nab^3\o\|_{L^{2}_{sc}(S_{u,\ub'})} d\ub'\\
&+\d^{\f14}((\M I^{(0)})^3+\M R^3+\underline{\M R}^3)\\
\leq&\f{\d^{\f12}}{|u|}\D_0\D_5+\d^{\f14}((\M I^{(0)})^3+\M R^3+\underline{\M R}^3).
\end{split}
\end{equation*}
Recalling the null structure equation
$$\div\chih=\f12\nab\tr\chi-\f12(\eta-\etb)\cdot(\chih-\f12\tr\chi)-\b,$$
we apply the elliptic estimates in Proposition \ref{ellipticthm2} to obtain
\begin{equation*}
\begin{split}
\|\nab^3\chih\|_{L^2_{sc}(\S)}\leq& \|\nab^3\tr\chi\|_{L^2_{sc}(\S)}+\|\nab^2\b\|_{L^2_{sc}(\S)}\\
&+\f{\d^{\f12}}{|u|}\|\nab^2(\eta,\etb)\|_{L^2_{sc}(\S)}\|\chi\|_{L^{\infty}_{sc}(\S)}+\f{\d^{\f12}}{|u|}\|\eta,\etb\|_{L^{\infty}_{sc}(\S)}\|\nab^2\chi\|_{L^2_{sc}(\S)}\\
&+\f{\d^{\f12}}{|u|}\|\nab\eta,\nab\etb\|_{L^4_{sc}(\S)}\|\nab\chi\|_{L^4_{sc}(\S)}\\
&+\d^{\f14}(\M I^{(0)}+\M R+\underline{\M R})^5\\
\leq& \|\nab^2\b\|_{L^2_{sc}(\S)}+\f{\d^{\f12}}{|u|}\D_0\D_5+\d^{\f14}(\M I^{(0)}+\M R+\underline{\M R})^5\\
&+\f{\d^{\f12}}{|u|}(\M O_{0,\infty}\M O_{2,2}+\M O^2_{1,4}).
\end{split}
\end{equation*}
\end{proof}

\begin{proposition}\label{O32eta}
 Under the assumption of Theorem \ref{main.thm} and the bootstrap assumption \eqref{BA.5}, we have
\begin{equation*}
\begin{split}
 \|\nab^3\eta\|_{L^2_{sc}(\S)}\leq& \|\nab^2\rho, \nab^2\sigma\|_{L^2_{sc}(\S)}+\f{\d^{\f12}}{|u|}\D_0\D_5+\d^{\f14}(\M I^{(0)}+\M R+\underline{\M R})^5.
\end{split}
\end{equation*}
\end{proposition}
\begin{proof}
The estimates derived for $\nab^2\T^{(s,s')}$ implies
\begin{equation*}
\begin{split}
\|\nab^2\mu\|_{L^2_{sc}(\S)}\leq&\int_0^{\ub}\f{\d^{-\f12}}{|u|}\|\chi\|_{L^{\infty}_{sc}(S_{u,\ub'})}\|\nab^3\eta,\nab^3\etb\|_{L^{2}_{sc}(S_{u,\ub'})} d\ub'\\
&+\d^{\f14}((\M I^{(0)})^3+\M R^3+\underline{\M R}^3)\\
\leq&\f{\d^{\f12}}{|u|}\D_0\D_5+\d^{\f14}((\M I^{(0)})^3+\M R^3+\underline{\M R}^3).
\end{split}
\end{equation*}
Thanks to the definition $\mu=-\div\eta-\rho$, $\eta$ obeys the following elliptic system:
$$\div\eta=-\mu-\rho,$$
$$\curl\eta=\sigma+\f12\chih\wedge\chibh.$$
We apply the elliptic estimates in Proposition \ref{ellipticthm2} to obtain
\begin{equation*}
\begin{split}
 \|\nab^3\eta\|_{L^2_{sc}(\S)}\leq& \|\nab^2\mu\|_{L^2_{sc}(\S)}+\|\nab^2\rho, \nab^2\sigma\|_{L^2_{sc}(\S)}\\
&+{\d^{\f12}}\|\nab^2\chih\|_{L^2_{sc}(\S)}\f{1}{|u|}\|\chibh\|_{L^{\infty}_{sc}(\S)}+{\d^{\f12}}\|\chih\|_{L^{\infty}_{sc}(\S)}\f{1}{|u|}\|\nab^2\chibh\|_{L^{2}_{sc}(\S)}\\
&+{\d^{\f12}}\|\nab\chih\|_{L^4_{sc}(\S)}\f{1}{|u|}\|\nab\chibh\|_{L^{4}_{sc}(\S)}\\
&+\d^{\f14}(\M I^{(0)}+\M R+\underline{\M R})^5\\
\leq&\|\nab^2\rho, \nab^2\sigma\|_{L^2_{sc}(\S)}+\f{\d^{\f12}}{|u|}\D_0\D_6+\d^{\f14}(\M I^{(0)}+\M R+\underline{\M R})^5\\
&+\d^{\f12}(\M O_{2,2}\M O_{0,\infty}+\M O^2_{1,4}).
\end{split}
\end{equation*}
\end{proof}

\begin{proposition}\label{O32omb}
 Under the assumption of Theorem \ref{main.thm} and the bootstrap assumption \eqref{BA.5}, we have
\begin{equation*}
\begin{split}
 \|\nab^3\omb,\nab^3\omb^{\dagger}\|_{L^2_{sc}(\S)}\leq \|\nab^2\beb\|_{L^2_{sc}(\S)}+\f{\d^{\f12}}{|u|}\D_0\D_5+\d^{\f14}(\M I^{(0)}+\M R+\underline{\M R})^5.
\end{split}
\end{equation*}
\end{proposition}

\begin{proof}
The following inequality follows from the estimates for $\nab^2\T^{(s,s')}$
\begin{equation*}
\begin{split}
\|\nab^2\underline{\kappa}\|_{L^2_{sc}(\S)}\leq&\int_0^{\ub}\f{\d^{-\f12}}{|u|}\|\chi,\o\|_{L^{\infty}_{sc}(S_{u,\ub'})}\|\nab^3\omb,\nab^3\omb^{\dagger}\|_{L^{2}_{sc}(S_{u,\ub'})} d\ub'\\
&+\int_0^{\ub}\f{\d^{-\f12}}{|u|}\|\omb\|_{L^{\infty}_{sc}(S_{u,\ub'})}\|\nab^3\o\|_{L^{2}_{sc}(S_{u,\ub'})} d\ub'\\
&+\int_0^{\ub}\f{\d^{-\f12}}{|u|}\|\eta,\etb\|_{L^{\infty}_{sc}(S_{u,\ub'})}\|\nab^3\eta,\nab^3\etb\|_{L^{2}_{sc}(S_{u,\ub'})} d\ub'\\
&+\d^{\f14}((\M I^{(0)})^3+\M R^3+\underline{\M R}^3)\\
\leq&\f{\d^{\f12}}{|u|}\D_0\D_5+\d^{\f14}((\M I^{(0)})^3+\M R^3+\underline{\M R}^3).
\end{split}
\end{equation*}
Recalling $\omb^{\dagger}$ is defined to be the solution of
$$\nab_4\omb^{\dagger}=\f12\sigma$$
with zero data. And $\underline{\kappa}$ is defined as
$$\underline{\kappa}:=-\nab\omb+{^{*}\nab}\omb^{\dagger}-\f12\beb.$$
Hence $\nab\omb$ and $\nab\omb^{\dagger}$ satisfy a $\div-\curl$ system:
$$\div\nab\omb=-\div\underline{\kappa}-\f12\div\beb,$$
$$\curl\nab\omb=0,$$
$$\curl\nab\omb^{\dagger}=\curl\underline{\kappa}+\f12\curl\beb,$$
$$\div\nab\omb^{\dagger}=0,$$
and elliptic estimates in Proposition \ref{ellipticthm}, we demonstrate
\begin{equation*}
\begin{split}
 \|\nab^3\omb,\nab^3\omb^{\dagger}\|_{L^2_{sc}(\S)}\leq& \|\nab^2\underline{\kappa}\|_{L^2_{sc}(\S)}+\|\nab^2\beb\|_{L^2_{sc}(\S)}\\
&+\d^{\f14}(\M I^{(0)}+\M R+\underline{\M R})^5\\
\leq&\|\nab^2\beb\|_{L^2_{sc}(\S)}+\f{\d^{\f12}}{|u|}\D_0\D_5+\d^{\f14}(\M I^{(0)}+\M R+\underline{\M R})^5.
\end{split}
\end{equation*}

\end{proof}

\begin{proposition}\label{O32etb}
 Under the assumption of Theorem \ref{main.thm} and the bootstrap assumption \eqref{BA.5}, we obtain
\begin{equation*}
\begin{split}
 \|\nab^3\etb\|_{L^2_{sc}(\S)}\leq& \|\nab^2\rho, \nab^2\sigma\|_{L^2_{sc}(\S)}+\f{\d^{\f12}}{|u|}\D_0\D_5+\d^{\f14}(\M I^{(0)}+\M R+\underline{\M R})^5\\
&+\d^{\f12}(\M O_{2,2}\M O_{0,\infty}+\M O^2_{1,4}).
\end{split}
\end{equation*}
\end{proposition}
\begin{proof}
From the estimates for $\nab^2\T^{(s,s')}$ above, we have
\begin{equation*}
\begin{split}
|u|^{-1}\|\nab^2\underline{\mu}\|_{L^2_{sc}(\S)}\leq&\int_{u_{\infty}}^{u}\f{\d^{\f12}}{|u|^4}\|\tr\chib\|_{L^{\infty}_{sc}(S_{u',\ub})}\|\nab^3\eta\|_{L^{2}_{sc}(S_{u',\ub})} du'\\
&+\int_{u_{\infty}}^{u}\f{\d^{\f12}}{|u|^4}\|\eta,\etb\|_{L^{\infty}_{sc}(S_{u,\ub'})}\|\nab^3\chib\|_{L^{2}_{sc}(S_{u,\ub'})} du'\\
&+\f{\d^{\f14}}{|u|}((\M I^{(0)})^3+\M R^3+\underline{\M R}^3)\\
\leq&\f{1}{|u|}\M O_{3,2}[\eta]+\f{\d^{\f12}}{|u|^2}\D_0\D_5+\f{\d^{\f14}}{|u|}((\M I^{(0)})^3+\M R^3+\underline{\M R}^3).
\end{split}
\end{equation*}

Hence the result in Proposition \ref{O32eta} implies
\begin{equation*}
\begin{split}
\|\nab^2\underline{\mu}\|_{L^2_{sc}(\S)}\leq& \|\nab^2\rho, \nab^2\sigma\|_{L^2_{sc}(\S)}+\f{\d^{\f12}}{|u|}\D_0\D_5+\d^{\f14}((\M I^{(0)})^3+\M R^3+\underline{\M R}^3)\\
&+\d^{\f12}(\M O_{2,2}\M O_{0,\infty}+\M O^2_{1,4}).
\end{split}
\end{equation*}

Thanks to the definition $\mub=-\div\etb-\rho$, $\eta$ obeys the following elliptic system:
$$\div\etb=-\mub-\rho,$$
$$\curl\etb=-\sigma-\f12\chih\wedge\chibh.$$
Employing the elliptic estimates in Proposition \ref{ellipticthm}, we obtain
\begin{equation*}
\begin{split}
 \|\nab^3\etb\|_{L^2_{sc}(\S)}\leq& \|\nab^2\mub\|_{L^2_{sc}(\S)}+\|\nab^2\rho, \nab^2\sigma\|_{L^2_{sc}(\S)}\\
&+{\d^{\f12}}\|\nab^2\chih\|_{L^2_{sc}(\S)}\f{1}{|u|}\|\chibh\|_{L^{\infty}_{sc}(\S)}+{\d^{\f12}}\|\chih\|_{L^{\infty}_{sc}(\S)}\f{1}{|u|}\|\nab^2\chibh\|_{L^{2}_{sc}(\S)}\\
&+{\d^{\f12}}\|\nab\chih\|_{L^4_{sc}(\S)}\f{1}{|u|}\|\nab\chibh\|_{L^{4}_{sc}(\S)}+\d^{\f14}(\M I^{(0)}+\M R+\underline{\M R})^5\\
\leq&\|\nab^2\rho, \nab^2\sigma\|_{L^2_{sc}(\S)}+\f{\d^{\f12}}{|u|}\D_0\D_5+\d^{\f14}(\M I^{(0)}+\M R+\underline{\M R})^5\\
&+\d^{\f12}(\M O_{2,2}\M O_{0,\infty}+\M O^2_{1,4}).
\end{split}
\end{equation*}

\end{proof}

\begin{proposition}
 Under the assumption of Theorem \ref{main.thm} and the bootstrap assumption \eqref{BA.5}, we have
\begin{equation*}
\begin{split}
 \|\nab^3\o,\nab^3\o^{\dagger}\|_{L^2_{sc}(\S)}\leq& \|\nab^2\b\|_{L^2_{sc}(\S)}+\f{\d^{\f12}}{|u|^2}\D_0\D_5+\d^{\f14}(\M I^{(0)}+\M R+\underline{\M R})^5.
\end{split}
\end{equation*}
\end{proposition}

\begin{proof}
Estimates obtained for $\nab^2\T^{(s,s')}$ lead to
\begin{equation*}
\begin{split}
|u|^{-1}\|\nab^2\kappa\|_{L^2_{sc}(\S)}\leq&\int_{u_{\infty}}^{u}\f{\d^{\f12}}{|u|^4}\|\chibh,\omb\|_{L^{\infty}_{sc}(S_{u',\ub})}\|\nab^3\o,\nab^3\o^{\dagger}\|_{L^{2}_{sc}(S_{u',\ub})} du'\\
&+\int_{u_{\infty}}^{u}\f{\d^{\f12}}{|u|^4}\|\o\|_{L^{\infty}_{sc}(S_{u',\ub})}\|\nab^3\omb\|_{L^{2}_{sc}(S_{u',\ub})} du'\\
&+\int_{u_{\infty}}^{u}\f{\d^{\f12}}{|u|^4}\|\eta,\etb\|_{L^{\infty}_{sc}(S_{u',\ub})}\|\nab^3\eta,\nab^3\etb\|_{L^{2}_{sc}(S_{u',\ub})} du'\\
&+\f{\d^{\f14}}{|u|}((\M I^{(0)})^3+\M R^3+\underline{\M R}^3).\\
\end{split}
\end{equation*}
Multiplying $|u|$ on both sides, we derive
$$\|\nab^2\kappa\|_{L^2_{sc}(\S)}\leq\f{\d^{\f12}}{|u|^2}\D_0\D_5+\d^{\f14}((\M I^{(0)})^3+\M R^3+\underline{\M R}^3).$$

Recalling $\o^{\dagger}$ solves
$$\nab_3\o^{\dagger}=\f12\sigma$$
with zero initial data on $H_{u_{\infty}}$. And ${\kappa}$ is defined:
$${\kappa}:=\nab\o+{^{*}\nab}\o^{\dagger}-\f12\beb.$$
Hence $\nab\o$ and $\nab\o^{\dagger}$ satisfy a $\div-\curl$ system:
$$\div\nab\o=\div{\kappa}+\f12\div\b,$$
$$\curl\nab\o=0,$$
$$\curl\nab\o^{\dagger}=\curl{\kappa}+\f12\curl\b,$$
$$\div\nab\o^{\dagger}=0,$$
and elliptic estimates in Proposition \ref{ellipticthm}, we demonstrate
\begin{equation*}
\begin{split}
 \|\nab^3\o,\nab^3\o^{\dagger}\|_{L^2_{sc}(\S)}\leq& \|\nab^2{\kappa}\|_{L^2_{sc}(\S)}+\|\nab^2\b\|_{L^2_{sc}(\S)}\\
&+\d^{\f14}(\M I^{(0)}+\M R+\underline{\M R})^5\\
\leq&\|\nab^2\b\|_{L^2_{sc}(\S)}+\f{\d^{\f12}}{|u|^2}\D_0\D_5+\d^{\f14}(\M I^{(0)}+\M R+\underline{\M R})^5.
\end{split}
\end{equation*}
\end{proof}

We are left to derive bounds for $\|\nab^3\tr\chib\|_{L^2_{sc}(\S)}$ and $\|\nab^3\chibh\|_{L^2_{sc}(\S)}$.

\begin{proposition}
 Under the assumption of Theorem \ref{main.thm} and the bootstrap assumption \eqref{BA.5}, we have
\begin{equation*}
\begin{split}
\|\nab^3\tr\chib\|_{L^2_{sc}(\S)}\leq \|\nab^2\beb\|_{L^2_{sc}(\S)}+{\d^{\f12}}\D_0\D_5+\d^{\f14}((\M I^{(0)})+\M R+\underline{\M R})^5,
\end{split}
\end{equation*}
and
\begin{equation*}
\begin{split}
&\f{1}{|u|}\|\nab^3\chibh\|_{L^2_{sc}(\S)}\\
\leq & \f{1}{|u|}\|\nab^2\beb\|_{L^2_{sc}(\S)}+\|\nab^2\rho, \nab^2\sigma\|_{L^2_{sc}(\S)}\\
&+\f{\d^{\f12}}{|u|}\D_0\D_5+\d^{\f14}(\M I^{(0)}+\M R+\underline{\M R})^5+\d^{\f12}(\M O_{2,2}\M O_{0,\infty}+\M O^2_{1,4}).\\
\end{split}
\end{equation*}


\end{proposition}

\begin{proof}
The derived estimates for $\nab^2\T^{(s,s')}$ guarantee that
\begin{equation*}
\begin{split}
&|u|^{-1}\|\nab^3(\oo)\|_{L^2_{sc}(\S)}\\
\leq&\int_{u_{\infty}}^{u}\f{\d^{\f12}}{|u|^4}\|\tr\chib\|_{L^{\infty}_{sc}(S_{u',\ub})}\|\nab^3\omb\|_{L^{2}_{sc}(S_{u',\ub})} du'\\
&+\int_{u_{\infty}}^{u}\f{\d^{\f12}}{|u|^4}\|\chibh,\omb, \oo\|_{L^{\infty}_{sc}(S_{u',\ub})}\|\nab^3\chib\|_{L^{2}_{sc}(S_{u',\ub})} du'\\
&+\d^{\f14}((\M I^{(0)})^3+\M R^3+\underline{\M R}^3)\\
\leq&\f{1}{|u|}\M O_{3,2}[\omb]+\f{\d^{\f12}}{|u|}\D_0\D_5+\d^{\f14}((\M I^{(0)})^3+\M R^3+\underline{\M R}^3).
\end{split}
\end{equation*}

Hence the result in Proposition \ref{O32omb} implies
\begin{equation*}
\begin{split}
\|\nab^3(\oo)\|_{L^2_{sc}(\S)}\leq &\M O_{3,2}[\omb]+{\d^{\f12}}\D_0\D_5+\d^{\f14}((\M I^{(0)})^3+\M R^3+\underline{\M R}^3)\\
\leq& \|\nab^2\beb\|_{L^2_{sc}(\S)}+{\d^{\f12}}\D_0\D_5+\d^{\f14}((\M I^{(0)})^3+\M R^3+\underline{\M R}^3).
\end{split}
\end{equation*}

To obtain estimate for $\nab^3\tr\chib$, we employ
\begin{equation*}
\begin{split}
&\|\nab^3\tr\chib\|_{L^2_{sc}(\S)}\\
\leq&  \|\nab^3(\oo)\|_{L^2_{sc}(\S)}+\f{\d^{\f12}}{|u|^2}\|\tr\chib\|_{L^{\infty}_{sc}(\S)}|u|\|\nab^3\O\|_{L^2_{sc}(\S)}\\
&+\f{\d^{\f12}}{|u|}\|\nab^2\O\|_{L^4_{sc}(\S)}\|\nab\tr\chib\|_{L^4_{sc}(\S)}+\f{\d^{\f12}}{|u|}\|\nab\O\|_{L^4_{sc}(\S)}\|\nab^2\tr\chib\|_{L^4_{sc}(\S)}\\
\leq&  \|\nab^3(\oo)\|_{L^2_{sc}(\S)}+\f{\d^{\f12}}{|u|^2}\|\tr\chib\|_{L^{\infty}_{sc}(\S)}|u|\|\nab^3\O\|_{L^2_{sc}(\S)}\\
&+\f{\d^{\f12}}{|u|}\|\nab\eta,\nab\etb\|_{L^4_{sc}(\S)}\|\nab\tr\chib\|_{L^4_{sc}(\S)}+\f{\d^{\f12}}{|u|}\|\eta,\etb\|_{L^4_{sc}(\S)}\|\nab^2\tr\chib\|_{L^4_{sc}(\S)}\\
&+\f{\d}{|u|^2}\|\eta,\etb\|_{L^{\infty}_{sc}(\S)}\|\eta,\etb\|_{L^4_{sc}(\S)}\|\nab\tr\chib\|_{L^4_{sc}(\S)},
\end{split}
\end{equation*}
where we adopt Proposition \ref{Omega} and the fact
$\nab\O=\O\ee.$

As a consequence, we need estimate for $\|\nab^3\O\|_{L^2_{sc}(\S)}$.
Using Propositions \ref{commute0}-\ref{commute3}, and
$$\nab_3\O=-2\omb,$$
it can be verified that for $\O$, we have
\begin{equation*}
\begin{split}
\nab_3\nab\O+\f12\tr\chib\nab\O=(\chibh,\omb)\nab\O+\O\nab\omb+\ee\O\omb,
\end{split}
\end{equation*}

\begin{equation*}
\begin{split}
&\nab_3\nab^2\O+\tr\chib\nab^2\O\\
=&\chibh\nab^2\O+(\nab\chib,\nab\omb)\nab\O+(\chibh,\omb)\nab^2\O+\O\nab^2\omb+(\nab\eta,\nab\etb)\O\omb\\
&+\ee\nab\O\omb+\ee\O\nab\omb+\chib\ee\nab\O+\beb\nab\O+\ee\ee\O\omb,
\end{split}
\end{equation*}
and
\begin{equation*}
\begin{split}
&\nab_3\nab^3\O+\f32\tr\chib\nab^3\O\\
=&\nab\chib\nab^2\O+(\chibh,\omb)\nab^3\O+\nab\omb\nab^2\O+(\nab^2\chib,\nab^2\omb)\nab\O+\O\nab^3\omb+(\nab^2\eta,\nab^2\etb)\O\omb\\
&+(\nab\eta,\nab\etb)\nab\O\omb+(\nab\eta,\nab\etb)\O\nab\omb+\ee\nab^2\O\omb+\ee\nab\O\nab\omb\\
&+\ee\O\nab^2\omb+\nab\chib\ee\nab\O+\chib(\nab\eta,\nab\etb)\nab\O+\chib\ee\nab^2\O+\nab\beb\nab\O+\beb\nab^2\O\\
&+(\nab\eta,\nab\etb)\ee\O\omb+\ee\ee\nab\O\omb+\ee\ee\O\nab\omb+\chib\ee\ee\nab\O\\
&+\beb\nab\O\ee+\ee\ee\ee\O\omb.
\end{split}
\end{equation*}

$\nab^3\O$ has signature $(\f32,\f32)$. With Gronwall's inequality and Proposition \ref{evolution lemma},
we obtain

\begin{equation*}
\begin{split}
&\f{1}{|u|}\|\nab^3\O\|_{L^2_{sc}(\S)}\\
\leq &\int_{u_{\infty}}^u\f{1}{|u'|^3}\|\nab^3\omb\|_{L^2_{sc}(S_{u',\ub})}du'+\int_{u_{\infty}}^u\f{\d^{\f12}}{|u'|^3}\f{1}{|u'|}\|\nab\chib,\nab\omb\|_{L^4_{sc}(S_{u',\ub})}\|\nab^2\O\|_{L^4_{sc}(S_{u',\ub})}du'\\
&+\int_{u_{\infty}}^u\f{\d^{\f12}}{|u'|^3}\f{1}{|u'|}\|\nab^2\chib,\nab^2\omb\|_{L^4_{sc}(S_{u',\ub})}\|\nab\O\|_{L^4_{sc}(S_{u',\ub})}du'\\
&+\int_{u_{\infty}}^u\f{\d^{\f12}}{|u'|^4}\|\nab^2\eta,\nab^2\etb\|_{L^4_{sc}(S_{u',\ub})}\|\omb\|_{L^4_{sc}(S_{u',\ub})}du'\\
&+\int_{u_{\infty}}^u\f{\d^{\f12}}{|u'|^4}\|\nab\eta,\nab\etb\|_{L^4_{sc}(S_{u',\ub})}\|\nab\omb\|_{L^4_{sc}(S_{u',\ub})}du'\\
&+\int_{u_{\infty}}^u\f{\d^{\f12}}{|u'|^4}\|\eta,\etb\|_{L^4_{sc}(S_{u',\ub})}\|\nab^2\omb\|_{L^4_{sc}(S_{u',\ub})}du'\\
&+\int_{u_{\infty}}^u\f{\d^{\f12}}{|u'|^4}\|\nab\beb\|_{L^4_{sc}(S_{u',\ub})}\|\nab\O\|_{L^4_{sc}(S_{u',\ub})}du'\\
&+\int_{u_{\infty}}^u\f{\d^{\f12}}{|u'|^4}\|\beb\|_{L^4_{sc}(S_{u',\ub})}\|\nab^2\O\|_{L^4_{sc}(S_{u',\ub})}du'\\
&+\int_{u_{\infty}}^u\f{\d}{|u'|^5}\|\nab\eta,\nab\etb\|_{L^4_{sc}(S_{u',\ub})}\|\nab\O\|_{L^4_{sc}(S_{u',\ub})}\|\omb\|_{L^{\infty}_{sc}(S_{u',\ub})}du'\\
&+\int_{u_{\infty}}^u\f{\d}{|u'|^5}\|\eta,\etb\|_{L^4_{sc}(S_{u',\ub})}\|\nab^2\O\|_{L^4_{sc}(S_{u',\ub})}\|\omb\|_{L^{\infty}_{sc}(S_{u',\ub})}du'\\
&+\int_{u_{\infty}}^u\f{\d}{|u'|^4}\|\eta,\etb\|_{L^{\infty}_{sc}(S_{u',\ub})}\|\nab\O\|_{L^4_{sc}(S_{u',\ub})}\f{1}{|u'|}\|\nab\omb,\nab\chib\|_{L^{4}_{sc}(S_{u',\ub})}du'\\
&+\int_{u_{\infty}}^u\f{\d^{\f12}}{|u'|^3}\f{\d^{\f12}}{|u'|^2}\|\chib\|_{L^{\infty}_{sc}(S_{u',\ub})}\|\nab\O\|_{L^4_{sc}(S_{u',\ub})}\|\nab\eta,\nab\etb\|_{L^{4}_{sc}(S_{u',\ub})}du'\\
&+\int_{u_{\infty}}^u\f{\d^{\f12}}{|u'|^3}\f{\d^{\f12}}{|u'|^2}\|\chib\|_{L^{\infty}_{sc}(S_{u',\ub})}\|\nab^2\O\|_{L^4_{sc}(S_{u',\ub})}\|\eta,\etb\|_{L^{4}_{sc}(S_{u',\ub})}du'\\
&+\int_{u_{\infty}}^u\f{\d}{|u'|^5}\|\nab\eta,\nab\etb\|_{L^4_{sc}(S_{u',\ub})}\|\eta,\etb\|_{L^4_{sc}(S_{u',\ub})}\|\omb\|_{L^{\infty}_{sc}(S_{u',\ub})}du'\\
&+\int_{u_{\infty}}^u\f{\d}{|u'|^5}\|\nab\omb\|_{L^4_{sc}(S_{u',\ub})}\|\eta,\etb\|_{L^4_{sc}(S_{u',\ub})}\|\eta,\etb\|_{L^{\infty}_{sc}(S_{u',\ub})}du'\\
&+\int_{u_{\infty}}^u\f{\d}{|u'|^5}\|\beb\|_{L^2_{sc}(S_{u',\ub})}\|\nab\O\|_{L^{\infty}_{sc}(S_{u',\ub})}\|\eta,\etb\|_{L^{\infty}_{sc}(S_{u',\ub})}du'\\
&+\int_{u_{\infty}}^u\f{\d}{|u'|^4}\f{\d^{\f12}}{|u'|^2}\|\chib,\omb\|_{L^{\infty}_{sc}(S_{u',\ub})}\|\nab\O, \eta, \etb\|_{L^{\infty}_{sc}(S_{u',\ub})}\|\eta,\etb\|_{L^{\infty}_{sc}(S_{u',\ub})}\|\eta,\etb\|_{L^{\infty}_{sc}(S_{u',\ub})}du'.\\
\end{split}
\end{equation*}

Therefore, we derive
$$\f{1}{|u|}\|\nab^3\O\|_{L^2_{sc}(\S)}\leq \f{1}{|u|^2}\M O_{3,2}[\omb]+\f{\d^{\f12}}{|u|^2}(\M I^{(0)}+\M R+\underline{\M R})^5.$$
Multiplying $|u|^2$ on both sides, we obtain
$$|u|\|\nab^3\O\|_{L^2_{sc}(\S)}\leq \M O_{3,2}[\omb]+{\d^{\f12}}(\M I^{(0)}+\M R+\underline{\M R})^5.$$

Putting the estimates above altogether and adopting the result in Proposition \ref{O32omb}, we derive
\begin{equation*}
\begin{split}
\|\nab^3\tr\chib\|_{L^2_{sc}(\S)}\leq &\M O_{3,2}[\omb]+{\d^{\f12}}\D_0\D_5+\d^{\f14}((\M I^{(0)})+\M R+\underline{\M R})^5\\
\leq& \|\nab^2\beb\|_{L^2_{sc}(\S)}+{\d^{\f12}}\D_0\D_5+\d^{\f14}((\M I^{(0)})+\M R+\underline{\M R})^5.
\end{split}
\end{equation*}

To obtain the estimates for $\nab^3\chibh$, we employ the Codazzi equation
$$\div\chibh=\beb+\f12\nab\tr\chib-\f12(\eta-\etb)\cdot(\chibh-\f12\tr\chib)$$
and elliptic estimates in Proposition \ref{ellipticthm2} to derive
\begin{equation*}
\begin{split}
&\f{1}{|u|}\|\nab^3\chibh\|_{L^2_{sc}(\S)}\\
\leq & \f{1}{|u|}\|\nab^3\tr\chib\|_{L^2_{sc}(\S)}+\f{1}{|u|}\|\nab^2\beb\|_{L^2_{sc}(\S)}+\|\nab^2\eta,\nab^2\etb\|_{L^2_{sc}(\S)}\f{\d^{\f12}}{|u|^2}\|\chib\|_{L^{\infty}_{sc}(\S)}\\
&+\f{\d^{\f12}}{|u|}\|\nab\eta,\nab\etb\|_{L^4_{sc}(\S)}\f{1}{|u|}\|\nab\chib\|_{L^{4}_{sc}(\S)}+\f{\d^{\f12}}{|u|}\|\eta,\etb\|_{L^{\infty}_{sc}(\S)}\f{1}{|u|}\|\nab^2\chib\|_{L^{2}_{sc}(\S)}\\
&+\d^{\f14}(\M I^{(0)}+\M R+\underline{\M R})^5\\
\leq & \f{1}{|u|}\|\nab^2\beb\|_{L^2_{sc}(\S)}+\|\nab^2\eta,\nab^2\etb\|_{L^2_{sc}(\S)}+\f{\d^{\f12}}{|u|}\D_0\D_5+\d^{\f14}(\M I^{(0)}+\M R+\underline{\M R})^5\\
&+\f{\d^{\f12}}{|u|}(\M O_{2,2}\M O_{0,\infty}+\M O_{1,4}\M O_{1,4}).
\end{split}
\end{equation*}

For $\|\nab^2\eta,\nab^2\etb\|_{L^2_{sc}(\S)}$, we have
$$\|\nab^2\eta,\nab^2\etb\|_{L^2_{sc}(\S)}\leq \d^{\f14}\|\nab^2\eta,\nab^2\etb\|_{L^4_{sc}(\S)}.$$

The conclusions in Proposition \ref{O14} immediately imply

\begin{equation*}
\begin{split}
&\f{1}{|u|}\|\nab^3\chibh\|_{L^2_{sc}(\S)}\\
\leq & \f{1}{|u|}\|\nab^2\beb\|_{L^2_{sc}(\S)}+\d^{\f14}\M O_{1,4}\\
&+\f{\d^{\f12}}{|u|}\D_0\D_5+\d^{\f14}(\M I^{(0)}+\M R+\underline{\M R})^5+\d^{\f12}(\M O_{2,2}\M O_{0,\infty}+\M O^2_{1,4}).\\
\end{split}
\end{equation*}
\end{proof}

Gathering all the estimates above, we obtain
\begin{equation*}
\begin{split}
 \M O_{3,2}\leq \|\nab^2\b,\nab^2\rho,\nab^2\sigma,\nab^2\beb\|_{L^2_{sc}(\S)}+\d^{\f14}(\M I^{(0)}+\M R+\underline{\M R})^5.
\end{split}
\end{equation*}

Let $$\D_5\gg \|\nab^2\b,\nab^2\rho,\nab^2\sigma,\nab^2\beb\|_{L^2_{sc}(\S)}+\d^{\f14}(\M I^{(0)}+\M R+\underline{\M R})^5.$$
Then we improve bootstrap assumption \eqref{BA.5} and derive
\begin{proposition}\label{O32}
 \begin{equation*}
\begin{split}
 \M O_{3,2}\leq C\l \|\nab^2\b,\nab^2\rho,\nab^2\sigma,\nab^2\beb\|_{L^2_{sc}(\S)}+\d^{\f14}(\M I^{(0)}+\M R+\underline{\M R})^5\r,
\end{split}
\end{equation*}
where $C$ is a large universal constant.
\end{proposition}

\subsection{$\M O_{2,4}$ ESTIMATES}
We recall the following norms:
\begin{equation*}
\begin{split}
&\M O_{2,4}\\
=&\sup_{u,\ub}\l\|\nab^2\chih, \nab^2\tr\chi, \nab^2\o, \nab^2\eta, \nab^2\etb, \nab^2\tr\chib, \nab^2\omb\|_{L^4_{sc}(\S)}\\
&\quad\quad\quad+\f{1}{|u|}\|\nab^2\chibh\|_{L^4_{sc}(\S)}\r.
\end{split}
\end{equation*}
For any $S_{u,\ub}$-tangent tensor $\phi$, we have the following Gagliardo-Nirenberg's inequality
$$\|\nab^2 \phi\|_{L^4_{sc}(\S)}\leq \|\nab^3 \phi\|^{\f23}_{L^2_{sc}(\S)}\|\nab \phi\|^{\f13}_{L^4_{sc}(\S)}.$$
With this inequality, for $\M O_{2,4}$ we deduce

\begin{proposition}\label{O24}
\begin{equation*}
 \M O_{2,4}\leq C\|\nab^2\b, \nab^2\rho, \nab^2\sigma, \nab^2\beb\|_{L^2_{sc}(\S)}+C(\M I^{(0)}+\M R+\underline{\M R})^5. 
\end{equation*}
\end{proposition}

Moreover, in the end of of this section we prove a useful proposition for energy estimate.
\begin{proposition}\label{O242}
Under the assumption of Theorem \ref{main.thm}, we have 
 \begin{equation*}
 \begin{split}
&\int_{u_{\infty}}^{u}\f{\d^{\f12}}{|u'|^2}\f{1}{|u'|}\|\nab^2\chi, \nab^2\o, \nab^2\eta, \nab^2\etb, \nab^2\chib, \nab^2\omb\|_{L^{4}_{sc}(S_{u',\ub})} du'\\ 
\leq&\f{\d^{\f12}}{|u|^{\f12}}C\|\nab^2\b, \nab^2\rho, \nab^2\sigma, \nab^2\beb\|_{L^2_{sc}(\Hbu)}+\f{\d^{\f12}}{|u|}C(\M I^{(0)}+\M R+\underline{\M R})^5,
\end{split}
\end{equation*}
and
\begin{equation*}
 \begin{split}
&\int_{0}^{\ub}\d^{-\f12}\f{1}{|u|}\|\nab^2\chi, \nab^2\o, \nab^2\eta, \nab^2\etb, \nab^2\chib, \nab^2\omb\|_{L^{4}_{sc}(S_{u,\ub'})} d\ub'\\ 
\leq&{\d^{\f12}}C\|\nab^2\b, \nab^2\rho, \nab^2\sigma, \nab^2\beb\|_{L^2_{sc}(\Hu)}+{\d^{\f12}}C(\M I^{(0)}+\M R+\underline{\M R})^5.
\end{split}
\end{equation*}
\end{proposition}

\begin{proof}
Employing the results in Proposition \ref{O32}, we derive
\begin{equation*}
 \begin{split}
&\int_{u_{\infty}}^{u}\f{\d^{\f12}}{|u'|^2}\f{1}{|u'|}\|\nab^2\chi, \nab^2\o, \nab^2\eta, \nab^2\etb, \nab^2\chib, \nab^2\omb\|_{L^{4}_{sc}(S_{u',\ub})} du'\\ 
\leq& \int_{u_{\infty}}^{u}\f{\d^{\f12}}{|u'|^2}C\l\|\nab\b, \nab\rho, \nab\sigma, \nab\beb\|_{L^2_{sc}(\S)}+(\M I^{(0)}+\M R+\underline{\M R})^5\r du'\\
\leq&\f{\d^{\f12}}{|u|^{\f12}}C\|\nab\b, \nab\rho, \nab\sigma, \nab\beb\|_{L^2_{sc}(\Hbu)}+\f{\d^{\f12}}{|u|}C(\M I^{(0)}+\M R+\underline{\M R})^5,
\end{split}
\end{equation*}
and
\begin{equation*}
 \begin{split}
&\int_{0}^{\ub}\d^{-\f12}\f{1}{|u|}\|\nab^2\chi, \nab^2\o, \nab^2\eta, \nab^2\etb, \nab^2\chib, \nab^2\omb\|_{L^{4}_{sc}(S_{u,\ub'})} d\ub'\\ 
\leq& \int_{0}^{\ub}\d^{-\f12}C\l\|\nab\b, \nab\rho, \nab\sigma, \nab\beb\|_{L^2_{sc}(\S)}+(\M I^{(0)}+\M R+\underline{\M R})^5\r\\
\leq&{\d^{\f12}}C\|\nab\b, \nab\rho, \nab\sigma, \nab\beb\|_{L^2_{sc}(\Hu)}+{\d^{\f12}}C(\M I^{(0)}+\M R+\underline{\M R})^5.
\end{split}
\end{equation*}

\end{proof}

\section{CURVATURE ESTIMATE}
Now we move to energy estimates (curvature estimate).
We will show in particular $\M R+\underline{\M R}\ls \M I^{(0)}$. Together with the estimates in the previous
sections, we will therefore improve all of the bootstrap assumptions \eqref{BA.0}, \eqref{BA.1}, \eqref{BA.2} and \eqref{BA.4} to obtain
Theorem \ref{main.thm}.

The estimates in this section are based on an {\bf key observation}:

Both curvature components $\Psi$ with signature $s(\Psi)=(s,s')$, and the kth-order angular derivatives of curvature components $\Psi$ with signature $s(\nab^k\Psi)=(s,s')$, satisfy the following transport equations:

\begin{equation} \label{tran3333}
\nab_3 \Psi^{(s,s')}+(\frac 12+s')\tr\chib\Psi^{(s,s')}=\M D\Psi^{(s-\frac12,s'+\frac12)}+\sum_{s_1+s_2=s, \atop s'_1+s'_2=s'+1}\psi^{(s_1,s'_1)}\cdot\Psi^{(s_2,s'_2)},
\end{equation}
or
\begin{equation} \label{tran4444}
\nab_4 \Psi^{(s-\frac 12,s'+\frac 12)}={^{*}{\M D}}\Psi^{(s,s')}+\sum_{\st_1+\st_2=s+\frac 12, \atop \st'_1+\st'_2=s'+\frac 12}\psi^{(\st_1,\st'_1)}\cdot\Psi^{(\st_2,\st'_2)},
\end{equation}
where $\Psi^{(s,s')}$ and $\psi^{(s,s')}$ stand for S-tangent tensor fields with signatures $(s,s')$.
$\M D$ and ${^{*}\M D}$ are Hodge operators listed below. And ${^{*}\M D}$ is the $L^2$ adjoint of ${\M D}$. 

\begin{itemize}
 \item The operator $\M D_1$ takes any 1-form $\phi$ into the pairs of functions $(\div \phi, \curl \phi)$.
 \item The operator $\M D_2$ takes any $S$ tangent symmetric traceless tensor $\phi$ into the $S$-tangent 
       one form $\div \phi$.
 \item The operator ${^*{\M D}}_1$ takes the pair of scalar functions $(\rho,\sigma)$ into the $S$-tangent
        1-form $-\nab\rho+{^*\nab\sigma}$.
 \item The operator ${^*{\M D}}_2$ takes 1-form $\phi$ on $S$ into the 2-covariant, symmetric, traceless tensors
       $-\f12\l\nab_b\phi_a+\nab_a\phi_b-(\div\phi)\gamma_{ab}\r$.
\end{itemize}

When $k=0$, we can simply check null Bianchi equations. 
When $k\geq1$, this observation can be verified by taking $k$th order angular derivatives of null Bianchi equations and using Propositions \ref{commute0}-\ref{commute3}.

For Hodge operators $\M D$ and ${^{*}\M D}$, we address a useful proposition:
\begin{proposition}
For  $\Psi^{(s,s')}$ and $\Psi^{(s-\frac 12,s'+\frac 12)}$ satisfying \eqref{tran3333} and \eqref{tran4444}, we have
\begin{equation}
 \int_{D_{u,\ub}}\l|u'|^{4s'}\Psi^{(s,s')}\M D\Psi^{(s-\frac12,s'+\frac12)}+|u'|^{4s'}\Psi^{(s-\frac 12,s'+\frac 12)}{^{*}{\M D}}\Psi^{(s,s')}\r du' d\ub'=0.
\end{equation}
\end{proposition}
\begin{proof}
 This follows from the definitions of $\M D, {^{*}\M D}$ and integrating by parts.
\end{proof}


\subsection{Curvature Estimates in the Scale Invariant Norms.}
Through integration by parts, we have the following formula:
\begin{proposition}\label{intbyparts4}
Give $r$ tensorfields $\phi_1$ and $\phi_2$, we have
\begin{equation*}
\begin{split}
&\int_{D_{u,\ub}} \phi_1 \nabla_4\phi_2+\int_{D_{u,\ub}}\phi_2\nabla_4\phi_1\\
=& \int_{\Hb_{\ub}^{(u_{\infty},u)}} \phi_1\phi_2-\int_{\Hb_0^{(u_{\infty},u)}} \phi_1\phi_2+\int_{D_{u,\ub}}(2\omega-\trch)\phi_1\phi_2.
\end{split}
\end{equation*}
\end{proposition}

\begin{proposition}\label{intbyparts3}
Given $r$ tensorfields $\phi_1$ and $\phi_2$, we have
\begin{equation*}
\begin{split}
&\int_{D_{u,\ub}} \phi_1 \nabla_3\phi_2+\int_{D_{u,\ub}}\phi_2\nabla_3\phi_1\\
=& \int_{\Hu} \phi_1\phi_2-\int_{H_{u_{\infty}}^{(0,\ub)}} \phi_1\phi_2+\int_{D_{u,\ub}}(2\omb-\tr\chib)\phi_1\phi_2.
\end{split}
\end{equation*}
\end{proposition}

\begin{proposition}\label{intbyparts12}
Given an $r$ tensorfield $^{(1)}\phi$ and an $r-1$ tensorfield $^{(2)}\phi$, we have
\begin{equation*}
\begin{split}
&\int_{D_{u,\ub}}{ }^{(1)}\phi^{A_1A_2...A_r}\nabla_{A_r}{ }^{(2)}\phi_{A_1...A_{r-1}}+\int_{D_{u,\ub}}\nabla^{A_r}{ }^{(1)}\phi_{A_1A_2...A_r}{ }^{(2)}\phi^{A_1...A_{r-1}}\\
=& -\int_{D_{u,\ub}}(\eta+\etab){ }^{(1)}\phi{ }^{(2)}\phi.
\end{split}
\end{equation*}
\end{proposition}

With these propositions above, we obtain
\begin{proposition}\label{energy1}
Assuming a pair of S-tangent tensor fields $\Psi^{(s,s')}$ and $\Psi^{(s-\frac 12,s'+\frac 12)}$ satisfy 
\begin{equation*} 
\nab_3 \Psi^{(s,s')}+(\frac 12+s')\tr\chib\Psi^{(s,s')}=\M D\Psi^{(s-\frac12,s'+\frac12)}+J_3,
\end{equation*}
and
\begin{equation*} 
\nab_4 \Psi^{(s-\frac 12,s'+\frac 12)}={^{*}{\M D}}\Psi^{(s,s')}+J_4,
\end{equation*}
then we have
\begin{equation*}
\begin{split}
&\|\Psi^{(s,s')}\|^2_{L^2_{sc}(H_u^{(0,\ub)})}+\|\Psi^{(s-\frac12,s'+\frac12)}\|^2_{L^2_{sc}(\Hb_{\ub}^{(u_{\infty},u)})}\\
\leq & \|\Psi^{(s,s')}\|^2_{L^2_{sc}(H_{u_{\infty}}^{(0,\ub)})}+\|\Psi^{(s-\frac12,s'+\frac12)}\|^2_{L^2_{sc}(\Hb_{0}^{(u_{\infty},u)})}\\
&+ \int_0^{\ub}\int_{u_{\infty}}^u \frac{\delta^{-\frac32}}{|u'|}\|(\eta+\etb)\cdot\Psi^{(s,s')}\cdot\Psi^{(s-\frac12,s'+\frac12)}\|_{L^1_{sc}(S_{u',\ub'})} du'd{\ub}'\\
&+ \int_0^{\ub}\int_{u_{\infty}}^u \frac{\delta^{-\frac32}}{|u'|}\|\Psi^{(s,s')}\cdot J_3\|_{L^1_{sc}(S_{u',\ub'})} du'd{\ub}'\\
&+ \int_0^{\ub}\int_{u_{\infty}}^u \frac{\delta^{-\frac32}}{|u'|}\|\Psi^{(s-\frac12,s'+\frac12)}\cdot J_4\|_{L^1_{sc}(S_{u',\ub'})}du'd{\ub'}.
\end{split} 
\end{equation*}
\end{proposition}

\begin{proof}
With the aid of equations and employing Propositions \ref{intbyparts4}, \ref{intbyparts3} and \ref{intbyparts12}, we compute
\begin{equation*}
\begin{split} 
&\int_{D_{u,\ub}}<|u'|^{2s'}\Psi^{(s-\f12,s'+\f12)},|u'|^{2s'}\nab_4\Psi^{(s-\f12,s'+\f12)}>_{\gamma}\\
=&\int_{D_{u,\ub}}<|u'|^{2s'}\Psi^{(s-\f12,s'+\f12)},|u'|^{2s'}{^*\M D}\Psi^{(s,s')}>_{\gamma}\\
&+\int_{D_{u,\ub}}<|u'|^{2s'}\Psi^{(s-\f12,s'+\f12)},|u'|^{2s'}J_4>_{\gamma}+\int_{D_{u,\ub}}|u'|^{4s'}(\eta+\etb)\Psi^{(s,s')}\Psi^{(s-\frac12,s'+\frac12)}\\
=&\int_{D_{u,\ub}}<-|u'|^{2s'}{\M D}\Psi^{(s-\f12,s'+\f12)},|u'|^{2s'}\Psi^{(s,s')}>_{\gamma}\\
&+\int_{D_{u,\ub}}<|u'|^{2s'}\Psi^{(s-\f12,s'+\f12)},|u'|^{2s'}J_4>_{\gamma}+\int_{D_{u,\ub}}|u'|^{4s'}(\eta+\etb)\Psi^{(s,s')}\Psi^{(s-\frac12,s'+\frac12)}\\
=&-\int_{D_{u,\ub}}<|u'|^{2s'}\nab_3\Psi^{(s,s')},|u'|^{2s'}\Psi^{(s,s')}>_{\gamma}\\
&-\int_{D_{u,\ub}}<|u'|^{2s'}(\f12+s')\tr\chib\Psi^{(s,s')},|u'|^{2s'}\Psi^{(s,s')}>_{\gamma}\\
&+\int_{D_{u,\ub}}<|u'|^{2s'}\Psi^{(s,s')},|u'|^{2s'}J_3>_{\gamma}+\int_{D_{u,\ub}}<|u'|^{2s'}\Psi^{(s-\f12,s'+\f12)},|u'|^{2s'}J_4>_{\gamma}\\
&+\int_{D_{u,\ub}}|u'|^{4s'}(\eta+\etb)\Psi^{(s,s')}\Psi^{(s-\frac12,s'+\frac12)}\\
=&-\int_{D_{u,\ub}}<\nab_3(|u'|^{2s'}\Psi^{(s,s')}),|u'|^{2s'}\Psi^{(s,s')}>_{\gamma}\\
&-\int_{D_{u,\ub}}<\f{2s'}{|u'|}\Omega^{-1}|u'|^{2s'}\Psi^{(s,s')},|u'|^{2s'}\Psi^{(s,s')}>_{\gamma}\\
&-\int_{D_{u,\ub}}<|u'|^{2s'}(\f12+s')\tr\chib\Psi^{(s,s')},|u'|^{2s'}\Psi^{(s,s')}>_{\gamma}\\
&+\int_{D_{u,\ub}}<|u'|^{2s'}\Psi^{(s,s')},|u'|^{2s'}J_3>_{\gamma}+\int_{D_{u,\ub}}<|u'|^{2s'}\Psi^{(s-\f12,s'+\f12)},|u'|^{2s'}J_4>_{\gamma}\\
&+\int_{D_{u,\ub}}|u'|^{4s'}(\eta+\etb)\Psi^{(s,s')}\Psi^{(s-\frac12,s'+\frac12)}\\
=&-\int_{D_{u,\ub}}<\nab_3(|u'|^{2s'}\Psi^{(s,s')}),|u'|^{2s'}\Psi^{(s,s')}>_{\gamma}\\
&-\int_{D_{u,\ub}}<\f{2s'}{|u'|}(\Omega^{-1}-1)|u'|^{2s'}\Psi^{(s,s')},|u'|^{2s'}\Psi^{(s,s')}>_{\gamma}\\
&+\int_{D_{u,\ub}}<\f{1}{|u'|}|u'|^{2s'}\Psi^{(s,s')},|u'|^{2s'}\Psi^{(s,s')}>_{\gamma}\\
&-\int_{D_{u,\ub}}<|u'|^{2s'}(\f12+s')(\tr\chib+\f{2}{|u'|})\Psi^{(s,s')},|u'|^{2s'}\Psi^{(s,s')}>_{\gamma}\\
&+\int_{D_{u,\ub}}<|u'|^{2s'}\Psi^{(s,s')},|u'|^{2s'}J_3>_{\gamma}+\int_{D_{u,\ub}}<|u'|^{2s'}\Psi^{(s-\f12,s'+\f12)},|u|^{2s'}J_4>_{\gamma}\\
&+\int_{D_{u,\ub}}|u'|^{4s'}(\eta+\etb)\Psi^{(s,s')}\Psi^{(s-\frac12,s'+\frac12)}.\\
\end{split}
\end{equation*}

On the other hand, applying Proposition \ref{intbyparts3} gives
\begin{equation*}
\begin{split} 
&-\int_{D_{u,\ub}}<\nab_3(|u'|^{2s'}\Psi^{(s,s')}),|u'|^{2s'}\Psi^{(s,s')}>_{\gamma}\\
=&-\f12\int_{\Hu}|u'|^{4s'}(\Psi^{(s,s')})^2+\f12 \int_{H_{u_{\infty}}^{(0,\ub)}}|u'|^{4s'}(\Psi^{(s,s')})^2\\
&-\int_{D_{u,\ub}}\omb|u'|^{4s'}(\Psi^{(s,s')})^2+\f12\int_{D_{u,\ub}}\tr\chib|u'|^{4s'}(\Psi^{(s,s')})^2\\
=&-\f12\int_{\Hu}|u'|^{4s'}(\Psi^{(s,s')})^2+\f12 \int_{H_{u_{\infty}}^{(0,\ub)}}|u'|^{4s'}(\Psi^{(s,s')})^2\\
&-\int_{D_{u,\ub}}\omb|u'|^{4s'}(\Psi^{(s,s')})^2+\f12\int_{D_{u,\ub}}(\tr\chib+\f{2}{|u'|})|u'|^{4s'}(\Psi^{(s,s')})^2\\
&-\int_{D_{u,\ub}}\f{1}{|u'|}|u'|^{4s'}(\Psi^{(s,s')})^2.
\end{split}
\end{equation*}

Combining the two identities above together, we arrive at 
\begin{equation*}
\begin{split} 
&\int_{D_{u,\ub}}<|u'|^{2s'}\Psi^{(s-\f12,s'+\f12)},|u'|^{2s'}\nab_4\Psi^{(s-\f12,s'+\f12)}>_{\gamma}\\
=&-\int_{D_{u,\ub}}<\f{2s'}{|u'|}(\Omega^{-1}-1)|u'|^{2s'}\Psi^{(s,s')},|u'|^{2s'}\Psi^{(s,s')}>_{\gamma}\\
&-\int_{D_{u,\ub}}<s'(\tr\chib+\f{2}{|u'|})|u'|^{2s'}\Psi^{(s,s')},|u'|^{2s'}\Psi^{(s,s')}>_{\gamma}\\
&+\int_{D_{u,\ub}}<|u'|^{2s'}\Psi^{(s,s')},|u'|^{2s'}J_3>_{\gamma}+\int_{D_{u,\ub}}<|u'|^{2s'}\Psi^{(s-\f12,s'+\f12)},|u'|^{2s'}J_4>_{\gamma}\\
&+\int_{D_{u,\ub}}|u'|^{4s'}(\eta+\etb)\Psi^{(s,s')}\Psi^{(s-\frac12,s'+\frac12)}-\int_{D_{u,\ub}}\omb|u'|^{4s'}(\Psi^{(s,s')})^2\\
&-\f12\int_{\Hu}|u'|^{4s'}(\Psi^{(s,s')})^2+\f12 \int_{H_{u_{\infty}}^{(0,\ub)}}|u'|^{4s'}(\Psi^{(s,s')})^2.\\
\end{split}
\end{equation*}
Meanwhile, adopting Proposition \ref{intbyparts4}, we get
\begin{equation*}
\begin{split} 
&\int_{D_{u,\ub}}<|u'|^{2s'}\Psi^{(s-\f12,s'+\f12)},|u'|^{2s'}\nab_4\Psi^{(s-\f12,s'+\f12)}>_{\gamma}\\
=&\f12\int_{\Hbu}|u'|^{4s'}(\Psi^{(s-\f12,s'+\f12)})^2-\f12\int_{H_0^{(u_{\infty},u)}}|u'|^{4s'}(\Psi^{(s-\f12,s'+\f12)})^2\\
&+\int_{D_{u,\ub}}(\o-\f12\tr\chi)|u'|^{4s'}(\Psi^{(s-\f12,s'+\f12)})^2.
\end{split}
\end{equation*}
Therefore, we conclude
\begin{equation*}
\begin{split} 
&\f12\int_{\Hbu}|u'|^{4s'}(\Psi^{(s-\f12,s'+\f12)})^2+\f12\int_{\Hu}|u'|^{4s'}(\Psi^{(s,s')})^2\\
=&\f12\int_{\Hb_0^{(u_{\infty},u)}}|u'|^{4s'}(\Psi^{(s-\f12,s'+\f12)})^2+\f12 \int_{H_{u_{\infty}}^{(0,\ub)}}|u'|^{4s'}(\Psi^{(s,s')})^2\\
&-\int_{D_{u,\ub}}<\f{2s'}{|u'|}(\Omega^{-1}-1)|u'|^{2s'}\Psi^{(s,s')},|u'|^{2s'}\Psi^{(s,s')}>_{\gamma}\\
&-\int_{D_{u,\ub}}<s'(\tr\chib+\f{2}{|u'|})|u'|^{2s'}\Psi^{(s,s')},|u'|^{2s'}\Psi^{(s,s')}>_{\gamma}\\
&+\int_{D_{u,\ub}}<|u'|^{2s'}\Psi^{(s,s')},|u'|^{2s'}J_3>_{\gamma}+\int_{D_{u,\ub}}<|u'|^{2s'}\Psi^{(s-\f12,s'+\f12)},|u'|^{2s'}J_4>_{\gamma}\\
&+\int_{D_{u,\ub}}|u'|^{4s'}(\eta+\etb)\Psi^{(s,s')}\Psi^{(s-\frac12,s'+\frac12)}-\int_{D_{u,\ub}}\omb|u'|^{4s'}(\Psi^{(s,s')})^2\\
&-\int_{D_{u,\ub}}(\o-\f12\tr\chi)|u'|^{4s'}(\Psi^{(s-\f12,s'+\f12)})^2.\\
\end{split}
\end{equation*}
Multiplying $\d^{2s-3}$ on both sides and using the Gronwall's inequality, we obtain desired estimate.
\end{proof}

\subsection{Curvature Estimate I}
With all the preparations, we are ready to prove energy estimates for curvature components.
The curvature component $\Psi^{(s,s')}$ satisfies
\begin{equation*} 
\nab_3 \Psi^{(s,s')}+(\frac 12+s')\tr\chib\Psi^{(s,s')}=\M D\Psi^{(s-\frac12,s'+\frac12)}+\sum_{s_1+s_2=s, \atop s'_1+s'_2=s'+1}\psi^{(s_1,s'_1)}\cdot\Psi^{(s_2,s'_2)},
\end{equation*}
and
\begin{equation*} 
\nab_4 \Psi^{(s-\frac 12,s'+\frac 12)}={^*\M D}\Psi^{(s,s')}+\sum_{\st_1+\st_2=s+\frac 12, \atop \st'_1+\st'_2=s'+\frac 12}\psi^{(\st_1,\st'_1)}\cdot\Psi^{(\st_2,\st'_2)}.
\end{equation*}
Applying Proposition \ref{energy1} and adopting H\"older's and Gronwall's inequality, we conclude
\begin{proposition}\label{energy11}
Assuming a pair of S-tangent tensor fields $\Psi^{(s,s')}$ and $\Psi^{(s-\frac 12,s'+\frac 12)}$ satisfy \eqref{tran3333} and \eqref{tran4444}, then we have
\begin{equation*}
\begin{split}
&\|\Psi^{(s,s')}\|^2_{L^2_{sc}(H_u^{(0,\ub)})}+\|\Psi^{(s-\frac12,s'+\frac12)}\|^2_{L^2_{sc}(\Hb_{\ub}^{(u_{\infty},u)})}\\
\leq & \|\Psi^{(s,s')}\|^2_{L^2_{sc}(H_{u_{\infty}}^{(0,\ub)})}+\|\Psi^{(s-\frac12,s'+\frac12)}\|^2_{L^2_{sc}(\Hb_{0}^{(u_{\infty},u)})}\\
&+  \frac{\delta^{\frac12}}{|u|}\sum_{s_1+s_2+s_3=2s, \atop s'_1+s'_2+s'_3=2s'+1}\sup_{u',\ub'}\frac{1}{|u'|}\|\psi^{(s_1,s'_1)}\|_{L^{\infty}_{sc}(S_{u',\ub'})} \|\Psi^{(s_2,s'_2)}\|_{L^2_{sc}(H_u^{(0,\ub)})}\\
&\times\|\Psi^{(s_3,s'_3)}\|_{L^2_{sc}(H_u^{(0,\ub)})},
\end{split}
\end{equation*}
or
\begin{equation*}
\begin{split}
&\|\Psi^{(s,s')}\|^2_{L^2_{sc}(H_u^{(0,\ub)})}+\|\Psi^{(s-\frac12,s'+\frac12)}\|^2_{L^2_{sc}(\Hb_{\ub}^{(u_{\infty},u)})}\\
\leq & \|\Psi^{(s,s')}\|^2_{L^2_{sc}(H_{u_{\infty}}^{(0,\ub)})}+\|\Psi^{(s-\frac12,s'+\frac12)}\|^2_{L^2_{sc}(\Hb_{0}^{(u_{\infty},u)})}\\
+&  \delta^{\frac12}\sum_{s_1+s_2+s_3=2s, \atop s'_1+s'_2+s'_3=2s'+1} \sup_{u',\ub'}\frac{1}{|u'|}\|\psi^{(s_1,s'_1)}\|_{L^{\infty}_{sc}(S_{u',\ub'})} \|\Psi^{(s_2,s'_2)}\|_{L^2_{sc}(\Hb_{\ub}^{(u_{\infty},u)})}\\
&\times\|\Psi^{(s_3,s'_3)}\|_{L^2_{sc}(\Hb_{\ub}^{(u_{\infty},u)})}.
\end{split}
\end{equation*}
\end{proposition}

With this, we arrive at
\begin{proposition}\label{energy a}
Under the assumption of Theorem \ref{main.thm} and the bootstrap assumption \eqref{BA.4}, we have
$$\M R[\a]+\underline{\M R}[\b]\leq \M R^{(0)}+\d^{\f14}\M R.$$
\end{proposition}

\begin{proof}
Using Proposition \ref{energy11} and H\"older's inequality, for the pair 
$$\Psi^{(s,s')}=\a, \quad \quad \Psi^{(s-\frac 12,s'+\frac 12)}=\b,$$
we have
\begin{equation*}
\begin{split}
&\d\|\Psi^{(s,s')}\|^2_{L^2_{sc}(H_u^{(0,\ub)})}+\d\|\Psi^{(s-\frac12,s'+\frac12)}\|^2_{L^2_{sc}(\Hb_{\ub}^{(u_{\infty},u)})}\\
\leq & \d\|\Psi^{(s,s')}\|^2_{L^2_{sc}(H_{u_{\infty}}^{(0,\ub)})}+\d\|\Psi^{(s-\frac12,s'+\frac12)}\|^2_{L^2_{sc}(\Hb_{0}^{(u_{\infty},u)})}\\
&+  \frac{\delta^{\frac12}}{|u|}\sum_{s_1+s_2+s_3=2s, \atop s'_1+s'_2+s'_3=2s'+1}\sup_{u',\ub'}\frac{1}{|u'|}\|\psi^{(s_1,s'_1)}\|_{L^{\infty}_{sc}(S_{u',\ub'})} \d^{\f12}\|\Psi^{(s_2,s'_2)}\|_{L^2_{sc}(H_u^{(0,\ub)})}\\
&\times\d^{\f12}\|\Psi^{(s_3,s'_3)}\|_{L^2_{sc}(H_u^{(0,\ub)})}.
\end{split}
\end{equation*}
We do not have $\ab$ term in the null Bianchi equations for $\nab_3\a$ and $\nab_4\b$. Hence all the curvature
components are bounded in $L^{2}_{sc}(\Hu)$ norms. The anomalous terms we have are due to $\chibh$ and $\a$. And they are bounded through the estimates above.
The conclusion of proposition follows.
\end{proof}

\begin{proposition}\label{energy98}
Under the assumption of Theorem \ref{main.thm} and the bootstrap assumption \eqref{BA.4}, we have
$$\M R[\rho,\sigma, \beb]+\underline{\M R}[\beb,\ab]\leq \M R^{(0)}+\d^{\f14}\underline{\M R}.$$
\end{proposition}

\begin{proof}
Applying Proposition \ref{energy11} and H\"older's inequality, for the pairs 
$$\Psi^{(s,s')}=\rho, \sigma, \quad \quad \Psi^{(s-\frac 12,s'+\frac 12)}=\beb,$$
$$\Psi^{(s,s')}=\beb, \quad \quad \Psi^{(s-\frac 12,s'+\frac 12)}=\ab,$$
we derive
\begin{equation*}
\begin{split}
&\|\Psi^{(s,s')}\|^2_{L^2_{sc}(H_u^{(0,\ub)})}+\|\Psi^{(s-\frac12,s'+\frac12)}\|^2_{L^2_{sc}(\Hb_{\ub}^{(u_{\infty},u)})}\\
\leq & \|\Psi^{(s,s')}\|^2_{L^2_{sc}(H_{u_{\infty}}^{(0,\ub)})}+\|\Psi^{(s-\frac12,s'+\frac12)}\|^2_{L^2_{sc}(\Hb_{0}^{(u_{\infty},u)})}\\
+&  \delta^{\frac12}\sum_{s_1+s_2+s_3=2s, \atop s'_1+s'_2+s'_3=2s'+1} \sup_{u',\ub'}\frac{1}{|u'|}\|\psi^{(s_1,s'_1)}\|_{L^{\infty}_{sc}(S_{u',\ub'})} \|\Psi^{(s_2,s'_2)}\|_{L^2_{sc}(\Hb_{\ub}^{(u_{\infty},u)})}\\
&\times\|\Psi^{(s_3,s'_3)}\|_{L^2_{sc}(\Hb_{\ub}^{(u_{\infty},u)})}.
\end{split}
\end{equation*}
We do not have $\a$ term in the null Bianchi equations for $\nab_3\rho, \nab_3\sigma, \nab_3\beb$ and $\nab_4\beb, \nab_4\ab$. Hence all the curvature
components are bounded in $L^{2}_{sc}(\Hbu)$ norms.  The anomalous terms we have are due to $\chibh$. And they are bounded through the estimates above.
The conclusion of proposition follows.
\end{proof}

\begin{proposition}
Under the assumption of Theorem \ref{main.thm} and the bootstrap assumption \eqref{BA.4}, we have
$$\M R[\b]+\underline{\M R}[\rho,\sigma]\ls (\M I^{(0)})+(\M R^{(0)})+\d^{\f18}(\M R+\underline{\M R})+(\M I^{(0)})^{\f12}(\M R^{(0)})^{\f14}\M R_1^{\f14}[\a].$$
\end{proposition}

\begin{proof}
Appealing to Propositions \eqref{L4}, \eqref{energy11} and H\"older's inequality, for the pair 
$$\Psi^{(s,s')}=\b, \quad \quad \Psi^{(s-\frac 12,s'+\frac 12)}=\rho, \sigma$$
we deduce
\begin{equation*}
\begin{split}
&\|\Psi^{(s,s')}\|^2_{L^2_{sc}(H_u^{(0,\ub)})}+\|\Psi^{(s-\frac12,s'+\frac12)}\|^2_{L^2_{sc}(\Hb_{\ub}^{(u_{\infty},u)})}\\
\leq & \|\Psi^{(s,s')}\|^2_{L^2_{sc}(H_{u_{\infty}}^{(0,\ub)})}+\|\Psi^{(s-\frac12,s'+\frac12)}\|^2_{L^2_{sc}(\Hb_{0}^{(u_{\infty},u)})}\\
&+\frac{\delta^{\frac12}}{|u|}\sum_{s_1+s_2+s_3=2s, \atop s'_1+s'_2+s'_3=2s'+1}\sup_{u',\ub'}\frac{1}{|u'|}\|\psi^{(s_1,s'_1)}\|_{L^{\infty}_{sc}(S_{u',\ub'})} \|\Psi^{(s_2,s'_2)}\|_{L^2_{sc}(H_u^{(0,\ub)})}\\
&\times\|\Psi^{(s_3,s'_3)}\|_{L^2_{sc}(H_u^{(0,\ub)})}\\
&+\f{1}{|u|}\sup_{u',\ub'}\f{\d^{\f14}}{|u'|}\|\chibh\|_{L^4_{sc}(S_{u',\ub'})}\|\rho,\sigma\|_{L^2_{sc}(\Hu)}\d^{\f14}\|\a\|^{\f12}_{L^2_{sc}(\Hu)}\|\nab\a\|^{\f12}_{L^2_{sc}(\Hu)}\\
&+\f{1}{|u|}\sup_{u',\ub'}\f{\d^{\f14}}{|u'|}\|\chibh\|_{L^4_{sc}(S_{u',\ub'})}\|\rho,\sigma\|_{L^2_{sc}(\Hu)}\d^{\f12}\|\a\|_{L^2_{sc}(\Hu)}.
\end{split}
\end{equation*}
For this pair, we encounter anomalous term $(\rho,\sigma)\chibh\a$, and it is bounded above.
Recalling for $\chibh$, from Proposition \ref{O04 3norm}, we have
\begin{equation*}
 \f{\d^{\f14}}{|u|}\|\chibh\|_{L^4_{sc}(\S)}\leq \f{\d^{\f14}}{|u_{\infty}|}\|\chibh\|_{L^4_{sc}(S_{u_{\infty},\ub})}+\f{\d^{\f14}}{|u|^{\f32}}\underline{\M R}^{\f12}_0 \underline{\M R}^{\f12}_1+\f{\d^{\f12}}{|u|^{\f32}} \underline{\M R}_0+\f{\d^{\f34}}{|u|^2}\D_0 \M O_{0,4}.
\end{equation*}
Using H\"older's inequality again, therefore we prove 
$$\M R^2[\b]+\underline{\M R}^2[\rho,\sigma]\leq (\M R^{(0)})^2+\d^{\f14}(\M R^2+\underline{\M R}^2)+\M I^{(0)}\M R_0^{\f12}[\a]\M R_1^{\f12}[\a].$$
With Proposition \ref{energy a}, we arrive at
$$\M R^2[\b]+\underline{\M R}^2[\rho,\sigma]\leq (\M I^{(0)})^2+(\M R^{(0)})^2+\d^{\f14}(\M R^2+\underline{\M R}^2)+\M I^{(0)}(\M R^{(0)})^{\f12}\M R_1^{\f12}[\a].$$
The desired estimate follows.
\end{proof}

Gathering all the estimates in this section, we conclude
\begin{proposition}
Under the assumption of Theorem \ref{main.thm} and bootstrap assumption \eqref{BA.4}, 
there exists $\d_0=\d_0(\M R, \underline{\M R})$, when $0<\d<\d_0$, 
we have 
$$\M R_0+\underline{\M R}_0\leq \M R^{(0)}+\M R_1[\a]+\d^{\f18}(\M I^{(0)}+\M R^{(0)}+\M R+\underline{\M R})^5.$$
\end{proposition}

\subsection{Curvature Estimate II}
The aim of this subsection is to derive energy estimate for first angular derivatives of curvature components.
And the pair $\nab\Psi^{(s,s')}$ and $\nab\Psi^{(s-\f12,s'+\f12)}$ satisfy
\begin{equation}\label{tran33333} 
\begin{split}
& \nab_3 \nab\Psi^{(s,s')}+(1+s')\tr\chib\nab\Psi^{(s,s')}\\
=&\nab\M D\Psi^{(s-\frac12,s'+\frac12)}+\sum_{s_1+s_2=s, \atop s'_1+s'_2=s'+1}\nab\psi^{(s_1, s'_1)}\cdot\Psi^{(s_2, s'_2)}+\sum_{s_1+s_2=s, \atop s'_1+s'_2=s'+1}\psi^{(s_1, s'_1)}\cdot\nab\Psi^{(s_2, s'_2)}\\
&+\sum_{s_3+s_4+s_5=s+\f12, \atop s'_3+s'_4+s'_5=s'+\f32}\psi^{(s_3, s'_3)}\cdot\psi^{(s_4, s'_4)}\cdot\Psi^{(s_5,s'_5)}+\sum_{s_6+s_7=s+\f12, \atop s'_6+s'_7=s'+\f32}\Psi^{(s_6,s'_6)}\cdot\Psi^{(s_7,s'_7)},
\end{split}
\end{equation}
or
\begin{equation}\label{tran44444} 
\begin{split}
& \nab_4 \nab\Psi^{(s-\f12,s'+\f12)}\\
=&\nab {^*\M D}\Psi^{(s,s')}+\sum_{s_1+s_2=s+\f12, \atop s'_1+s'_2=s'+\f12}\nab\psi^{(s_1, s'_1)}\cdot\Psi^{(s_2, s'_2)}+\sum_{s_1+s_2=s+\f12, \atop s'_1+s'_2=s'+\f12}\psi^{(s_1, s'_1)}\cdot\nab\Psi^{(s_2, s'_2)}\\
&+\sum_{s_3+s_4+s_5=s+1, \atop s'_3+s'_4+s'_5=s'+1}\psi^{(s_3, s'_3)}\cdot\psi^{(s_4, s'_4)}\cdot\Psi^{(s_5,s'_5)}+\sum_{s_6+s_7=s+1, \atop s'_6+s'_7=s'+1}\Psi^{(s_6,s'_6)}\cdot\Psi^{(s_7,s'_7)}.
\end{split}
\end{equation}

Adopting equations and H\"older's inequality, we arrive at 
\begin{proposition}
For curvature component $\nab\Psi^{(s,s')}$ and $\nab\Psi^{(s-\frac12,s'+\frac12)}$ obeying \eqref{tran33333} and \eqref{tran44444}, we have
\begin{equation*}
\begin{split}
&\|\nab\Psi^{(s,s')}\|^2_{L^2_{sc}(H_u^{(0,\ub)})}+\|\nab\Psi^{(s-\frac12,s'+\frac12)}\|^2_{L^2_{sc}(\Hb_{\ub}^{(u_{\infty},u)})}\\
\leq & \|\nab\Psi^{(s,s')}\|^2_{L^2_{sc}(H_{u_{\infty}}^{(0,\ub)})}+\|\nab\Psi^{(s-\frac12,s'+\frac12)}\|^2_{L^2_{sc}(\Hb_{0}^{(u_{\infty},u)})}\\
&\quad+M_1+M_2+M_3+M_4+M_5+M_6+M_7+M_8,
\end{split} 
\end{equation*}
where $M_1-M_8$ are listed below:
\begin{equation*}
\begin{split}
M_1=&\int_0^{\ub}\int_{u_{\infty}}^u \frac{\delta^{-\frac32}}{|u'|}\sum_{s_1+s_2=s, \atop s'_1+s'_2=s'+1}\|\nab\psi^{(s_1, s'_1)}\cdot\Psi^{(s_2, s'_2)}\cdot\nab\Psi^{(s,s')}\|_{L^1_{sc}(S_{u',\ub'})} du'd{\ub}',\\
M_2=&\int_0^{\ub}\int_{u_{\infty}}^u \frac{\delta^{-\frac32}}{|u'|}\sum_{s_1+s_2=s, \atop s'_1+s'_2=s'+1}\|\psi^{(s_1, s'_1)}\cdot\nab\Psi^{(s_2, s'_2)}\cdot\nab\Psi^{(s,s')}\|_{L^1_{sc}(S_{u',\ub'})} du'd{\ub}',\\
M_3=&\int_0^{\ub}\int_{u_{\infty}}^u \frac{\delta^{-\frac32}}{|u'|}\sum_{s_1+s_2=s+\f12, \atop s'_1+s'_2=s'+\f12}\|\nab\psi^{(s_1, s'_1)}\cdot\Psi^{(s_2, s'_2)}\cdot\nab\Psi^{(s-\f12,s'+\f12)}\|_{L^1_{sc}(S_{u',\ub'})} du'd{\ub}',\\
M_4=&\int_0^{\ub}\int_{u_{\infty}}^u \frac{\delta^{-\frac32}}{|u'|}\sum_{s_1+s_2=s+\f12, \atop s'_1+s'_2=s'+\f12}\|\psi^{(s_1, s'_1)}\cdot\nab\Psi^{(s_2, s'_2)}\cdot\nab\Psi^{(s-\f12,s'+\f12)}\|_{L^1_{sc}(S_{u',\ub'})} du'd{\ub}',\\
M_5=&\int_0^{\ub}\int_{u_{\infty}}^u \frac{\delta^{-\frac32}}{|u'|}\sum_{s_3+s_4+s_5=s+\f12, \atop s'_3+s'_4+s'_5=s'+\f32}\|\psi^{(s_3, s'_3)}\cdot\psi^{(s_4, s'_4)}\cdot\Psi^{(s_5,s'_5)}\cdot\nab\Psi^{(s,s')}\|_{L^1_{sc}(S_{u',\ub'})} du'd{\ub}',\\
M_6=&\int_0^{\ub}\int_{u_{\infty}}^u \frac{\delta^{-1}}{|u'|^2}\sum_{s_3+s_4+s_5=s+1, \atop s'_3+s'_4+s'_5=s'+1}\|\psi^{(s_3, s'_3)}\cdot\psi^{(s_4, s'_4)}\cdot\Psi^{(s_5,s'_5)}\|_{L^2_{sc}(S_{u',\ub'})}\\
&\quad\times\|\nab\Psi^{(s-\f12,s'+\f12)}\|_{L^2_{sc}(S_{u',\ub'})} du'd{\ub}',\\
M_7=&\int_0^{\ub}\int_{u_{\infty}}^u \frac{\delta^{-\frac32}}{|u'|}\sum_{s_1+s_2=s+\f12, \atop s'_1+s'_2=s'+\f32}\|\Psi^{(s_1, s'_1)}\cdot\Psi^{(s_2, s'_2)}\cdot\nab\Psi^{(s,s')}\|_{L^1_{sc}(S_{u',\ub'})} du'd{\ub}',\\
M_8=&\int_0^{\ub}\int_{u_{\infty}}^u \frac{\delta^{-\frac32}}{|u'|}\sum_{s_1+s_2=s+1, \atop s'_1+s'_2=s'+1}\|\Psi^{(s_1, s'_1)}\cdot\Psi^{(s_2, s'_2)}\cdot\nab\Psi^{(s-\f12,s'+\f12)}\|_{L^1_{sc}(S_{u',\ub'})} du'd{\ub}'.\\
\end{split}
\end{equation*}
\end{proposition}

We will do estimate term by term. For $M_1$,
it is easy to verify that for equations \eqref{tran33333} and \eqref{tran44444}, $\a$ and $\ab$ terms will not appear at the same time.
If there is no $\ab$ term, we have 
\begin{equation*}
\begin{split}
M_1=&\int_0^{\ub}\int_{u_{\infty}}^u \frac{\delta^{-\frac32}}{|u'|}\sum_{s_1+s_2=s, \atop s'_1+s'_2=s'+1}\|\nab\psi^{(s_1, s'_1)}\cdot\Psi^{(s_2, s'_2)}\cdot\nab\Psi^{(s,s')}\|_{L^1_{sc}(S_{u',\ub'})} du'd{\ub}'\\
\leq&\int_0^{\ub}\int_{u_{\infty}}^u \frac{\delta^{-\frac12}}{|u'|^3}\sum_{s_1+s_2=s, \atop s'_1+s'_2=s'+1}\|\nab\psi^{(s_1, s'_1)}\|_{L^4_{sc}(\s)}\|\Psi^{(s_2, s'_2)}\|_{L^4_{sc}(\s)}\\
&\quad\times\|\nab\Psi^{(s,s')}\|_{L^2_{sc}(S_{u',\ub'})} du'd{\ub}'\\
\leq&\int_0^{\ub}\int_{u_{\infty}}^u \frac{\delta^{-\frac12}}{|u'|^3}\sum_{s_1+s_2=s, \atop s'_1+s'_2=s'+1}\|\nab\psi^{(s_1, s'_1)}\|_{L^4_{sc}(\s)}\|\Psi^{(s_2, s'_2)}\|^{\f12}_{L^2_{sc}(\s)}\\
&\quad\times\|\nab\Psi^{(s_2, s'_2)}\|^{\f12}_{L^2_{sc}(\s)}\|\nab\Psi^{(s,s')}\|_{L^2_{sc}(S_{u',\ub'})} du'd{\ub}'\\
&+\int_0^{\ub}\int_{u_{\infty}}^u \frac{\delta^{-\frac14}}{|u'|^3}\sum_{s_1+s_2=s, \atop s'_1+s'_2=s'+1}\|\nab\psi^{(s_1, s'_1)}\|_{L^4_{sc}(\s)}\|\Psi^{(s_2, s'_2)}\|_{L^2_{sc}(\s)}\\
&\quad\times\|\nab\Psi^{(s,s')}\|_{L^2_{sc}(S_{u',\ub'})} du'd{\ub}'\\
\leq&\int_0^{\ub}\int_{u_{\infty}}^u \frac{\delta^{\frac12}}{|u'|^2}\sum_{s_1+s_2=s, \atop s'_1+s'_2=s'+1}\f{1}{|u'|}\|\nab\psi^{(s_1, s'_1)}\|_{L^4_{sc}(\s)}\d^{-\f14}\|\Psi^{(s_2, s'_2)}\|^{\f12}_{L^2_{sc}(\s)}\\
&\quad\times\d^{-\f14}\|\nab\Psi^{(s_2, s'_2)}\|^{\f12}_{L^2_{sc}(\s)}\d^{-\f12}\|\nab\Psi^{(s,s')}\|_{L^2_{sc}(S_{u',\ub'})} du'd{\ub}'\\
&+\int_0^{\ub}\int_{u_{\infty}}^u \frac{\delta^{\frac34}}{|u'|^2}\sum_{s_1+s_2=s, \atop s'_1+s'_2=s'+1}\f{1}{|u'|}\|\nab\psi^{(s_1, s'_1)}\|_{L^4_{sc}(\s)}\d^{-\f12}\|\Psi^{(s_2, s'_2)}\|_{L^2_{sc}(\s)}\\
&\quad\times\d^{-\f12}\|\nab\Psi^{(s,s')}\|_{L^2_{sc}(S_{u',\ub'})} du'd{\ub}'\\
\leq&\f{\d^{\f14}}{|u|}\sum_{s_1+s_2=s, \atop s'_1+s'_2=s'+1}\sup_{u'\ub'}\f{1}{|u'|}\|\nab\psi^{(s_1, s'_1)}\|_{L^4_{sc}(\s)}\d^{\f14}\|\Psi^{(s_2, s'_2)}\|^{\f12}_{L^2_{sc}(\Hu)}\\
&\quad\times\|\nab\Psi^{(s_2, s'_2)}\|^{\f12}_{L^2_{sc}(\Hu)}\|\nab\Psi^{(s,s')}\|_{L^2_{sc}(\Hu)}\\
&+\f{\d^{\f14}}{|u|}\sum_{s_1+s_2=s, \atop s'_1+s'_2=s'+1}\sup_{u'\ub'}\f{1}{|u'|}\|\nab\psi^{(s_1, s'_1)}\|_{L^4_{sc}(\s)}\\
&\quad\times\d^{\f12}\|\Psi^{(s_2, s'_2)}\|_{L^2_{sc}(\Hu)}\|\nab\Psi^{(s,s')}\|_{L^2_{sc}(\Hu)}.\\
\end{split} 
\end{equation*}
For the last step, we use definition of scale invariant norm and H\"older's inequality.
 
Similarly, if there is no $a$ term in $M_1$, we derive 
\begin{equation*}
\begin{split}
M_1=&\int_0^{\ub}\int_{u_{\infty}}^u \frac{\delta^{-\frac32}}{|u'|}\sum_{s_1+s_2=s, \atop s'_1+s'_2=s'+1}\|\nab\psi^{(s_1, s'_1)}\cdot\Psi^{(s_2, s'_2)}\cdot\nab\Psi^{(s,s')}\|_{L^1_{sc}(S_{u',\ub'})} du'd{\ub}'\\
\leq&\int_0^{\ub}\int_{u_{\infty}}^u \frac{\delta^{-\frac12}}{|u'|^3}\sum_{s_1+s_2=s, \atop s'_1+s'_2=s'+1}\|\nab\psi^{(s_1, s'_1)}\|_{L^4_{sc}(\s)}\|\Psi^{(s_2, s'_2)}\|_{L^4_{sc}(\s)}\\
&\quad\times\|\nab\Psi^{(s,s')}\|_{L^2_{sc}(S_{u',\ub'})} du'd{\ub}'\\
\leq&\int_0^{\ub}\int_{u_{\infty}}^u \frac{\delta^{-\frac12}}{|u'|^3}\sum_{s_1+s_2=s, \atop s'_1+s'_2=s'+1}\|\nab\psi^{(s_1, s'_1)}\|_{L^4_{sc}(\s)}\|\Psi^{(s_2, s'_2)}\|^{\f12}_{L^2_{sc}(\s)}\\
&\quad\times\|\nab\Psi^{(s_2, s'_2)}\|^{\f12}_{L^2_{sc}(\s)}\|\nab\Psi^{(s,s')}\|_{L^2_{sc}(S_{u',\ub'})} du'd{\ub}'\\
&+\int_0^{\ub}\int_{u_{\infty}}^u \frac{\delta^{-\frac14}}{|u'|^3}\sum_{s_1+s_2=s, \atop s'_1+s'_2=s'+1}\|\nab\psi^{(s_1, s'_1)}\|_{L^4_{sc}(\s)}\|\Psi^{(s_2, s'_2)}\|_{L^2_{sc}(\s)}\\
&\quad\times\|\nab\Psi^{(s,s')}\|_{L^2_{sc}(S_{u',\ub'})} du'd{\ub}'\\
\leq&\int_0^{\ub}\int_{u_{\infty}}^u {\delta^{-\frac12}}\sum_{s_1+s_2=s, \atop s'_1+s'_2=s'+1}\f{1}{|u'|}\|\nab\psi^{(s_1, s'_1)}\|_{L^4_{sc}(\s)}\f{1}{|u'|^{\f12}}\|\Psi^{(s_2, s'_2)}\|^{\f12}_{L^2_{sc}(\s)}\\
&\quad\times\f{1}{|u'|^{\f12}}\|\nab\Psi^{(s_2, s'_2)}\|^{\f12}_{L^2_{sc}(\s)}\f{1}{|u'|}\|\nab\Psi^{(s,s')}\|_{L^2_{sc}(S_{u',\ub'})} du'd{\ub}'\\
&+\int_0^{\ub}\int_{u_{\infty}}^u \d^{-\f14}\sum_{s_1+s_2=s, \atop s'_1+s'_2=s'+1}\f{1}{|u'|}\|\nab\psi^{(s_1, s'_1)}\|_{L^4_{sc}(\s)}\f{1}{|u'|}\|\Psi^{(s_2, s'_2)}\|_{L^2_{sc}(\s)}\\
&\quad\times\f{1}{|u'|}\|\nab\Psi^{(s,s')}\|_{L^2_{sc}(S_{u',\ub'})} du'd{\ub}'\\
\leq&{\d^{\f12}}\sum_{s_1+s_2=s, \atop s'_1+s'_2=s'+1}\sup_{u'\ub'}\f{1}{|u'|}\|\nab\psi^{(s_1, s'_1)}\|_{L^4_{sc}(\s)}\|\Psi^{(s_2, s'_2)}\|^{\f12}_{L^2_{sc}(\Hbu)}\\
&\quad\times\|\nab\Psi^{(s_2, s'_2)}\|^{\f12}_{L^2_{sc}(\Hbu)}\|\nab\Psi^{(s,s')}\|_{L^2_{sc}(\Hbu)}\\
&+{\d^{\f34}}\sum_{s_1+s_2=s, \atop s'_1+s'_2=s'+1}\sup_{u'\ub'}\f{1}{|u'|}\|\nab\psi^{(s_1, s'_1)}\|_{L^4_{sc}(\s)}\\
&\quad\times\|\Psi^{(s_2, s'_2)}\|_{L^2_{sc}(\Hbu)}\|\nab\Psi^{(s,s')}\|_{L^2_{sc}(\Hbu)}.\\
\end{split} 
\end{equation*}
For the last step, we use definition of scale invariant norm and H\"older's inequality.
The anomalous terms of $M_1$ type are $(\nab\chib,\nab\omb, \nab\eta)\a\nab\Psi$ and $\nab\chib(\b, \rho, \sigma)\nab\Psi$, where $\Psi$ is a curvature component.
These terms are bounded through the estimates.

In the same fashion, we bound $M_2, M_3, M_4, M_5$ and $M_6$.
For $M_2$, we have either 
\begin{equation*}
\begin{split}
M_2\leq &\f{\d^{\f12}}{|u|}\sum_{s_1+s_2=s, \atop s'_1+s'_2=s'+1}\sup_{u'\ub'}\f{1}{|u'|}\|\psi^{(s_1, s'_1)}\|_{L^4_{sc}(\s)}\|\nab\Psi^{(s_2, s'_2)}\|^{\f12}_{L^2_{sc}(\Hu)}\\
&\quad\times\|\nab^2\Psi^{(s_2, s'_2)}\|^{\f12}_{L^2_{sc}(\Hu)}\|\nab\Psi^{(s,s')}\|_{L^2_{sc}(\Hu)}\\
&+\f{\d^{\f34}}{|u|}\sum_{s_1+s_2=s, \atop s'_1+s'_2=s'+1}\sup_{u'\ub'}\f{1}{|u'|}\|\psi^{(s_1, s'_1)}\|_{L^4_{sc}(\s)}\\
&\quad\times\|\nab\Psi^{(s_2, s'_2)}\|_{L^2_{sc}(\Hu)}\|\nab\Psi^{(s,s')}\|_{L^2_{sc}(\Hu)},\\
\end{split} 
\end{equation*}
or
\begin{equation*}
\begin{split}
M_2\leq&{\d^{\f12}}\sum_{s_1+s_2=s, \atop s'_1+s'_2=s'+1}\sup_{u'\ub'}\f{1}{|u'|}\|\psi^{(s_1, s'_1)}\|_{L^4_{sc}(\s)}\|\nab\Psi^{(s_2, s'_2)}\|^{\f12}_{L^2_{sc}(\Hbu)}\\
&\quad\times\|\nab^2\Psi^{(s_2, s'_2)}\|^{\f12}_{L^2_{sc}(\Hbu)}\|\nab\Psi^{(s,s')}\|_{L^2_{sc}(\Hbu)}\\
&+{\d^{\f34}}\sum_{s_1+s_2=s, \atop s'_1+s'_2=s'+1}\sup_{u'\ub'}\f{1}{|u'|}\|\psi^{(s_1, s'_1)}\|_{L^4_{sc}(\s)}\\
&\quad\times\|\nab\Psi^{(s_2, s'_2)}\|_{L^2_{sc}(\Hbu)}\|\nab\Psi^{(s,s')}\|_{L^2_{sc}(\Hbu)}.\\
\end{split} 
\end{equation*}
We encounter $\chibh(\nab\b,\nab\rho, \nab\sigma \nab\beb)\nab\Psi$ as anomalous term of $M_2$ type. They are bounded
through the estimates.

$M_3$ and $M_4$ can be bounded in the same manner as for $M_1$ and $M_2$, respectively.

The term $M_5$ satisfies either
\begin{equation*}
\begin{split}
M_5\leq &\f{\d^{\f14}}{|u|}\sum_{s_3+s_4+s_5=s+\f12, \atop s'_3+s'_4+s'_5=s'+\f32}\sup_{u'\ub'}\f{\d^{\f12}}{|u'|^2}\|\psi^{(s_3, s'_3)}\|_{L^{\infty}_{sc}(\s)}\|\psi^{(s_4, s'_4)}\|_{L^4_{sc}(\s)}\d^{\f14}\|\Psi^{(s_5, s'_5)}\|^{\f12}_{L^2_{sc}(\Hu)}\\
&\quad\times\|\nab\Psi^{(s_5, s'_5)}\|^{\f12}_{L^2_{sc}(\Hu)}\|\nab\Psi^{(s,s')}\|_{L^2_{sc}(\Hu)}\\
&+\f{\d^{\f14}}{|u|}\sum_{s_3+s_4+s_5=s+\f12, \atop s'_3+s'_4+s'_5=s'+\f32}\sup_{u'\ub'}\f{\d^{\f12}}{|u'|^2}\|\psi^{(s_3, s'_3)}\|_{L^{\infty}_{sc}(\s)}\|\psi^{(s_4, s'_4)}\|_{L^4_{sc}(\s)}\\
&\quad\times\d^{\f12}\|\Psi^{(s_5, s'_5)}\|_{L^2_{sc}(\Hu)}\|\nab\Psi^{(s,s')}\|_{L^2_{sc}(\Hu)},\\
\end{split} 
\end{equation*}
or
\begin{equation*}
\begin{split}
M_5\leq&{\d^{\f12}}\sum_{s_3+s_4+s_5=s+\f12, \atop s'_3+s'_4+s'_5=s'+\f32}\sup_{u'\ub'}\f{\d^{\f12}}{|u'|^2}\|\psi^{(s_3, s'_3)}\|_{L^{\infty}_{sc}(\s)}\|\psi^{(s_4, s'_4)}\|_{L^4_{sc}(\s)}\|\Psi^{(s_5, s'_5)}\|^{\f12}_{L^2_{sc}(\Hbu)}\\
&\quad\times\|\nab\Psi^{(s_5, s'_5)}\|^{\f12}_{L^2_{sc}(\Hbu)}\|\nab\Psi^{(s,s')}\|_{L^2_{sc}(\Hbu)}\\
&+{\d^{\f34}}\sum_{s_3+s_4+s_5=s+\f12, \atop s'_3+s'_4+s'_5=s'+\f32}\sup_{u'\ub'}\f{\d^{\f12}}{|u'|^2}\|\psi^{(s_3, s'_3)}\|_{L^{\infty}_{sc}(\s)}\|\psi^{(s_4, s'_4)}\|_{L^4_{sc}(\s)}\\
&\quad\times\|\Psi^{(s_5, s'_5)}\|_{L^2_{sc}(\Hbu)}\|\nab\Psi^{(s,s')}\|_{L^2_{sc}(\Hbu)}.\\
\end{split} 
\end{equation*}

We control $M_6$ in the same method.
With these estimates, $M_5$ and $M_6$ type anomalous terms $(\eta,\etb)(\chib,\omb, \eta,\etb)\a\nab\Psi, (\eta,\etb)\chib(\b, \rho, \sigma, \beb)\nab\Psi$ are clearly bounded.

We bound $M_7$ by
\begin{equation*}
\begin{split}
M_7\leq&\int_0^{\ub}\int_{u_{\infty}}^u \frac{\delta^{\frac12}}{|u'|^2}\sum_{s_1+s_2=s+\f12, \atop s'_1+s'_2=s'+\f32}\f{1}{|u'|^{\f12}}\|\Psi^{(s_1, s'_1)}\|^{\f12}_{L^2_{sc}(\s)}\f{1}{|u'|^{\f12}}\|\nab\Psi^{(s_1, s'_1)}\|^{\f12}_{L^2_{sc}(\s)}\\
&\quad\times\d^{-\f14}\|\Psi^{(s_2, s'_2)}\|^{\f12}_{L^2_{sc}(\s)}\d^{-\f14}\|\nab\Psi^{(s_2, s'_2)}\|^{\f12}_{L^2_{sc}(\s)}\\
&\quad\times\d^{-\f12}\|\nab\Psi^{(s,s')}\|_{L^2_{sc}(S_{u',\ub'})} du'd{\ub}'\\
&+\int_0^{\ub}\int_{u_{\infty}}^u \frac{\delta^{\frac34}}{|u'|^2}\sum_{s_1+s_2=s+\f12, \atop s'_1+s'_2=s'+\f32}\f{1}{|u'|}\|\Psi^{(s_1, s'_1)}\|_{L^2_{sc}(\s)}\\
&\quad\times\d^{-\f14}\|\Psi^{(s_2, s'_2)}\|^{\f12}_{L^2_{sc}(\s)}\d^{-\f14}\|\nab\Psi^{(s_2, s'_2)}\|^{\f12}_{L^2_{sc}(\s)}\\
&\quad\times\d^{-\f12}\|\nab\Psi^{(s,s')}\|_{L^2_{sc}(S_{u',\ub'})} du'd{\ub}'\\
&+\int_0^{\ub}\int_{u_{\infty}}^u \frac{\delta^{\frac34}}{|u'|^2}\sum_{s_1+s_2=s+\f12, \atop s'_1+s'_2=s'+\f32}\f{1}{|u'|^{\f12}}\|\Psi^{(s_1, s'_1)}\|^{\f12}_{L^2_{sc}(\s)}\f{1}{|u'|^{\f12}}\|\nab\Psi^{(s_1, s'_1)}\|^{\f12}_{L^2_{sc}(\s)}\\
&\quad\times\d^{-\f12}\|\Psi^{(s_2, s'_2)}\|_{L^2_{sc}(\s)}\d^{-\f12}\|\nab\Psi^{(s,s')}\|_{L^2_{sc}(S_{u',\ub'})} du'd{\ub}'\\
&+\int_0^{\ub}\int_{u_{\infty}}^u \frac{\delta}{|u'|^2}\sum_{s_1+s_2=s+\f12, \atop s'_1+s'_2=s'+\f32}\f{1}{|u'|}\|\Psi^{(s_1, s'_1)}\|_{L^2_{sc}(\s)}\\
&\quad\times\d^{-\f12}\|\Psi^{(s_2, s'_2)}\|_{L^2_{sc}(\s)}\d^{-\f12}\|\nab\Psi^{(s,s')}\|_{L^2_{sc}(S_{u',\ub'})} du'd{\ub}'.\\
\end{split} 
\end{equation*}
Hence, we have
\begin{equation*}
\begin{split}
M_7\leq&\f{\d^{\f12}}{|u|^{\f32}}\sum_{s_1+s_2=s+\f12, \atop s'_1+s'_2=s'+\f32}\|\Psi^{(s_1, s'_1)}\|^{\f12}_{L^2_{sc}(\Hbu)}\|\nab\Psi^{(s_1, s'_1)}\|^{\f12}_{L^2_{sc}(\Hbu)}\\
&\quad\times\|\Psi^{(s_2, s'_2)}\|^{\f12}_{L^2_{sc}(\Hu)}\|\nab\Psi^{(s_2, s'_2)}\|^{\f12}_{L^2_{sc}(\Hu)}\|\nab\Psi^{(s, s')}\|_{L^2_{sc}(\Hu)}\\
&+\f{\d^{\f34}}{|u|^{\f32}}\sum_{s_1+s_2=s+\f12, \atop s'_1+s'_2=s'+\f32}\|\Psi^{(s_1, s'_1)}\|_{L^2_{sc}(\Hbu)}\|\Psi^{(s_2, s'_2)}\|^{\f12}_{L^2_{sc}(\Hu)}\\
&\quad\times\|\nab\Psi^{(s_2, s'_2)}\|^{\f12}_{L^2_{sc}(\Hu)}\|\nab\Psi^{(s, s')}\|_{L^2_{sc}(\Hu)}\\
&+\f{\d^{\f34}}{|u|^{\f32}}\sum_{s_1+s_2=s+\f12, \atop s'_1+s'_2=s'+\f32}\|\Psi^{(s_1, s'_1)}\|^{\f12}_{L^2_{sc}(\Hbu)}\|\nab\Psi^{(s_1, s'_1)}\|^{\f12}_{L^2_{sc}(\Hbu)}\\
&\quad\times\|\Psi^{(s_2, s'_2)}\|_{L^2_{sc}(\Hu)}\|\nab\Psi^{(s, s')}\|_{L^2_{sc}(\Hu)}\\
&+\f{\d}{|u|^{\f32}}\sum_{s_1+s_2=s+\f12, \atop s'_1+s'_2=s'+\f32}\|\Psi^{(s_1, s'_1)}\|_{L^2_{sc}(\Hbu)}\|\Psi^{(s_2, s'_2)}\|_{L^2_{sc}(\Hu)}\|\nab\Psi^{(s, s')}\|_{L^2_{sc}(\Hu)}.\\
\end{split} 
\end{equation*}

Alternatively $M_7$ is bounded by
\begin{equation*}
\begin{split}
M_7\leq&\int_0^{\ub}\int_{u_{\infty}}^u \frac{1}{|u'|}\sum_{s_1+s_2=s+\f12, \atop s'_1+s'_2=s'+\f32}\d^{-\f14}\|\Psi^{(s_1, s'_1)}\|^{\f12}_{L^2_{sc}(\s)}\d^{-\f14}\|\nab\Psi^{(s_1, s'_1)}\|^{\f12}_{L^2_{sc}(\s)}\\
&\quad\times\f{1}{|u'|^{\f12}}\|\Psi^{(s_2, s'_2)}\|^{\f12}_{L^2_{sc}(\s)}\f{1}{|u'|^{\f12}}\|\nab\Psi^{(s_2, s'_2)}\|^{\f12}_{L^2_{sc}(\s)}\\
&\quad\times\f{1}{|u'|}\|\nab\Psi^{(s,s')}\|_{L^2_{sc}(S_{u',\ub'})} du'd{\ub}'\\
&+\int_0^{\ub}\int_{u_{\infty}}^u \frac{\delta^{\frac14}}{|u'|}\sum_{s_1+s_2=s+\f12, \atop s'_1+s'_2=s'+\f32}\d^{-\f12}\|\Psi^{(s_1, s'_1)}\|_{L^2_{sc}(\s)}\\
&\quad\times\f{1}{|u'|^{\f12}}\|\Psi^{(s_2, s'_2)}\|^{\f12}_{L^2_{sc}(\s)}\f{1}{|u'|^{\f12}}\|\nab\Psi^{(s_2, s'_2)}\|^{\f12}_{L^2_{sc}(\s)}\\
&\quad\times\f{1}{|u'|}\|\nab\Psi^{(s,s')}\|_{L^2_{sc}(S_{u',\ub'})} du'd{\ub}'\\
&+\int_0^{\ub}\int_{u_{\infty}}^u \frac{\delta^{\frac14}}{|u'|}\sum_{s_1+s_2=s+\f12, \atop s'_1+s'_2=s'+\f32}\d^{-\f14}\|\Psi^{(s_1, s'_1)}\|^{\f12}_{L^2_{sc}(\s)}\d^{-\f14}\|\nab\Psi^{(s_1, s'_1)}\|^{\f12}_{L^2_{sc}(\s)}\\
&\quad\times\f{1}{|u'|}\|\Psi^{(s_2, s'_2)}\|_{L^2_{sc}(\s)}\f{1}{|u'|}\|\nab\Psi^{(s,s')}\|_{L^2_{sc}(S_{u',\ub'})} du'd{\ub}'\\
&+\int_0^{\ub}\int_{u_{\infty}}^u \frac{\delta^{\f12}}{|u'|}\sum_{s_1+s_2=s+\f12, \atop s'_1+s'_2=s'+\f32}\d^{-\f12}\|\Psi^{(s_1, s'_1)}\|_{L^2_{sc}(\s)}\\
&\quad\times\f{1}{|u'|}\|\Psi^{(s_2, s'_2)}\|_{L^2_{sc}(\s)}\f{1}{|u'|}\|\nab\Psi^{(s,s')}\|_{L^2_{sc}(S_{u',\ub'})} du'd{\ub}'.\\
\end{split} 
\end{equation*}
Therefore, we derive
\begin{equation*}
\begin{split}
M_7\leq&\f{\d^{\f14}}{|u|}\sum_{s_1+s_2=s+\f12, \atop s'_1+s'_2=s'+\f32}\d^{\f14}\|\Psi^{(s_1, s'_1)}\|^{\f12}_{L^2_{sc}(\Hu)}\|\nab\Psi^{(s_1, s'_1)}\|^{\f12}_{L^2_{sc}(\Hu)}\\
&\quad\times\|\Psi^{(s_2, s'_2)}\|^{\f12}_{L^2_{sc}(\Hbu)}\|\nab\Psi^{(s_2, s'_2)}\|^{\f12}_{L^2_{sc}(\Hbu)}\|\nab\Psi^{(s, s')}\|_{L^2_{sc}(\Hbu)}\\
&+\f{\d^{\f14}}{|u|}\sum_{s_1+s_2=s+\f12, \atop s'_1+s'_2=s'+\f32}\d^{\f12}\|\Psi^{(s_1, s'_1)}\|_{L^2_{sc}(\Hu)}\|\Psi^{(s_2, s'_2)}\|^{\f12}_{L^2_{sc}(\Hbu)}\\
&\quad\times\|\nab\Psi^{(s_2, s'_2)}\|^{\f12}_{L^2_{sc}(\Hbu)}\|\nab\Psi^{(s, s')}\|_{L^2_{sc}(\Hbu)}\\
&+\f{\d^{\f12}}{|u|}\sum_{s_1+s_2=s+\f12, \atop s'_1+s'_2=s'+\f32}\d^{\f14}\|\Psi^{(s_1, s'_1)}\|^{\f12}_{L^2_{sc}(\Hu)}\|\nab\Psi^{(s_1, s'_1)}\|^{\f12}_{L^2_{sc}(\Hu)}\\
&\quad\times\|\Psi^{(s_2, s'_2)}\|_{L^2_{sc}(\Hbu)}\|\nab\Psi^{(s, s')}\|_{L^2_{sc}(\Hbu)}\\
&+\f{\d^{\f12}}{|u|}\sum_{s_1+s_2=s+\f12, \atop s'_1+s'_2=s'+\f32}\d^{\f12}\|\Psi^{(s_1, s'_1)}\|_{L^2_{sc}(\Hu)}\|\Psi^{(s_2, s'_2)}\|_{L^2_{sc}(\Hbu)}\|\nab\Psi^{(s, s')}\|_{L^2_{sc}(\Hbu)}.\\
\end{split} 
\end{equation*}

In the same manner, $M_8$ is controlled as well.

Putting all the estimates in this section altogether, we arrive at
\begin{proposition}\label{energy912}
Under the assumption of Theorem \ref{main.thm} and bootstrap assumption \eqref{BA.4}, 
there exists $\d_0=\d_0(\M R, \underline{\M R})$, when $0<\d<\d_0$, 
we have 
$$\M R_1+\underline{\M R}_1\leq \M R^{(0)}+\d^{\f18}(\M I^{(0)}+\M R^{(0)}+\M R+\underline{\M R})^5.$$
\end{proposition}

\subsection{Curvature Estimate III}
Our goal in this subsection is to derive estimates for the second angular derivatives of curvature components and 
the pair $\nab^2\Psi^{(s,s')}$ and $\nab^2\Psi^{(s-\f12,s'+\f12)}$ obey
\begin{equation}\label{tran3333333} 
\begin{split}
& \nab_3 \nab^2\Psi^{(s,s')}+(\f32+s')\tr\chib\nab^2\Psi^{(s,s')}\\
=&\nab^3\Psi^{(s-\frac12,s'+\frac12)}+\sum_{s_1+s_2=s, \atop s'_1+s'_2=s'+1}\nab^2\psi^{(s_1, s'_1)}\cdot\Psi^{(s_2, s'_2)}+\sum_{s_1+s_2=s, \atop s'_1+s'_2=s'+1}\nab\psi^{(s_1, s'_1)}\cdot\nab\Psi^{(s_2, s'_2)}\\
&+\sum_{s_3+s_4+s_5=s+\f12, \atop s'_3+s'_4+s'_5=s'+\f32}\nab\psi^{(s_3, s'_3)}\cdot\psi^{(s_4, s'_4)}\cdot\Psi^{(s_5,s'_5)}+\sum_{s_3+s_4+s_5=s+\f12, \atop s'_3+s'_4+s'_5=s'+\f32}\psi^{(s_3, s'_3)}\cdot\psi^{(s_4, s'_4)}\cdot\nab\Psi^{(s_5,s'_5)}\\
&+\sum_{s_1+s_2=s+\f12, \atop s'_1+s'_2=s'+\f32}\Psi^{(s_1, s'_1)}\cdot\nab\Psi^{(s_2, s'_2)}+\sum_{s_1+s_2=s, \atop s'_1+s'_2=s'+1}\psi^{(s_1, s'_1)}\cdot\nab^2\Psi^{(s_2, s'_2)}\\
&+\sum_{s_6+s_7+s_8+s_9=s+1, \atop s'_6+s'_7+s'_8+s'_9=s'+2}\psi^{(s_6,s'_6)}\cdot\psi^{(s_7, s'_7)}\cdot\psi^{(s_8, s'_8)}\cdot\Psi^{(s_9,s'_9)}\\
&+\sum_{s_{10}+s_{11}+s_{12}=s+1, \atop s'_{10}+s'_{11}+s'_{12}=s'+2}\psi^{(s_{10},s'_{10})}\cdot\Psi^{(s_{11},s'_{11})}\cdot\Psi^{(s_{12},s'_{12})},
\end{split}
\end{equation}
or 
\begin{equation}\label{tran4444444} 
\begin{split}
& \nab_4 \nab^2\Psi^{(s-\f12, s'+\f12)}\\
=&\nab^3\Psi^{(s, s')}+\sum_{s_1+s_2=s+\f12, \atop s'_1+s'_2=s'+\f12}\nab^2\psi^{(s_1, s'_1)}\cdot\Psi^{(s_2, s'_2)}+\sum_{s_1+s_2=s+\f12, \atop s'_1+s'_2=s'+\f12}\nab\psi^{(s_1, s'_1)}\cdot\nab\Psi^{(s_2, s'_2)}\\
&+\sum_{s_3+s_4+s_5=s+1, \atop s'_3+s'_4+s'_5=s'+1}\nab\psi^{(s_3, s'_3)}\cdot\psi^{(s_4, s'_4)}\cdot\Psi^{(s_5,s'_5)}+\sum_{s_3+s_4+s_5=s+1, \atop s'_3+s'_4+s'_5=s'+1}\psi^{(s_3, s'_3)}\cdot\psi^{(s_4, s'_4)}\cdot\nab\Psi^{(s_5,s'_5)}\\
&+\sum_{s_1+s_2=s+1, \atop s'_1+s'_2=s'+1}\Psi^{(s_1, s'_1)}\cdot\nab\Psi^{(s_2, s'_2)}+\sum_{s_1+s_2=s+\f12, \atop s'_1+s'_2=s'+\f12}\psi^{(s_1, s'_1)}\cdot\nab^2\Psi^{(s_2, s'_2)}\\
&+\sum_{s_6+s_7+s_8+s_9=s+\f32, \atop s'_6+s'_7+s'_8+s'_9=s'+\f32}\psi^{(s_6,s'_6)}\cdot\psi^{(s_7, s'_7)}\cdot\psi^{(s_8, s'_8)}\cdot\Psi^{(s_9,s'_9)}\\
&+\sum_{s_{10}+s_{11}+s_{12}=s+\f32, \atop s'_{10}+s'_{11}+s'_{12}=s'+\f32}\psi^{(s_{10},s'_{10})}\cdot\Psi^{(s_{11},s'_{11})}\cdot\Psi^{(s_{12},s'_{12})}.
\end{split}
\end{equation}

With equations above and H\"older's inequality, we arrive at 

\begin{proposition}
For curvature component $\nab^2\Psi^{(s,s')}$, $\nab^2\Psi^{(s-\f12, s'+\f12)}$ satisfying \eqref{tran3333333} and \eqref{tran4444444}, we have
\begin{equation*}
\begin{split}
&\|\nab^2\Psi^{(s,s')}\|^2_{L^2_{sc}(H_u^{(0,\ub)})}+\|\nab^2\Psi^{(s-\frac12,s'+\frac12)}\|^2_{L^2_{sc}(\Hb_{\ub}^{(u_{\infty},u)})}\\
\leq & \|\nab^2\Psi^{(s,s')}\|^2_{L^2_{sc}(H_{u_{\infty}}^{(0,\ub)})}+\|\nab^2\Psi^{(s-\frac12,s'+\frac12)}\|^2_{L^2_{sc}(\Hb_{0}^{(u_{\infty},u)})}\\
&+N_1+N_2+N_3+N_4+N_5+N_6+N_7+N_8+N_9\\
&+N_{10}+N_{11}+N_{12}+N_{13}+N_{14}+N_{15}+N_{16},
\end{split}
\end{equation*}
where $N_1-N_{16}$ are given as follows:
\begin{equation*}
\begin{split}
N_1=&\int_0^{\ub}\int_{u_{\infty}}^u \frac{\delta^{-\frac32}}{|u'|}\sum_{s_1+s_2=s, \atop s'_1+s'_2=s'+1}\|\nab^2\psi^{(s_1, s'_1)}\cdot\Psi^{(s_2, s'_2)}\cdot\nab^2\Psi^{(s,s')}\|_{L^1_{sc}(S_{u',\ub'})} du'd{\ub}',\\
N_2=&\int_0^{\ub}\int_{u_{\infty}}^u \frac{\delta^{-\frac32}}{|u'|}\sum_{s_1+s_2=s, \atop s'_1+s'_2=s'+1}\|\nab\psi^{(s_1, s'_1)}\cdot\nab\Psi^{(s_2, s'_2)}\cdot\nab^2\Psi^{(s,s')}\|_{L^1_{sc}(S_{u',\ub'})} du'd{\ub}',\\
N_3=&\int_0^{\ub}\int_{u_{\infty}}^u \frac{\delta^{-\frac32}}{|u'|}\sum_{s_1+s_2=s+\f12, \atop s'_1+s'_2=s'+\f32}\|\Psi^{(s_1, s'_1)}\cdot\nab\Psi^{(s_2, s'_2)}\cdot\nab^2\Psi^{(s,s')}\|_{L^1_{sc}(S_{u',\ub'})} du'd{\ub}',\\
N_4=&\int_0^{\ub}\int_{u_{\infty}}^u \frac{\delta^{-\frac32}}{|u'|}\sum_{s_1+s_2=s, \atop s'_1+s'_2=s'+1}\|\psi^{(s_1, s'_1)}\cdot\nab^2\Psi^{(s_2, s'_2)}\cdot\nab^2\Psi^{(s,s')}\|_{L^1_{sc}(S_{u',\ub'})} du'd{\ub}',\\
N_5=&\int_0^{\ub}\int_{u_{\infty}}^u \frac{\delta^{-\frac32}}{|u'|}\sum_{s_1+s_2=s+\f12, \atop s'_1+s'_2=s'+\f12}\|\nab^2\psi^{(s_1, s'_1)}\cdot\Psi^{(s_2, s'_2)}\cdot\nab^2\Psi^{(s-\f12,s'+\f12)}\|_{L^1_{sc}(S_{u',\ub'})} du'd{\ub}',\\
\end{split} 
\end{equation*}
\begin{equation*}
\begin{split}
N_6=&\int_0^{\ub}\int_{u_{\infty}}^u \frac{\delta^{-\frac32}}{|u'|}\sum_{s_1+s_2=s+\f12, \atop s'_1+s'_2=s'+\f12}\|\nab\psi^{(s_1, s'_1)}\cdot\nab\Psi^{(s_2, s'_2)}\cdot\nab^2\Psi^{(s-\f12,s'+\f12)}\|_{L^1_{sc}(S_{u',\ub'})} du'd{\ub}',\\
N_7=&\int_0^{\ub}\int_{u_{\infty}}^u \frac{\delta^{-\frac32}}{|u'|}\sum_{s_1+s_2=s+1, \atop s'_1+s'_2=s'+1}\|\Psi^{(s_1, s'_1)}\cdot\nab\Psi^{(s_2, s'_2)}\cdot\nab^2\Psi^{(s-\f12,s'+\f12)}\|_{L^1_{sc}(S_{u',\ub'})} du'd{\ub}',\\
N_8=&\int_0^{\ub}\int_{u_{\infty}}^u \frac{\delta^{-\frac32}}{|u'|}\sum_{s_1+s_2=s+\f12, \atop s'_1+s'_2=s'+\f12}\|\psi^{(s_1, s'_1)}\cdot\nab^2\Psi^{(s_2, s'_2)}\cdot\nab^2\Psi^{(s-\f12,s'+\f12)}\|_{L^1_{sc}(S_{u',\ub'})} du'd{\ub}',\\
N_9=&\int_0^{\ub}\int_{u_{\infty}}^u \frac{\delta^{-\frac32}}{|u'|}\sum_{s_3+s_4+s_5=s+\f12, \atop s'_3+s'_4+s'_5=s'+\f32}\|\nab\psi^{(s_3, s'_3)}\cdot\psi^{(s_4, s'_4)}\cdot\Psi^{(s_5,s'_5)}\cdot\nab^2\Psi^{(s,s')}\|_{L^1_{sc}(S_{u',\ub'})} du'd{\ub}',\\
N_{10}=&\int_0^{\ub}\int_{u_{\infty}}^u \frac{\delta^{-\frac32}}{|u'|}\sum_{s_3+s_4+s_5=s+\f12, \atop s'_3+s'_4+s'_5=s'+\f32}\|\psi^{(s_3, s'_3)}\cdot\psi^{(s_4, s'_4)}\cdot\nab\Psi^{(s_5,s'_5)}\cdot\nab^2\Psi^{(s,s')}\|_{L^1_{sc}(S_{u',\ub'})} du'd{\ub}',\\
N_{11}=&\int_0^{\ub}\int_{u_{\infty}}^u \frac{\delta^{-\frac32}}{|u'|}\sum_{s_3+s_4+s_5=s+1, \atop s'_3+s'_4+s'_5=s'+1}\|\nab\psi^{(s_3, s'_3)}\cdot\psi^{(s_4, s'_4)}\cdot\Psi^{(s_5,s'_5)}\\
&\quad\quad\quad\times\nab^2\Psi^{(s-\f12,s'+\f12)}\|_{L^1_{sc}(S_{u',\ub'})} du'd{\ub}',\\
N_{12}=&\int_0^{\ub}\int_{u_{\infty}}^u \frac{\delta^{-\frac32}}{|u'|}\sum_{s_3+s_4+s_5=s+1, \atop s'_3+s'_4+s'_5=s'+1}\|\psi^{(s_3, s'_3)}\cdot\psi^{(s_4, s'_4)}\cdot\nab\Psi^{(s_5,s'_5)}\\
&\quad\quad\quad\times\nab^2\Psi^{(s-\f12,s'+\f12)}\|_{L^1_{sc}(S_{u',\ub'})} du'd{\ub}',\\
N_{13}=&\int_0^{\ub}\int_{u_{\infty}}^u \frac{\delta^{-\frac32}}{|u'|}\f{\d}{|u'|^2}\sum_{s_6+s_7+s_8+s_9=s+1, \atop s'_6+s'_7+s'_8+s'_9=s'+2}\|\psi^{(s_6,s'_6)}\cdot\psi^{(s_7, s'_7)}\cdot\psi^{(s_8, s'_8)}\|_{L^{4}_{sc}(S_{u',\ub'})}\\
&\quad\times\|\Psi^{(s_9,s'_9)}\|_{L^4_{sc}(S_{u',\ub'})}\|\nab^2\Psi^{(s,s')}\|_{L^2_{sc}(S_{u',\ub'})} du'd{\ub}',\\
N_{14}=&\int_0^{\ub}\int_{u_{\infty}}^u \frac{\delta^{-\frac32}}{|u'|}\f{\d}{|u'|^2}\sum_{s_6+s_7+s_8+s_9=s+\f32, \atop s'_6+s'_7+s'_8+s'_9=s'+\f32}\|\psi^{(s_6,s'_6)}\cdot\psi^{(s_7, s'_7)}\cdot\psi^{(s_8, s'_8)}\|_{L^{4}_{sc}(S_{u',\ub'})}\\
&\quad\times\|\Psi^{(s_9,s'_9)}\|_{L^4_{sc}(S_{u',\ub'})}\|\nab^2\Psi^{(s-\f12,s'+\f12)}\|_{L^2_{sc}(S_{u',\ub'})} du'd{\ub}',\\
N_{15}=&\int_0^{\ub}\int_{u_{\infty}}^u \frac{\delta^{-\frac32}}{|u'|}\sum_{s_{10}+s_{11}+s_{12}=s+1, \atop s'_{10}+s'_{11}+s'_{12}=s'+2}\|\psi^{(s_{10}, s'_{10})}\cdot\Psi^{(s_{11}, s'_{11})}\cdot\Psi^{(s_{12},s'_{12})}\\
&\quad\quad\quad\times\nab^2\Psi^{(s,s')}\|_{L^1_{sc}(S_{u',\ub'})} du'd{\ub}',\\
N_{16}=&\int_0^{\ub}\int_{u_{\infty}}^u \frac{\delta^{-\frac32}}{|u'|}\sum_{s_{10}+s_{11}+s_{12}=s+\f32, \atop s'_{10}+s'_{11}+s'_{12}=s'+\f32}\|\psi^{(s_{10}, s'_{10})}\cdot\Psi^{(s_{11}, s'_{11})}\cdot\Psi^{(s_{12},s'_{12})}\\
&\quad\quad\quad\times\nab^2\Psi^{(s-\f12,s'+\f12)}\|_{L^1_{sc}(S_{u',\ub'})} du'd{\ub}'.\\
\end{split} 
\end{equation*}
\end{proposition}

%
We will bound these term by term.

In the same fashion as in previous subsection, with Proposition \ref{L4} and H\"older's inequality, we have the following bounds.

We control $N_1$ either by

\begin{equation*}
\begin{split}
N_1\leq&\int_0^{\ub}\int_{u_{\infty}}^u\f{\d^{-\f12}}{|u'|^3}\sum_{s_1+s_2=s, \atop s'_1+s'_2=s'+1}\|\nab^2\psi^{(s_1,s'_1)}\|_{L^4_{sc}(S_{u',\ub'})}\|\Psi^{(s_2,s'_2)}\|_{L^4_{sc}(S_{u',\ub'})}\|\nab^2\Psi^{(s,s')}\|_{L^2_{sc}(S_{u',\ub'})}du'd\ub'\\
\leq&\int_{u_{\infty}}^u\f{\d^{\f12}}{|u'|^3}\sum_{s_1+s_2=s, \atop s'_1+s'_2=s'+1}\|\nab^2\psi^{(s_1,s'_1)}\|_{L^4_{sc}(S_{u',\ub'})}\l\int_0^{\ub}\d^{-\f12}\|\Psi^{(s_2,s'_2)}\|_{L^4_{sc}(S_{u',\ub'})}\\
&\quad\quad\quad\times\d^{-\f12}\|\nab^2\Psi^{(s,s')}\|_{L^2_{sc}(S_{u',\ub'})}d\ub'\r du'\\
\leq&\l\f{\d^{\f14}}{|u|^{\f12}}C\|\nab^2\b,\nab^2\rho,\nab^2\sigma,\nab^2\beb\|_{L^2_{sc}(\Hbu)}+\f{\d^{\f14}}{|u|}C(\M I^{(0)}+\M R+\underline{\M R})^5\r\\
&\quad\times\l\d^{\f14}\|\Psi^{(s_2,s'_2)}\|^{\f12}_{L^2_{sc}(\Hu)}\|\nab\Psi^{(s_2,s'_2)}\|^{\f12}_{L^2_{sc}(\Hu)}\\
&\quad\quad\quad+\d^{\f12}\|\Psi^{(s_2,s'_2)}\|_{L^2_{sc}(\Hu)}\r\|\nab^2\Psi^{(s,s')}\|_{L^2_{sc}(\Hu)},
\end{split} 
\end{equation*}
where we use Proposition \ref{O242}, or by

\begin{equation*}
\begin{split}
N_1\leq&\int_0^{\ub}\int_{u_{\infty}}^u\f{\d^{-\f12}}{|u'|^3}\sum_{s_1+s_2=s, \atop s'_1+s'_2=s'+1}\|\nab^2\psi^{(s_1,s'_1)}\|_{L^4_{sc}(S_{u',\ub'})}\|\Psi^{(s_2,s'_2)}\|_{L^4_{sc}(S_{u',\ub'})}\|\nab^2\Psi^{(s,s')}\|_{L^2_{sc}(S_{u',\ub'})}du'd\ub'\\
\leq&\int_{0}^{\ub}\f{\d^{-\f12}}{|u'|}\sum_{s_1+s_2=s, \atop s'_1+s'_2=s'+1}\|\nab^2\psi^{(s_1,s'_1)}\|_{L^4_{sc}(S_{u',\ub'})}\l\int_{u_{\infty}}^{u}\f{1}{|u'|}\|\Psi^{(s_2,s'_2)}\|_{L^4_{sc}(S_{u',\ub'})}\\
&\quad\quad\quad\times\f{1}{|u'|}\|\nab^2\Psi^{(s,s')}\|_{L^2_{sc}(S_{u',\ub'})}du'\r d\ub'\\
\leq&\l\d^{\f12}C\|\nab^2\b,\nab^2\rho,\nab^2\sigma,\nab^2\beb\|_{L^2_{sc}(\Hu)}+\d^{\f12}C(\M I^{(0)}+\M R+\underline{\M R})^5\r\\
&\quad\times\l\|\Psi^{(s_2,s'_2)}\|^{\f12}_{L^2_{sc}(\Hbu)}\|\nab\Psi^{(s_2,s'_2)}\|^{\f12}_{L^2_{sc}(\Hbu)}\\
&\quad\quad\quad+\d^{\f14}\|\Psi^{(s_2,s'_2)}\|_{L^2_{sc}(\Hbu)}\r\|\nab^2\Psi^{(s,s')}\|_{L^2_{sc}(\Hbu)},
\end{split} 
\end{equation*}
where we use Proposition \ref{O242}.

Anomalous terms $(\nab^2\chib,\nab^2\omb, \nab^2\eta, \nab^2\etb)\a\nab^2\Psi$, $\nab^2\chibh(\b, \rho, \sigma)\nab^2\Psi$ and $(\nab^2\chi,\nab^2\o)\ab\nab^2\Psi$ of $N_1$ 
type can be treated through the estimate above.

For $N_2$, we derive
\begin{equation*}
\begin{split}
N_2\leq&\f{\d^{\f12}}{|u|}\sum_{s_1+s_2=s, \atop s'_1+s'_2=s'+1}\sup_{u'\ub'}\f{1}{|u'|}\|\nab\psi^{(s_1, s'_1)}\|_{L^4_{sc}(\s)}\|\nab\Psi^{(s_2, s'_2)}\|^{\f12}_{L^2_{sc}(\Hu)}\\
&\quad\times\|\nab^2\Psi^{(s_2, s'_2)}\|^{\f12}_{L^2_{sc}(\Hu)}\|\nab^2\Psi^{(s,s')}\|_{L^2_{sc}(\Hu)}\\
&+\f{\d^{\f34}}{|u|}\sum_{s_1+s_2=s, \atop s'_1+s'_2=s'+1}\sup_{u'\ub'}\f{1}{|u'|}\|\nab\psi^{(s_1, s'_1)}\|_{L^4_{sc}(\s)}\\
&\quad\times\|\nab\Psi^{(s_2, s'_2)}\|_{L^2_{sc}(\Hu)}\|\nab^2\Psi^{(s,s')}\|_{L^2_{sc}(\Hu)},\\
\end{split} 
\end{equation*}
or
\begin{equation*}
\begin{split}
N_2\leq&{\d^{\f12}}\sum_{s_1+s_2=s, \atop s'_1+s'_2=s'+1}\sup_{u'\ub'}\f{1}{|u'|}\|\nab\psi^{(s_1, s'_1)}\|_{L^4_{sc}(\s)}\|\nab\Psi^{(s_2, s'_2)}\|^{\f12}_{L^2_{sc}(\Hbu)}\\
&\quad\times\|\nab^2\Psi^{(s_2, s'_2)}\|^{\f12}_{L^2_{sc}(\Hbu)}\|\nab^2\Psi^{(s,s')}\|_{L^2_{sc}(\Hbu)}\\
&+{\d^{\f34}}\sum_{s_1+s_2=s, \atop s'_1+s'_2=s'+1}\sup_{u'\ub'}\f{1}{|u'|}\|\nab\psi^{(s_1, s'_1)}\|_{L^4_{sc}(\s)}\\
&\quad\times\|\nab\Psi^{(s_2, s'_2)}\|_{L^2_{sc}(\Hbu)}\|\nab^2\Psi^{(s,s')}\|_{L^2_{sc}(\Hbu)}.\\
\end{split} 
\end{equation*} 
The anomalous term of $N_{2}$ type is $\nab\chibh(\nab\a, \nab\b,\nab\rho, \nab\sigma, \nab\beb)\nab^2\Psi$ and it is bounded through the estimate above.

We bound $N_3$ through

\begin{equation}\label{N3}
\begin{split}
N_3=&\int_0^{\ub}\int_{u_{\infty}}^u \frac{\delta^{-\frac32}}{|u'|}\sum_{s_1+s_2=s+\f12, \atop s'_1+s'_2=s'+\f32}\|\Psi^{(s_1, s'_1)}\cdot\nab\Psi^{(s_2, s'_2)}\cdot\nab^2\Psi^{(s,s')}\|_{L^1_{sc}(S_{u',\ub'})} du'd{\ub}'\\
\leq&\int_0^{\ub}\int_{u_{\infty}}^u \frac{\delta^{-\frac12}}{|u'|^3}\sum_{s_1+s_2=s+\f12, \atop s'_1+s'_2=s'+\f32}\|\Psi^{(s_1, s'_1)}\|_{L^4_{sc}(\s)}\|\nab\Psi^{(s_2, s'_2)}\|_{L^4_{sc}(\s)}\\
&\quad\times\|\nab^2\Psi^{(s,s')}\|_{L^2_{sc}(S_{u',\ub'})} du'd{\ub}'\\
\leq&\int_0^{\ub}\int_{u_{\infty}}^u \frac{\delta^{-\frac12}}{|u'|^3}\sum_{s_1+s_2=s+\f12, \atop s'_1+s'_2=s'+\f32}\|\Psi^{(s_1, s'_1)}\|^{\f12}_{L^2_{sc}(\s)}\|\nab\Psi^{(s_1, s'_1)}\|^{\f12}_{L^2_{sc}(\s)}\\
&\quad\times\|\Psi^{(s_2, s'_2)}\|^{\f12}_{L^2_{sc}(\s)}\|\nab\Psi^{(s_2, s'_2)}\|^{\f12}_{L^2_{sc}(\s)}\|\nab^2\Psi^{(s,s')}\|_{L^2_{sc}(S_{u',\ub'})} du'd{\ub}'\\
&+\int_0^{\ub}\int_{u_{\infty}}^u \frac{\delta^{-\frac12}}{|u'|^3}\sum_{s_1+s_2=s+\f12, \atop s'_1+s'_2=s'+\f32}\d^{\f14}\|\Psi^{(s_1, s'_1)}\|_{L^2_{sc}(\s)}\\
&\quad\times\|\nab\Psi^{(s_2, s'_2)}\|^{\f12}_{L^2_{sc}(\s)}\|\nab^2\Psi^{(s_2, s'_2)}\|^{\f12}_{L^2_{sc}(\s)}\|\nab^2\Psi^{(s,s')}\|_{L^2_{sc}(S_{u',\ub'})} du'd{\ub}'\\
&+\int_0^{\ub}\int_{u_{\infty}}^u \frac{\delta^{-\frac12}}{|u'|^3}\sum_{s_1+s_2=s+\f12, \atop s'_1+s'_2=s'+\f32}\|\Psi^{(s_1, s'_1)}\|^{\f12}_{L^2_{sc}(\s)}\|\nab\Psi^{(s_1, s'_1)}\|^{\f12}_{L^2_{sc}(\s)}\\
&\quad\times\d^{\f14}\|\nab\Psi^{(s_2, s'_2)}\|_{L^2_{sc}(\s)}\|\nab^2\Psi^{(s,s')}\|_{L^2_{sc}(S_{u',\ub'})} du'd{\ub}'\\
&+\int_0^{\ub}\int_{u_{\infty}}^u \frac{\delta^{-\frac12}}{|u'|^3}\sum_{s_1+s_2=s+\f12, \atop s'_1+s'_2=s'+\f32}\d^{\f14}\|\Psi^{(s_1, s'_1)}\|_{L^2_{sc}(\s)}\\
&\quad\times\d^{\f14}\|\nab\Psi^{(s_2, s'_2)}\|_{L^2_{sc}(\s)}\|\nab^2\Psi^{(s,s')}\|_{L^2_{sc}(S_{u',\ub'})} du'd{\ub}'.\\
\end{split} 
\end{equation}
Furthermore, starting with \eqref{N3}, we have
\begin{equation*}
\begin{split}
N_3\leq&\int_0^{\ub}\int_{u_{\infty}}^u \frac{\delta^{\frac12}}{|u'|^2}\sum_{s_1+s_2=s+\f12, \atop s'_1+s'_2=s'+\f32}\f{1}{|u'|^{\f12}}\|\Psi^{(s_1, s'_1)}\|^{\f12}_{L^2_{sc}(\s)}\f{1}{|u'|^{\f12}}\|\nab\Psi^{(s_1, s'_1)}\|^{\f12}_{L^2_{sc}(\s)}\\
&\quad\times\d^{-\f14}\|\Psi^{(s_2, s'_2)}\|^{\f12}_{L^2_{sc}(\s)}\d^{-\f14}\|\nab\Psi^{(s_2, s'_2)}\|^{\f12}_{L^2_{sc}(\s)}\\
&\quad\times\d^{-\f12}\|\nab^2\Psi^{(s,s')}\|_{L^2_{sc}(S_{u',\ub'})} du'd{\ub}'\\
&+\int_0^{\ub}\int_{u_{\infty}}^u \frac{\delta^{\frac34}}{|u'|^2}\sum_{s_1+s_2=s+\f12, \atop s'_1+s'_2=s'+\f32}\f{1}{|u'|}\|\Psi^{(s_1, s'_1)}\|_{L^2_{sc}(\s)}\\
&\quad\times\d^{-\f14}\|\nab\Psi^{(s_2, s'_2)}\|^{\f12}_{L^2_{sc}(\s)}\d^{-\f14}\|\nab^2\Psi^{(s_2, s'_2)}\|^{\f12}_{L^2_{sc}(\s)}\\
&\quad\times\d^{-\f12}\|\nab^2\Psi^{(s,s')}\|_{L^2_{sc}(S_{u',\ub'})} du'd{\ub}'\\
&+\int_0^{\ub}\int_{u_{\infty}}^u \frac{\delta^{\frac34}}{|u'|^2}\sum_{s_1+s_2=s+\f12, \atop s'_1+s'_2=s'+\f32}\f{1}{|u'|^{\f12}}\|\Psi^{(s_1, s'_1)}\|^{\f12}_{L^2_{sc}(\s)}\f{1}{|u'|^{\f12}}\|\nab\Psi^{(s_1, s'_1)}\|^{\f12}_{L^2_{sc}(\s)}\\
&\quad\times\d^{-\f12}\|\nab\Psi^{(s_2, s'_2)}\|_{L^2_{sc}(\s)}\d^{-\f12}\|\nab^2\Psi^{(s,s')}\|_{L^2_{sc}(S_{u',\ub'})} du'd{\ub}'\\
&+\int_0^{\ub}\int_{u_{\infty}}^u \frac{\delta}{|u'|^2}\sum_{s_1+s_2=s+\f12, \atop s'_1+s'_2=s'+\f32}\f{1}{|u|}\|\Psi^{(s_1, s'_1)}\|_{L^2_{sc}(\s)}\\
&\quad\times\d^{-\f12}\|\nab\Psi^{(s_2, s'_2)}\|_{L^2_{sc}(\s)}\d^{-\f12}\|\nab^2\Psi^{(s,s')}\|_{L^2_{sc}(S_{u',\ub'})} du'd{\ub}'.\\
\end{split} 
\end{equation*}
Therefore, we deduce
\begin{equation*}
\begin{split}
N_3\leq&\f{\d^{\f12}}{|u|^{\f32}}\sum_{s_1+s_2=s+\f12, \atop s'_1+s'_2=s'+\f32}\|\Psi^{(s_1, s'_1)}\|^{\f12}_{L^2_{sc}(\Hbu)}\|\nab\Psi^{(s_1, s'_1)}\|^{\f12}_{L^2_{sc}(\Hbu)}\\
&\quad\times\|\nab\Psi^{(s_2, s'_2)}\|^{\f12}_{L^2_{sc}(\Hu)}\|\nab^2\Psi^{(s_2, s'_2)}\|^{\f12}_{L^2_{sc}(\Hu)}\|\nab^2\Psi^{(s, s')}\|_{L^2_{sc}(\Hu)}\\
&+\f{\d^{\f34}}{|u|^{\f32}}\sum_{s_1+s_2=s+\f12, \atop s'_1+s'_2=s'+\f32}\|\Psi^{(s_1, s'_1)}\|^{\f12}_{L^2_{sc}(\Hbu)}\|\nab\Psi^{(s_1, s'_1)}\|^{\f12}_{L^2_{sc}(\Hbu)}\\
&\quad\times\|\nab\Psi^{(s_2, s'_2)}\|_{L^2_{sc}(\Hu)}\|\nab^2\Psi^{(s, s')}\|_{L^2_{sc}(\Hu)}\\
&+\f{\d^{\f34}}{|u|^{\f32}}\sum_{s_1+s_2=s+\f12, \atop s'_1+s'_2=s'+\f32}\|\Psi^{(s_1, s'_1)}\|_{L^2_{sc}(\Hbu)}\\
&\quad\times\|\nab\Psi^{(s_2, s'_2)}\|^{\f12}_{L^2_{sc}(\Hu)}\|\nab^2\Psi^{(s_2, s'_2)}\|^{\f12}_{L^2_{sc}(\Hu)}\|\nab^2\Psi^{(s, s')}\|_{L^2_{sc}(\Hu)}\\
&+\f{\d}{|u|^{\f32}}\sum_{s_1+s_2=s+\f12, \atop s'_1+s'_2=s'+\f32}\|\Psi^{(s_1, s'_1)}\|_{L^2_{sc}(\Hbu)}\|\nab\Psi^{(s_2, s'_2)}\|_{L^2_{sc}(\Hu)}\|\nab^2\Psi^{(s, s')}\|_{L^2_{sc}(\Hu)}.\\
\end{split} 
\end{equation*}
Alternatively, with \eqref{N3} we get 
\begin{equation*}
\begin{split}
N_3\leq&\int_0^{\ub}\int_{u_{\infty}}^u \frac{1}{|u'|}\sum_{s_1+s_2=s+\f12, \atop s'_1+s'_2=s'+\f32}\d^{-\f14}\|\Psi^{(s_1, s'_1)}\|^{\f12}_{L^2_{sc}(\s)}\d^{-\f14}\|\nab\Psi^{(s_1, s'_1)}\|^{\f12}_{L^2_{sc}(\s)}\\
&\quad\times\f{1}{|u'|^{\f12}}\|\Psi^{(s_2, s'_2)}\|^{\f12}_{L^2_{sc}(\s)}\f{1}{|u'|^{\f12}}\|\nab\Psi^{(s_2, s'_2)}\|^{\f12}_{L^2_{sc}(\s)}\\
&\quad\times\f{1}{|u'|}\|\nab^2\Psi^{(s,s')}\|_{L^2_{sc}(S_{u',\ub'})} du'd{\ub}'\\
&+\int_0^{\ub}\int_{u_{\infty}}^u \frac{\delta^{\frac14}}{|u'|}\sum_{s_1+s_2=s+\f12, \atop s'_1+s'_2=s'+\f32}\d^{-\f12}\|\Psi^{(s_1, s'_1)}\|_{L^2_{sc}(\s)}\\
&\quad\times\f{1}{|u'|^{\f12}}\|\nab\Psi^{(s_2, s'_2)}\|^{\f12}_{L^2_{sc}(\s)}\f{1}{|u'|^{\f12}}\|\nab^2\Psi^{(s_2, s'_2)}\|^{\f12}_{L^2_{sc}(\s)}\\
&\quad\times\f{1}{|u'|}\|\nab^2\Psi^{(s,s')}\|_{L^2_{sc}(S_{u',\ub'})} du'd{\ub}'\\
&+\int_0^{\ub}\int_{u_{\infty}}^u \frac{\delta^{\frac14}}{|u'|}\sum_{s_1+s_2=s+\f12, \atop s'_1+s'_2=s'+\f32}\d^{-\f14}\|\Psi^{(s_1, s'_1)}\|^{\f12}_{L^2_{sc}(\s)}\d^{-\f14}\|\nab\Psi^{(s_1, s'_1)}\|^{\f12}_{L^2_{sc}(\s)}\\
&\quad\times\f{1}{|u'|}\|\nab\Psi^{(s_2, s'_2)}\|_{L^2_{sc}(\s)}\f{1}{|u'|}\|\nab^2\Psi^{(s,s')}\|_{L^2_{sc}(S_{u',\ub'})} du'd{\ub}'\\
&+\int_0^{\ub}\int_{u_{\infty}}^u \frac{\delta^{\f12}}{|u'|}\sum_{s_1+s_2=s+\f12, \atop s'_1+s'_2=s'+\f32}\d^{-\f12}\|\Psi^{(s_1, s'_1)}\|_{L^2_{sc}(\s)}\\
&\quad\times\f{1}{|u'|}\|\nab\Psi^{(s_2, s'_2)}\|_{L^2_{sc}(\s)}\f{1}{|u'|}\|\nab^2\Psi^{(s,s')}\|_{L^2_{sc}(S_{u',\ub'})} du'd{\ub}'.\\
\end{split} 
\end{equation*}
Therefore, we obtain
\begin{equation*}
\begin{split}
N_3\leq&\f{\d^{\f12}}{|u|}\sum_{s_1+s_2=s+\f12, \atop s'_1+s'_2=s'+\f32}\|\Psi^{(s_1, s'_1)}\|^{\f12}_{L^2_{sc}(\Hu)}\|\nab\Psi^{(s_1, s'_1)}\|^{\f12}_{L^2_{sc}(\Hu)}\\
&\quad\times\|\Psi^{(s_2, s'_2)}\|^{\f12}_{L^2_{sc}(\Hbu)}\|\nab\Psi^{(s_2, s'_2)}\|^{\f12}_{L^2_{sc}(\Hbu)}\|\nab^2\Psi^{(s, s')}\|_{L^2_{sc}(\Hbu)}\\
&+\f{\d^{\f34}}{|u|}\sum_{s_1+s_2=s+\f12, \atop s'_1+s'_2=s'+\f32}\|\Psi^{(s_1, s'_1)}\|^{\f12}_{L^2_{sc}(\Hu)}\|\nab\Psi^{(s_1, s'_1)}\|^{\f12}_{L^2_{sc}(\Hu)}\\
&\quad\times\|\Psi^{(s_2, s'_2)}\|_{L^2_{sc}(\Hbu)}\|\nab^2\Psi^{(s, s')}\|_{L^2_{sc}(\Hbu)}\\
&+\f{\d^{\f34}}{|u|}\sum_{s_1+s_2=s+\f12, \atop s'_1+s'_2=s'+\f32}\|\Psi^{(s_1, s'_1)}\|_{L^2_{sc}(\Hu)}\\
&\quad\times\|\Psi^{(s_2, s'_2)}\|^{\f12}_{L^2_{sc}(\Hbu)}\|\nab\Psi^{(s_2, s'_2)}\|^{\f12}_{L^2_{sc}(\Hbu)}\|\nab^2\Psi^{(s, s')}\|_{L^2_{sc}(\Hbu)}\\
&+\f{\d^{\f12}}{|u|}\sum_{s_1+s_2=s+\f12, \atop s'_1+s'_2=s'+\f32}\d^{\f12}\|\Psi^{(s_1, s'_1)}\|_{L^2_{sc}(\Hu)}\|\nab\Psi^{(s_2, s'_2)}\|_{L^2_{sc}(\Hbu)}\|\nab^2\Psi^{(s, s')}\|_{L^2_{sc}(\Hbu)}.\\
\end{split} 
\end{equation*}
The anomalous term of $N_{3}$ type is $\nab\beb\a\nab^2\Psi$, and it is bounded through the estimate above.

The term $N_4$ is controlled through
\begin{equation*}
\begin{split}
N_4\leq&\f{\d^{\f12}}{|u|}\sum_{s_1+s_2=s, \atop s'_1+s'_2=s'+1}\sup_{u'\ub'}\f{1}{|u'|}\|\psi^{(s_1, s'_1)}\|_{L^{\infty}_{sc}(\s)}\|\nab^2\Psi^{(s_2, s'_2)}\|_{L^2_{sc}(\Hu)}\\
&\quad\times\|\nab^2\Psi^{(s,s')}\|_{L^2_{sc}(\Hu)},\\
\end{split} 
\end{equation*}
or
\begin{equation*}
\begin{split}
N_4\leq&{\d^{\f12}}\sum_{s_1+s_2=s, \atop s'_1+s'_2=s'+1}\sup_{u'\ub'}\f{1}{|u'|}\|\psi^{(s_1, s'_1)}\|_{L^4_{sc}(\s)}\|\nab^2\Psi^{(s_2, s'_2)}\|_{L^2_{sc}(\Hbu)}\\
&\quad\times\|\nab^2\Psi^{(s,s')}\|_{L^2_{sc}(\Hbu)}.\\
\end{split} 
\end{equation*} 
The anomalous term of $N_{4}$ type is $\chibh(\nab^2\a, \nab^2\b,\nab^2\rho, \nab^2\sigma, \nab^2\beb)\nab^2\Psi$, which is bounded through the estimate above.

$N_5$, $N_6$, $N_7$ and $N_8$ are treated in the same manner as for $N_1$, $N_2$, $N_3$ and $N_4$, respectively.

In $N_9-N_{14}$, $\chib$ appears at most once in a term. Hence we derive the following estimates.
For $N_9$ we have
\begin{equation*}
\begin{split}
N_9\leq &\d^{\f14}\sum_{s_3+s_4+s_5=s+\f12, \atop s'_3+s'_4+s'_5=s'+\f32}\sup_{u'\ub'}\f{\d^{\f12}}{|u'|^2}\|\psi^{(s_3, s'_3)}\|_{L^{\infty}_{sc}(\s)}\f{1}{|u'|}\|\nab\psi^{(s_4, s'_4)}\|_{L^4_{sc}(\s)}\\
&\quad\times\d^{\f14}\|\Psi^{(s_5, s'_5)}\|^{\f12}_{L^2_{sc}(\Hu)}\|\nab\Psi^{(s_5, s'_5)}\|^{\f12}_{L^2_{sc}(\Hu)}\|\nab\Psi^{(s,s')}\|_{L^2_{sc}(\Hu)}\\
&+\f{\d^{\f14}}{|u|}\sum_{s_3+s_4+s_5=s+\f12, \atop s'_3+s'_4+s'_5=s'+\f32}\sup_{u'\ub'}\f{\d^{\f12}}{|u'|^2}\|\psi^{(s_3, s'_3)}\|_{L^{\infty}_{sc}(\s)}\f{1}{|u'|}\|\nab\psi^{(s_4, s'_4)}\|_{L^4_{sc}(\s)}\\
&\quad\times\d^{\f12}\|\Psi^{(s_5, s'_5)}\|_{L^2_{sc}(\Hu)}\|\nab\Psi^{(s,s')}\|_{L^2_{sc}(\Hu)},\\
\end{split} 
\end{equation*}
or
\begin{equation*}
\begin{split}
N_9\leq&{\d^{\f12}}\sum_{s_3+s_4+s_5=s+\f12, \atop s'_3+s'_4+s'_5=s'+\f32}\sup_{u'\ub'}\f{\d^{\f12}}{|u'|^2}\|\psi^{(s_3, s'_3)}\|_{L^{\infty}_{sc}(\s)}\|\nab\psi^{(s_4, s'_4)}\|_{L^4_{sc}(\s)}\|\Psi^{(s_5, s'_5)}\|^{\f12}_{L^2_{sc}(\Hbu)}\\
&\quad\times\|\nab\Psi^{(s_5, s'_5)}\|^{\f12}_{L^2_{sc}(\Hbu)}\|\nab\Psi^{(s,s')}\|_{L^2_{sc}(\Hbu)}\\
&+{\d^{\f34}}\sum_{s_3+s_4+s_5=s+\f12, \atop s'_3+s'_4+s'_5=s'+\f32}\sup_{u'\ub'}\f{\d^{\f12}}{|u'|^2}\|\psi^{(s_3, s'_3)}\|_{L^{\infty}_{sc}(\s)}\|\nab\psi^{(s_4, s'_4)}\|_{L^4_{sc}(\s)}\\
&\quad\times\|\Psi^{(s_5, s'_5)}\|_{L^2_{sc}(\Hbu)}\|\nab\Psi^{(s,s')}\|_{L^2_{sc}(\Hbu)}.\\
\end{split} 
\end{equation*}
The anomalous terms of $N_{9}$ type are 
$$(\chib,\eta,\etb,\omb)(\nab\chib,\nab\omb, \nab\eta,\nab\etb)\a\nab^2\Psi,$$
$$(\nab\eta,\nab\etb)\chib(\a, \b, \rho, \sigma, \beb)\nab^2\Psi,$$ 
and 
$$(\eta,\etb)\nab\chibh(\a, \b, \rho, \sigma, \beb)\nab^2\Psi.$$ 
All of them are bounded through the estimate above.

We bound $N_{10}$ by
\begin{equation*}
\begin{split}
N_{10}\leq &\f{\d^{\f12}}{|u|}\sum_{s_3+s_4+s_5=s+\f12, \atop s'_3+s'_4+s'_5=s'+\f32}\sup_{u'\ub'}\f{\d^{\f12}}{|u'|^2}\|\psi^{(s_3, s'_3)}\|_{L^{\infty}_{sc}(\s)}\|\psi^{(s_4, s'_4)}\|_{L^4_{sc}(\s)}\\
&\quad\times\|\nab\Psi^{(s_5, s'_5)}\|^{\f12}_{L^2_{sc}(\Hu)}\|\nab^2\Psi^{(s_5, s'_5)}\|^{\f12}_{L^2_{sc}(\Hu)}\|\nab\Psi^{(s,s')}\|_{L^2_{sc}(\Hu)}\\
&+\f{\d^{\f34}}{|u|}\sum_{s_3+s_4+s_5=s+\f12, \atop s'_3+s'_4+s'_5=s'+\f32}\sup_{u'\ub'}\f{\d^{\f12}}{|u'|^2}\|\psi^{(s_3, s'_3)}\|_{L^{\infty}_{sc}(\s)}\|\psi^{(s_4, s'_4)}\|_{L^4_{sc}(\s)}\\
&\quad\times\|\nab\Psi^{(s_5, s'_5)}\|_{L^2_{sc}(\Hu)}\|\nab\Psi^{(s,s')}\|_{L^2_{sc}(\Hu)},\\
\end{split} 
\end{equation*}
or
\begin{equation*}
\begin{split}
N_{10}\leq&{\d^{\f12}}\sum_{s_3+s_4+s_5=s+\f12, \atop s'_3+s'_4+s'_5=s'+\f12}\sup_{u'\ub'}\f{\d^{\f12}}{|u'|^2}\|\psi^{(s_3, s'_3)}\|_{L^{\infty}_{sc}(\s)}\|\psi^{(s_4, s'_4)}\|_{L^4_{sc}(\s)}\\
&\quad\times\|\nab\Psi^{(s_5, s'_5)}\|^{\f12}_{L^2_{sc}(\Hbu)}\|\nab^2\Psi^{(s_5, s'_5)}\|^{\f12}_{L^2_{sc}(\Hbu)}\|\nab\Psi^{(s,s')}\|_{L^2_{sc}(\Hbu)}\\
&+{\d^{\f34}}\sum_{s_3+s_4+s_5=s+\f12, \atop s'_3+s'_4+s'_5=s'+\f12}\sup_{u'\ub'}\f{\d^{\f12}}{|u'|^2}\|\psi^{(s_3, s'_3)}\|_{L^{\infty}_{sc}(\s)}\|\psi^{(s_4, s'_4)}\|_{L^4_{sc}(\s)}\\
&\quad\times\|\nab\Psi^{(s_5, s'_5)}\|_{L^2_{sc}(\Hbu)}\|\nab\Psi^{(s,s')}\|_{L^2_{sc}(\Hbu)}.\\
\end{split} 
\end{equation*}
The anomalous term of $N_{10}$ type is $(\eta,\etb)\chib(\nab\a, \nab\b,\nab\rho, \nab\sigma, \nab\beb)\nab^2\Psi$ and it is bounded through the estimate above.

The terms $N_{11}$ and $N_{12}$ are bounded in the same manner as for $N_9$ and $N_{10}$, respectively.

The term $N_{13}$ obeys
\begin{equation*}
\begin{split}
N_{13}\leq &\f{\d}{|u|^2}\sum_{s_6+s_7+s_8+s_9=s+1, \atop s'_6+s'_7+s'_8+s'_9=s'+2}\sup_{u'\ub'}\f{\d^{\f12}}{|u'|^2}\|\psi^{(s_6, s'_6)}\|_{L^{\infty}_{sc}(\s)}\\
&\quad\times\|\psi^{(s_7, s'_7)}\|_{L^{\infty}_{sc}(\s)}\|\psi^{(s_8, s'_8)}\|_{L^4_{sc}(\s)}\\
&\quad\times\|\Psi^{(s_9, s'_9)}\|^{\f12}_{L^2_{sc}(\Hu)}\|\nab\Psi^{(s_9, s'_9)}\|^{\f12}_{L^2_{sc}(\Hu)}\|\nab\Psi^{(s,s')}\|_{L^2_{sc}(\Hu)}\\
&+\f{\d^{\f54}}{|u|^2}\sum_{s_6+s_7+s_8+s_9=s+1, \atop s'_6+s'_7+s'_8+s'_9=s'+2}\sup_{u'\ub'}\f{\d^{\f12}}{|u'|^2}\|\psi^{(s_6, s'_6)}\|_{L^{\infty}_{sc}(\s)}\|\psi^{(s_7, s'_7)}\|_{L^{\infty}_{sc}(\s)}\\
&\quad\times\|\psi^{(s_8, s'_8)}\|_{L^4_{sc}(\s)}\|\Psi^{(s_9, s'_9)}\|_{L^2_{sc}(\Hu)}\|\nab\Psi^{(s,s')}\|_{L^2_{sc}(\Hu)},\\
\end{split} 
\end{equation*}
or
\begin{equation*}
\begin{split}
N_{13}\leq&\f{\d}{|u|}\sum_{s_6+s_7+s_8+s_9=s+1, \atop s'_6+s'_7+s'_8+s'_9=s'+2}\sup_{u'\ub'}\f{\d^{\f12}}{|u'|^2}\|\psi^{(s_6, s'_6)}\|_{L^{\infty}_{sc}(\s)}\\
&\quad\times\|\psi^{(s_7, s'_7)}\|_{L^{\infty}_{sc}(\s)}\|\psi^{(s_8, s'_8)}\|_{L^4_{sc}(\s)}\\
&\quad\times\|\Psi^{(s_9, s'_9)}\|^{\f12}_{L^2_{sc}(\Hbu)}\|\nab\Psi^{(s_9, s'_9)}\|^{\f12}_{L^2_{sc}(\Hbu)}\|\nab\Psi^{(s,s')}\|_{L^2_{sc}(\Hbu)}\\
&+\f{\d^{\f54}}{|u|}\sum_{s_6+s_7+s_8+s_9=s+1, \atop s'_6+s'_7+s'_8+s'_9=s'+2}\sup_{u'\ub'}\f{\d^{\f12}}{|u'|^2}\|\psi^{(s_6, s'_6)}\|_{L^{\infty}_{sc}(\s)}\|\psi^{(s_7, s'_7)}\|_{L^{\infty}_{sc}(\s)}\\
&\quad\times\|\psi^{(s_8, s'_8)}\|_{L^4_{sc}(\s)}\|\Psi^{(s_9, s'_9)}\|_{L^2_{sc}(\Hbu)}\|\nab\Psi^{(s,s')}\|_{L^2_{sc}(\Hbu)}.\\
\end{split} 
\end{equation*}
The anomalous terms of $N_{13}$ type are $\ee\ee(\eta,\etb,\chib,\omb)\a$ and $\ee\ee\chib(\b,\rho,\sigma)\nab^2\Psi$,  which are bounded through the estimate above.

We use the same method to control $N_{14}$.

For $N_{15}$, since there is no $\psi\a\a, \psi\a\ab$ or $\psi\ab\,\ab$ term. Employing Proposition \ref{R04}, one of the curvature components can 
be bounded in $\|\cdot\|_{L^4_{sc}(\S)}$ norm, hence we derive
\begin{equation*}
\begin{split}
N_{15}\leq &\f{\d^{\f14}}{|u|}\sum_{s_{10}+s_{11}+s_{12}=s+1, \atop s'_{10}+s'_{11}+s'_{12}=s+2}\sup_{u'\ub'}\f{\d^{\f12}}{|u'|^2}\|\psi^{(s_{10}, s'_{10})}\|_{L^{\infty}_{sc}(\s)}\d^{\f14}\|\Psi^{(s_{11}, s'_{11})}\|_{L^4_{sc}(\s)}\\
&\quad\times\|\Psi^{(s_{12}, s'_{12})}\|^{\f12}_{L^2_{sc}(\Hu)}\|\nab\Psi^{(s_{12}, s'_{12})}\|^{\f12}_{L^2_{sc}(\Hu)}\|\nab\Psi^{(s,s')}\|_{L^2_{sc}(\Hu)}\\
&+\f{\d^{\f14}}{|u|}\sum_{s_{10}+s_{11}+s_{12}=s+1, \atop s'_{10}+s'_{11}+s'_{12}=s+2}\sup_{u'\ub'}\f{\d^{\f12}}{|u'|^2}\|\psi^{(s_{10}, s'_{10})}\|_{L^{\infty}_{sc}(\s)}\|\Psi^{(s_{11}, s'_{11})}\|_{L^4_{sc}(\s)}\\
&\quad\times\d^{\f12}\|\Psi^{(s_{12}, s'_{12})}\|_{L^2_{sc}(\Hu)}\|\nab\Psi^{(s,s')}\|_{L^2_{sc}(\Hu)},\\
\end{split} 
\end{equation*}
or
\begin{equation*}
\begin{split}
N_{15}\leq&{\d^{\f12}}\sum_{s_{10}+s_{11}+s_{12}=s+1, \atop s'_{10}+s'_{11}+s'_{12}=s+2}\sup_{u'\ub'}\f{\d^{\f12}}{|u'|^2}\|\psi^{(s_{10}, s'_{10})}\|_{L^{\infty}_{sc}(\s)}\|\Psi^{(s_{11}, s'_{11})}\|_{L^4_{sc}(\s)}\\
&\quad\times\|\Psi^{(s_{12}, s'_{12})}\|^{\f12}_{L^2_{sc}(\Hbu)}\|\nab\Psi^{(s_{12}, s'_{12})}\|^{\f12}_{L^2_{sc}(\Hbu)}\|\nab\Psi^{(s,s')}\|_{L^2_{sc}(\Hbu)}\\
&+{\d^{\f34}}\sum_{s_{10}+s_{11}+s_{12}=s+1, \atop s'_{10}+s'_{11}+s'_{12}=s+2}\sup_{u'\ub'}\f{\d^{\f12}}{|u'|^2}\|\psi^{(s_{10}, s'_{10})}\|_{L^{\infty}_{sc}(\s)}\|\Psi^{(s_{11}, s'_{11})}\|_{L^4_{sc}(\s)}\\
&\quad\times\|\Psi^{(s_{12}, s'_{12})}\|_{L^2_{sc}(\Hbu)}\|\nab\Psi^{(s,s')}\|_{L^2_{sc}(\Hbu)}.\\
\end{split} 
\end{equation*}
The anomalous term of $N_{15}$ type is $\ee\b\a\nab^2\Psi$, and it is bounded through the estimate above.

We use the same method to control $N_{16}$.

Gathering all the estimates in this subsection, we conclude
\begin{proposition}
Under the assumption of Theorem \ref{main.thm} and bootstrap assumption \eqref{BA.4}, 
there exists $\d_0=\d_0(\M R, \underline{\M R})$, when $0<\d<\d_0$, 
we have 
$$\M R_2+\underline{\M R}_2\leq \M R^{(0)}+\d^{\f18}(\M I^{(0)}+\M R^{(0)}+\M R+\underline{\M R})^5.$$
\end{proposition}


Therefore, we arrive at
\begin{proposition}\label{energy final}
Under the assumption of Theorem \ref{main.thm}, we get
$$\M R+\underline{\M R}\leq C(\M R^{(0)}+\M I^{(0)}),$$
where $C$ is a large universal constant.
\end{proposition}
\begin{proof}
By choosing $\d$, such that $\d^{-\f{1}{200}}\gg R$ holds, we have improved  the bootstrap assumption \eqref{BA.4}. 
The conclusion follows.\\
\end{proof}

With the estimates in this section, we can prove another useful proposition.

Denote $\epsilon$ a small constant satisfying $\d\ll\epsilon\ll1$. 
Let ${\tilde{\M I}}^{(0)}=\epsilon^{-1} \M I^{(0)}$.
We have
\begin{proposition}\label{epsilon}
Under the assumption of Theorem \ref{main.thm2}, we obtain
$$\M R+\underline{\M R}\leq \epsilon {\tilde{\M I}}^{(0)}.$$ 
\end{proposition}

\begin{proof}
Multiplying $\epsilon^{-2}$ on both sides
of all the estimates in this section, we derive 
$$\epsilon^{-2}\M R^2+\epsilon^{-2}\underline{\M R}^2\leq \epsilon^{-2}\M R_0^2+\d^{\f14}\epsilon^{-2}({\M I}^{(0)})^5 \leq ({\tilde{\M I}}^{(0)})^2+\d^{\f14}\epsilon^{3}({\tilde{\M I}}^{(0)})^5.$$
From the assumptions of Theorem \ref{main.thm2}, we conclude that 
${\tilde{\M I}}^{(0)}=\epsilon^{-1} \M I^{(0)}\leq C,$
where $C$ is a universal large constant.
Times $\epsilon^2$ on both sides, we deduce the desired estimate.
\end{proof}

%
%
%
%
%

\section{FORMATION OF TRAPPED SURFACES}
In the section, we will prove\\
{\bf Theorem 4.2.}(Formation of Trapped Surfaces from Past Null Infinity)

Given $\M I^{(0)}$ and $c$, there exist $\d_0=\d_0(\M I^{(0)}, c)$ sufficiently large, such that for $0<\d<\d_0$, with initial data:
   \begin{itemize}
            \item $\sum_{i\leq 5, k\leq 3}\delta^{k} \d^{\frac12}\|\nab^{k}_4(|u_{\infty}|\nab)^{i}\chih\|_{L^{\infty}(S(1,\ub))}\leq \M I^{(0)}$ \quad along $u=u_{\infty}$
            \item $\sum_{2\leq j\leq 7}\delta^{\frac12}\|(\delta^{\frac12}|u_{\infty}|\nab)^j\chih_0\|_{L^2(S(u_{\infty},\ub))}\leq \epsilon$ \quad along $u=u_{\infty}$
            \item Minkowski initial data \quad \quad \quad \quad \quad \quad \quad \quad \quad along $\ub=0$
            \item $\int_0^{\delta}|u_{\infty}|^2|\chih|^2\geq 4c$ for every direction  \quad \quad along $u=u_{\infty}$
    \end{itemize}
Then $S_{c,\delta}$ is a trapped surface.

\begin{proof}
We first derive pointwise estimates for $|\chih|^2_{\gamma}$. 
Fix $(\theta^1, \theta^2)\in S^2$. We consider the following null structure equation 
\begin{equation*}
 \nab_3\chih+\frac 12 \tr\chib \chih=\nab\widehat{\otimes} \eta+2\omegab \chih-\frac 12 \tr\chi \chibh +\eta\widehat{\otimes} \eta.
\end{equation*}
We contract this two tensor with $\chih$ and get
\begin{equation*}
\f12 \nab_3|\chih|^2_{\gamma}+\f12\tr\chib |\chih|^2_{\gamma}-2\omb|\chih|^2_{\gamma}=\chih(\nab\widehat{\otimes}\eta-\f12\tr\chi\chibh+\eta\widehat{\otimes}\eta).
\end{equation*}
Rewriting this equality in coordinates, we obtain
\begin{equation*}
\begin{split} 
&\f{1}{2\Omega}(\f{\partial}{\partial u}+b^{A}\f{\partial}{\partial \theta^{A}})|\chih|^2_{\gamma}-\f{1}{\Omega|u|}|\chih|^2_{\gamma}+\f12(\tr\chib+\f{2}{\Omega|u|})|\chih|^2_{\gamma}-2\omb|\chih|^2_{\gamma}\\
=&\chih(\nab\widehat{\otimes}\eta-\f12\tr\chi\chibh+\eta\widehat{\otimes}\eta).
\end{split}
\end{equation*}
Employing the fact $\omb=-\f12\nab_3(\log \Omega)$, we get
\begin{equation*}
\begin{split} 
&\Omega^2 \exp(-\int_1^{u}\Omega \tr\chib du')\f{\partial}{\partial u}\l \exp(-\int_1^{u}\Omega \tr\chib du')\Omega^{-2}|\chih|^2_{\gamma}\r\\
=&-b^{A}\f{\partial}{\partial \theta^{A}}|\chih|^2_{\gamma}-b^{A}\f{\partial \Omega}{\partial \theta^{A}}+2\Omega\chih(\nab\widehat{\otimes}\eta-\Omega \tr\chi \chih+2\Omega \eta\widehat{\otimes}\eta).
\end{split}
\end{equation*}
Applying the identity
 $$\Omega \tr\chib=\Omega(\tr\chib+\f{2}{|u|})-\Omega\f{2}{|u|}=\Omega(\tr\chib+\f{2}{|u|})-(\Omega-1)\f{2}{|u|}-\f{2}{|u|}$$
and Propositions \ref{Omega} and \ref{O0infty trchib}, we deduce
\begin{equation*}
\f{\partial}{\partial u}(\Omega^{-1}u^2|\chih|^2_{\gamma})\leq 2u^2\l-b^{A}\f{\partial}{\partial \theta^{A}}|\chih|^2_{\gamma}-b^{A}\f{\partial \Omega}{\partial \theta^{A}}+2\Omega\chih(\nab\widehat{\otimes}\eta-\Omega \tr\chi \chibh+2\Omega \eta\widehat{\otimes}\eta)\r, 
\end{equation*}
when $b$ is sufficiently large. 

For $b^{A}$, we have equation
\begin{equation*}
\f{\partial b^{A}}{\partial \ub}=-4\Omega^2\zeta^{A},
\end{equation*}
which is from 
\begin{equation*}
[L,\Lb]=\f{\partial b^{A}}{\partial \ub}\f{\partial}{\partial \theta^{A}}.
\end{equation*}
Applying Propositions \ref{Omega} and \ref{O0infty} implies in $D_{u,\ub}$
\begin{equation*}
\|b^{A}\|_{L^{\infty}(\S)}\leq \f{\d}{|u|^2}.
\end{equation*}

For $\nab\chih$, we have

\begin{equation*}
\begin{split}
\|\nab\chih\|_{L^{\infty}(S_{u,\ub})}\leq &\|\nab\chih\|^{\frac 12}_{L^{4}(S_{u,\ub})}\|\nab^2\chih\|^{\frac 12}_{L^{4}(S(u,\ub))}+\frac{1}{|u|^{\frac 12}}\|\nab\chih\|_{L^{4}(S_{u,\ub})}\\
\leq& (\f{\d^{-\f34}}{|u|^{\f32}})^{\f12}(\f{\d^{-\f54}}{|u|^{\f52}})^{\f12}(I^{(0)})+\f{\d^{-\f34}}{|u|^{\f32}}I^{(0)}\\
=&\f{\d^{-1}}{|u|^2}(I^{(0)})+\f{\d^{-\f34}}{|u|^{\f32}}I^{(0)}.
\end{split}
\end{equation*}
Hence, we derive
\begin{equation*}
\begin{split}
\|\f{\partial}{\partial \theta^{A}}\chih\|_{L^{\infty}(S_{u,\ub})}\leq |u|\|\nab\chih\|_{L^{\infty}(S_{u,\ub})}\leq \f{\d^{-1}}{|u|}(I^{(0)})+\f{\d^{-\f34}}{|u|^{\f12}}I^{(0)}.
\end{split}
\end{equation*}

For $\nab\eta$, appealing to Propositions \ref{L4} and \ref{Linfty}, we obtain
\begin{equation*}
\begin{split}
\|\nab\eta\|_{L^{\infty}_{sc}(\S)}\leq & \|\nab\eta\|^{\f12}_{L^{4}_{sc}(\S)}\|\nab^2\eta\|^{\f12}_{L^{4}_{sc}(\S)}+\d^{\f14}\|\nab\eta\|_{L^{4}_{sc}(\S)}\\
\leq & \|\nab\eta\|^{\f23}_{L^{4}_{sc}(\S)}\|\nab^3\eta\|^{\f13}_{L^{2}_{sc}(\S)}+\d^{\f14}\|\nab\eta\|_{L^{4}_{sc}(\S)}.\\
\end{split}
\end{equation*}
From Proposition \ref{O32eta}, we have
$$\|\nab^3\eta\|_{L^{2}_{sc}(\S)}\leq \|\nab^2\rho, \nab^2\sigma\|_{L^{2}_{sc}(\S)}+\d^{\f14}(\M I^{(0)})^5.$$
With this bound, we derive
\begin{equation*}
\begin{split}
&\int_{u_{\infty}}^u u^2\|\chih\|_{L^{\infty}(S_{u',\ub})}\|\nab\eta\|_{L^{\infty}(S_{u',\ub})}du'\\
\leq&\int_{u_{\infty}}^u \f{\d^{-1}}{|u'|^2}\|\chih\|_{L^{\infty}_{sc}(S_{u',\ub})}\|\nab\eta\|_{L^{\infty}_{sc}(S_{u',\ub})}du'\\
\leq&\int_{u_{\infty}}^u \f{\d^{-1}}{|u'|^2}\|\chih\|_{L^{\infty}_{sc}(S_{u',\ub})}\|\nab\eta\|^{\f23}_{L^{4}_{sc}(S_{u',\ub})}\|\nab^3\eta\|^{\f13}_{L^{2}_{sc}(S_{u',\ub})}du'\\
&+\int_{u_{\infty}}^u \f{\d^{-1}}{|u'|^2}\|\chih\|_{L^{\infty}_{sc}(S_{u',\ub})}\d^{\f14}\|\nab\eta\|_{L^{4}_{sc}(S_{u',\ub})}du'\\
\leq&\int_{u_{\infty}}^u \f{\d^{-1}}{|u'|^2}\|\chih\|_{L^{\infty}_{sc}(S_{u',\ub})}\|\nab\eta\|^{\f23}_{L^{4}_{sc}(S_{u',\ub})}\|\nab^2\rho, \nab^2\sigma\|^{\f13}_{L^{2}_{sc}(S_{u',\ub})}du'\\
&+\int_{u_{\infty}}^u \f{\d^{-1}}{|u'|^2}\|\chih\|_{L^{\infty}_{sc}(S_{u',\ub})}\|\nab\eta\|^{\f23}_{L^{4}_{sc}(S_{u',\ub})}\d^{\f{1}{12}}(\M I^{(0)})^{\f53}du'\\
&+\int_{u_{\infty}}^u \f{\d^{-1}}{|u'|^2}\|\chih\|_{L^{\infty}_{sc}(S_{u',\ub})}\d^{\f14}\|\nab\eta\|_{L^{4}_{sc}(S_{u',\ub})}du'\\
\leq&\f{\d^{-1}}{|u|^{\f56}}\sup_{u,\ub}\|\chih\|_{L^{\infty}_{sc}(S_{u,\ub})}\|\nab\eta\|^{\f23}_{L^{4}_{sc}(\S)}\|\nab^2\rho, \nab^2\sigma\|^{\f13}_{L^{2}_{sc}(\Hbu)}\\
&+\f{\d^{-1}\d^{\f{1}{12}}}{|u|}\sup_{u,\ub}\|\chih\|_{L^{\infty}_{sc}(S_{u,\ub})}\|\nab\eta\|^{\f23}_{L^{4}_{sc}(S_{u,\ub})}(\M I^{(0)})^{\f53}\\
&+\f{\d^{-1}\d^{\f14}}{|u|}\sup_{u,\ub}\|\chih\|_{L^{\infty}_{sc}(S_{u,\ub})}\|\nab\eta\|_{L^{4}_{sc}(S_{u,\ub})}\\
\leq&\f{\d^{-1}}{|u|^{\f56}}\sup_{u,\ub}\|\chih\|_{L^{\infty}_{sc}(S_{u,\ub})}\|\nab\eta\|^{\f23}_{L^{4}_{sc}(\S)}\epsilon^{\f13}({\tilde{\M I}}^{(0)})^{\f13}\\
&+\f{\d^{-1}\d^{\f{1}{12}}}{|u|}\sup_{u,\ub}\|\chih\|_{L^{\infty}_{sc}(S_{u,\ub})}\|\nab\eta\|^{\f23}_{L^{4}_{sc}(S_{u,\ub})}(\M I^{(0)})^{\f53}\\
&+\f{\d^{-1}\d^{\f14}}{|u|}\sup_{u,\ub}\|\chih\|_{L^{\infty}_{sc}(S_{u,\ub})}\|\nab\eta\|_{L^{4}_{sc}(S_{u,\ub})},\\
\end{split}
\end{equation*}
where we employ Proposition \ref{epsilon} in the last inequality.

Combining estimates in Proposition \ref{O0infty}, for fixed $(\theta^1,\theta^2)\in S^2$ we conclude
\begin{equation*}
\Omega^{-1}|u|^2|\chih|^2_{\gamma}(u,\ub,\theta^1,\theta^2)\geq|u_{\infty}|^2|\chih|^2_{\gamma}(u_{\infty},\ub,\theta^1,\theta^2)-\f{\d^{-1}\epsilon^{\f13}}{|u|^{\f56}}({\tilde{\M I}}^{(0)})^{\f13}-\f{\d^{-1}\d^{\f{1}{12}}}{|u|}(\M I^{(0)})^5. 
\end{equation*}
This yields
\begin{equation*}
\begin{split}
&\int_0^{\delta}\Omega^{-1}|u|^2|\chih|^2_{\gamma}(u,\ub',\theta^1,\theta^2)d\ub'\\
\geq &\int_0^{\d}|u_{\infty}|^2|\chih|^2_{\gamma}(u_{\infty},\ub',\theta^1,\theta^2)d\ub'-\f{\epsilon^{\f13}}{|u|^{\f56}}({\tilde{\M I}}^{(0)})^{\f13}(\M I^{(0)})^5-\f{\d^{\f{1}{12}}}{|u|}(\M I^{(0)})^5\geq 3c.  
\end{split}
\end{equation*}
Considering another null structure equation
$$\nab_4 \tr\chi+\f12(\tr\chi)^2=-|\chih|^2-2\o\tr\chi,$$
and after employing $\o=-\f12\nab_4(\log \Omega)$, we have
\begin{equation*}
\begin{split}
\nab_4 \tr\chi+\f12(\tr\chi)^2=&-|\chih|^2-2\o\tr\chi\\
=&-|\chih|^2+\nab_4(\log\Omega)\tr\chi\\
=&-|\chih|^2+\f{1}{\Omega}\nab_4\Omega \tr\chi.
\end{split}
\end{equation*}
This is equivalent to 
\begin{equation*}
\begin{split}
\nab_4(\Omega^{-1} \tr\chi)=&-\Omega^{-2}\nab_4\Omega\cdot \tr\chi+\Omega^{-1}\nab_4\tr\chi\\
=&\Omega^{-1}(\nab_4\tr\chi-\Omega^{-1}\nab_4\Omega\cdot \tr\chi)\\
=&\Omega^{-1}(-\f12(\tr\chi)^2-|\chih|^2).
\end{split}
\end{equation*}
Applying the fact $e_4=\Omega^{-1}\f{\partial}{\partial \ub}$, for every $(\theta^1,\theta^2)\in S^2$ we conclude
\begin{equation*}
\begin{split}
&\Omega^{-1}c^2\tr\chi(c,\d, \theta^1, \theta^2)\\
\leq & \Omega^{-1}c^2\tr\chi(c,0, \theta^1, \theta^2)-\int_0^{\d}\Omega^{-1}c^2|\chih|^2(c,\ub',\theta^1,\theta^2)d\ub'\\
\leq & 2c-3c\\
< & 0.
\end{split}
\end{equation*}
Recall that in $D_{u,\ub}$ the following inequality holds
\begin{equation*}
\|\tr\chib+\f{2}{|u|}\|_{L^{\infty}(S_{u,\ub})}\leq \f{\d^{\f12}}{|u|^2}.
\end{equation*}
Therefore $S_{c,\d}$ is a trapped surface.
\end{proof}

\section{IN THE PURSUIT OF CHRISTODOULOU'S RESULTS}
In this section, we claim that given the same initial data as in \cite{Chr:book}, we get consistent bounds for curvature components and null Ricci coefficients as in \cite{Chr:book}.
The author will write an additional note on this. And here we outline the proof.\\

Comparing with the norms in \cite{Chr:book}, the norms we employ in previous sections are weaker norms.
So far we have proved a semi-global existence result in these weaker norms. 
With initial data in \cite{Chr:book}, we promote the results in weaker norms to stronger norms.

\subsection{Initial data}

The prescribed initial data along $H_{u_{\infty}}$ in \cite{Chr:book} are demonstrated below:

$$ \omega=0, \quad (\|\omega\|_{L^{\infty}(S_{u_{\infty},\ub})}\leq \frac{1}{|u_{\infty}|^2})$$ 
$$\|\chih\|_{L^{\infty}(S_{u_{\infty},\ub})}\leq \frac{\delta^{-\frac 12}}{|u_{\infty}|}, \quad \|\tr\chi-\frac{2}{|u_{\infty}|}\|_{L^{\infty}(S_{u_{\infty},\ub})}\leq \frac{1}{|u_{\infty}|^2}, \quad \|\eta,\etb\|_{L^{\infty}(S_{u_{\infty},\ub})}\leq \frac{\delta^{\frac 12}}{|u_{\infty}|^2},$$
$$\|\chibh\|_{L^{\infty}(S_{u_{\infty},\ub})}\leq \frac{\delta^{\frac 12}}{|u_{\infty}|^2},\quad \|\tr\chib+\frac{2}{|u_{\infty}|}\|_{L^{\infty}{S(u_{\infty},\ub)}}\leq \frac{\delta}{|u_{\infty}|^2},\quad \|\omb\|_{L^{\infty}(S_{u_{\infty},\ub})}\leq \frac{\delta}{|u_{\infty}|^3},$$
$$\|K-\frac{1}{|u_{\infty}|^2}\|_{L^{\infty}(S_{u_{\infty},\ub})}\leq \frac{\delta^{\frac 12}}{|u_{\infty}|^3}, \quad \|\alpha\|_{L^{\infty}(S_{u_{\infty},\ub})}\leq \frac{\delta^{-\frac 32}}{|u_{\infty}|}, \quad \|\beta\|_{L^{\infty}(S_{u_{\infty},\ub})}\leq \frac{\delta^{-\frac 12}}{|u_{\infty}|^2},$$
$$\|\rho,\sigma\|_{L^{\infty}(S_{u_{\infty},\ub})}\leq \frac{1}{|u_{\infty}|^3}, \quad \|\beb\|_{L^{\infty}(S_{u_{\infty},\ub})}\leq \frac{\delta}{|u_{\infty}|^4}, \quad \|\ab\|_{L^{\infty}(S_{u_{\infty},\ub})}\leq \frac{\delta^{\frac 32}}{|u_{\infty}|^5}.$$

%
%
%
%

\subsection{Energy Estimates}
The curvature norms $\M R^*$ and $\underline{\M R}^*$ adopted in \cite{Chr:book} are as below:

\begin{equation*}
\begin{split}
\M R^*_0+\underline{\M R}^*_0:=&\d^{\f12}\|\a\|_{L^2_{sc}(\Hu)}+\|\b,\rho, \sigma\|_{L^2_{sc}(\Hu)}+\d^{-\f12}\|\beb\|_{L^2_{sc}(\Hu)}\\
&+\d^{\f12}\|\b\|_{L^2_{sc}(\Hbu)}+\|\rho, \sigma, \beb\|_{L^2_{sc}(\Hbu)}+\d^{-\f12}\|\ab\|_{L^2_{sc}(\Hbu)},
\end{split} 
\end{equation*}

\begin{equation*}
\begin{split}
\M R^*_1+\underline{\M R}^*_1:=&\|\nab\a\|_{L^2_{sc}(\Hu)}+\d^{-\f12}\|\nab\b,\nab\rho, \nab\sigma\|_{L^2_{sc}(\Hu)}+\d^{-1}\|\nab\beb\|_{L^2_{sc}(\Hu)}\\
&+\|\nab\b\|_{L^2_{sc}(\Hbu)}+\d^{-\f12}\|\nab\rho, \nab\sigma, \nab\beb\|_{L^2_{sc}(\Hbu)}+\d^{-1}\|\nab\ab\|_{L^2_{sc}(\Hbu)},
\end{split} 
\end{equation*}

\begin{equation*}
\begin{split}
\M R^*_2+\underline{\M R}^*_2:=&\d^{-\f12}\|\nab^2\a\|_{L^2_{sc}(\Hu)}+\d^{-1}\|\nab^2\b,\nab^2\rho, \nab^2\sigma\|_{L^2_{sc}(\Hu)}\\
&+\d^{-\f32}\|\nab^2\beb\|_{L^2_{sc}(\Hu)}+\d^{-\f12}\|\nab^2\b\|_{L^2_{sc}(\Hbu)}\\
&+\d^{-1}\|\nab^2\rho, \nab^2\sigma, \nab^2\beb\|_{L^2_{sc}(\Hbu)}+\d^{-\f32}\|\nab^2\ab\|_{L^2_{sc}(\Hbu)}.
\end{split} 
\end{equation*}

With the results in previous sections and these new norms, we can prove
\begin{proposition}\label{Chr1}
Under the assumption of Theorem \ref{main.thm}, we have
$${\M R_0^*}^2+{\underline{\M R}_0^*}^2\leq {({\M R^*}^{(0)})}^2+C(\M I^{(0)})^{20},$$
where $C$ is a universal large constant.
\end{proposition}

\begin{proof}
Since for 
$$\Psi^{(s,s')}=\a, \quad \quad \Psi^{(s-\frac 12,s'+\frac 12)}=\b,$$
$$\Psi^{(s,s')}=\b, \quad \quad \Psi^{(s-\frac 12,s'+\frac 12)}=\rho,\sigma,$$
$$\Psi^{(s,s')}=\rho,\sigma, \quad \quad \Psi^{(s-\frac 12,s'+\frac 12)}=\beb,$$
the desired estimates hold trivially. We are left to consider pairs 
$$\Psi^{(s,s')}=\beb, \quad \quad \Psi^{(s-\frac 12,s'+\frac 12)}=\ab.$$
From Proposition \eqref{energy98}, we have
\begin{equation*}
\begin{split}
&\|\beb\|^2_{L^2_{sc}(H_u^{(0,\ub)})}+\|\ab\|^2_{L^2_{sc}(\Hb_{\ub}^{(u_{\infty},u)})}\\
\leq & \|\beb\|^2_{L^2_{sc}(H_{u_{\infty}}^{(0,\ub)})}+\|\ab\|^2_{L^2_{sc}(\Hb_{0}^{(u_{\infty},u)})}\\
+& \d^{\f12}\sup_{u',\ub'}\frac{1}{|u'|}\|\chibh,\eta,\etb, \omb\|_{L^{\infty}_{sc}(S_{u',\ub'})} \|\rho,\sigma,\beb,\ab\|_{L^2_{sc}(\Hb_{\ub}^{(u_{\infty},u)})}\|\beb,\ab\|_{L^2_{sc}(\Hb_{\ub}^{(u_{\infty},u)})}\\
+& \f{\d^{\f12}}{|u|}\sup_{u',\ub'}\frac{1}{|u'|}\|\chibh,\eta,\etb, \omb\|_{L^{\infty}_{sc}(S_{u',\ub'})} \|\rho,\sigma,\beb\|_{L^2_{sc}(\Hu)}\d^{-\f12}\|\beb\|_{L^2_{sc}(\Hu)}.
\end{split}
\end{equation*}
Multiplying $\d^{-1}$ on both sides, we have
\begin{equation*}
\begin{split}
&\d^{-1}\|\beb\|^2_{L^2_{sc}(H_u^{(0,\ub)})}+\d^{-1}\|\ab\|^2_{L^2_{sc}(\Hb_{\ub}^{(u_{\infty},u)})}\\
\leq & \d^{-1}\|\beb\|^2_{L^2_{sc}(H_{u_{\infty}}^{(0,\ub)})}+\d^{-1}\|\ab\|^2_{L^2_{sc}(\Hb_{0}^{(u_{\infty},u)})}\\
+& \sup_{u',\ub'}\frac{1}{|u'|}\|\chibh,\eta,\etb, \omb\|_{L^{\infty}_{sc}(S_{u',\ub'})} \|\rho,\sigma,\beb,\ab\|_{L^2_{sc}(\Hb_{\ub}^{(u_{\infty},u)})}\d^{-\f12}\|\ab\|_{L^2_{sc}(\Hb_{\ub}^{(u_{\infty},u)})}\\
+& \f{1}{|u|}\sup_{u',\ub'}\frac{1}{|u'|}\|\chibh,\eta,\etb, \omb\|_{L^{\infty}_{sc}(S_{u',\ub'})} \|\rho,\sigma,\beb\|_{L^2_{sc}(\Hu)}\d^{-\f12}\|\beb\|_{L^2_{sc}(\Hu)}.
\end{split}
\end{equation*}
Thus, we obtain
$$\M R^*[\beb]^2+\underline{\M R}^*[\ab]^2\leq ({\M R^*}^{(0)})^2+\M O_{0,\infty}\underline{\M R}\,\underline{\M R}^*[\ab]+\M O_{0,\infty}{\M R}{\M R}^*[\beb].$$
Using H\"older's inequality, we arrive at
$${\M R_0^*}^2+{\underline{\M R}_0^*}^2\leq {({\M R^*}_0^{(0)})}^2+\M O^2_{0,\infty}{\underline{\M R}}^2+\M O^2_{0,\infty}{\M R}^2.$$
Applying the conclusion in Proposition \ref{energy final}, we have proved desired estimate.
\end{proof}

For $\M R_1^*$ and $\underline{\M R}_1^*$, we get
\begin{proposition}
Under the assumption of Theorem \ref{main.thm}, the following inequality holds
$${\M R_1^*}^2+{\underline{\M R}_1^*}^2\leq {({\M R^*}^{(0)})}^2+C(\M I^{(0)})^{20},$$
where $C$ is a universal large constant.
\end{proposition}

\begin{proof}

For $\a$, we have ${\M R^*}^2_1[\a]+{\underline{\M R}^*}^2_1[\b]={\M R}^2_1[\a]+{\underline{\M R}}^2_1[\b]$.

For $\b, \rho, \sigma$, 
using estimates in Proposition \ref{energy912} and similar methods in Proposition \ref{Chr1}, we derive
\begin{equation*}
\begin{split}
{\M R_1^*}^2[\b,\rho,\sigma]+{\underline{\M R}_1^*}^2[\rho,\sigma,\beb]\leq &{({\M R^*}_0^{(0)})}^2+C(\M I^{(0)}+\M R+\underline{\M R})^5{\underline{\M R}}\,{\underline{\M R}_1^*}[\rho,\sigma,\beb]\\
&+C(\M I^{(0)}+\M R+\underline{\M R})^5{\M R}{\M R}_1^*[\b,\rho,\sigma].
\end{split}
\end{equation*}
With H\"older's inequality, we obtain desired estimates.

Similarly, for $\beb$ we deduce
\begin{equation*}
\begin{split}
{\M R_1^*}^2[\beb]+{\underline{\M R}_1^*}^2[\ab]\leq &{({\M R^*}_0^{(0)})}^2+C(\M I^{(0)}+\M R+\underline{\M R})^5{\underline{\M R}_1^*}[\rho,\sigma,\beb]{\underline{\M R}_1^*}[\ab]\\
&+C(\M I^{(0)}+\M R+\underline{\M R})^5{\M R}_1^*[\rho,\sigma]{\M R}_1^*[\beb].
\end{split}
\end{equation*}
Since, we have already estimated ${\M R}_1^*[\rho,\sigma]$ and ${\underline{\M R}_1^*}[\rho,\sigma,\beb]$. After employing
H\"older's inequality and Proposition \ref{energy final}, we finish the proof.
\end{proof}

In the same fashion, we conclude

\begin{proposition}
Under the assumption of Theorem \ref{main.thm}, we get
$${\M R_2^*}^2+{\underline{\M R}_2^*}^2\leq {({\M R^*}^{(0)})}^2+C(\M I^{(0)})^{20},$$
where $C$ is a universal large constant.
\end{proposition}

\subsection{Coefficients Estimates}

We define $\M R^{*}_{0,\i}$ as 
\begin{equation}
\begin{split}
\M R^{*}_{0,\i}:=&\delta^{\frac 32}|u|\|\alpha\|_{L^{\infty}(\S)}+\delta^{\frac 12}|u|^2\|\beta\|_{L^{\infty}(\S)}+|u|^3\|\rho,\sigma\|_{L^{\infty}(\S)}\\
&+\delta^{-1}|u|^4\|\beb\|_{L^{\infty}(\S)}+\delta^{-\frac 32}|u|^{5}\|\ab\|_{L^{\infty}(\S)}.
\end{split}
\end{equation}

With the energy estimates obtained above, it can be verified that
$$\M R^{*}_{0,\i}\leq C(\M I^{(0)})^{20},$$
where $C$ is a universal large constant. With the aid of this bound, we improve several $L^{\infty}(\S)$ estimates for Ricci coefficients.

For $\eta$, using $\nab_4\eta=-\chi\cdot(\eta-\etb)-\beta$, we have $\|\eta\|_{L^{\infty}(\S)}\leq \frac{\delta^{\frac 12}}{|u|^2}C(\M I^{(0)})^{20}$.\\  
For $\etb$, with $\nab_3\etb=-\chib\cdot(\etb-\eta)-\beb$, we obtain $\|\etb\|_{L^{\infty}(\S)}\leq \frac{\delta^{\frac 12}}{|u|^2}C(\M I^{(0)})^{20}$.\\  
For $\omega$, employing $\nab_3 \omega=2\omega\omb+(\eta,\etb)(\eta,\etb)+\frac 12 \rho$, we derive  $\|\omega\|_{L^{\infty}(\S)}\leq \frac{1}{|u|^2}C(\M I^{(0)})^{20}$.\\
For $\omb$, adopting $\nab_4 \omb=2\omega\omb+(\eta,\etb)(\eta,\etb)+\frac 12 \rho$, we conclude  $\|\omb\|_{L^{\infty}(S)}\leq \frac{\delta}{|u|^3}C(\M I^{(0)})^{20}$.\\

In summary, starting with initial data in \cite{Chr:book} by employing an alternative strategy, which is outlined in this section,
we arrive at
\begin{theorem}
Given Minkowskian initial data on $\Hb_{\ub}^{(u_{\infty},u)}$ and Christodoulou's initial data on $H_{u_{\infty}}^{(0,\ub)}$ as in \cite{Chr:book}, we can derive estimates that are consistent with \cite{Chr:book}.  
\end{theorem}

\appendix

\section{Norms in standard $L^p$ form}
For convenience, in this section we list the norms in the standard $L^p$ form:

\begin{equation}\label{O_{0,infty}}
\begin{split}
\mathcal{O}_{0,\infty}(u,\underline{u})=&\delta^{\frac{1}{2}}(|u|\|\hat{\chi}\|_{L^{\infty}(\S)}+|u|\|\omega\|_{L^{\infty}(\S)})+\d^{\f12}|u|\|\tr\chi\|_{L^{\infty}(\S)}\\
&+|u|^2\|\eta,\underline{\eta}\|_{L^{\infty}(\S)}+\delta^{-\frac{1}{2}}|u|^2\|\hat{\underline{\chi}}\|_{L^{\infty}(\S)}\\
&+\delta^{-\f12}|u|^2\|\tr\underline{\chi}+\frac{2}{|u|}\|_{L^{\infty}(\S)}+\delta^{-\frac{1}{2}}|u|^3\|\underline{\omega}\|_{L^{\infty}(\S)},
\end{split}
\end{equation}

\begin{equation}\label{O_{0,4}}
\begin{split}
\mathcal
{O}_{0,4}(u,\underline{u})=&\delta^{\frac{1}{2}}|u|^{\frac{1}{2}}\|\hat{\chi}\|_{L^{4}(\S)}+\delta^{\frac{1}{4}}|u|^{\frac{1}{2}}\|\omega\|_{L^{4}(\S)}+\d^{\f14}|u|^{\frac{1}{2}}\|\tr\chi\|_{L^{4}(\S)}\\
&+\delta^{-\frac{1}{4}}|u|^{\frac{3}{2}}\|\eta,\underline{\eta}\|_{L^{4}(\S)}+\delta^{-\frac{1}{2}}|u|^{\frac{3}{2}}\|\hat{\underline{\chi}}\|_{L^{4}(\S)}\\
&+|u|^{\f12}\|\tr\underline{\chi}\|_{L^{4}(\S)}+\delta^{-\frac{3}{4}}|u|^{\frac{5}{2}}\|\underline{\omega}\|_{L^{4}(\S)},
\end{split}
\end{equation}

\begin{equation}\label{O_{0,2}}
\begin{split}
\mathcal
{O}_{0,2}(u,\underline{u})=&\delta^{\frac{1}{2}}\|\hat{\chi}\|_{L^{2}(\S)}+\|\omega\|_{L^{2}(\S)}+\|\tr\chi\|_{L^{2}(\S)}+\delta^{-\frac{1}{2}}|u|\|\eta,\underline{\eta}\|_{L^{2}(\S)}\\
&+\delta^{-\frac{1}{2}}|u|\|\hat{\underline{\chi}}\|_{L^{2}(\S)}+\|\tr\underline{\chi}\|_{L^{2}(\S)}+\delta^{-1}|u|^{2}\|\underline{\omega}\|_{L^{2}(\S)},
\end{split}
\end{equation}

\begin{equation}\label{O_{1,4}}
\begin{split}
\mathcal
{O}_{1,4}(u,\underline{u})=&\delta^{\frac{3}{4}}|u|^{\frac{3}{2}}\|\nabla\hat{\chi}\|_{L^{4}(\S)}+\delta^{\frac{3}{4}}|u|^{\frac{3}{2}}\|\nabla{\tr\chi}\|_{L^{4}(\S)}+\delta^{\frac{3}{4}}|u|^{\frac{3}{2}}\|\nabla\omega\|_{L^{4}(\S)}\\
&+\delta^{\frac{1}{4}}|u|^{\frac{5}{2}}\|\nabla(\eta,\underline{\eta})\|_{L^{4}(\S)}+\delta^{-\frac{1}{4}}|u|^{\frac{5}{2}}\|\nabla\hat{\underline{\chi}}\|_{L^{4}(\S)}\\
&+\delta^{-\frac 14}|u|^{\frac{7}{2}}\|\nabla{\tr\underline{\chi}}\|_{L^{4}(\S)}+\delta^{-\frac 14}|u|^{\frac{7}{2}}\|\nabla\underline{\omega}\|_{L^{4}(\S)},
\end{split}
\end{equation}

\begin{equation}\label{O_{1,2}}
\begin{split}
\mathcal
{O}_{1,2}(u,\underline{u})=&\delta^{\frac{1}{2}}|u|\|\nabla\hat{\chi}\|_{L^{2}(\S)}+\delta^{\frac{1}{2}}|u|\|\nabla{\tr\chi}\|_{L^{2}(\S)}+\delta^{\frac{1}{2}}|u|\|\nabla\omega\|_{L^{2}(\S)}\\
&+|u|^{2}\|\nabla(\eta,\underline{\eta})\|_{L^{2}(\S)}+\delta^{-\frac{1}{2}}|u|^{2}\|\nabla\hat{\underline{\chi}}\|_{L^{2}(\S)}\\
&+\delta^{-\frac{1}{2}}|u|^{3}\|\nabla{\tr\underline{\chi}}\|_{L^{2}(\S)}+\delta^{-\frac{1}{2}}|u|^{3}\|\nabla\underline{\omega}\|_{L^{2}(\S)},
\end{split}
\end{equation}

\begin{equation}\label{R_0(H)}
\begin{split}
\mathcal
{R}_0(u,\underline{u})&=\delta\|\alpha\|_{L^2(H_{u}^{(0,\underline{u})})}+|u|\|\beta\|_{L^2(H_{u}^{(0,\underline{u})})}+\delta^{-\frac{1}{2}}|u|^2\|\rho,\sigma\|_{L^2(H_{u}^{(0,\underline{u})})}\\
&+\delta^{-1}|u|^3\|\underline{\beta}\|_{L^2(H_{u}^{(0,\underline{u})})},
\end{split}
\end{equation}

\begin{equation}\label{R_0(Hb)}
\begin{split}
\mathcal
{\underline{R}}_0(u,\underline{u})=&\delta\|\beta\|_{L^2(\underline{H}_{\underline{u}}^{(u_{\infty},u)})}+\||u'|(\rho,\sigma)\|_{L^2(\underline{H}_{\underline{u}}^{(u_{\infty},u)})}+\delta^{-\frac{1}{2}}\||u'|^2\underline{\beta}\|_{L^2(\underline{H}_{\underline{u}}^{(u_{\infty},u)})}\\
&+\delta^{-1}|\||u'|^3\underline{\alpha}\|_{L^2(\underline{H}_{\underline{u}}^{(u_{\infty},u)})},
\end{split}
\end{equation}

\begin{equation}\label{R_1(H)}
\begin{split}
\mathcal
{R}_1(u,\underline{u})=&\delta|u|\|\nabla\alpha\|_{L^2(H_{u}^{(0,\underline{u})})}+\delta^{\frac{1}{2}}|u|^2\|\nabla\beta\|_{L^2(H_{u}^{(0,\underline{u})})}
+|u|^3\|\nabla(\rho,\sigma)\|_{L^2(H_{u}^{(0,\underline{u})})}\\
&+\delta^{-\frac{1}{2}}|u|^4\|\nabla\underline{\beta}\|_{L^2(H_{u}^{(0,\underline{u})})},
\end{split}
\end{equation}

\begin{equation}\label{R_1(Hb)}
\begin{split}
\mathcal
{\underline{R}}_1(u,\underline{u})=&\delta\||u'|\nabla\beta\|_{L^2(\underline{H}_{\underline{u}}^{(u_{\infty},u)})}+\delta^{\frac{1}{2}}\||u'|^2\nabla(\rho,\sigma)\|_{L^2(\underline{H}_{\underline{u}}^{(u_{\infty},u)})}
+\||u'|^3\nabla\underline{\beta}\|_{L^2(\underline{H}_{\underline{u}}^{(u_{\infty},u)})}\\
&+\delta^{-\frac{1}{2}}||u'|^4{\nabla\ab}\|_{L^2(\underline{H}_{\underline{u}}^{(u_{\infty},u)})}.
\end{split}
\end{equation}

\section{Transport Equations for Elliptic Estimates}
In this section, we write down the transport equations for $\T\in\{\nab\tr\chi, \nab\tr\chib, \mu, \mub, \kappa, \underline{\kappa}\}.$

For $\nab\tr\chi$, we have
\begin{equation*}
 \begin{split}
  \nab_4\nab\tr\chi=(\nab\chi,\nab\o)(\chi,\o)+(\eta,\etb)(\o,\chi)(\o,\chi),
 \end{split}
\end{equation*}

\begin{equation*}
 \begin{split}
  &\nab_4\nab^2\tr\chi\\
=&(\nab^2\chi,\nab^2\o)(\chi,\o)+(\nab\chi,\nab\o)(\nab\chi,\nab\o)+(\nab\eta,\nab\etb)(\o,\chi)(\o,\chi)\\
&+(\eta,\etb)(\nab\o,\nab\chi)(\o,\chi)+\b\nab\tr\chi+\ee\ee(\o,\chi)(\o,\chi),
 \end{split}
\end{equation*}

\begin{equation*}
 \begin{split}
  &\nab_4\nab^3\tr\chi\\
=&(\nab^3\chi,\nab^3\o)(\chi,\o)+(\nab^2\chi,\nab^2\o)(\nab\chi,\nab\o)+(\nab^2\eta,\nab^2\etb)(\o,\chi)(\o,\chi)\\
&+(\nab\eta,\nab\etb)(\nab\o,\nab\chi)(\o,\chi)+(\eta,\etb)(\nab^2\o,\nab^2\chi)(\o,\chi)+\nab\b\nab\tr\chi+\b\nab^2\chi\\
&+(\nab\eta,\nab\etb)\ee(\o,\chi)(\o,\chi)+\ee\ee(\nab\o,\nab\chi)(\o,\chi)+\b\ee\nab\tr\chi\\
&+\ee(\nab\chi,\nab\o)(\nab\chi,\nab\o)+\ee\ee\ee(\o,\chi)(\o,\chi).
 \end{split}
\end{equation*}

%
%
%
%

For $\mu$, we obtain
\begin{equation*}
 \begin{split}
\nab_4\mu=\chi\mu+\chi(\nab\eta,\nab\etb)+\ee\nab\chi+\chibh\a+\ee\b+\chi\rho+\chi\ee\ee,
 \end{split}
\end{equation*}

\begin{equation*}
 \begin{split}
&\nab_4\nab\mu\\
=&\chi\nab\mu+\nab\chi\mu+\nab\chi(\nab\eta,\nab\etb)+\chi(\nab^2\eta,\nab^2\etb)+\ee\nab^2\chi+\chibh\nab\a\\
&+\nab\chibh\a+(\nab\eta,\nab\etb)\b+\ee\nab\b+\nab\chi\rho+\chi\nab\rho+\nab\chi\ee\ee\\
&+\chi(\nab\eta,\nab\etb)\ee+\ee\mu\chi+\ee\chibh\a+\ee\ee\b\\
&+\ee\chi\rho+\chi\ee\ee\ee,
 \end{split}
\end{equation*}

\begin{equation*}
 \begin{split}
&\nab_4\nab^2\mu\\
=&\chi\nab^2\mu+\nab\chi\nab\mu+\nab^2\chi\mu+\nab^2\chi(\nab\eta,\nab\etb)+\nab\chi(\nab^2\eta,\nab^2\etb)\\
&+\chi(\nab^3\eta,\nab^3\etb)+\nab\chibh\nab\a+\chibh\nab^2\a+\nab^2\chibh\a+(\nab^2\eta,\nab^2\etb)\b+(\nab\eta,\nab\etb)\nab\b\\
&+\ee\nab^2\b+\nab^2\chi\rho+\nab\chi\nab\rho+\chi\nab^2\rho+\nab^2\chi\ee\ee+\nab\chi(\nab\eta,\nab\etb)\ee\\
&+\chi(\nab^2\eta,\nab^2\etb)\ee+(\nab\eta,\nab\etb)\mu\chi+\ee\nab\mu\chi+\ee\mu\nab\chi+\ee\ee\chibh\a\\
&+(\nab\eta,\nab\etb)\chibh\a+\ee\nab\chibh\a+\ee\chibh\nab\a+(\nab\eta,\nab\etb)\ee\b+\ee\ee\nab\b\\
&+(\nab\eta,\nab\etb)\chi\rho+\ee\nab\chi\rho+\ee\chi\nab\rho+\nab\chi\ee\ee\ee+\chi(\nab\eta,\nab\etb)\ee\ee\\
&+\eta\nab\mu\chi+\b\nab\mu+\ee\ee\ee\b+\ee\ee\chi\rho+\chi\ee\ee\ee\ee.
 \end{split}
\end{equation*}

For $\mub$, we demonstrate
\begin{equation*}
 \begin{split}
\nab_3\mub+\tr\chib\,\mub=\tr\chib\nab\eta+\chibh(\nab\eta,\nab\etb)+\ee\nab\chib+\chih\ab+\ee\beb+\chib\rho+\chib\ee\ee,
 \end{split}
\end{equation*}

\begin{equation*}
 \begin{split}
&\nab_3\nab\mub+\f32\tr\chib\nab\mub\\
=&\chibh\nab\mub+\tr\chib\nab^2\eta+\nab\chib(\nab\eta,\nab\etb)+\ee\nab^2\chib\\
&+\nab\chih\ab+\chih\nab\ab+(\nab\eta,\nab\etb)\beb+\ee\nab\beb+\nab\chib\rho+\chib\nab\rho+\nab\chib\ee\ee\\
&+\chib(\nab\eta,\nab\etb)\ee+\ee\mub\,\chib+\ee\chih\ab+\nab\chibh(\nab\eta,\nab\etb)+\chibh(\nab^2\eta,\nab^2\etb)\\
&+\ee\ee\beb+\chib\ee\rho+\chib\ee\ee\ee+\ee\tr\chib\nab\eta+\ee\ee\nab\chib,
 \end{split}
\end{equation*}

\begin{equation*}
 \begin{split}
&\nab_3\nab^2\mub+2\tr\chib\nab^2\mub\\
=&\chibh\nab^2\mub+\tr\chib\nab^3\eta+\nab\chib(\nab^2\eta,\nab^2\etb)+\nab^2\chib(\nab\eta,\nab\etb)\\
&+\ee\nab^3\chib+\nab\chih\nab\ab+\nab^2\chih\ab+\chih\nab^2\ab+(\nab^2\eta,\nab^2\etb)\beb+(\nab\eta,\nab\etb)\nab\beb+\ee\nab^2\beb\\
&+\nab^2\chibh\rho+\nab\chib\nab\rho+\chib\nab^2\rho+\nab^2\chibh\ee\ee+\nab\chib(\nab\eta,\nab\etb)\ee+\ee\tr\chib\nab^2\eta\\
&+\chib(\nab^2\eta,\nab^2\etb)\ee+(\nab\eta,\nab\etb)\mub\,\chib+\ee\nab\mub\,\chib+\ee\mub\nab\chib+\ee\ee\tr\chib\nab\eta\\
&+(\nab\eta,\nab\etb)\chih\ab+\ee\nab\chih\ab+\ee\chih\nab\ab+(\nab\eta,\nab\etb)\ee\beb+\ee\ee\nab\beb+\nab\chib\ee\rho\\
&+\chib(\nab\eta,\nab\etb)\rho+\chib\ee\nab\rho+\nab\chib\ee\ee\ee+\chib(\nab\eta,\nab\etb)\ee\ee+\beb\nab\mub\\
&+\ee\ee\mub\,\chib+\ee\ee\chih\ab+\ee\ee\ee\beb\\
&+\chib\ee\ee\rho+\chib\ee\ee\ee\ee.
\end{split}
\end{equation*}

For $\underline{\kappa}$, we derive
\begin{equation*}
 \begin{split}
\nab_4\underline{\kappa}=&\chi\underline{\kappa}+\o\beb+\chibh\b+\ee(\rho,\sigma)+\chi(\nab\omb,\nab\omb^{\dagger})+\nab\o\omb+\o\nab\omb+(\nab\eta,\nab\etb)\ee\\
&+\ee\o\omb+\ee\ee\ee,
\end{split}
\end{equation*}

\begin{equation*}
 \begin{split}
&\nab_4\nab\underline{\kappa}\\
=&\chi\nab\underline{\kappa}+\nab\chi\underline{\kappa}+\nab\o\beb+\o\nab\beb+\nab\chibh\b+\chibh\nab\b\\
&+(\nab\eta,\nab\etb)(\rho,\sigma)+\ee(\nab\rho,\nab\sigma)+\nab\chih(\nab\o,\nab\o^{\dagger})+\chih(\nab^2\omb,\nab^2\omb^{\dagger})\\
&+\nab^2\o\omb+\nab\o\nab\omb+\o\nab^2\omb+(\nab^2\eta,\nab^2\etb)\ee+(\nab\eta,\nab\etb)(\nab\eta,\nab\etb)\\
&+(\nab\eta,\nab\etb)\o\omb+\ee\nab\o\omb+\ee\o\nab\omb+(\nab\eta,\nab\etb)\ee\ee+\ee\underline{\kappa}\chi+\b\underline{\kappa}\\
&+\ee\o\beb+\ee\chibh\b+\ee\ee(\rho,\sigma)+\ee\chih(\nab\omb,\nab\omb^{\dagger})+\ee\ee\o\omb\\
&+\ee\ee\ee\ee,
\end{split}
\end{equation*}

\begin{equation*}
 \begin{split}
&\nab_4\nab^2\underline{\kappa}\\
=&\chi\nab^2\underline{\kappa}+\nab\chi\nab\underline{\kappa}+\nab^2\chi\underline{\kappa}+\nab^2\o\beb\\
&+\nab\o\nab\beb+\o\nab^2\beb+\nab^2\chibh\b+\nab\chibh\nab\b+\chibh\nab^2\b+(\nab^2\eta,\nab^2\etb)(\rho,\sigma)\\
&+(\nab\eta,\nab\etb)(\nab\rho,\nab\sigma)+(\eta,\etb)(\nab^2\rho,\nab^2\sigma)+\nab^2\chih(\nab\o,\nab\o^{\dagger})\\
&+\nab\chih(\nab^2\omb,\nab^2\omb^{\dagger})+\chih(\nab^3\omb,\nab^3\omb^{\dagger})+\nab^3\o\omb+\nab^2\o\nab\omb+\nab\o\nab^2\omb\\
&+\o\nab^3\omb+(\nab^3\eta,\nab^3\etb)(\eta,\etb)+(\nab^2\eta,\nab^2\etb)(\nab\eta,\nab\etb)+(\nab^2\eta,\nab^2\etb)\o\omb\\
&+(\nab\eta,\nab\etb)\nab\o\omb+(\nab\eta,\nab\etb)\o\nab\omb+\ee\nab^2\o\omb+\ee\o\nab^2\omb+(\nab^2\eta,\nab^2\eta)\ee\ee\\
&+(\nab\eta,\nab\eta)(\nab\eta,\nab\etb)\ee+(\nab\eta,\nab\etb)\underline{\kappa}\chi+(\eta,\etb)\nab\underline{\kappa}\chi+\ee\nab\o\nab\omb\\
&+(\eta,\etb)\underline{\kappa}\nab\chi+\nab\b\underline{\kappa}+\b\nab\underline{\kappa}+(\nab\eta,\nab\etb)\o\beb\\
&+\ee\nab\o\beb+\ee\o\nab\beb+(\nab\eta,\nab\etb)\chibh\b+\ee\nab\chibh\b+\ee\chibh\nab\b+(\nab\eta,\nab\etb)\ee(\rho,\sigma)\\
&+\ee\ee(\nab\rho,\nab\sigma)+(\nab\eta,\nab\etb)\chih(\nab\omb,\nab\omb^{\dagger})+\ee\nab\chih(\nab\omb,\nab\omb^{\dagger})\\
&+\ee\chih(\nab^2\omb,\nab^2\omb^{\dagger})+(\nab\eta,\nab\etb)\ee\o\omb+\ee\ee\nab\o\omb+\ee\ee\o\nab\omb\\
&+(\nab\eta,\nab\etb)\ee\ee\ee+\ee\nab\underline{\kappa}\chi+\b\nab\underline{\kappa}+\ee\ee\underline{\kappa}\chi\\
&+\ee\b\underline{\kappa}+\ee\ee\o\beb+\ee\chibh\b+\ee\ee\ee(\rho,\sigma)+\ee\ee\chih(\nab\omb,\nab\omb^{\dagger})\\
&+\ee\ee\ee\o\omb+\ee\ee\ee\ee\ee.
\end{split}
\end{equation*}

For $\kappa$, we have
\begin{equation*}
\begin{split}
\nab_3\kappa+\f12\tr\chib\kappa=&\omb\b+\chih\beb+\ee(\rho,\sigma)+\chibh(\nab\o,\nab\o^{\dagger})+\nab\o\omb+\o\nab\omb\\
&+(\nab\eta,\nab\etb)\ee+\ee\o\omb+\ee\ee\ee,
\end{split}
\end{equation*}

\begin{equation*}
\begin{split}
&\nab_3\nab\kappa+\tr\chib\nab\kappa\\
=&\chibh\nab\kappa+\nab\tr\chib\kappa+\nab\omb\b+\omb\nab\b+\nab\chih\beb+\chih\nab\beb+\ee\ee(\rho,\sigma)\\
&+(\nab\eta,\nab\etb)(\rho,\sigma)+\ee(\nab\rho,\nab\sigma)+\nab\chibh(\nab\o,\nab\o^{\dagger})+\chibh(\nab^2\o,\nab^2\o^{\dagger})\\
&+\nab^2\o\omb+\nab\o\nab\omb+\o\nab^2\omb+(\nab^2\eta,\nab^2\etb)\ee+(\nab\eta,\nab\etb)(\nab\eta,\nab\etb)\\
&+(\nab\eta,\nab\etb)\o\omb+\ee\nab\o\omb+\ee\o\nab\omb+(\nab\eta,\nab\etb)\ee\ee+\ee\kappa\chib+\beb\kappa\\
&+\ee\omb\b+\ee\chih\beb+\ee\chibh(\nab\o,\nab\o^{\dagger})+\ee\ee\o\omb+\ee\ee\ee\ee,
\end{split}
\end{equation*}

\begin{equation*}
\begin{split}
&\nab_3\nab^2\kappa+\f32\tr\chib\nab^2\kappa\\
=&\chibh\nab^2\kappa+\nab\chibh\nab\kappa+\nab^2\tr\chib\kappa+\nab\tr\chib\nab\kappa\\
&+\nab^2\omb\b+\nab\omb\nab\b+\omb\nab^2\b+\nab^2\chih\beb+\nab\chih\nab\beb+\chih\nab^2\beb+(\nab^2\eta,\nab^2\etb)(\rho,\sigma)\\
&+(\nab\eta,\nab\etb)(\nab\rho,\nab\sigma)+\ee(\nab^2\rho,\nab^2\sigma)+\nab^2\chibh(\nab\o,\nab\o^{\dagger})\\
&+\nab\chibh(\nab^2\o,\nab^2\o^{\dagger})+\chibh(\nab^3\o,\nab^3\o^{\dagger})+\nab^3\o\omb+\nab^2\o\nab\omb\\
&+\nab\o\nab^2\omb+\o\nab^3\omb+(\nab^3\eta,\nab^3\etb)(\eta,\etb)+(\nab^2\eta,\nab^2\etb)(\nab\eta,\nab\etb)\\
&+(\nab^2\eta,\nab^2\etb)\o\omb+(\nab\eta,\nab\etb)\nab\o\omb+(\nab\eta,\nab\etb)\o\nab\omb+(\eta,\etb)\nab^2\o\omb+(\eta,\etb)\o\nab^2\omb\\
&+(\eta,\etb)\nab\o\nab\omb+(\nab^2\eta,\nab^2\etb)\ee\ee+(\nab\eta,\nab\etb)(\nab\eta,\nab\etb)\ee+(\nab\eta,\nab\etb)\kappa\chib\\
&+\ee\nab\kappa\chib+\ee\kappa\nab\chib+\nab\beb\kappa+\beb\nab\kappa+(\nab\eta,\nab\etb)\omb\b+(\eta,\etb)\nab\omb\b\\
&+(\eta,\etb)\omb\nab\b+(\nab\eta,\nab\etb)\chih\beb+(\eta,\etb)\nab\chih\beb+(\eta,\etb)\chih\nab\beb\\
&+(\nab\eta,\nab\etb)\chibh(\nab\o,\nab\o^{\dagger})+(\eta,\etb)\nab\chibh(\nab\o,\nab\o^{\dagger})+(\eta,\etb)\chibh(\nab^2\o,\nab^2\o^{\dagger})\\
&+(\nab\eta,\nab\etb)\ee\o\omb+\ee\ee\nab\o\omb+\ee\ee\o\nab\omb\\
&+(\nab\eta,\nab\etb)\ee\ee\ee+\ee(\nab\eta,\nab\etb)(\rho,\sigma)\\
&+\ee\ee(\nab\rho,\nab\sigma)+\ee\ee\kappa\chib+\ee\beb\kappa+\ee\ee\omb\b\\
&+\ee\ee\chih\beb+\ee\ee\chibh(\nab\o,\nab\o^{\dagger})+\ee\ee\ee\o\omb\\
&+\ee\ee\ee\ee\ee.
\end{split}
\end{equation*}

For $\oo$, we obtain

\begin{equation}\label{Omega trchi 3}
\begin{split}
\nab_3(\Omega\tr\chib-\f{2}{u})+\tr\chib(\Omega\tr\chib-\f{2}{u})=-4\Omega\omb\tr\chib-\Omega|\chibh|^2+\f{\Omega^{-1}}{2}(\Omega\tr\chib-\f{2}{u})^2,\\
\end{split}
\end{equation}

\begin{equation*}
\begin{split}
&\nab_3\nab(\Omega \tr\chib-\f{2}{u})+\f32\tr\chib\nab(\Omega \tr\chib-\f{2}{u})\\
=&\chibh\nab(\Omega \tr\chib-\f{2}{u})+\nab\tr\chib(\Omega \tr\chib-\f{2}{u})+\nab\Omega\omb\tr\chib+\Omega\nab\omb\tr\chib+\Omega\omb\nab\tr\chib\\
&+\nab\Omega\chibh\,\chibh+\Omega\chibh\nab\chibh+\nab\Omega\Omega^{-2}(\Omega \tr\chib-\f{2}{u})^2+\Omega^{-1}\nab(\Omega \tr\chib-\f{2}{u})(\Omega \tr\chib-\f{2}{u})\\
&+\ee\tr\chib(\Omega \tr\chib-\f{2}{u})+\ee\Omega\omb\tr\chib+\ee\Omega\chibh\,\chibh+\ee\Omega^{-1}(\Omega \tr\chib-\f{2}{u})^2,
\end{split}
\end{equation*}

\begin{equation*}
\begin{split}
&\nab_3\nab^2(\oo)+2\tr\chib\nab^2(\oo)\\
=&\chibh\nab^2(\oo)+\nab\chibh\nab(\oo)+\nab^2\tr\chib(\oo)+\nab\tr\chib\nab(\oo)\\
&+\nab^2\O\omb\tr\chib+\nab\O\nab\o\tr\chib+\nab\O\omb\nab\tr\chib+\O\nab^2\omb\tr\chib+\O\nab\omb\nab\tr\chib\\
&+\O\omb\nab^2\tr\chib+\nab^2\O\chibh\,\chibh+\nab\O\nab\chib\,\chibh+\O\chibh\nab^2\chibh+\nab^2\O\O^{-2}(\oo)^2\\
&+\nab\O\nab\O\O^{-3}(\oo)^2+\nab\O\O^{-2}\nab(\oo)(\oo)\\
&+\O^{-1}\nab^2(\oo)(\oo)+(\nab\eta,\nab\etb)\tr\chib(\oo)\\
&+\ee\nab\tr\chib(\oo)+\O^{-1}\nab(\oo)\nab(\oo)\\
&+\ee\tr\chib\nab(\oo)+(\nab\eta,\nab\etb)\O\omb\tr\chib+\ee\nab\O\omb\tr\chib\\
&+\ee\O\nab\omb\tr\chib+\ee\O\omb\nab\tr\chib+(\nab\eta,\nab\etb)\O\chibh\,\chibh+\ee\nab\O\chibh\,\chibh\\
&+\ee\O\chibh\nab\chibh+(\nab\eta,\nab\etb)\O^{-1}(\oo)^2+\ee\nab\O\O^{-2}(\oo)^2\\
&+\ee\O^{-1}\nab(\oo)(\oo)+\ee\chib\nab(\oo)+\beb\nab(\oo)\\
&+\ee\chibh\nab(\oo)+\ee\nab\tr\chib(\oo)+\ee\nab\O\omb\tr\chib\\
&+\ee\O\nab\omb\tr\chib+\ee\O\omb\nab\tr\chib+\ee\nab\O\chibh\,\chibh+\ee\O\chibh\nab\chibh\\
&+\ee\nab\O\O^{-2}(\oo)^2+\ee\ee\tr\chib(\oo)+\ee\ee\O\omb\tr\chib\\
&+\ee\ee\O\chibh\,\chibh+\ee\ee\O^{-1}(\oo)^2,
\end{split}
\end{equation*}

\begin{equation*}
\begin{split}
&\nab_3\nab^3(\oo)+\f52\tr\chib\nab^3(\oo)\\
=&\chibh\nab^3(\oo)+\nab^2\chibh\nab(\oo)+\nab\chibh\nab^2(\oo)\\
&+\nab^3\tr\chib(\oo)+\nab^2\tr\chib\nab(\oo)\\
&+\nab\tr\chib\nab^2(\oo)+\nab^3\O\omb\tr\chib+\nab^2\O\nab\omb\tr\chib+\nab^2\O\omb\nab\tr\chib+\nab\O\nab^2\omb\tr\chib\\
&+\nab\O\nab\omb\nab\tr\chib+\nab\O\omb\nab^2\tr\chib+\O\nab^3\omb\tr\chib+\O\nab^2\omb\nab\tr\chib+\O\nab\omb\nab^2\tr\chib\\
&+\O\omb\nab^3\tr\chib+\nab^3\O\chibh\,\chibh+\nab^2\O\nab\chibh\,\chibh+\nab\O\nab^2\chibh\,\chibh+\nab\O\nab\chibh\nab\chibh\\
&+\O\nab\chibh\nab^2\chibh+\O\chibh\nab^3\chibh+\nab^3\O\O^{-2}(\oo)^2\\
&+\nab^2\O\nab\O\O^{-3}(\oo)^2+\nab^2\O\O^{-2}\nab(\oo)(\oo)\\
&+\nab^2\O\nab\O\O^{-3}(\oo)^2+\nab\O\nab\O\nab\O\O^{-4}(\oo)^2\\
&+\nab\O\nab\O\O^{-3}\nab(\oo)(\oo)+\nab^2\O\O^{-2}\nab(\oo)(\oo)\\
&+\nab\O\O^{-2}\nab^2(\oo)(\oo)+\ee\nab^2\O\O^{-2}(\oo)^2\\
&+\nab\O\O^{-2}\nab(\oo)\nab(\oo)+\nab\O\O^{-2}\nab^2(\oo)(\oo)\\
&+\O^{-1}\nab^3(\oo)(\oo)+(\nab^2\eta,\nab^2\etb)\tr\chib(\oo)\\
&+(\nab\eta,\nab\etb)\nab\tr\chib(\oo)+\O^{-1}\nab^2(\oo)\nab(\oo)\\
&+\nab\O\O^{-2}\nab(\oo)\nab(\oo)+\O^{-1}\nab^2(\oo)\nab(\oo)\\
&+(\nab\eta,\nab\etb)\tr\chib\nab(\oo)+\ee\nab^2\tr\chib(\oo)+\ee\nab\chib\nab(\oo)\\
&+\ee\chib\nab^2(\oo)+(\nab^2\eta,\nab^2\etb)\O\omb\tr\chib+(\nab\eta,\nab\etb)\nab\O\omb\tr\chib+(\nab\eta,\nab\etb)\O\nab\omb\tr\chib\\
&+(\nab\eta,\nab\etb)\O\omb\nab\tr\chib+\ee\nab\O\nab\omb\tr\chib+\ee\nab\O\omb\nab\tr\chib+(\nab^2\eta,\nab^2\etb)\O\omb\tr\chib\\
&+\ee\nab^2\O\omb\tr\chib+(\nab\eta,\nab\etb)\O\nab\omb\tr\chib+\ee\O\nab^2\omb\tr\chib+\ee\O\nab\omb\nab\tr\chib\\
&+\ee\O\omb\nab^2\tr\chib+(\nab^2\eta,\nab^2\etb)\O\chibh\,\chibh+(\nab\eta,\nab\etb)\nab\O\chibh\,\chibh+(\nab\eta,\nab\etb)\O\nab\chibh\,\chibh\\
&+\ee\nab^2\O\chibh\,\chibh+\ee\nab\O\nab\chibh\,\chibh+\ee\O\nab\chibh\nab\chibh+(\nab\eta,\nab\etb)\nab\O\O^{-2}(\oo)^2\\
&+\ee\O\chibh\nab^2\chibh+(\nab^2\eta,\nab^2\etb)\O^{-1}(\oo)^2+(\nab\eta,\nab\etb)\O^{-1}\nab(\oo)(\oo)\\
&+(\nab\eta,\nab\etb)\nab\O\O^{-2}(\oo)^2\\
\end{split}
\end{equation*}

\begin{equation*}
\begin{split}
&+\ee\nab\O\nab\O\O^{-3}(\oo)^2+\ee\nab\O\O^{-2}\nab(\oo)(\oo)\\
&+(\nab\eta,\nab\etb)\O^{-1}\nab(\oo)(\oo)+\ee\nab\O\O^{-2}\nab(\oo)(\oo)\\
&+\ee\O^{-1}\nab^2(\oo)(\oo)+\ee\O^{-1}\nab(\oo)\nab(\oo)\\
&+(\nab\eta,\nab\etb)\chib\nab(\oo)+\ee\nab\chib\nab(\oo)+\ee\chib\nab^2(\oo)\\
&+\nab\beb\nab(\oo)+\beb\nab^2(\oo)+(\nab\eta,\nab\etb)\chibh\nab(\oo)+\ee\nab\chibh\nab(\oo)\\
&+\ee\chibh\nab^2(\oo)+(\nab\eta,\nab\etb)\nab\tr\chib(\oo)+\ee\nab^2\tr\chib(\oo)\\
&+\ee\nab\tr\chib\nab(\oo)+(\nab\eta,\nab\etb)\nab\O\omb\tr\chib+\ee\nab^2\O\omb\tr\chib+\ee\nab\O\nab\omb\tr\chib\\
&+\ee\nab\O\omb\nab\tr\chib+(\nab\eta,\nab\etb)\O\nab\omb\tr\chib+\ee\O\nab^2\omb\tr\chib+\ee\O\nab\omb\nab\tr\chib\\
&+(\nab\eta,\nab\etb)\O\omb\nab\tr\chib+\ee\O\nab\omb\nab\tr\chib+\ee\O\omb\nab^2\tr\chib\\
&+(\nab\eta,\nab\etb)\nab\O\chibh\,\chibh+\ee\nab^2\O\chibh\,\chibh+\ee\nab\O\nab\chibh\,\chibh+(\nab\eta,\nab\etb)\O\chibh\nab\chibh\\
&+\ee\O\nab\chibh\nab\chibh+\ee\O\chibh\nab^2\chibh+(\nab\eta,\nab\etb)\nab\O\O^{-2}(\oo)^2\\
&+\ee\nab^2\O\O^{-2}(\oo)^2+\ee\nab\O\nab\O\O^{-3}(\oo)^2\\
&+\ee\nab\O\O^{-2}\nab(\oo)(\oo)+(\nab\eta,\nab\etb)\nab\O\O^{-2}(\oo)^2\\
&+\ee\nab^2\O\O^{-2}(\oo)^2+\ee\nab\O\nab\O\O^{-3}(\oo)^2\\
&+\ee\nab\O\O^{-2}\nab(\oo)(\oo)+(\nab\eta,\nab\etb)\ee\tr\chib(\oo)\\
&+\ee\ee\nab\tr\chib(\oo)+\ee\ee\chib\nab(\oo)\\
&+(\nab\eta,\nab\etb)\ee\O\omb\tr\chib+\ee\ee\nab\O\omb\tr\chib+\ee\ee\O\nab\omb\tr\chib+\ee\ee\O\omb\nab\tr\chib\\
&+(\nab\eta,\nab\etb)\ee\O\chibh\,\chibh+\ee\ee\nab\O\chibh\,\chibh+\ee\ee\O\nab\chibh\,\chibh\\
&+(\nab\eta,\nab\etb)\ee\O^{-1}(\oo)^2+\ee\ee\nab\O\O^{-2}(\oo)^2+\ee\beb\nab(\oo)\\
&+\ee\ee\O^{-1}\nab(\oo)(\oo)+\ee\chib\nab^2(\oo)+\beb\nab^2(\oo)\\
&+\ee\ee\ee\tr\chib(\oo)+\ee\ee\ee\O\omb\tr\chib\\
&+\ee\ee\ee\O\chibh\,\chibh+\ee\ee\ee\O^{-1}(\oo)^2.
\end{split}
\end{equation*}

\section{Transport Equations for Energy Estimates}
In this section, we demonstrate the null Bianchi equations for angular derivatives of curvature components. For $\a,\b$, we have

\begin{equation*}
\begin{split}
&\nab_3\nab\a+\tr\chib\nab\a\\
=&\nab\nab\hat{\otimes}\b+\omb\nab\a+(\eta,\etb)\nab\b+\chibh\nab\a+\beb\a\\
&+\nab\tr\chib\a+\nab\omb\a+\nab\chih(\rho,\sigma)+(\nab\eta,\nab\etb)\b\\
&+(\eta,\etb)\chib\a+(\eta,\etb)\omb\a+(\eta,\etb)\chih(\rho,\sigma)+(\eta,\etb)(\eta,\etb)\b,
\end{split} 
\end{equation*}

\begin{equation*}
\begin{split}
&\nab_3\nab^2\a+\f32\tr\chib\nab\a\\
=&\nab^2\nab\hat{\otimes}\b+(\eta,\etb)\nab^2\b+\chibh\nab^2\a+\omb\nab^2\a+\a\nab\beb+\beb\nab\a\\
&+\nab\chib\nab\a+\nab\omb\nab\a+\nab\chih\nab(\rho,\sigma)+(\nab\eta,\nab\etb)\nab\b+(\eta,\etb)\chib\nab\a\\
&+(\eta,\etb)\omb\nab\a+(\eta,\etb)\chih(\nab\rho,\nab\sigma)+(\eta,\etb)(\eta,\etb)\nab\b\\
&+\nab^2\tr\chib\a+\nab^2\omb\a+\nab^2\chih(\rho,\sigma)+(\nab^2\eta,\nab^2\etb)\b\\
&+(\nab\eta,\nab\etb)\chib\a+\ee\nab\chib\a+(\nab\eta,\nab\etb)\omb\a+(\eta,\etb)\nab\omb\a\\
&+(\nab\eta,\nab\etb)\chih(\rho,\sigma)+(\eta,\etb)\nab\chih(\rho,\sigma)+(\nab\eta,\nab\etb)(\eta,\etb)\b\\
&+(\eta,\etb)\beb\a+(\eta,\etb)(\eta,\etb)(\chib,\omb)\a+(\eta,\etb)(\eta,\etb)\chih(\rho,\sigma)\\
&+(\eta,\etb)(\eta,\etb)(\eta,\etb)\b,
\end{split} 
\end{equation*}

\begin{equation*}
\begin{split}
&\nab_4\nab\b\\
=&\nab \div\a+\o\nab\b+\chi\nab\b+(\eta,\etb)\nab\a+\b\b\\
&+\nab\tr\chi\b+\nab\o\b+\nab\eta\a\\
&+\chi\etb\b+(\eta,\etb)\tr\chi\b+(\eta,\etb)\o\b+(\eta,\etb)\eta\a,
\end{split} 
\end{equation*}

\begin{equation*}
\begin{split}
&\nab_4\nab^2\b\\
=&\nab^2 \div\a+(\eta,\etb)\nab^2\a+(\o,\chi)\nab^2\b+\b\nab\b\\
&+(\nab\chi,\nab\o)\nab\b+(\nab\eta,\nab\etb)\nab\a+(\eta,\etb)(\chi,\o)\nab\b+(\eta,\etb)(\eta,\etb)\nab\a\\
&+(\nab^2\tr\chi,\nab^2\o)\b+(\nab^2\eta,\nab^2\etb)\a+(\nab\eta,\nab\etb)\chi\b+(\eta,\etb)\nab\chi\b\\
&+(\nab\eta,\nab\etb)\o\b+(\eta,\etb)\nab\o\b+(\nab\eta,\nab\etb)(\eta,\etb)\a\\
&+(\eta,\etb)\b\b+(\o,\chi)(\eta,\etb)(\eta,\etb)\b+(\eta,\etb)(\eta,\etb)(\eta,\etb)\a.
\end{split} 
\end{equation*}

For $\b,\rho,\sigma$, we obtain

\begin{equation*}
\begin{split}
&\nab_3\nab\b+\f32\tr\chib\nab\b\\
=&\nab^2\rho+\nab^{*}\nab\sigma+\chibh\nab\b+\nab\omb\b+\omb\nab\b+\nab\chih\beb+\chih\nab\beb+\nab\tr\chib\b+\ee\tr\chib\b\\
&+(\nab\eta,\nab\etb)(\rho,\sigma)+\ee(\nab\rho,\nab\sigma)+\eta\b\chib+\b\beb+\ee\omb\b+\ee\chih\beb+\ee\ee(\rho,\sigma),
\end{split} 
\end{equation*}

\begin{equation*}
\begin{split}
&\nab_3\nab^2\b+2\tr\chib\nab^2\b\\
=&\nab^3\rho+\nab\nab^{*}\nab\sigma+\nab\chibh\nab\b+\chibh\nab^2\b+\nab^2\omb\b+\nab\omb\nab\b+\omb\nab^2\b\\
&+\nab^2\chih\beb+\nab\chih\nab\beb+\chih\nab^2\beb+(\nab^2\eta,\nab^2\etb)(\rho,\sigma)+(\nab\eta,\nab\etb)(\nab\rho,\nab\sigma)+\nab\tr\chib\nab\b+\nab^2\tr\chib\b\\
&+\ee(\nab^2\rho,\nab^2\sigma)+(\nab\eta,\nab\etb)\chib\b+\ee\nab\chib\b+\ee\chib\nab\b+\nab\b\beb+\b\nab\beb\\
&+(\nab\eta,\nab\etb)\omb\b+\ee\nab\omb\b+\ee\omb\nab\b+(\nab\eta,\nab\etb)\chih\beb+\ee\nab\chih\beb+\ee\chih\nab\beb\\
&+(\nab\eta,\nab\etb)\ee(\rho,\sigma)+\ee\ee(\nab\rho,\nab\sigma)+\ee\nab\b\chib+\ee\ee\chib\b+\ee\b\beb\\
&+\ee\ee\omb\b+\ee\ee\chih\beb+\ee\ee\ee(\rho,\sigma),
\end{split} 
\end{equation*}

\begin{equation*}
\begin{split}
&\nab_4\nab\rho\\
=&\nab \div \b+\nab\chibh\a+\chibh\nab\a+\chi(\nab\rho,\nab\sigma)+\nab\chi(\rho,\sigma)+(\nab\eta,\nab\etb)\b\\
&+\ee\nab\b+\ee\chi(\rho,\sigma)+\b(\rho,\sigma)+\ee\chibh\a+\ee\ee\b,
\end{split} 
\end{equation*}

\begin{equation*}
\begin{split}
&\nab_4\nab\sigma\\
=&-\nab \div^{*}\b+\nab\chibh\a+\chibh\nab\a+\chi(\nab\rho,\nab\sigma)+\nab\chi(\rho,\sigma)+(\nab\eta,\nab\etb)\b\\
&+\ee\nab\b+\ee\chi(\rho,\sigma)+\b(\rho,\sigma)+\ee\chibh\a+\ee\ee\b,
\end{split} 
\end{equation*}

\begin{equation*}
\begin{split}
&\nab_4\nab^2\rho\\
=&\nab^2 \div\b+\nab^2\chibh\a+\nab\chibh\nab\a+\chibh\nab^2\a+\chibh\nab^2\a+\chi(\nab^2\rho,\nab^2\sigma)\\
&+\nab\chi(\nab\rho,\nab\sigma)+\nab^2\chi(\rho,\sigma)+(\nab^2\eta,\nab^2\etb)\b+(\nab\eta,\nab\etb)\nab\b+\ee\nab^2\b\\
&+(\nab\eta,\nab\etb)\chi(\rho,\sigma)+(\eta,\etb)\nab\chi(\rho,\sigma)+(\eta,\etb)\chi(\nab\rho,\nab\sigma)\\
&+\nab\b(\rho,\sigma)+\b(\nab\rho,\nab\sigma)+(\nab\eta,\nab\etb)\chibh\a+\ee\nab\chibh\a+\ee\chibh\nab\a\\
&+(\nab\eta,\nab\etb)\ee\b+\ee\ee\nab\b+\ee(\nab\rho,\nab\sigma)\chi+\chi(\nab^2\rho,\nab^2\sigma)+\ee\ee\nab\b\\
&+\ee\ee\chi(\rho,\sigma)+\ee\b(\rho,\sigma)+\ee\ee\chibh\a+\ee\ee\ee\b,
\end{split} 
\end{equation*}

\begin{equation*}
\begin{split}
&\nab_4\nab^2\sigma\\
=&-\nab^2 \div^{*}\b+\nab^2\chibh\a+\nab\chibh\nab\a+\chibh\nab^2\a+\chibh\nab^2\a+\chi(\nab^2\rho,\nab^2\sigma)\\
&+\nab\chi(\nab\rho,\nab\sigma)+\nab^2\chi(\rho,\sigma)+(\nab^2\eta,\nab^2\etb)\b+(\nab\eta,\nab\etb)\nab\b+\ee\nab^2\b\\
&+(\nab\eta,\nab\etb)\chi(\rho,\sigma)+(\eta,\etb)\nab\chi(\rho,\sigma)+(\eta,\etb)\chi(\nab\rho,\nab\sigma)\\
&+\nab\b(\rho,\sigma)+\b(\nab\rho,\nab\sigma)+(\nab\eta,\nab\etb)\chibh\a+\ee\nab\chibh\a+\ee\chibh\nab\a\\
&+(\nab\eta,\nab\etb)\ee\b+\ee\ee\nab\b+\ee(\nab\rho,\nab\sigma)\chi+\chi(\nab^2\rho,\nab^2\sigma)+\ee\ee\nab\b\\
&+\ee\ee\chi(\rho,\sigma)+\ee\b(\rho,\sigma)+\ee\ee\chibh\a+\ee\ee\ee\b.
\end{split} 
\end{equation*}

For $\rho,\sigma,\beb$, we derive

\begin{equation*}
\begin{split}
&\nab_3\nab\rho+2\tr\chib\nab\rho\\
&=-\nab \div\beb+\nab\tr\chib(\rho, \sigma)+\nab\chih\ab+\chih\nab\ab+(\nab\eta,\nab\etb)\beb+\ee\nab\beb+\chibh(\nab\rho, \nab\sigma)\\
&+\ee\chib(\rho, \sigma)+\ee\chih\ab+\ee\ee\beb,
\end{split} 
\end{equation*}

\begin{equation*}
\begin{split}
&\nab_3\nab\sigma+2\tr\chib\nab\sigma\\
&=-\nab \div^{*}\beb+\nab\tr\chib(\rho, \sigma)+\nab\chih\ab+\chih\nab\ab+(\nab\eta,\nab\etb)\beb+\ee\nab\beb+\chibh(\nab\rho, \nab\sigma)\\
&+\ee\chib(\rho,\sigma)+\ee\chih\ab+\ee\ee\beb,
\end{split} 
\end{equation*}

\begin{equation*}
\begin{split}
&\nab_3\nab^2\sigma+\f52\tr\chib\nab^2\sigma\\
&=-\nab^2 \div^{*}\beb+\nab^2\tr\chib(\rho,\sigma)+\nab\tr\chib(\nab\rho, \nab\sigma)+\nab^2\chih\ab+\nab\chih\nab\ab+\chih\nab^2\ab\\
&+(\nab^2\eta,\nab^2\etb)\beb+(\nab\eta,\nab\etb)\nab\beb+\ee\nab^2\beb+\nab\chibh(\nab\rho,\nab\sigma)\\
&+\chibh(\nab^2\rho,\nab^2\sigma)+(\nab\eta,\nab\etb)\chib(\rho,\sigma)+\ee\nab\chib(\rho,\sigma)+\ee\chib(\nab\rho,\nab\sigma)\\
&+(\nab\eta,\nab\etb)\chih\ab+\ee\nab\chih\ab+\ee\chih\nab\ab+(\nab\eta,\nab\etb)\ee\beb+\ee\ee\nab\beb+\chibh(\nab^2\rho,\nab^2\sigma)\\
&+\beb(\nab\rho,\nab\sigma)+\chib(\eta,\etb)(\nab\rho,\nab\sigma)+\ee\chibh(\nab\rho,\nab\sigma)\\
&+\ee\ee\chib(\rho,\sigma)+\ee\ee\chih\ab+\ee\ee\ee\beb,
\end{split} 
\end{equation*}

\begin{equation*}
\begin{split}
&\nab_4\nab\beb\\
=&-\nab^2\rho+\nab^{*}\nab\sigma+\nab\tr\chi\beb+\tr\chi\nab\beb+\nab\o\beb+\o\nab\beb+\nab\chibh\b+\chibh\nab\b\\
&+\ee(\nab\rho,\nab\sigma)+(\nab\eta,\nab\etb)(\rho,\sigma)+\chi\nab\beb+\b\beb+\chi\ee\beb+\ee\o\beb\\
&+\ee\chibh\b+\ee\ee(\rho,\sigma),
\end{split} 
\end{equation*}

\begin{equation*}
\begin{split}
&\nab_4\nab^2\beb\\
=&-\nab^3\rho+\nab\nab^{*}\nab\sigma+\nab^2\tr\chi\beb+\nab\tr\chi\nab\beb+\tr\chi\nab^2\beb+\nab^2\o\beb\\
&+\nab\o\nab\beb+\o\nab^2\beb+\nab^2\chibh\b+\nab\chibh\nab\b+\chibh\nab^2\b+(\nab\eta,\nab\etb)(\nab\rho,\nab\sigma)\\
&+\ee(\nab^2\rho,\nab^2\sigma)+(\nab^2\eta,\nab^2\etb)(\rho,\sigma)+\nab\chi\nab\beb+\chi\nab^2\beb+\nab\b\beb+\b\nab\beb\\
&+\nab\chi\ee\beb+\chi(\nab\eta,\nab\etb)\beb+\chib\ee\nab\beb+(\nab\eta,\nab\etb)\o\beb+\ee\nab\o\beb+\ee\o\nab\beb\\
&+(\nab\eta,\nab\etb)\chibh\b+\ee\nab\chibh\b+\ee\chibh\nab\b+(\nab\eta,\nab\etb)\ee(\rho,\sigma)+\ee\ee(\nab\rho,\nab\sigma)\\
&+\ee\nab\beb\chi+\chi\nab^2\beb+\ee\ee(\nab\rho,\nab\sigma)+\ee(\nab\eta,\nab\etb)(\rho,\sigma)\\
&+\ee\chi\nab\beb+\ee\b\beb+\chi\ee\ee\beb+\ee\ee\o\beb\\
&+\ee\ee\chibh\b+\ee\ee\ee(\rho,\sigma).
\end{split} 
\end{equation*}

For $\beb,\ab$, we have

\begin{equation*}
\begin{split}
&\nab_3\nab\beb+\f52\tr\chib\nab\beb\\
=&-\nab \div\ab+\chibh\nab\beb+\nab\omb\beb+\omb\nab\beb+(\nab\eta,\nab\etb)\ab+\ee\nab\ab+\nab\tr\chib\,\beb\\
&+\beb\,\beb+\chibh\nab\beb+\chib\ee\beb+\ee\omb\beb+\ee\ee\ab,
\end{split} 
\end{equation*}

\begin{equation*}
\begin{split}
&\nab_3\nab^2\beb+3\tr\chib\nab^2\beb\\
=&-\nab^2 \div\ab+\chibh\nab^2\beb+\nab\chibh\nab\beb+\nab^2\omb\beb+\nab\omb\nab\beb+\omb\nab^2\beb+\nab^2\tr\chib\,\beb+\nab\tr\chib\nab\beb\\
&+(\nab^2\eta,\nab^2\etb)\ab+(\nab\eta,\nab\etb)\nab\ab+\ee\nab^2\ab+\nab\beb\,\beb+\nab\chibh\nab\beb\\
&+\chibh\nab^2\beb+\nab\chib\ee\beb+\chib(\nab\eta,\nab\etb)\beb+\chib\ee\nab\beb+(\nab\eta,\nab\etb)\omb\beb\\
&+\ee\nab\omb\beb+\ee\omb\nab\beb+(\nab\eta,\nab\etb)\ee\ab+\ee\ee\nab\ab+\chib\ee\nab\beb+\chibh\nab^2\beb\\
&+\ee\tr\chib\nab\beb+\ee\beb\,\beb+\ee\chibh\nab\beb+\chib\ee\ee\beb\\
&+\ee\ee\omb\beb+\ee\ee\ee\ab,
\end{split} 
\end{equation*}

\begin{equation*}
\begin{split}
&\nab_4\nab\ab\\
=&-\nab\nab\widehat{\otimes}\beb+\nab\tr\chi\ab+\chi\nab\ab+\nab\o\ab+\o\nab\ab+\nab\chibh(\rho,\sigma)+\chibh(\nab\rho,\nab\sigma)\\
&+(\nab\eta,\nab\etb)\beb+\ee\nab\beb+\ee\chi\ab+\ee\o\ab+\ee\chibh(\rho,\sigma)+\ee\ee\beb,
\end{split} 
\end{equation*}

\begin{equation*}
\begin{split}
&\nab_4\nab^2\ab\\
=&-\nab^2\nab\widehat{\otimes}\beb+\nab^2\tr\chi\ab+\nab\tr\chi\nab\ab+\chi\nab^2\ab+\nab^2\o\ab+\nab\o\nab\ab+\o\nab^2\ab\\
&+\nab^2\chibh(\rho,\sigma)+\nab\chibh(\nab\rho,\nab\sigma)+\chibh(\nab^2\rho,\nab^2\sigma)+(\nab^2\eta,\nab^2\etb)\beb\\
&+(\nab\eta,\nab\etb)\nab\beb+\ee\nab^2\beb+(\nab\eta,\nab\etb)\chi\ab+\ee\nab\chi\ab+\ee\chi\nab\ab\\
&+(\nab\eta,\nab\etb)\o\ab+\ee\nab\o\ab+\ee\o\nab\ab+(\nab\eta,\nab\etb)\chibh(\rho,\sigma)+\ee\nab\chibh(\rho,\sigma)\\
&+\ee\chibh(\nab\rho,\nab\sigma)+(\nab\eta,\nab\etb)\ee\beb+\ee\ee\nab\beb\\
&+\ee\chi\nab\ab+\b\nab\ab+\chi\nab^2\ab+\ee\ee\chi\ab+\ee\ee\o\ab\\
&+\ee\ee\chibh(\rho,\sigma)+\ee\ee\ee\beb.
\end{split} 
\end{equation*}

\end{document}